\documentclass[a4paper,11pt]{report}
\pdfoutput=1 % to allow compilation in JCAP

\usepackage{amsmath}
\usepackage{bm}
\usepackage{graphicx}
\usepackage{physics}
\usepackage{amssymb}
\usepackage{gensymb}
\usepackage{enumitem}
\usepackage{geometry}
\usepackage{xspace}
\usepackage{pdflscape}
\usepackage{placeins}
\usepackage{epstopdf}
\usepackage{comment}
\usepackage[breaklinks=true, urlcolor  = blue]{hyperref}
\usepackage{comment}
\usepackage{collectbox}
\usepackage{pdfpages}
\usepackage[english,french]{babel}
\newcommand{\ie}{{i.e.~}}

\newcommand{\eg}{e.g.~}

\newcommand{\OmegaGW}{\Omega_{\mathrm{GW}}}
\newcommand{\rhoGW}{\rho_{\mathrm{GW}}}

\let\oldsqrt\sqrt
\def\sqrt{\mathpalette\DHLhksqrt}
\def\DHLhksqrt#1#2{%
\setbox0=\hbox{$#1\oldsqrt{#2\,}$}\dimen0=\ht0
\advance\dimen0-0.2\ht0
\setbox2=\hbox{\vrule height\ht0 depth -\dimen0}%
{\box0\lower0.4pt\box2}}

\newcommand{\sss}[1]{{\scriptscriptstyle{#1}}}
\newcommand{\boldmathsymbol}[1]{{\ensuremath{\boldsymbol{#1}}}}

\newcommand{\uPl}{\mathrm{Pl}}

\newcommand{\ueff}{\mathrm{eff}}

\newcommand{\usssPl}{\sss{\uPl}}

\newcommand{\bmk}{\boldmathsymbol{k}}
\newcommand{\bmx}{\boldmathsymbol{x}}

\newcommand{\calH}{\mathcal{H}}

\newcommand{\g}{\mathrm{g}}

\newcommand{\Mpc}{\mathrm{Mpc}}

\newcommand{\cs}{c_{_\mathrm{S}}}

\newcommand{\Mp}{M_\usssPl}

\newcommand{\efolds}{$e$-folds}

\newcommand{\beq}{\begin{equation}}
\newcommand{\eeq}{\end{equation}}
\newcommand{\bea}{\begin{equation}\begin{aligned}}
\newcommand{\eea}{\end{aligned}\end{equation}}

\newlength{\wsingfig}
\newlength{\wdblefig}
\newlength{\wquadfig}
\newlength{\wtriplefig}
\newcommand{\Eq}[1]{Eq.~(\ref{#1})}

\newcommand{\Fig}[1]{Fig.~{\ref{#1}}}

\newcommand{\Sec}[1]{Sec.~\ref{#1}}
\newcommand{\Ch}[1]{Ch.~\ref{#1}}

\newcommand{\App}[1]{Appendix~\ref{#1}}

\newcommand{\mr}{\mathrm{r}}
\newcommand{\mt}{\mathrm{t}}

\begin{document}
\sloppy
%\pagenumbering{gobble}
\date{}
%\maketitle

\vspace{-2cm}
\begin{center}
    
    {\Large \scshape Université de Paris}

    \vspace{2em}

    \includegraphics[width=2.2cm]{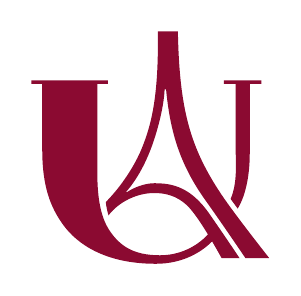} \hspace{1cm}
    \includegraphics[height=2.5cm]{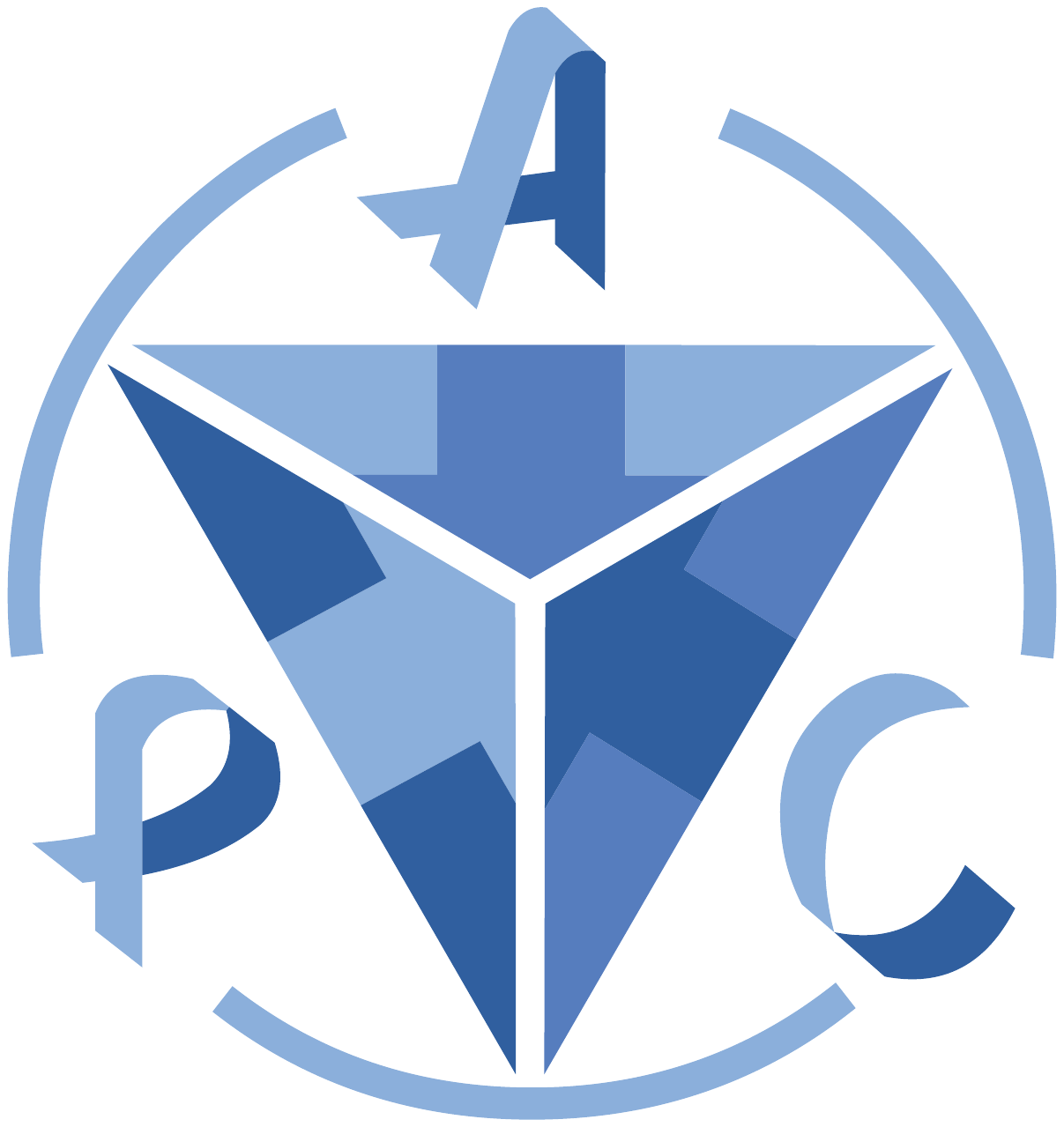} \hspace{1cm}
    \includegraphics[height=2.5cm]{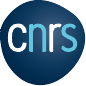} \hspace{1cm}
    \includegraphics[height=2.5cm]{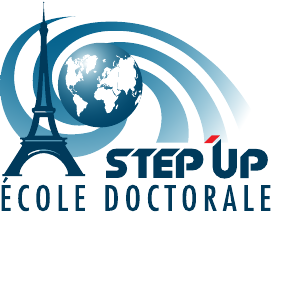}

    \vspace{.5em} 

    École Doctorale des Sciences de la Terre et de l'Environnement et Physique de l'Univers - ED 560 
    \vspace{.5em} 

    Laboratoire AstroParticules et Cosmologie (APC) - Groupe Théorie

    \vspace{2em}

    \rule{\linewidth}{1pt} 
    \vspace{.1em}

    {\bfseries \LARGE Studying Aspects of the Early Universe with Primordial Black Holes}

    \vspace{.5em}
    \rule{\linewidth}{1pt} 

    \vspace{1em}

    {\large Thèse de Doctorat de Physique de l'Univers}

    \vspace{.5em}

     de Theodoros {\scshape Papanikolaou}

    \vspace{.5em}

    dirigée par Vincent {\scshape Vennin } 

    \vspace{1.5em}

    {\itshape présentée et soutenue publiquement le 20 septembre 2021 \\ devant le jury composé de :}

    \vspace{1em}

    %\begin{table}
        %\centering
        \begin{tabular}{r l}
        
        	   Christian Thomas {\scshape Byrnes} & Rapporteur  \\
            Senior Lecturer (University of Sussex) & \\
            
            S\'ebastien {\scshape Clesse} & Rapporteur \\
            Assistant Professor (Universit\'e Libre de Bruxelles) & \\
            
            Antonio {\scshape Riotto} & Examinateur \\
            Full Professor (Universit\'e de Gen\`eve) & \\

            Danièle {\scshape Steer} & Présidente du jury \\
            Professeur des Universités (Université de Paris, APC) & \\

            Vincent {\scshape Vennin} & Directeur de thèse \\
            Chargé de Recherche (CNRS, APC) & \\

        \end{tabular}
\thispagestyle{empty}
%\newpage
%\includepdf[pages=-]{Papers/UP_vd_liste_sous_droits.pdf}
\newpage
\pagenumbering{arabic} 
\end{center}
\newpage
\begin{Large}
\textbf{Acknowledgements}
\end{Large}
\vspace{0.5cm}

First of all, I want to express my deepest gratitude to Vincent Vennin who gave me the opportunity to conduct research in a motivating and pleasant environment within the Laboratoire AstroParticule et Cosmologie (APC). As a PhD advisor, he introduced me to the field of early universe cosmology and gave me motivations to contuct research on the field of primordial black holes. I want to thank him as well for his psychological support and encouragement to endure and continue the journey towards the PhD.

In addition, I am delighted to thank my collaborators David Langlois, J\'er\^ ome Martin and Lucas Pinol with I whom I worked closely. I learnt a lot from their pedagogy, scientific rigour, physical intuitiveness and scientific integrity.

A special thankfulness goes to Ilia Musco, with whom I collaborated a lot on the field of the gravitational collapse of primordial black holes. I want to thank him together with Antonio Riotto for the kind hospitality in the Departement of Theoretical Physics of the University of Geneva as well as for the fruitful scientific discussions we had from which I learnt a lot.

I want to thank as well S\'ebastien Clesse, Chris Thomas Byrnes, Antonio Riotto and Dani\`ele Steer for their scientific interest to my work and for having accepted being members of my PhD thesis jury and review my PhD manuscript. 

Furthermore, I want to thank the directors of APC, Stavros Katsanevas and Antoine Kouchner and the director of the theory group of APC Dimitri Semikoz for providing me with the best conditions in order to carry out research as well as the directors of the STEP'UP Doctoral School Yannick Giraud-H\'eraud and Alessandra Tonazzo for their patience and for their guidance to find the necessary funding resources for my PhD. 

At this point, I have to acknowledge thankfulness to the Fondation CFM pour la
Recherche in France, the Alexander S. Onassis Public Benefit Foundation in Greece, the Foundation for Education
and European Culture in Greece and the A.G. Leventis Foundation for their interest to my PhD research field as well as for their funding support during the three years of my PhD studies.

Moreover, it's a pleasure to give a special thanks to my office mates C\'elia, Gabriel, Jan and Am\'elie who contributed to forming a nice working environment for me as well as to my PhD colleagues in APC Pierre, Jewel and Hugo for the nice discussions we had during launch and coffee times.

In addition, I am grateful to my spiritual fathers, archimandrite father Gabriel, archpriest father Panagiotis and the Bishop Ir\'en\'ee of Reggio for their psychological support and kindness as well as for providing me with all the necessary spiritual arms during my PhD studies.

Last but not least, I want to thank my parents Nikolaos and Regina, my brother Christos and my fianc\'ee Adela-Maria for their love, patience and support during these years and all the upcoming ones.

\newpage
\begin{Large}
\textbf{Abstract}
\end{Large}
\vspace{0.5cm}

This thesis by publication is devoted to the study of aspects of the early universe in the context of primordial black hole (PBH) physics. Since the early '70s, when PBHs were initially proposed, PBHs have been attracting an increasing interest within the scientific community given the fact that they can address a number of fundamental issues of modern cosmology and at the same time give access to different physical phenomena depending on their mass. Interestingly, with low mass PBHs one can probe and constrain the physics of the early universe, such as inflation and reheating whereas with high mass PBHs one can probe gravitational physics phenomena like the large-scale structure formation and the origin of dark matter. %In particular, the recent developments in the field of gravitational wave astronomy make PBHs potentially detected by gravitational wave experiments, a fact which renders these astrophysical objects and the theories associated to PBHs testable against the advent of gravitational wave observational data.

In the following PhD thesis, we firstly review the fundamentals of the early universe cosmology and we recap the basics of the PBHs physics covering both theoretical and observational aspects. In particular, we propose a refinement in the determination of the PBH formation threshold, a fundamental quantity in PBH physics, in the context of a time-dependent equation-of-state parameter. Afterwards, we briefly present the theory of inflationary perturbations, which is the theoretical framework within which PBHs are studied in this thesis.

Then, in the second part of the thesis,  we review the core of the research conducted within my PhD, in which aspects of the early universe and the gravitational wave physics are combined with the physics of PBHs. Moreover, aspects of the PBH gravitational collapse process are studied in the presence of anisotropies. Specifically, we study PBHs produced from the preheating instability in the context of single-field inflation. In particular, we find that PBHs produced during preheating can potentially dominate the universe's content and drive reheating through their evaporation. Then, we focus on the scalar induced second-order stochastic gravitational wave background (SGWB) produced during an era before BBN in which ultralight PBHs dominate the energy budget of the universe. By taking then into account gravitational wave backreaction effects we set model-independent constraints on the initial abundance of ultralight PBHs as a function of their mass. Afterwards, we study in a covariant way the anisotropic spherical gravitational collapse of PBHs during a radiation-dominated era in which one can compute the PBH formation threshold as a function of the anisotropy.

Finally, we summarize our research results by discussing future prospects opened up as a result of the work we have done within the PhD. In particular, we emphasize the fact that one can narrow down the CMB observational predictions by studying PBHs produced from the single-field inflation preheating instability as well as the potential detectability of ultralight PBHs by future gravitational experiments such as LISA, Einstein Telescope and SKA.
\vspace{0.5cm}

\newpage
\selectlanguage{french}
\begin{Large}
\textbf{Résumé}
\end{Large}
\vspace{0.5cm}

Cette thèse sur articles est dédiée à l'étude des aspects de l'univers primordial par le biais des trous noirs primordiaux (TNP). Depuis que les TNP ont été initialement proposés dans les années 70, ils attirent de plus en plus l'intérêt de la communauté scientifique étant donné le fait que ces objets astrophysiques apportent un éclairage sur un grand nombre de problèmes de la cosmologie contemporaine et en parallèle peuvent donner accès à une grande variété de phénomènes physiques en fonction de de leur masse. En particulier, les TNP de petite masse peuvent donner accès à la physique de l'univers primordial comme la physique de l'inflation et du rechauffement tandis qu' avec les TNP de grande masse on peut explorer des phénomènes de la physique gravitationnelle comme la formation des structures de grande échelle et l'origine de la matière noire.

Dans cette thèse, on rapelle tout d'abord les fondements de la cosmologie de l'univers primordial et les essentiels de la physique des TNP en couvrant à la fois des aspects théoriques et observationnelles. En particulier, on propose un raffinement des méthodes sur la détermination du seuil de formation des TNP, une quantité fondamentale dans le domaine de recherche des TNP, dans le contexte d'un paramètre d'équation d'état dépendant du temps. Ensuite, on se réfère brièvement à la théorie des perturbations inflationnaires, qui constitue le cadre théorique dans lequel les TNP sont étudiés dans cette thèse.

Dans une deuxième partie, on présente la recherche effectuée au sein de mes études doctorales, dans laquelle des aspects de la physique de l'univers primordial se combinent avec la physique des ondes gravitationnelles. De plus, des facettes de l'effondrement gravitationnel des TNP en présence des anisotropies sont étudiées. Plus spécifiquement, on étudie les TNP produits de l'instabilité du préchauffement dans le contexte de la théorie de l'inflation avec un champ scalaire. En particulier, on trouve que les TNP produits pendant la période du préchauffement peuvent potentiellement dominer le contenu énergétique de l'univers et conduire au réchauffement de l'univers à travers leur évaporation.  

Ensuite, on se concentre sur le fond stochastique d'ondes gravitationnelles induites par perturbations scalaires à travers des effets gravitationnels de second ordre pendant une période cosmique avant l'époque de la nucléosynthèse du Big Bang, où des trous noirs primordiaux ultralégers constituent la composante principale du budget énergétique de l’univers. En demandant alors que ces ondes gravitationnelles induites ne se produisent pas en excès à la fin de la période de domination énergétique des TNP,  on impose des contraintes indépedentes du modèle de production de TNP sur leur abondance au moment de leur formation en fonction de leur masse. Puis, on étudie d'une manière covariante l'effondrement gravitationnel sphérique et anisotrope des TNP se produisant pendant une époque cosmique dominée par la radiation.

Enfin, on résume les résultats de notre recherche en discutant les perspectives qu'ouvre le travail effectué au sein du doctorat. En particulier, nous insistons sur le fait que les prédictions observationnelles des modèles d'inflation à un champ scalaire concernant les anisotropies du fonds diffus cosmologiques peuvent être affinées par la prise en compte des TNP produits pendant la phase de préchauffement. De plus, on souligne la detectabilité potentielle des TNP ultralégers par des futures expériences gravitationnelles comme LISA, Einstein Telescope et SKA.

\vspace{0.5cm}

\textbf{Mots Clés:} trous noirs primordiaux, inflation, préchauffement, ondes gravitationnelles induites, effondrement gravitationnel de trous noirs primordiaux.
\selectlanguage{english}

\newpage
\chapter*{Introduction}
Primordial black holes (PBHs), firstly proposed more than 50 years ago ~\cite{1967SvA....10..602Z}, are attracting increasing attention given that they can address a number of issues of modern cosmology. According to theoretical arguments they may indeed constitute a part or all of the dark matter~\cite{Chapline:1975ojl} and they may explain the generation of large-scale structures through Poisson fluctuations~\cite{Meszaros:1975ef,Afshordi:2003zb}. Furthermore, they may provide seeds for supermassive black holes in galactic nuclei~\cite{1984MNRAS.206..315C, Bean:2002kx} as well as account for the progenitors of the black-hole merging events recently detected by the LIGO/VIRGO collaboration~\cite{LIGOScientific:2018mvr} through their gravitational wave (GW) emission.

The idea for their existence dates back in 1967 when Novikov and Zeldovich ~\cite{1967SvA....10..602Z} proposed that black holes can  form in the early universe through accretion of the surrounding radiation. Some years later,  Stephen Hawking in 1971 ~\cite{1971MNRAS.152...75H} and his PhD student Bernard Carr in 1974 ~\cite{1974MNRAS.168..399C}, who pioneered the field of PBHs, considered also formation of PBHs establishing the modern way of viewing the PBH formation mechanism. In particular, they claimed that PBHs form out of the gravitational collapse of high overdensity regions whose energy density exceeds a critical threshold value, which in general depends on the characteristic scale and the shape of the overdensity region as well as on the time at which the gravitational collapse is taking place ~\cite{Carr:1975qj} and on the details of the surroundings.

This type of black holes is different from the astrophysical black holes in the sense that they do not form out of the collapse of a star, an astrophysical process which imposes a lower bound on the mass of the forming black hole at around 3 solar masses ~\cite{Rhoades:1974fn}. On the contrary, PBHs can form at whichever epoch of the cosmic history when an overdensity region is highly compressed and collapses to a black hole under an extremely strong gravitational force. As realized very early by Hawking~\cite{1971MNRAS.152...75H} the mass of a PBH is roughly equal to the mass inside the cosmic horizon at the time of formation, a fact which makes the mass range of PBHs very wide given the time dependence of the cosmic horizon scale. 
In particular, one can produce super-massive black holes like the ones residing in the center of galaxies, with typical PBH masses $m_\mathrm{PBH}\sim 10^6 M_\odot$ ~\cite{Bernal:2017nec}, where $M_\odot$ stands for the solar mass,  as well as ultra-light PBHs with $m_\mathrm{PBH}\sim 10^{-15}M_\odot$ ~\cite{Poulter:2019ooo} [See ~\cite{Carr:2005zd} and the references therein]. 

This last fact that black holes can acquire a very small mass of the order of the elementary particles or of the Planck mass initiated the idea of Hawking that black holes should be strongly affected by quantum phenomena, an idea which led to his famous work in 1974 ~\cite{Hawking:1974sw} showing that the mass-energy of a black hole is evaporated away with a thermal radiation spectrum and that the time of evaporation of a black hole depends cubicly on their mass, namely ~\cite{Hawking:1974rv}
\beq\label{Delta t_evap}
\Delta t_\mathrm{evap} = \frac{160}{\pi g_\ueff}\frac{m_\mathrm{BH}^3}{\Mp^4},
\eeq
where $g_\ueff$ is the effective number of relativistic degrees of freedom at the time of the black evaporation, $m_\mathrm{BH}$ is the black hole mass and $\Mp \sim 4.34\times 10^{-6}\mathrm{g}$ is the reduced Plack mass. Therefore, black holes with masses less than $10^{15}\mathrm{g}$ have evaporated by now. This critical mass $M_\mathrm{c} = 10^{15}\mathrm{g}$ is very important since with it one can divide PBHs in three categories depending on their mass, and these categories are related to different physical phenomena.

Specifically, the small mass PBHs ($m_\mathrm{PBH}\leq 10^{15}$g) which have evaporated by now can give access to the early universe physics such as the physics of  inflation and the primordial cosmological perturbations ~\cite{Kalaja:2019uju}, the Big Bang Nucleosynthesis (BBN) physics ~\cite{2010arXiv1006.5342S,Keith:2020jww}, the physics of the cosmic microwave background (CMB) ~\cite{Ali-Haimoud:2016mbv}, the primordial gravitational wave physics ~\cite{Clesse:2018ogk} and primordial phase transitions ~\cite{Jedamzik:1999am}. On the other hand, with the intermediate mass PBHs which evaporate in our era we can probe high energy astrophysical phenomena like the cosmic ray background through PBH Hawking evaporation ~\cite{Carr:2016hva}. Finally, the higher mass PBHs which still exist today, ($m_\mathrm{PBH}> 10^{15}$g), can give access to gravitational physics phenomena like gravitational lensing ~\cite{Niikura:2019kqi,Zumalacarregui:2017qqd}, large scale structure (LSS) formation ~\cite{Carr:2018rid} as well as to the physics of the dark sector of the universe, namely the dark matter (DM) ~\cite{Carr:2020xqk} and the dark energy (DE) ~\cite{Nesseris:2019fwr}.

Given all this motivation for the research in the area of PBH physics, there has been initiated during the last years a research interest on setting constraints on the abundance of PBHs depending on their mass. These constraints range from micro-lensing constraints, dynamical constraints (such as constraints from the abundance of wide dwarfs in our local galaxy, or from the existence of a star cluster near the centers of ultra-faint dwarf galaxies), constraints from the cosmic microwave background due to the radiation released in PBH accretion, constraints from the primordial power spectrum as well as from the nature of the statistics of the cosmological fluctuations and constraints from the extragalactic gamma-ray background to which Hawking evaporation of PBHs contributes.  For a recent review, see ~\cite{Carr:2020gox}. 

During my PhD I focused on PBHs produced during the metric preheating instability phase in the context of single-field inflation as well as on the scalar induced gravitational waves produced from a universe filled with primordial black holes. I also engaged myself in studying the anisotropic gravitational collapse of PBHs formed in a radiation dominated era. 

This thesis is organized as follows. In \Ch{sec:early universe} we recap the fundamentals of the early universe cosmology, by presenting the basics of a homogeneous and isotropic universe and by reviewing briefly its thermal history as well as the shortcomings of the Hot-Big Bang theory which initiated the theory of inflation.

In \Ch{sec:PBH formation}, after providing the reader with the fundamental notions of PBH physics as well as with the current observational status in the PBH field we propose  a refined way of calculation of the PBH formation threshold in the context of  a time-dependent equation-of-state parameter. We highlight as well the implications of PBHs in cosmology.

In \Ch{sec:Inflation Theory and PBH production from Preheating}, after introducing the theory of inflationary perturbations, which is the fundamental theoretical framework within which PBHs were studied throughout this thesis, we review the literature related to preheating and describe the results of our research regarding PBHs produced from metric preheating in the context of single-field inflation~\cite{Martin:2019nuw, Martin:2020fgl}.

In \Ch{sec:PBHs and Induced Gravitational Waves}, we recapitulate briefly the various ways with which PBHs can be connected with gravitational waves and we give the fundamentals of the calculation of the stochastic background of induced gravitational waves. Then, we present the results of our research concerning induced gravitational waves produced from a universe filled with ultralight PBHs ~\cite{Papanikolaou:2020qtd}.

In \Ch{sec:Anisotropic Collapse of PBHs}, after introducing the hydrodynamic equations describing the PBH gravitational collapse we propose a covariant formulation for the equation of state of a spherically symmetric anisotropic radiation fluid which can potentially collapse and form a PBH. Then, by making use of a gradient expansion perturbative scheme we extract the initial conditions of the hydrodynamic and metric perturbations and investigate how the PBH formation threshold depends on the anisotropic character of the collapse.

Finally, we summarize our research results and discuss future prospects opened up as a result of the work we have done within the PhD.

\newpage
\tableofcontents
\chapter{Early Universe Cosmology}\label{sec:early universe}
In this chapter, we present the fundamental elements necessary for the description of the early universe, when PBHs are assumed to be formed. Very briefly, we adduce firstly the basic notions and the theoretical framework describing a homogeneous and isotropic universe. Then, we give a brief description of the thermal history and the composition of the universe during the different cosmic epochs. Finally, we recap the shortcomings of the Hot Big Bang theory which gave rise to inflation, the ``standard theory'' for the description of the very early moments of the cosmic history  and which generated  the primordial cosmological perturbations seeding the large scale structures observed today as well as the relic cosmic microwave background radiation. 

\section{The Homegeneous and Isotropic Universe}
\subsection{The Hubble parameter and the redshift}\label{sec:Hubble parameter + z}
As it is well established, the universe is expanding and the rate of this expansion can be described through a universal scale factor, $a(t)$, which encodes all the information about the expansion ``history'' of the universe. This quantity depends on the cosmic time $t$, which is the time measured by a local comoving observer. From the point of view of this observer, the distances measured can be written as
 \beq\label{physical distances}
 R(t) = a(t) r,
 \eeq
where $R(t)$ is the physical distance and $r$ is the comoving distance. In a similar way, one can define a useful time variable $\eta$ defined as 
\beq\label{conformal time}
\mathrm{d}\eta \equiv \frac{\mathrm{d}t}{a(t)},
\eeq
known as the conformal time.
One then can naturally define the rate of the universe expansion, known as the Hubble parameter, $H$, as
\beq\label{Hubble parameter}
H\equiv \frac{1}{a}\frac{\mathrm{d}a}{\mathrm{d}t} = \frac{\dot{a}}{a}
\eeq
This parameter appears as a proportionality factor in the famous Hubble law, relating the expansion velocity $U$ with the physical distance between two points in the universe,
\beq\label{Hubble Law}
U = \frac{\mathrm{d}R}{\mathrm{d}t} = HR.
\eeq
The Hubble parameter, $H$, has dimensions of inverse time and is very important since it gives an order of magnitude prediction for the age of the universe at the time one measures it. On the contrary, the Hubble radius, $cH^{-1}$, where $c$ is the speed of light, determines the size of the observable universe at the time one measures it, i.e. the scale of causal contact within our universe.

From the point of view of observations, the expansion of the universe is measured with the redshift variable, $z$ which is the relative change of the wavelength of a photon, $\mathrm{d}\lambda/\lambda$, which travels between the emission source and the observer. This relative change is due to the expansion of the universe and reads as
\beq\label{redshift}
1+z \equiv \frac{a(t_\mathrm{obs})}{a(t_\mathrm{em})}, 
\eeq
where $a(t_\mathrm{obs})$  and $a(t_\mathrm{em})$ are the scale factors at the times of observation and emission of the photons respectively. By measuring redshifts $z$ one then  can reconstruct the expansion rate of the universe, $H$. The current value of the Hubble parameter as measured by Planck satellite, which captured and analysed the CMB radiation, is ~\cite{Aghanim:2018eyx}
\beq\label{H_0}
H_0 = 67.4\pm 0.5 \mathrm{km s^{-1}Mpc^{-1}}.
\eeq
However, different experiments probing late-universe cosmology phenomena based on different astrophysical measurements are finding different values with the tension between different probes being quite intriguing ~\cite{Riess:2016jrr,Aghanim:2018eyx}. [See ~\cite{DiValentino:2020zio} for a review.]
\subsection{The FRLW metric}
The standard Hot Big Bang paradigm for the universe is based on the \textit{cosmological principle} which states that the universe is spatially homogeneous and isotropic in large scales ($\sim$ 100Mpc). This principle has observational evidences and the most astonishing one is the nearly identical temperature of the CMB radiation coming from different parts of the sky. Adopting thus the cosmological principle, one inevitably constrain the form of the metric, i.e the infinitesimal line element between two points in the universe,  which should describe a homogeneous and isotropic universe. The general form of this metric, known as Friedmann- Lemaitre-Robertson-Walker (FLRW) metric, reads as ~\cite{Friedmann:1924bb,Lemaitre:1927zz,Robertson:1935zz,1935MNRAS..95..263W}:
 \beq \label{FRLW metric - cosmic time}
 \mathrm{d}s^2\equiv g_{\mathrm{\mu\nu}}\mathrm{d}x^\mathrm{\mu}\mathrm{d}x^\mathrm{\nu} = -\mathrm{d}t^2+a^{2}(t)\left[ \frac{\mathrm{d}r^2}{1-K r^2}+r^2\left(\mathrm{d}\theta^2+\sin^2\theta\mathrm{d}\phi^2\right)\right] 
 \eeq
 where $g_{\mathrm{\mu\nu}}$ is the metric tensor, $a(t)$ is the scale factor with dimensions of length and $r$,$\theta$,$\phi$ are the comoving coordinates which are dimensionless. Finally, $K$ is the spatial curvature signature of the metric ($K=0$: flat univese, $K=\pm1$: closed (spherical) and opened (hyperbolic) universe respectively). In terms of the conformal time defined in \Eq{conformal time}, the metric takes the following form which is very useful since it simplifies as we will see later the calculations,
  \beq \label{FRLW metric - conformal time}
  \mathrm{d}s^2=-a^2(\eta)\left[ \mathrm{d}\eta^2+\frac{\mathrm{d}r^2}{1-Kr^2}+r^2(\mathrm{d}\theta^2+\sin^2\theta\mathrm{d}\phi^2)\right].
  \eeq
 
\subsection{The Einstein Equations}
Having determined then the infinitesimal line element in an expanding homogeneous and isotropic universe one can relate the expansion of the universe, a manifestation of the curvature of the space-time, to the energy-mass content of the universe through the Einstein's equations of General Relativity (GR) which can be derived from the variation of the action $\mathcal{S}$ describing the Universe.This action can decomposed in two parts, a part $\mathcal{S}_\mathrm{grav}$  which describes the gravitational sector of the universe and a part $\mathcal{S}_\mathrm{matter}$ which describes the matter content in the universe. These two parts read as \footnote{Hereafter, unless stated otherwise, we work in units where $c=\hbar=k_\mathrm{B}=1$.} ~\cite{Einstein:1915ca,Einstein:1916vd,Hilbert:1915tx}
\begin{eqnarray}\label{Actions}
\mathcal{S}_\mathrm{grav} = \frac{1}{16\pi G} \int \mathrm{d}^4x \sqrt{-g}\left(R - 2\Lambda\right) \\
\mathcal{S}_\mathrm{matter} = \int\mathcal{L}_\mathrm{matter}\sqrt{-g}\mathrm{d}^4x, 
\end{eqnarray}
where $G$ is the Newton constant, $\Lambda$ is a cosmological constant, $\mathcal{L}_\mathrm{matter}$ is the Lagrangian of matter in the universe, $g$ is the determinant of the metric $g_{\mathrm{\mu\nu}}$, $R$ is the Ricci scalar defined as a contraction of the Ricci tensor $R_{\mathrm{\mu\nu}}$, i.e. $R\equiv g_{\mathrm{\mu\nu}}R^{\mathrm{\mu\nu}}$ . The Ricci tensor reads as $R_{\mathrm{\mu\nu}}\equiv \partial_\mathrm{\rho} \Gamma^\mathrm{\rho}_{\mathrm{\mu\nu}} -  \partial_\mathrm{\nu} \Gamma^\mathrm{\rho}_{\mathrm{\mu\rho}} + \Gamma^\mathrm{\rho}_{\mathrm{\mu\nu}}\Gamma^\mathrm{\lambda}_{\mathrm{\rho\lambda}} -  \Gamma^\mathrm{\rho}_{\mathrm{\mu\lambda}}\Gamma^\mathrm{\lambda}_{\mathrm{\nu\rho}}$, where the Christoffel symbols are given by $\Gamma^\mathrm{\rho}_{\mathrm{\mu\nu}} = \frac{g^{\mathrm{\rho\lambda}}}{2}\left(\partial_\mathrm{\nu} g_{\mathrm{\lambda\mu}} +\partial_\mathrm{\mu} g_{\mathrm{\lambda\nu}}-\partial_\lambda g_{\mathrm{\mu\nu}}\right)$. 

By varying then these two parts of the action one obtains that 
\begin{eqnarray}
\frac{16\pi G}{\sqrt{-g}} \frac{\partial\mathcal{S}_\mathrm{grav}}{\partial g_{\mathrm{\mu\nu}}} = R_{\mathrm{\mu\nu}} - \frac{1}{2}Rg_\mathrm{\mathrm{\mu\nu}} + \Lambda g_{\mathrm{\mu\nu}} \\
-\frac{2}{\sqrt{-g}}\frac{\partial\mathcal{S}_\mathrm{matter}}{\partial g_{\mathrm{\mu\nu}}} = g_{\mathrm{\mu\nu}}\mathcal{L}_\mathrm{matter} - 2 \frac{\delta \mathcal{L}_\mathrm{matter}}{\delta g^{\mathrm{\mu\nu}}}
\end{eqnarray}
Demanding then that $\frac{\partial\mathcal{S}}{\partial g_{\mathrm{\mu\nu}}} =  \frac{\partial\mathcal{S}_\mathrm{grav}}{\partial g_{\mathrm{\mu\nu}}} + \frac{\partial\mathcal{S}_\mathrm{matter}}{\partial g_{\mathrm{\mu\nu}}}=0$ one obtains the Einstein equations given by ~\cite{Einstein:1915ca}
 \beq\label{Einstein's Equations}
G_{\mathrm{\mu\nu}} + \Lambda g_{\mathrm{\mu\nu}}= 8\pi G T_{\mathrm{\mu\nu}}, 
\eeq
where we have defined the Einstein tensor as $G_{\mathrm{\mu\nu}} \equiv R_{\mathrm{\mu\nu}} - \frac{1}{2}Rg_\mathrm{\mathrm{\mu\nu}}$. The stress-energy tensor for matter is defined as
\beq\label{stress-energy tensor}
T_{\mathrm{\mu\nu}}\equiv -\frac{2}{\sqrt{-g}}\frac{\partial\mathcal{S}_\mathrm{matter}}{\partial g_{\mathrm{\mu\nu}}}.
\eeq
With the Einstein equations \Eq{Einstein's Equations} one can relate the curvature of space-time described in terms the geometrical quantities $G_{\mathrm{\mu\nu}}$ or $R_{\mathrm{\mu\nu}}$, $R$ and $g_{\mathrm{\mu\nu}}$ to the energy-mass content of the universe described with the energy-momentum tensor $T_{\mathrm{\mu\nu}}$.
 \subsection{Dynamics of an Expanding Universe}
Now, by taking into consideration the cosmological principle and treating the universe background medium as a perfect fluid, the energy-momentum tensor can be written in a general form as 
\beq\label{T_munu - perfect fluid}
T^{\mathrm{\mu\nu}} = - pg^{\mathrm{\mu\nu}} +(p+\rho)u^\mathrm{\mu} u^\mathrm{\nu}
\eeq
where $p$ and $\rho$ are the pressure and energy densities of the fluid respectively and $u^\mu$ is the four velocity of a comoving observer for whom space is homogeneous and isotropic. One thus has that $u^\mu=\delta^{\mu,0}$, where $\delta$ is the Kr\"{o}necker delta. Therefore, by solving the Einstein equations for a homogeneous and isotropic universe described with the FRLW metric \Eq{FRLW metric - cosmic time} and filled with a perfect fluid described in terms of the stress-energy tensor in \Eq{T_munu - perfect fluid},  one can extract the following equations, which govern the evolution of the scale factor $a(t)$ in a homogeneous and isotropic universe ~\cite{Friedman:1922kd,Raychaudhuri:1953yv}.
\beq \label{Friedmann Equation}
H^2=\left( \frac{\dot{a}}{a}\right)^2=\frac{\rho}{3\Mp^2}-\frac{k}{a^2} +\frac{\Lambda}{3}  \rm{\;\;\;\;Friedmann\;Equation}
\eeq
\beq \label{Acceleration Equation}
\frac{\ddot{a}}{a}=-\frac{4\pi G}{3}(\rho+3p) +\frac{\Lambda}{3}, \rm{\;\;\;\;\;\;\;\;\;\;\;\;\;Raychaudhuri\;Equation}
\eeq
where we have used the definition of the reduced Planck mass as $\Mp^2\equiv\frac{1}{8\pi G}$.

At this point, one should point out that combining the Friedmann and the Raychaudhuri equations one can obtain the continuity equation which reads as 
\beq \label{Continuity Equation}
\dot{\rho}+3H(\rho+p)=0. \rm{\;\;\;\;\;\;\;\;\;\;\;Continuity\;Equation}
\eeq
The above equation can be also obtained from the covariant conservation of the energy-momentum tensor, i.e.  $\nabla_{\mu}T^{\mu\nu}=0$, where $\nabla_{\mu}$ is the covariant derivative and can be seen as the first law of thermodynamics, $\mathrm{d}U_\mathrm{th} + p\mathrm{d}V=0$, describing an adiabatic expansion, where the thermal energy density $U_\mathrm{th}$ can be defined as $U_\mathrm{th}\equiv\rho V$ and the volume $V$ as $V\equiv a^3$.

Here, we should stress out that one can write the Friedmann equation in a more compact form introducing the dimensionless variable $\Omega$ such as $\Omega=\frac{\rho}{\rho_\mathrm{c}}$, where $\rho_\mathrm{c}=\frac{3H^{2}}{8\pi G}$,  which quantifies the deviation of the energy density of the universe from the critical energy density, $\rho_\mathrm{c}$, of a spatial flat universe. Thus, straightforwardly one obtains that \Eq{Friedmann Equation} can be recast as
 \beq \label{Friedman Equation - Omega form}
 \Omega-1=\frac{K}{a^{2}H^{2}}
 \eeq
and one can see that for $K=-1$ (hyperbolic geometry), $\Omega<1$ whereas for $K=+1$ (spherical geometric), $\Omega>1$. Regarding the case in which $K=0$ (Euclidean geometry), $\Omega=1$ and the universe is spatiallly flat.

\subsection{The constant equation of state}\label{The constant w}
 At this point, we can find the evolution of the scale factor for different components of the energy budget of the universe which may dominate in different periods of the cosmic history. By viewing the dominant component of the energy content of the universe as a perfect fluid, the universe thermal state can be described by the following equation of state
 \beq\label{equation of state}
 p=w\rho,
 \eeq
 where $w$ is the equation-of-state parameter determining the nature of the fluid. The case where $w$ is constant, the most commonly studied one, describes quite well the universe's thermal state in the different periods of the cosmic history. Assuming then a constant equation-of-state parameter $w=p/\rho$ for the dominant component of the universe one can work out from \Eq{Continuity Equation} and \Eq{Friedmann Equation} the dynamics of the space expansion as well as of the energy density of the universe. In particular, one can straightforwardly find that
\beq\label{rho in a w environment}
\rho= \rho_\mathrm{ini}\left(\frac{a}{a_\mathrm{ini}}\right)^{-3(1+w)}
\eeq

 \beq\label{scale factor in a w environment}
a = \left\{ 
 \begin{array}{ll}
  a_\mathrm{ini} \left[1 \pm\frac{3}{2}\sqrt{\frac{\rho_\mathrm{ini}}{3}}\frac{t-t_\mathrm{ini}}{\Mp}\right]^\frac{2}{3(1+w)} & w\neq -1 \\
  a_\mathrm{ini}\exp{\left(\pm\sqrt{\frac{\rho_\mathrm{ini}}{3}}\frac{t-t_\mathrm{ini}}{\Mp}\right)} & w=-1 \\
  \end{array}
  \right.
\eeq
where the index $\mathrm{ini}$ denotes an initial time. The $+$ sign accounts for an expanding universe, $H>0$, whereas the $-$ sign for a contracting universe, $H<0$.
 
Below, we refer to some characteristic values of $w$ which can describe the universe thermal state at different cosmic epochs. The case of $w=0$ describes a fluid of non-relativistic particles (matter domination era) where $\rho_\mathrm{m}\sim a^{-3}$ whereas when $w=1/3$ one can identify a fluid of relativistic particles (radiation domination era) where $\rho_\mathrm{r}\sim a^{-4}$. The case $w=-1$ describes a thermal state of negative pressure in which the vacuum energy dominates the universe energy content. This is the case for a $\Lambda$ domination era where one can assign from the Friedmann equation \Eq{Friedmann Equation} an energy density to the cosmological constant $\Lambda$, namely $\rho_\mathrm{\Lambda}=\Lambda\Mp^2 = \mathrm{constant}$.  With the same reasoning one can associate an energy density to the spatial curvature, $\rho_\mathrm{K}=-\frac{3K}{a^2}\Mp^2$ which can be viewed as the energy density of a perfect fluid with  $w=-1/3$. Thus, taking into account the above discussion one can rewrite the Friedmann equation \Eq{Friedmann Equation} in the following form
\beq\label{Friedmann Equation in terms of rhos}
H^2 = \frac{1}{3\Mp^2}\left(\rho_\mathrm{matter}+\rho_\mathrm{K}+\rho_\mathrm{\Lambda}\right) = \frac{\rho_\mathrm{t}}{3\Mp^2},
\eeq
where $\rho_\mathrm{matter}$ accounts for the sum of the energy densities of ordinary matter, dark matter, radiation and any other constituent of the universe and $\rho_\mathrm{t}$ is the total energy density in the universe. 
\subsection{The horizon scale}\label{sec:horizon scales}
The concept of the horizon is fundamental in cosmology. Below we discriminate between the notion of cosmological/Hubble horizon or Hubble radius and that of the particle horizon. The Hubble horizon or Hubble radius is defined as
\beq\label{R_H}
R_\mathrm{H} \equiv H^{-1}
\eeq
and is the distance at which the galaxy recession velocity is equal to the speed of light. Galaxies outside a sphere of a radius equal to the Hubble radius recede from us at a speed faster than the speed of light. This does not violate the special relativity postulate that the maximum speed in the universe is $c$, because it is spacetime itself that is expanding faster than the speed of light, not objects within that spacetime.  In a more formal way, the fact that galaxies can recede from us with a speed faster than the speed of light is not a problem given the fact that Lorentz symmetry is a local symmetry.

The particle horizon on the other side is defined as the region where causal contact has been established through photon interactions. More precisely, at a specific time the particle horizon is the extent of our light cone in the the past at $t=0$ . From \Eq{FRLW metric - cosmic time} taking $d\theta=d\phi=0$ the infinitesimal distances traveled by photons is $dr=dt/a(t)$. Therefore, the particle horizon, $L_\mathrm{P}$ is defined as
 \beq\label{R_P}
 R_\mathrm{P}\equiv a(t)\int_{0}^{t} \frac{dt'}{a(t')}
 \eeq
Assuming a polynomial behavior of $a(t)$, i.e. $a(t)\propto t^n$ with $n<1$, which is the case for the majority of the cosmic epochs, one finds that $R_\mathrm{P} =\frac{nH^{-1}}{1-n} \sim H^{-1}$. One then can see that the Hubble horizon and the particle horizon are of the same order and hereafter they can be used interchangeably as the horizon scale unless stated differently. Therefore, the horizon scale, $H^{-1}$, gives a very good estimate of the region within which causal contact has been established and is identified as well with the scale at which general relativistic effects become important. An important relevant quantity is the horizon mass, $M_\mathrm{H}$, defined as the mass inside the horizon:
\beq\label{M_H}
M_\mathrm{H}\equiv\frac{4\pi}{3}\rho_tR^3_\mathrm{H} 
\eeq
Combining then \Eq{M_H} and \Eq{Friedmann Equation in terms of rhos} one can infer that
\beq\label{cosmological horizon}
R_\mathrm{H}=2GM_\mathrm{H}.
\eeq
The above expression which relates the mass inside the horizon and the horizon scale is the same expression used for the definition of the black hole apparent horizon in spherical symmetry, a fact which reflects the common physical nature of the cosmological horizon and the black hole apparent horizon. Both of them can be viewed as trapped surfaces in the context of the theory of general relativity ~\cite{Faraoni:2016xgy}.
      
 Finally, it is important to distinguish between physical lengths inside and outside the horizon which will give us below critical behaviors. Therefore, a length scale $\lambda$ related to its wave number is the comoving scale associated to $\lambda$ times the scale factor. Thus, $\lambda=\frac{2\pi a}{k}$ and our conditions take the following form:
 $$\frac{k}{aH}\ll1 \Rightarrow{\rm Scale \; \lambda \; outside \; the \; horizon}$$
 $$\frac{k}{aH}\gg1 \Rightarrow {\rm Scale \; \lambda \; inside \; the \;horizon}$$
\section{The Thermal History of the Universe}
\label{sec:Thermal History}
The Cosmic Microwave Background radiation was firstly detected in 1965 by Penzias and Wilson and later confirmed by the sattelite probes COBE, WMAP and Planck. As it was found, CMB constitues a nearly uniform signal at microwave frequencies coming from all directions in the sky with a high degree of isotropy. It is interpreted as the black-body radiation emitted at the moment of the last scattering of photons with matter at around 380.000 years after the Big Bang singularity. Today, the present temperature of this black-body spectrum is $T_\mathrm{CMB,0}
\simeq 2.725K$ while the high degree of isotropy, namely  $\Delta T/T \sim 10^{-5}$, strongly suggests a homogeneous universe on sufficiently large scales $\sim 100\mathrm{Mpc}$. 

Accounting therefore for the cosmological redshift presented in \Sec{sec:Hubble parameter + z} and for the adiabatic expansion of the universe (no heat transfer) one can infer that a black-body state stays as a black-body state with a temperature decreasing with the expansion as
\beq\label{T_r}
T_\mathrm{r}\sim 1/a
\eeq
Therefore, as we go deeply in the radiation dominated universe the temperature increases as $1/a$ and at the time when universe begins it becomes infinite. This leads to the standard cosmological picture of the Hot Big Bang universe: One has initially an initial state at some finite time in the past when the universe was infinitely hot, followed by a radiation era during which the universe is gradually cooling down as $T_\mathrm{r}\sim 1/a$. During this period of radiation, photons strongly interact with matter and at the end of this period when the universe is cold enough, the first atoms form and photons can travel freely in the universe without interacting with matter. This triggers the onset of a matter dominated era during which large scale structures such as galaxies, stars and planets form. Finally at some point, the vaccum energy, largely quoted as ``dark energy", described above in terms of the cosmological constant $\Lambda$, inevitably dominates the energy content of the universe driving in its turn an accelerated expansion as the one we observe today. Let us now describe in more detail the different epochs of the thermal history of the universe ~\cite{Kolb:1990vq}:

\begin{itemize}
\item{\textbf{$100\mathrm{GeV}<T<T_\mathrm{Pl}=10^{19}\mathrm{GeV}$}

This is the cosmic epoch of the very early moments of the cosmic history during which inflation is assumed to take place at some point ~\cite{Starobinsky:1980te,Guth:1980zm,Linde:1981mu,Albrecht:1982wi,Linde:1983gd}. During this inflationary epoch,  the universe undergoes an accelerated expansion where physical lengths are stretched out so much that they become larger than the horizon scale conserving however the isotropy. This inflationary period is supposed to be driven by one (inflaton field) or more scalar fields which at the end of inflation decay to relativistic particles which thermalise by reaching a common temperature, quoted as the reheating temperature [For more details on reheating see ~\cite{Turner:1983he,Shtanov:1994ce,Kofman:1994rk,Kofman:1997yn}].\footnote{After inflation, the inflaton or/and the other scalar fields oscillates at the bottom of his/their potential, a fact which sources a parametric instability in the equation of motion of the metric perturbations, that are enhanced on small scales. These enhanced perturbations depending on the details of their collapse dynamics can constitute the seeds either for the formation of virialised objects or for the formation of PBHs.} When reheating is over, the universe is dominated by relativistic particles which increase the entropic degrees of freedom and which lead to the transition to the radiation era (Hot Big Bang phase). 
}
\item{\textbf{$T\sim 100 \mathrm{GeV}$}

At a temperature $T$ around $100 \mathrm{GeV}$ the electroweak phase transition ~\cite{Kirzhnits:1972ut,Weinberg:1974hy,Dine:1992wr} takes place in which the $SU(2)\times U(1)$ electroweak symmetry breaks into the $U(1)$ symmetry of electromagnetism. During this phase transition, the weak nuclear and electromagnetic forces separate and the physics of the universe at this time is described by the Standard Model (SM) or some extension. The electroweak symmetry breaking time corresponds as well to the last time at which it possible to generate a matter/antimatter asymmetry through a process of baryogenesis ~\cite{Sakharov:1967dj,Farrar:1993hn}. 
}
\item{\textbf{$T\sim 100 \mathrm{MeV}$}

At $T\sim 100 \mathrm{MeV}$ the Quantum ChromoDynamic (QCD) phase transition takes place during which the  plasma of quarks and gluons become bound leading to the formation of hadrons ~\cite{Olive:1980dy,Fuller:1987ue}. This transition is associated to a chiral symmetry breaking mechanism and it is considered to play a significant role to the generation of primordial magnetic fields ~\cite{Cheng:1994yr}.
}

\item{\textbf{$1\mathrm{MeV}<T<100 \mathrm{MeV}$}

During this era, all the elementary particles ($\mathrm{\gamma, \nu, e, \bar{e}, n}$ and $\mathrm{p}$) interact with each other and form a bath of thermal equilibrium. At a temperature $T\sim 1 \mathrm{MeV}$ the neutrinos decouple from the thermal bath and primordial nucleosynthesis of light elements (mostly $\mathrm{H,D,He,Li}$ and $\mathrm{Be}$) starts taking place already at a temperature $T\sim 10 \mathrm{MeV}$ and end at a temperature of $T\sim 100 \mathrm{keV}$ ~\cite{Alpher:1948ve,Wagoner:1966pv,Sarkar:1995dd}. Heavier elements are formed later in the interior of the stars through stellar nucleosynthesis or through star explosions.
}
\item{\textbf{$T\sim 100  \mathrm{keV}$}

 At a temperature $T\sim 100  \mathrm{keV}$ the onset of the matter domination era takes place during which through the recombination process ~\cite{Peebles:1968ja} the first atoms form when free electrons bind with nucleons. Given the dynamical nature of the recombination process initially some electrons are free and can interact with photons which remain coupled to the unbound electrons during the first stages of recombination. However, soon after the end of the recombination process photons decouple from matter and are free to travel in the universe producing a black-body radiation spectrum, the well studied CMB radiation. This relic radiation is the oldest ``snapshot" of the universe one can get ~\cite{Penzias:1965wn}.}
 
 \item{\textbf{$1 \mathrm{meV}<T<100  \mathrm{keV}$}
 
 After photon decoupling at $T\sim 100 \mathrm{keV}$ the different thermal processes present in the earliest epochs of the cosmic history stop taking place and the universe gradually cools down entering the so called ``dark" ages during which structure formation takes place through gravitational processes ~\cite{Gunn:1972sv}. However, at a temperature of around $T\sim 1  \mathrm{meV}$ reionisation processes occur when energetic objects inside the already formed galaxies ionize the neutral hydrogen creating again, as during the eras before recombination, the conditions for an ionized plasma in the intergalactic medium ~\cite{Gunn:1965hd,Becker:2001ee,Barkana:2000fd}. However, due to the expansion of the universe the matter is so much diluted that interactions are much less frequent explaining in this way the transparency of the universe in the subsequent cosmic epochs ~\cite{Vishniac:1987wm}. 
 }
 \item{\textbf{$T<1 \mathrm{meV}$}
 
 After reionisation, at a temperature $T\sim 0.33  \mathrm{meV}$, equivalent to $9$  billions years after the Big Bang singularity, dark energy dominates and the universe enters the era of its accelerated expansion continuing to cooling down ~\cite{Perlmutter:1998np,Riess:1998cb,Schmidt:1998ys}. Today, its temperature, namely the temperature of the CMB relic radiation is $T\simeq 2.725K$.}
\end{itemize}
\section{The Composition of the Universe}\label{sec:universe composition}
Having described in a concise way before the thermal history of the universe, we will see here how the energy content of the universe evolves with time. In particular, one can consider that each constituent of the energy content of the universe can be described in terms of a perfect fluid and assuming for simplicity that there is no considerable energy transfer between the different constituents the total energy density of the universe can be read as the sum of the energy density of the different energy components [See \Eq{rho in a w environment}],
\beq\label{rho_total}
\rho_\mathrm{t} = \sum_i\rho_\mathrm{ini,i}\left(\frac{a}{a_\mathrm{ini}}\right)^{-3(1+w_i)}.
\eeq
where the index $i$ denotes the different constituents which are dominant during the different epochs of the cosmic history. 
One then can specify the energy density of every energy component at a specific time and then from \Eq{rho_total} they can infer the dynamics of $\rho_\mathrm{tot}$. To do so in a ``compact" way, we introduce the dimensionless parameters $\Omega_{i}$ which measure the energy contribution of the different components of the universe in its energy budget and are defined as
\beq\label{Omega_i}
\Omega_{i}\equiv \frac{\rho_i}{\rho_\mathrm{c}},
\eeq
where $\rho_\mathrm{c}$ is the critical energy density required for a flat universe [See the discussion above \Eq{Friedman Equation - Omega form}].
Consequently, one can easily deduce that \Eq{Friedmann Equation in terms of rhos} can be recast as
\beq\label{Omega_total}
\Omega_\mathrm{tot}=\sum_i \Omega_i = 1.
\eeq
Following, the results of the Planck satellite which captured and studied the CMB relic radiation we give below the the $\Omega$ parameters for the different constituents of the universe today. Then, from \Eq{rho_total}, \Eq{Omega_i} and \Eq{Omega_total} we can reconstruct the time evolution of the composition of the universe.
\begin{itemize}
\item{\textbf{Baryonic Matter}

In this constituent of the universe counts the ordinary matter in form of cold baryons, which are composite subatomic particles made up of quarks, like the protons and the neutrons and which are heavier than the leptons, namely the three generations of electrons and neutrinos. Their contribution according to Planck 2018 results  ~\cite{Aghanim:2018eyx} is 
$\Omega^{(0)}_\mathrm{b} \simeq 0.049$ \footnote{With the index (0) we refer to today.}.
}
\item{\textbf{Radiation}

In this constituent, we account for relativistic particles, namely photons of the CMB and neutrinos. Their overall contribution is  extremely tiny and account for $\Omega^{(0)}_\mathrm{r} \simeq 10^{-4}$ ~\cite{Aghanim:2018eyx}.
}
\item{\textbf{Dark Matter}

This constituent of the universe was postulated to exist in order to explain many observations findings related to galaxy rotation curves and large scale structure formation. This non-relativistic form of matter, described in terms of a fluid with $w=0$, is of non baryonic form and therefore its unknown nature is an active field of study. Its current contribution is $\Omega^{(0)}_\mathrm{DM}=0.265$ ~\cite{Aghanim:2018eyx} and as one can infer its energy contribution is more than five times bigger than that of the ordinary baryonic matter.
}
\item{
\textbf{Curvature}

From the Friedman equation in the form of \Eq{Friedman Equation - Omega form}  and taking into account the definitions of the $\Omega$ parameters [See \Eq{Omega_i}] one can identify and $\Omega$ parameter associated to the spatial curvature which in the case where the universe is not flat, namely when $K=\pm 1$, can be viewed as explained above \Eq{Friedmann Equation in terms of rhos} as a fluid with $w=-1/3$. However, the observations made so far are still consistent with a spatially flat universe with $\Omega^{(0)}_K\simeq 0$. The current constraints on $\Omega_K$ read as $\Omega^{(0)}_K = -0.001 \pm 0.002$ at $ 95\%$ confidence level ~\cite{Aghanim:2018eyx}.
}
\item{\textbf{Dark Energy}

This constituent of the universe was postulated to exist like dark matter to balance the missing bulk part of the total energy density of the universe. It was also introduced to explain the acceleration in the expansion of the universe observed in '90s which points towards the existence of a fluid with $w\simeq -1$. This is why the cosmological constant $\Lambda$ is considered one of the main candidates for the dark energy. Similarly to dark matter, dark energy constitutes an active field of research and its current energy contribution is calculated to be $\Omega^{(0)}_\mathrm{DE}\simeq 0.685$ ~\cite{Aghanim:2018eyx}.}
\end{itemize}
In \Fig{fig:Universe Composition} below, we see in the left panel the current composition of the universe displayed in a pie chart. In left panel on the other hand, we show the evolution of the energy contribution of the different constituents of the universe as a function of the scale factor normalized with respect to the scale factor today, $a_0$. As one may see, by assuming this simple picture of non interacting fluids for the different constituents of the universe we reproduce quite well the thermal history of the universe presented in \Sec{sec:Thermal History}. We clearly see an initial radiation domination epoch for $a<a_\mathrm{eq}$ during which different phase transitions take place and thermal processes lead to the primordial nucleosynthesis of the light elements and nuclei. Then, a subsequent matter domination era for $a_\mathrm{eq}<a<a_\mathrm{acc}$ drives the universe cosmic history during which the large scale structures form and finally a late dark energy era for $a>a_\mathrm{acc}$ during which the universe expands in a  accelerated way.
\begin{figure}[h!]
\begin{center}
\includegraphics[width=0.496\textwidth, clip=true]{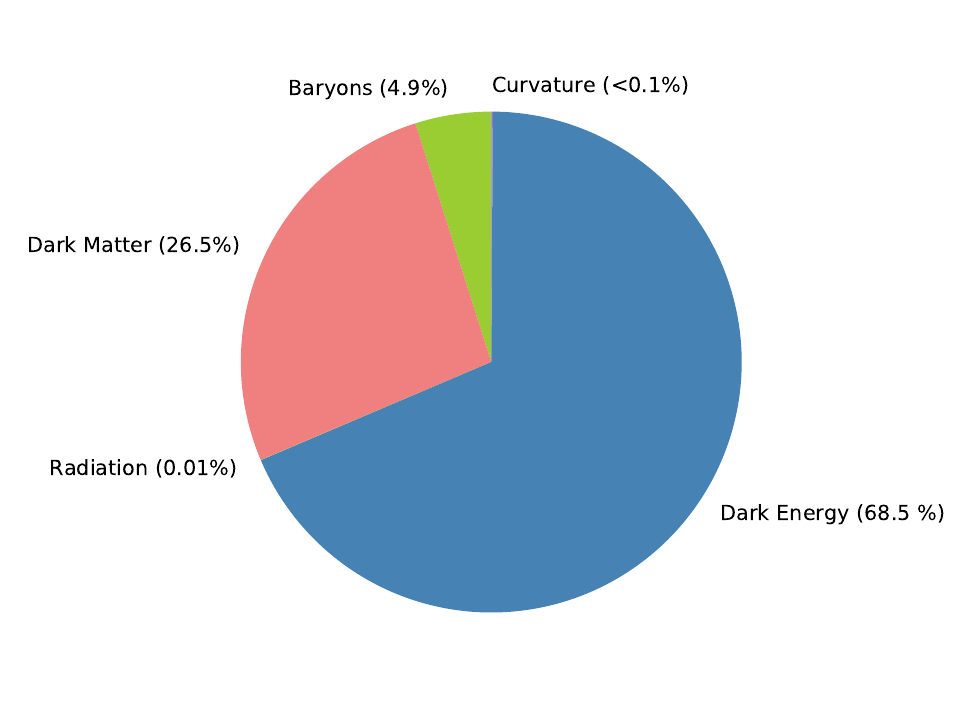}
\includegraphics[width=0.496\textwidth, clip=true] {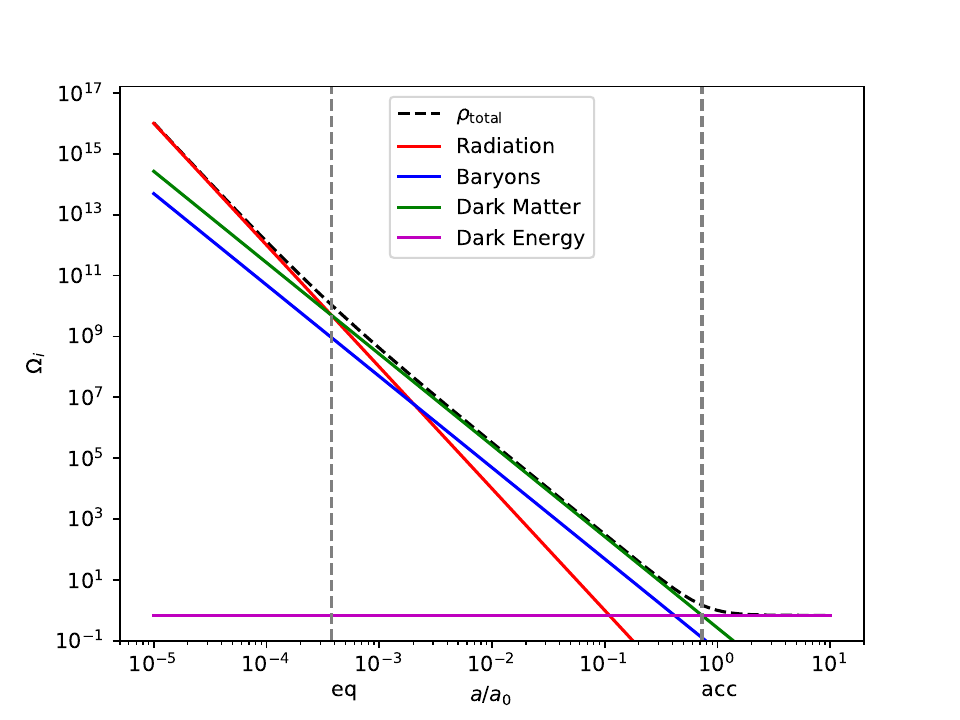}
\caption{ Left Panel: The composition of the universe today. Right Panel: The dynamics of the energy density contribution of the different constituents of the universe. The left dashed vertical line denotes the time at matter-radiation equality $(\mathrm{eq})$ when the first atoms form while the right dashed vertical one denotes the time when the dark energy dominated the universe energy budget driving an accelerated expansion $(\mathrm{acc})$.}
\label{fig:Universe Composition}
\end{center}
\end{figure}

\section{The Problems of the Hot Big Bang Universe}\label{subsec:HBB Problems}
After having presented the fundamental notions and the theoretical framework of a homogeneous and an isotropic universe and giving a brief thermal history of our universe dictated by the Hot Big Bang theory we feature here the basic shortcomings of the standard Hot Big Bang cosmology that motivated inflation ~\cite{Starobinsky:1980te,Guth:1980zm,Linde:1981mu,Albrecht:1982wi,Linde:1983gd}. Actually, we will refer to two of them: \textbf{the horizon problem} and \textbf{the flatness problem}.
 
 \subsection{The horizon problem}
 The horizon problem ~\cite{Rindler:1956yx,Misner:1967uu} or large scale homogeneity problem is qualitatively the fact that regions separated by distances greater than the speed of light times the age of the universe (no causal connected regions) are observed to have similar density and temperature fluctuations up to $10^{-5}$, a fact which is contradicting since they should not know each other due to the principle of special relativity for the finitude of the speed of light. Therefore, there should have been an information exchange between these regions in the past. More quantitatively, we can see the evolution of the horizon and of physical lengths during radiation and matter dominating epochs. In particular, from equation \Eq{scale factor in a w environment} one obtains for radiation domination (RD) ($w=1/3$) and matter domination (MD) ($w=0$) that 
\beq
R_{H}=
\begin{cases}
\frac{4}{3}\Mp\sqrt{\frac{3}{\rho_\mathrm{ini}}}\left(1 + \frac{3}{2}\sqrt{\frac{\rho_\mathrm{ini}}{3}}\frac{t-t_\mathrm{ini}}{\Mp}\right), \mathrm{ \; for \; RD} \\
\Mp\sqrt{\frac{3}{\rho_\mathrm{ini}}}\left(1 + \frac{3}{2}\sqrt{\frac{\rho_\mathrm{ini}}{3}}\frac{t-t_\mathrm{ini}}{\Mp}\right), \mathrm{ \; for \; MD}
\end{cases}
\eeq
 On the other hand, the physical distances evolve as we showed before as
 $$
R_{\mathrm{phys}} = a(t) r = 
\begin{cases}
a_\mathrm{ini}\left(1 + \frac{3}{2}\sqrt{\frac{\rho_\mathrm{ini}}{3}}\frac{t-t_\mathrm{ini}}{\Mp}\right)^{1/2},  \mathrm{ \; for \; RD}\\
a_\mathrm{ini}\left(1 + \frac{3}{2}\sqrt{\frac{\rho_\mathrm{ini}}{3}}\frac{t-t_\mathrm{ini}}{\Mp}\right)^{2/3},  \mathrm{ \; for \; MD}
\end{cases}
$$
Thus, one can infer that the horizon scale, $R_\mathrm{H}$ grows faster than physical distances both in the RD and MD eras. Consequently, in the past there should have been regions which were causally disconnected. However, our universe is extremely homogeneous and isotropic (e.g. CMB temperature fluctuations $\frac{\delta T}{T}\sim10^{-5}$ on angular scales larger than $1\deg$, which corresponds to the horizon scale at the time of the emission of CMB). 

 \subsection{The Flatness Problem}
 Regarding the flatness problem one can see how the $\Omega$ parameter, related to the spatial curvature of the universe through \Eq{Friedman Equation - Omega form}, evolves in time.  From \Eq{Friedman Equation - Omega form} we see that if the universe is perfectly flat today then $\Omega=1$ at all times. If however there is a small curvature $K\neq0$ then $\Omega$ will depend on time. Below, we consider the case where $K=+1\neq 0$ for the RD and MD eras. Knowing then that  $\rho_{r}\propto a^{-4}$ for the RD era and $\rho_{m}\propto a^{-3}$ for the MD era and taking into account that $H^{2}\propto\rho_\mathrm{t}$,  \Eq{Friedman Equation - Omega form} is equivalent to 
 \beq\label{Omega-1 in RD and MD}
 \Omega -1 \propto 
 \begin{cases}
 a^{2}, \mathrm{\;for\;RD} \\
 a, \mathrm{\;for\;MD} 
 \end{cases}
 \eeq
Consequently, knowing that $\Omega \sim 0.01$ today and that $T_\mathrm{r}\sim 1/a$ one can easily infer that at the beginning of the radiation dominated era, which here we identify with the epoch of BBN where $T_\mathrm{BBN}\sim 1\mathrm{MeV}$, 
 \beq \label{Omega -1:flatness problem}
 \frac{|\Omega-1|_{T=T_{\mathrm{BBN}}}}{|\Omega-1|_{T=T_{0}}}\approx \left( \frac{a^{2}_\mathrm{BBN}}{a^{2}_{0}}\right) \approx \left( \frac{T^{2}_{0}}{T^{2}_{\mathrm{BBN}}}\right)\approx O(10^{-16})
 \eeq
 where $T_{0}\sim10^{-13}eV\sim2.7K$ is the present temperature of CMB.
 Therefore, in order to recover the value $\Omega_{0}-1\sim 1$ today we must assume that the value of $\Omega-1$ at early times (Planck era) is perfectly fine-tuned to a value very close to zero $10^{-16}$ but not exactly zero! That is the flatness problem, also dubbed as the ``fine-tuning" problem and lies in understanding the mysterious mechanism which led the universe to start its expansion with almost spatially flat initial conditions ~\cite{1970grun.conf.....D,Hawking:1979ig}. 
 
 \subsubsection{The flatness problem and the entropy conservation}\leavevmode\\
 
Let us see here, how the flatness problem is related to the assumption of the adiabatic expansion. Equation (\ref{Friedman Equation - Omega form}) can be recast in the RD era, where $\rho_{r}=\frac{\pi^2}{30}g_{*}(T)T^{4}$ with $g_{*}$ being the number of relativistic degrees of freedom when the universe's temperature is $T$, as follows
\beq\label{Omega-1:entropy problem 1} 
\Omega-1=\frac{90}{\pi^2g_{*}(T)}\frac{k\Mp^{2}}{a^{2}T^{4}}=\left[\frac{1440}{\pi^2 g_{*}(T)}\right]^{1/3} \frac{kM^{2}_{Pl}}{S^{\frac{2}{3}}T^{2}},
\eeq
 where in the last equality we use the fact the entropy $S$ is defined as $S\equiv sV$, where the entropy density $s$ reads as $s=\frac{2\pi^2}{45}g_{*}(T) T^3$ and the volume $V=a^3$. Thus,  \Eq{Omega-1:entropy problem 1} at the BBN time reads as
 \beq \label{Omega-1:entropy problem 2}
 |\Omega-1|_{T=T_\mathrm{BBN}}=\left[\frac{1440}{\pi^2 g_{*}(T)}\right]^{1/3}\frac{\Mp^2}{S^{2/3}_\mathrm{BBN}T^{2}_{\mathrm{BBN}}} \approx 10^{-16}
 \eeq
In the last step, we used the fact that the entropy in a comoving volume is conserved and it is equal to $10^{90}$ according to observational evidence from the matter-antimatter asymmetry \cite{Canetti:2012zc} and that the number of relativistic degrees of freedom at BBN time, where $T_\mathrm{BBN}\sim 1\mathrm{MeV}$, is $\g_{*}(T_\mathrm{BBN})=106.75$, having accounted only for the SM particles. Evidently, we find again the same ``fine-tuning" problem as before in which the universe should have started with almost spatially flat initial conditions. However, now this ``fine-tuning" problem arises because we have adopted the assumption of entropy conservation. 

 \subsection{Solving the problems}
Regarding the horizon problem mentioned above, in order to solve it, the universe has to pass through a primordial period in which physical lengths $R_\mathrm{phys}$ grow faster than the horizon scale  $H^{-1}$. Specifically, if there is a period in which physical lengths grow faster than the horizon then the photons that appear to be causally disconnected in the time of last scattering (when CMB was emitted) where $\lambda>H^{-1}$ had the chance to ``talk" to each other in a primordial cosmic era where $\lambda<H^{-1}$. In this way, we recover the homogeneity and isotropy of CMB solving the horizon problem. This last condition can be expressed in terms of the evolution of the scale factor $a(t)$. Thus, since $\lambda\propto a$ and $H^{-1}=a/\dot{a}$ we should impose a period in the cosmic history where
\beq
\left(\frac{\lambda}{H^{-1}}\right)^{.}=\ddot{a}>0.
\eeq
This equation can be recast, using \Eq{Acceleration Equation} and the fact that during this early cosmic era the universe's energy content is dominated by a fluid $X$ with an equation-of-state parameter $w_\mathrm{X}=p/\rho$, in the following form
\beq\label{Inflation Condition:Horizon Probelm}
w_\mathrm{X}<-1/3.
\eeq
In order to solve now the flatness/entropy problem, we should demand 
 %In order to solve the above mentioned problems we should have an expansion period of the universe in which non adiabaticity should have been occurred. This will explain the flatness and entropy problem  and will lead to the large values of  entropy $S$ we observe today. 
that in an initial era of the cosmic history before the onset of the radiation era, the parameter $\Omega-1$ should decrease allowing in this way to obtain very low values of the order of $10^{-16}$. To ensure this, one can assume that the universe during this early era is prevailed by a fluid $X$ with equation of state $w_\mathrm{X}$. Combining therefore \Eq{Friedman Equation - Omega form}, \Eq{rho in a w environment} and \Eq{Friedmann Equation in terms of rhos} one straightforwardly obtains that 
 \beq
 \Omega-1\propto a^{1+3w_\mathrm{X}}
 \eeq
 Consequently, in order for $\Omega -1 $ to decrease one requires that
 \beq\label{Inflation Condition:Flatness/Entropy Probelm}
 w_\mathrm{X}<-1/3.
 \eeq
This last condition, $w_\mathrm{X}<-1/3$ for the solution of the flatness/entropy problem is the same as the condition to address the horizon problem and defines the inflationary period in the cosmic expansion. During this period, the universe expands in an accelerated way, i.e. $\ddot{a}>0$ and the $\Omega$ parameter at the end of inflation is forced to take a value very close to one, but not exactly one, independently of its initial value. 

%At this point, we should stress out that during inflation the universe expands in an adiabatic way, which means that the local state of matter at some spacetime point $(t,\boldmathsymbol{x})$ of the perturbed universe is the same as in the background but at slightly different time $t+\delta t(\boldmathsymbol{x})$ with the time shift $\delta t(\boldmathsymbol{x})$ being a local quantity which does not depend on the different energy components of the universe. Quantitatively, this means that for two different energy component of the universe $i$ and $j$ their energy density perturbations should satisfy the following relation
%\beq\label{adiabatic perturbations}
%\frac{\delta\rho_\mathrm{i}}{\dot{\rho}_\mathrm{i}} = \frac{\delta\rho_\mathrm{j}}{\dot{\rho}_\mathrm{j}} = \delta t(\boldmathsymbol{x}).
%\eeq

At this point, we should stress out that during inflation the universe expands in an adiabatic way, i.e. there is no entropy production. More rigorously, this means that one should ensure the covariant conservation of the stress energy tensor, i.e. $\nabla_\mathrm{\mu}T^{\mathrm{\mu\nu}} = 0$. To ensure however the transition to the radiation era the universe should pass through a non adiabatic period of reheating during which an enormous amount of entropy is generated through relativistic degrees of freedom, solving in this way naturally the flatness/entropy problem. This early phase transition era is broadly quoted as (pre)reheating and was one of the topics studied within my PhD where we studied together with by collaborators the production of PBHs during the period of preheating in the context of single-field inflationary models. 
 
\newpage
\chapter{PBH Formation}\label{sec:PBH formation}
In this chapter, we  introduce the fundamentals of PBH physics. Firstly, we give the basic theoretical framework in the field of PBH research by introducing the notions of the PBH mass function, the PBH characteristic scale and the PBH threshold. Then, we  present briefly the current observational status in the domain of  PBH physics by describing the different observational constraints on the abundance of PBHs as a function of their mass. Finally, we underlie the implications of PBHs in cosmology.
\section{PBH Basics}\label{sec:PBH basics}
\subsection{The PBH Mass}\label{sec:PBH mass}
As we saw in the discussion after \Eq{M_H} the expression which relates the mass inside the Hubble radius and the Hubble radius is the same expression used for the definition of the black hole apparent horizon in spherical symmetry. This fact, as mentioned in \Sec{sec:horizon scales}, reflects the common physical nature of the cosmological horizon and the black hole apparent horizon from the point of view of general relativity. In particular, the black hole apparent horizon is the asymptotic location of the outermost trapped surface for outgoing light-rays whereas the cosmological horizon is the innermost trapped surface for incoming light rays. One then expects that the mass of a PBH is the same with the mass inside the horizon at PBH formation epoch, which is considered roughly as the time at which the PBH characteristic scale crosses the Hubble radius. However, more accurate analysis shows that the mass of a PBH is a fraction of the mass inside the horizon at the time of PBH formation and reads as 
\beq\label{M_PBH}
m_\mathrm{PBH} = \gamma M_\mathrm{H},
\eeq
where $\gamma \sim O(1)$  is an efficiency parameter encapsulating the details of the gravitational collapse. 

At this point, one should stress out the importance of scaling laws in PBH formation process firstly noted by Jedmazik and Niemeyer ~\cite{Niemeyer:1997mt} and further investigated by Musco et al. ~\cite{Musco:2008hv,Musco:2012au} which can refine the computation of the PBH mass. Specifically, when the local/mean energy density excess  is sufficiently close to the critical threshold $\delta_\mathrm{c}$, i.e. $|\delta-\delta_\mathrm{c}| \ll 1$, then the refined PBH mass is given by the following scaling law,
\beq\label{critical collapse}
m_\mathrm{PBH} \propto M_\mathrm{H}\left(\delta-\delta_\mathrm{c}\right)^{p},
\eeq
where $p\simeq 0.37$ is a universal exponent. The above critical scaling behavior was already found in the context of spherical symmetric collapse of a massless scalar field firstly studied by Choptuik ~\cite{Choptuik:1992jv} and further explored by subsequent studies ~\cite{Evans:1994pj,Gundlach:1999cu}. Here, it is important to mention that the scaling law in \Eq{critical collapse} breaks down when one approaches very small values of the difference $\delta-\delta_\mathrm{c}$ due to generation of shock waves in nearly critical collapse, imposing in this way a minimum mass for $m_\mathrm{PBH}$ at the order of $10^{-4}$ of the mass inside the horizon ~\cite{Hawke:2002rf}.

\subsection{The PBH characteristic scale}\label{sec:PBH scale}
\begin{figure}[h!]
\begin{center}
\includegraphics[width = 0.7\textwidth, height = 0.6\textwidth, clip=true]{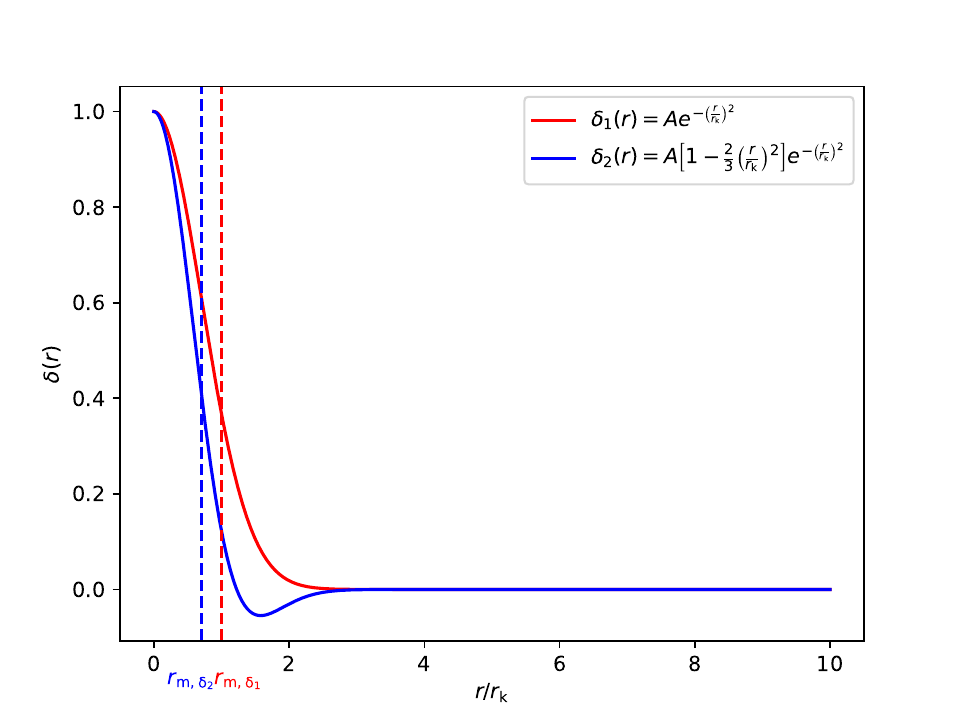}
\caption{ The PBH scale for a compensated and a non-compensated energy density profile $\delta(r)$. The blue solid line represents a Gaussian (non-compensated) energy density profile, denoted as $\delta_1(r)$, whereas the red solid line stands for a Mexican-hat (compensated) profile, denoted as $\delta_2(r)$. In both profiles, the parameters $A$ and $r_\mathrm{k}$ are chosen to be $A=r_\mathrm{k}=1$. By making use of \Eq{r_m definition}, we plot with the vertical dashed blue line the PBH scale in the case of the Mexican-hat profile and with the vertical dashed red line the PBH scale for the Gaussian profile. }
\label{fig:PBH_scale}
\end{center}
\end{figure}

Having defined before the PBH mass as the horizon mass at the time of PBH formation, approximately equal to the time at which the PBH characteristic scale crosses the Hubble radius, one will inevitably ask the question what is this characteristic scale. In general, assuming spherical symmetry, it is considered to be roughly equal to the scale at which the local energy density excess of the overdensity/energy density profile, $\delta(r)$, \footnote{The  local energy density excess is defined as $\delta(r,t) \equiv \frac{\rho(r,t)-\rho_\mathrm{b}(t)}{\rho_\mathrm{b}(t)}$, where $\rho_\mathrm{b}(t)$ is the energy density of the background and $\rho(r,t)$ is the energy density of the overdensity. The energy density profile $\delta(r)$ can be viewed as well as the time-independent part of the local energy density excess in the superhorizon regime where one can perform a gradient expansion approximation ~\cite{Musco:2018rwt}.} at PBH formation time is zero. However, there are energy density profiles which are always positive, such as the Gaussian one, which give an infinite PBH scale. These profiles are called non-compensated profiles whereas profiles in which $\delta(r)$ becomes negative at some point are the compensated ones. See \Fig{fig:PBH_scale} in which a compensated and a non-compensated energy density profiles are shown  together with the respective PBH scales.

For this reason, the PBH characteristic scale is usually defined in a more refined way for any type of energy density profiles by using the notion of the compaction function, firstly introduced by Shibata $\&$ Sasaki in ~\cite{Shibata_1999} and then recently used by Musco ~\cite{Musco:2018rwt}. The compaction function $\mathcal{C}$ is defined in a similar fashion as the Schwarzschild condition for the formation of a black hole apparent horizon, $R = 2GM$ and can be seen as a  local measure of the gravitational potential. In particular, it is defined as twice the local mass-excess over the areal radius and reads as
\beq\label{C definition}
\mathcal{C}(r,t) \equiv 2 \frac{\delta M(r,t)}{R(r,t)}, 
\eeq
where $\delta M(r,t)$ is the mass excess of a local overdense region and $R(r,t)=a(t)r$ is the areal radius of this region. Then, the characteristic comoving scale, $r_\mathrm{m}$ of the overdense region, is defined as the the position of the maximum of the compaction function, usually computed on the superhorizon regime, where the compaction function becomes time-independent ~\cite{Musco:2018rwt}.
\beq\label{r_m definition}
\mathcal{C}^\prime(r_\mathrm{m}) = 0.
\eeq
In \Fig{fig:PBH_scale}, one clearly sees that if using the condition \ref{r_m definition}, they can clearly determine a finite PBH scale even for non-compensated energy density profiles, like the Gaussian one.

\subsection{The PBH Mass Function}\label{sec:PBH mass function}
We consider here the standard PBH formation scenario during which PBHs form out of primordial energy density fluctuations when the local/mean energy density excess of an overdense region is larger than a critical threshold, $\delta_\mathrm{c}$. In this case, when $\delta>\delta_\mathrm{c}$ the overdense region stops expanding and collapses against the pressure of the background medium, forming in this way a PBH. Consequently, in the context of Press-Schechter formalism ~\cite{Press:1973iz}, the PBH mass function is defined as the probability that the local/mean energy density excess of an overdense region of mass $M$  is larger than a critical threshold $\delta_\mathrm{c}$: \footnote{In the standard Press-Schechter approach, the density contrast of an overdense region is smoothed using a window function. In this way, it is introduced a smoothing scale $R$ and the smoothed density contrast becomes $\delta(\boldmathsymbol{x},R)\equiv \int \mathrm{d}^3 \boldmathsymbol{x}^\prime W(|\boldmathsymbol{x}^\prime -\boldmathsymbol{x}|,R) \delta(\boldmathsymbol{x}^\prime)$, where $ W(|\boldmathsymbol{x}^\prime -\boldmathsymbol{x}|,R)$ is the window function. %However, due to uncertainties introduced by the window function at the level of the PBH abundances we define here the PBH mass function without smoothing the energy density excess ~\cite{Ando:2018qdb,Young:2019osy}. In this case, the PBH mass is taken as being a fraction of the mass inside the horizon at the time of PBH formation. See the discussion in \Sec{sec:PBH mass}.
}
\beq\label{PBH mass function - Press-Schecter}
\beta(M) \equiv P\left[\delta > \delta_\mathrm{c}\right] = \int_{\delta_\mathrm{c}}^{\infty}P(M,\delta)\mathrm{d}\delta,
\eeq
where $P(M,\delta)$ is the probability density function (PDF) of the density fluctuations which can potentially collapse and form PBHs. The PBH mass function is a very important quantity since it is the one constrained by observational probes. See ~\cite{Carr:2020gox} for a review about the constraints on $\beta(M)$. 

Regarding the limitations of the Press-Schechter formalism, which render approximate this approach in some regimes one should mention the well known cloud-in-cloud problem ~\cite{Jedamzik:1994nr}  in which small overdense regions which are parts of larger overdensities and collapse to form PBHs are not taken into account leading in this way to an underestimation of the PBH abundance. In addition, the Press-Schechter approach assumes an underlying Gaussian density field which is not only the case. To address thus these problems, the excursion-set formalism was introduced initially by ~\cite{Peacock:1990zz} and further developed by ~\cite{Bower:1991kf,Bond:1990iw} to tackle mainly the cloud-in-cloud problem, in which one should treat the density fluctuation, $\delta$, as a random variable and solve stochastic model equations to obtain analytically ~\cite{Maggiore:2009rv,Maggiore:2009rw,Maggiore:2009rx} or numerically ~\cite{Bond:1990iw,Percival:2001nv} the mass function. Regarding now the limitations on the Gaussian nature of the underlying density fields, there have been proposed  some extensions of the Press-Schechter formalism in the context of non-Gaussian regimes ~\cite{Matarrese:2000iz,LoVerde:2007ri} as well as studies of non-Gaussian initial conditions in the context of the excursion set theory regarding the halo mass functions ~\cite{Achitouv:2011sq,Achitouv:2012ux}. 

At this point, one should stress out that the PBH mass function is also often calculated in the context of peak theory ~\cite{Bardeen:1985tr} which studies the statistics of the peaks of a Gaussian density field and which assumes that a PBH is formed when a local density peak exceeds a certain threshold value. The peak theory approach similarly to the Press-Schechter formalism suffers as well from the Gaussian assumption for the underlying density field. 

Here, one should point out that the Press-Press-Schechter forrmalism, the peak theory as well as the excursion set theory are not related to the companction function method introduced before to compute the PBH characteristic scale.

\subsection{The PBH Threshold}\label{sec:PBH Threshold}
As we saw before, in order to determine the PBH mass function one should have an expression for the PDF of the density fluctuations which can collapse and form PBHs as well as an expression of the critical threshold value, $\delta_\mathrm{c}$. The PDF of the density fluctuations is rather model dependent and one cannot say much more about it without specifying the specific model which can give rise to PBH formation. The critical threshold however, in most cases, depends on the characteristic scale and the shape of the collapsing overdensity region, the time at which the gravitational collapse is taking place ~\cite{Carr:1975qj} as well as on the details of the surroundings. In what follows, we will try to give a brief summary in a historical order of the analytic and numerical works done so far for the determination of the critical PBH formation threshold $\delta_\mathrm{c}$.
\subsubsection{Early Approaches} \leavevmode\\
The first historical attempt for the determination of the PBH formation threshold was done by Bernard Carr and Stephen Hawking between 1974 and 1975 ~\cite{1974MNRAS.168..399C,Carr:1975qj} where they used a simplified Jeans instability criterion in the context of Newtonian gravity to determine $\delta_\mathrm{c}$. Specifically, they required that an overdense region in the early Universe can collapse to  form a PBH if its characteristic scale is larger than the Jeans length at maximum expansion. This led B. Carr to his famous result that $\delta_\mathrm{c}\sim w$ at horizon crossing time, where $w$ is the equation-of-state parameter defined in \Sec{The constant w}. Afterwards, the PBH formation threshold was studied  for the first time numerically through hydrodynamic simulations by some pioneering works from Nadezhin, Novikov $\&$ Polnarev in 1978 ~\cite{1978SvA....22..129N}, Bicknell $\&$ Henriksen in 1979 ~\cite{1979ApJ...232..670B} and Novikov $\&$ Polnarev in 1980 ~\cite{1980SvA....24..147N}. 

Then, after  a break of almost 20 years, the PBH formation threshold was studied again by highly sophisticated simulations this time performed in 1999 by Niemeyer $\&$ Jedmazik ~\cite{Niemeyer:1997mt} and Shibata $\&$ Sasaki ~\cite{Shibata_1999} which expressed the PBH formation threshold in terms of the energy density and curvature perturbation and which gave the same range for $\delta_\mathrm{c}$ varying between $0.3$ and $0.5$ depending on the shape of the energy density/curvature profiles considered. %The later conclusion was reached by Green et al. in 2004 ~\cite{Green:2004wb} which made use of the relation of the energy density and curvature perturbations in the linear regime and was quantified with numerical simulations performed by Musco et al. in 2005 ~\cite{Musco_2005}.

\subsubsection{Contemporary Approaches} \leavevmode\\

In the last decades, a lot of progress has been made in the research for the determination of the PBH formation threshold  both at the analytic as well as at the numerical level. 

In particular, T.Harada, C-M. Yoo $\&$ K. Kohri in 2013 ~\cite{Harada_2013} considered a ``three zone" spherical symmetric model for the description of the energy density field in which an initially sharply peaked overdense region is modeled as a homogeneous core (closed universe) surrounded by an underdense shell which separates the overdense region from the expanding background universe. In the end, after comparing the time at which the pressure sound wave crosses the overdensity with the onset time of the gravitational collapse they updated the PBH formation threshold value obtained by Carr in 1975 and in the uniform Hubble gauge their expression for $\delta_\mathrm{c}$ as a function of the equation-of-state parameter $w$ reads as:
\beq\label{delta_c-HYK}
\delta_\mathrm{c}= \sin^2\left(\frac{\pi\sqrt{w}}{1+3w}\right).
\eeq
At this point, it is important to stress out that the above mentioned expression for $\delta_\mathrm{c}$ is valid for at least the cases where $w\ll 1$ where one expects negligible pressure gradients which can not break up the homogeneity of the overdense region. 

Some years later, knowing the dependence of $\delta_\mathrm{c}$ on the shape of the initial energy density perturbation which collapses to a PBH already since the early numerical works in 1999 from Niemeyer $\&$ Jedmazik ~\cite{Niemeyer:1997mt} and Shibata $\&$ Sasaki ~\cite{Shibata_1999}, the authors of~\cite{Escriva:2019phb} quantified this effect by introducing a shape parameter in terms of the compaction function defined in \Eq{C definition} through which one can describe the shape of the initial density perturbation around the peak of the collapsing overdensity.  With ``shape'' here,  one refers to the broadness or sharpness of the energy density perturbation around its peak. In particular, the shape parameter is related to the second derivative of the compaction function at the comoving characteristic scale of the perturbation and it is defined on superhorizon scales where the compaction function is time independent as
\beq\label{shape parameter definition}
\alpha \equiv -\frac{r^2_\mathrm{m}\mathcal{C}^{\prime\prime}(r_\mathrm{m})}{4\mathcal{C}(r_\mathrm{m})}
\eeq
Here, it is important to mention that the compaction function computed at $r_\mathrm{m}$ is equal, as it can be straightforwardly checked, to the average energy density excess over a volume of radius $r_\mathrm{m}$, 
\beq\label{delta_m definition}
\delta_\mathrm{m} \equiv \frac{1}{V}\int^{r_\mathrm{m}}_0 4\pi r^2 \delta(r) = \mathcal{C}(r_\mathrm{m}),
\eeq
where $V=4\pi r^3_\mathrm{m}/3$ and $\delta(r)$ is the superhorizon time independent energy density perturbation. Consequently, one can formulate the PBH formation criterion by requiring that a PBH forms when the compaction function at $r_\mathrm{m}$, $\mathcal{C}(r_\mathrm{m})$ or equivalently the average perturbation amplitude, $\delta_\mathrm{m}$ is greater than a critical threshold which depends on the shape of the initial energy density profile as well as on the characteristic scale, $r_\mathrm{m}$ of the collapsing overdense region. This threshold was studied numerically in ~\cite{Nakama:2013ica} and recently in ~\cite{Musco:2018rwt}.

Furthermore,  it is important to point out here that the authors of ~\cite{Escriva:2019phb}, by making use of an effective basis for the initial curvature profile which can reproduce any realistic curvature for the calculation of the PBH formation threshold, deduced in the case of PBH formation during a radiation era a universal analytic threshold for the average compaction function as a function of the shape parameter defined in \Eq{shape parameter definition}. Their analytic expression for the threshold reads as
\beq\label{delta_c - Germani - Escriva - Sheth}
\delta_\mathrm{c} =\frac{4}{15}e^{-\frac{1}{\alpha}}\frac{\alpha^{1-\frac{5}{2\alpha}}}{\Gamma\left(\frac{5}{2\alpha}\right)-\Gamma\left(\frac{5}{2\alpha},\frac{1}{\alpha}\right)},
\eeq
where $\Gamma(x)$ is the gamma function and $\Gamma(x,y)$ is the incomplete gamma function and $\alpha	$ is the shape parameter given by \Eq{shape parameter definition}. The work of ~\cite{Escriva:2019phb} was generalized for an arbitrary equation-of-state parameter $w$ and it was found that for $w>1/3$ one can find an analytic formula for $\delta_\mathrm{c}$ as a function of $\alpha$ and $w$. We do not give here the full expression since it is quite complicated. For $w<1/3$ the determination of an analytic PBH formation threshold remains an open issue given that in this regime the full shape of the compaction function is necessary. 

At this point it is very important to underlie the huge interest raised recently in the role of non-linearities ~\cite{Kawasaki:2019mbl,Young:2019yug,Germani:2019zez,Young:2020xmk} and non-Gaussianities ~\cite{Young:2013oia,Young:2015cyn, Franciolini:2018vbk, DeLuca:2019qsy,Yoo:2019pma} for the determination of the PBH formation threshold  as well as the dependence of the PBH abundance ~\cite{Young:2014ana} on the details of the initial power spectrum of curvature perturbations which gave rise to PBHs ~\cite{Germani:2018jgr,Yoo:2018kvb,Yoo:2020dkz}. In addition, we should mention that the majority of the research work conducted in the literature assumes spherical collapse of the initial perturbations which leads to the production of non rotating PBHs. However, in a more realistic case, one can in principle expect non spherical collapse of the initial overdense regions which in general induces velocity field generation and therefore rotation effects. This last aspect was studied both analytically ~\cite{Kuhnel:2016exn} and numerically ~\cite{Yoo:2020lmg} showing that in principle a non-spherical collapse can make harder the PBH formation leading to the increase of the PBH formation threshold. Finally, regarding rotation, which has not necessarily generated due to non-spherical gravitational collapse, there has been been done a lot of analytic ~\cite{He:2019cdb,DeLuca:2019buf} and numerical work  ~\cite{Baumgarte:2016xjw,Gundlach:2017tqq} pointing out that the PBH formation threshold in general increases with the angular momentum which in its turn prevents the gravitational collapse.

\subsection{The PBH formation threshold in a  time-dependent $w$ background}
Having reviewed early and contemporary approaches about the determination of the PBH formation threshold, we extract here for the first time, to the best of our knowledge, the PBH formation threshold, $\delta_\mathrm{c}$, in the case of a time-dependent equation-of-state parameter, $w$. To do that, we follow closely and generalize the analytic treatment of ~\cite{Harada_2013}, in which one can compute $\delta_\mathrm{c}$ in the uniform Hubble gauge defined in the next subsections, by considering the ``three-zone" model adopted in ~\cite{Harada_2013}. In particular, in this subsection, we initially introduce the ``three-zone" model, then we compute the energy density perturbation of the overdensity region in the uniform Hubble gauge and finally we present a scheme to compute the PBH formation threshold in the case of a time-dependent equation-of-state parameter. 

\subsubsection{The ``three-zone" model} \leavevmode\\

\begin{figure}[h!]
\begin{center}
\includegraphics[width=0.896\textwidth, clip=true]{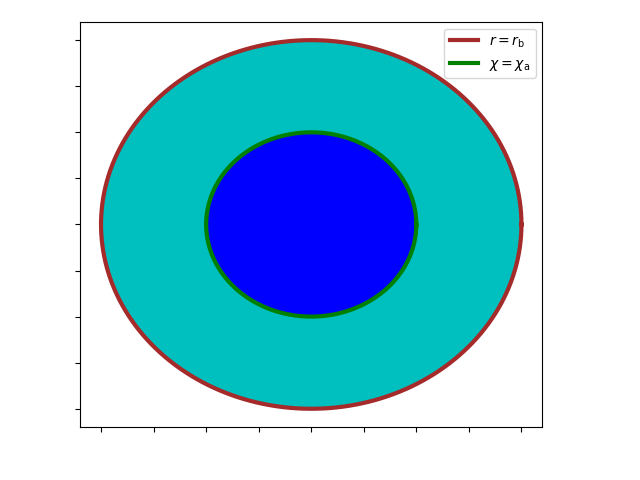}
\caption{The spherical ``three-zone" model: The overdensity region is shown in blue and it is surrounded by a spherical underdense layer depicted with cyan. The boundary between the overdensity region and the spherical underdense layer is shown with the green circumference at $\chi=\chi_\mathrm{a}$ whereas the boundary between the underdense layer and the FLRW flat background is depicted with the brown circumference at $r=r_\mathrm{b}$.}
\label{fig:three_zone_model}
\end{center}
\end{figure}

 In the spherically symmetric ``three-zone" model,  the overdense region is a homogeneous core (closed universe) surrounded by a thin underdense spherical shell which compensates the overdensity and separates the overdense region from the expanding background universe. See below \Fig{fig:three_zone_model}.

On the one hand, the background metric corresponds to a flat FLRW universe and reads as
\beq\label{background metric}
\mathrm{d}s^2 = - \mathrm{d}t^2 + a^2_\mathrm{b}(t)\left(\mathrm{d}r^2+r^2\mathrm{d}\Omega^2\right),
\eeq
where $\mathrm{d}\Omega^2$ is the line element of a unit two-sphere and $a_\mathrm{b}(t)$ is the scale factor of the background universe. The respective Friedmann equation reads as
\beq\label{Friedmann equation - background universe}
H^2_\mathrm{b} = \left(\frac{\dot{a}_\mathrm{b}}{a_\mathrm{b}}\right)^2 = \frac{\rho_\mathrm{b}}{3\Mp^2},
\eeq
where $\rho_\mathrm{b}$ and $H_\mathrm{b}$ is the energy density and the Hubble parameter of the background universe. 

On the other hand, the overdense region corresponds to a close ($K=1$) FLRW universe with a metric
\beq\label{overdense region metric}
\mathrm{d}s^2 = - \mathrm{d}t^2 + a^2(t)\left(\mathrm{d}\chi^2+\sin^2\chi \mathrm{d}\Omega^2\right)
\eeq
and a Friedmann equation 
\beq\label{Friedmann equation - overdense region}
\left(\frac{\dot{a}}{a}\right)^2 = \frac{\rho}{3\Mp^2} - \frac{1}{a^2},
\eeq
where $\rho$ is the energy density of the overdense region.

The underdense spherical shell is matched to the closed FLRW universe describing the overdensity at $\chi=\chi_\mathrm{a}$ while the flat FLRW background universe is matched to the the compensating underdense layer at $r=r_\mathrm{b}$. Therefore, the areal radius at the edge of the overdense region, $R_\mathrm{a}$ as well as that a the edge of the surrounding underdense spherical shell read as
\beq\label{matching conditions}
R_\mathrm{a} = a\sin\chi_\mathrm{a}, \quad R_\mathrm{b}=a_\mathrm{b}r_\mathrm{b}
\eeq

\subsubsection{Defining the energy density perturbation on the uniform Hubble gauge} \leavevmode\\

Then, having introduced the spherical ``three-zone'' model, we extract here the energy density perturbation at horizon crossing time on the uniform Hubble gauge, in which the Hubble parameters of the overdensity and that of the background are the same, i.e. $H=H_\mathrm{b}$. To do so, we firstly introduce the energy density parameter, $\Omega$ of the overdense region defined as 
\beq\label{Omega overdensity definition}
\Omega \equiv \frac{\rho}{3\Mp^2H^2} = 1 + \frac{1}{a^2H^2},
\eeq
where in the last equality we have used \Eq{Friedmann equation - overdense region}. Then, using the expression for the areal radius at $\chi = \chi_\mathrm{a}$, i.e.  $R_\mathrm{a} = a\sin\chi_\mathrm{a}$, as well as the definition of the horizon scale, i.e. $R_\mathrm{H}=H^{-1}$, one can find straightforwardly that 
\beq\label{Omega - sinxa relation}
(\Omega-1)\left(\frac{R_\mathrm{a}}{R_\mathrm{H}}\right)^2 = \sin^2\chi_\mathrm{a},
\eeq
an expression which relates $\Omega$ with the scale of the overdensity. In addition, one can relate $\Omega$ with the energy density perturbation of the overdense region with respect to the background defined as 
\beq\label{delta overdensity definition}
\delta \equiv \frac{\rho-\rho_\mathrm{b}}{\rho_\mathrm{b}}.
\eeq
Specifically, by solving \Eq{delta overdensity definition} for $\rho$ and substituting $\rho$ in  \Eq{Omega overdensity definition} one can obtain that 
\beq\label{Omega-delta relation}
\Omega = (1+\delta) \left(\frac{H_\mathrm{b}}{H}\right)^2, 
\eeq
where $\rho_\mathrm{b}$ has been expressed in terms of $H_\mathrm{b}$ through \Eq{Friedmann equation - background universe}.
Then, one can extract the energy density perturbation at horizon crossing time, $\delta_\mathrm{H}$, at the time when $R_\mathrm{a} = H^{-1}_\mathrm{b}$, by solving for $\delta$ \Eq{Omega-delta relation} and substituting $\Omega$ from \Eq{Omega - sinxa relation}. Finally, one gets that
\beq\label{delta at horizon crossing}
\delta_\mathrm{H} = \left(\frac{H}{H_\mathrm{b}}\right)^2 - \cos^2\chi_\mathrm{a}.
\eeq
In the uniform Hubble time slicing, in which $H=H_\mathrm{b}$, \Eq{delta at horizon crossing} becomes
\beq\label{delta at horizon crossing in the UH gauge}
\delta^\mathrm{UH}_\mathrm{H} = \sin^2\chi_\mathrm{a},
\eeq 
where $\delta^\mathrm{UH}_\mathrm{H}$ denotes $\delta$ in the uniform Hubble gauge at horizon crossing time. We should note here that the above expression for $\delta^\mathrm{UH}_\mathrm{H}$ does not depend on the equation of state of the universe at PBH formation time.

\subsubsection{The PBH formation threshold refined} \leavevmode\\

After expressing the energy density perturbation in the uniform Hubble gauge we compute now the PBH formation threshold in the case of a time dependent equation-of-state parameter. In particular, we compute the threshold by comparing the pressure and the gravitational force or equivalently the sound crossing time over the radius of the overdensity and the free fall time from the maximum expansion to complete collapse. To do so, we redefine the scale factor $a$ and the cosmic time $t$ such as that the Friedmann equation for the overdensity, \Eq{Friedmann equation - overdense region} takes the Tolman-Bondi form, valid for the dust case, which has an analytic parametric solution. Specifically, we redefine $a$ and $t$ as follows
\begin{eqnarray}\label{a and t transforms}
\tilde{a}=a e^{3I(a)} \\
\mathrm{d}\tilde{t}=\mathrm{d}t e^{3I(a)} \left[1+3w(a)\right]
\end{eqnarray}
where $I\left(a\right)\equiv\int_{a_\mathrm{ini}}^{a} \frac{w(x)}{x}\mathrm{d}x$ and the index $\mathrm{ini}$ denotes the initial time. Then, solving the continuity equation (\ref{Continuity Equation}) for a time-dependent equation-of-state parameter and using the coordinate transformation of \Eq{a and t transforms}, the Friedmann equation of the overdensity region (\ref{Friedmann equation - overdense region}) can be written in a dust form as
\beq\label{Friedmann equation - overdense region - dust form}
\left(\frac{\mathrm{d}\tilde{a}}{\mathrm{d}\tilde{t}}\right)^2=\frac{A}{\tilde{a}} - 1,
\eeq
where $A=\frac{\rho_\mathrm{ini}a^3_\mathrm{ini}}{3\Mp^2}$ and we have used the fact that $\frac{\mathrm{d}\tilde{a}}{\mathrm{d}\tilde{t}} = \frac{\mathrm{d}a}{\mathrm{d}t}$. The above equation can be integrated and gives a parametric solution of the form
\beq\label{dust parametric solution}
\tilde{a}=\tilde{a}_\mathrm{max}\frac{1-\cos\eta}{2}, \quad  \tilde{t}=\tilde{t}_\mathrm{max}\frac{\eta-\sin\eta}{\pi},
\eeq
with $\eta\in \left[0,2\pi\right]$. In the above parametric solution, $\eta$ is the conformal time defined  in terms of redefined scale factor and cosmic time, i.e. $\mathrm{d}\tilde{t}\equiv \tilde{a}\mathrm{d}\eta$,  and $\tilde{a}_\mathrm{max}$ and $\tilde{t}_\mathrm{max}$ are the redefined scale factor and cosmic time at the maximum expansion time  respectively and are given as follows:
\beq\label{a_max - t_max}
\tilde{a}_\mathrm{max}=\frac{\Omega_\mathrm{ini}}{\Omega_\mathrm{ini}-1}\tilde{a}_\mathrm{ini}, \quad \tilde{t}_\mathrm{max}=\frac{\pi}{2}\tilde{a}_\mathrm{max}.
\eeq

%Then, in terms of $(\eta,\chi)$ coordinates the the line element of the overdensity can be written as
%\beq\label{overdense region metric redefined}
%\mathrm{d}s^2 = \tilde{a}^2(\eta)\bigg\{ -\frac{\mathrm{d}\eta^2}{e^{6I(\eta)}\left[1+3w(\eta)\right]^2} +\mathrm{d}\chi^2 +\sin^2\chi \mathrm{d}\Omega^2\bigg\}.
%\eeq

%Regarding not the PBH formation, an apparent horizon is formed when a marginally trapped surface is formed, i.e. when $\frac{2M}{R} = 1$ where $M$ is the Misner-Sharp mass in spherical symmetric spacetimes [See in ~\cite{Misner:1964je, Hayward:1994bu} for more details]. In the case of a close Friedmann universe one has that ~\cite{Harada_2013}
%\beq\label{apparent horizon}
%\frac{2M}{R} = \left(1+\dot{a}^2\right)\sin^2\chi, 
%\eeq
%where the $^{.}$ means derivation with respect to cosmic time either the initial one or the redefined one since as pointed out before $\frac{d\tilde{a}}{d\tilde{t}} = \frac{da}{dt}$. 
%Consequently, in the case of dust universe as the one described with \Eq{Friedmann equation - overdense region - dust form} and \Eq{dust parametric solution} there is a marginally trapped surface when 
%\beq\label{trapped surface}
%\sin^2\chi = \frac{1-\cos\eta}{2}.
%\eeq
%Thus, an apparent horizon is formed when $\eta=2\chi$ and $\eta = 2\pi -2\chi$. For perturbations with $0<\chi_\mathrm{a}<\pi/2$, in which we concentrate here, the apparent horizon corresponds to 
%\beq\label{apparent horizon eta-chi level}
%\eta = 2\pi -2\chi.
%\eeq
Concerning now the sound wave propagation in a close Friedman geometry, the latter is dictated by the following equation
\beq
a\frac{\mathrm{d}\chi}{\mathrm{d}t}=c_\mathrm{s}(t),
\eeq
where $c^2_\mathrm{s}$ is the sound speed of an adiabatic fluid with a time-dependent equation-of-state parameter, $w$ computed in the \App{app:sound speed calculation}. Using now the conformal time $\eta$ introduced before with the use of the redefined variables $\tilde{a}$ and $\mathrm{d}\tilde{t}$ and \Eq{dust parametric solution} the above equation becomes
\beq\label{sound wave equation}
\frac{\mathrm{d}\chi}{\mathrm{d}\eta}=\frac{c_\mathrm{s}(\eta)}{1+3w(\eta)}
\eeq

One then can establish the PBH formation criterium by demanding that the time at which the sound wave crosses the radius of the overdensity, i.e. $\eta(\chi_\mathrm{a})$ is larger than the time at which the overdensity reaches the maximum expansion, i.e. $\eta_\mathrm{max}=\pi$. In this way, the pressure gradient will not have time to prevent the gravitational collapse whose onset time is considered here as the time of maximum expansion. To do so, in contrast with the treatment of ~\cite{Harada_2013}  one should solve numerically \Eq{sound wave equation} and demand that 
\beq\label{PBH formation criterion}
\eta_\mathrm{num}(\chi_\mathrm{a}) =\pi, 
\eeq
where $\eta_\mathrm{num}(\chi)$ is the numerical solution of  \Eq{sound wave equation} and $\chi_\mathrm{a}$ is the comoving scale at which the sound wave crosses the overdensity at the time of the maximum expansion. 
Therefore, from \Eq{delta at horizon crossing in the UH gauge} one can see that in the uniform Hubble slice gauge, the PBH formation threshold for a time dependent equation-of state parameter reads as 
\beq\label{delta_c w time dependent}
\delta_\mathrm{c}=\sin^2\chi_\mathrm{a},
\eeq 
with $\chi_\mathrm{a}$ being the solution of $\eta_\mathrm{num}(\chi_\mathrm{a}) =\pi$. 

At this point, one should stress out that the black hole apparent horizon should form after the onset of the gravitational collapse, i.e. the time of the maximum expansion. Thus, one should demand as well that $\eta_\mathrm{h}>\eta_\mathrm{max}=\pi$ where $\eta_\mathrm{h}$ is the time of formation of the apparent horizon which is obtained when $\frac{2M}{R} = 1$ where $M$ is the Misner-Sharp mass in spherical symmetric spacetimes [See in ~\cite{Misner:1964je, Hayward:1994bu} for more details]. A rigorous analysis shows that in the case of a closed FLRW universe, the condition $\frac{2M}{R} = 1$ gives that 
\beq
\eta_\mathrm{h} =2\chi_\mathrm{a} \quad \mathrm{or} \quad 2\pi - 2\chi_\mathrm{a}.
\eeq
Given the fact that the coordinates in \Eq{overdense region metric} cannot cover entirely the overdense region of perturbations for which $\pi/2<\chi_\mathrm{a}<\pi$ we focus here on perturbations for which  $0<\chi_\mathrm{a}<\pi/2$ and therefore $\eta_\mathrm{h}=2\pi - 2\chi_\mathrm{a}$. Demanding then that $\eta_\mathrm{h}>\eta_\mathrm{max}=\pi$ one has that $\chi_\mathrm{a}<\pi/2$. Here, we should stress out that in the case $w$ is constant then $c^2_\mathrm{s}=w$, \Eq{sound wave equation} can be solved analytically and the requirement that $\eta(\chi_\mathrm{a}) = \pi$ with $0<\chi_\mathrm{a}<\pi/2$ leads to the formula for $\delta_\mathrm{c}$ obtained in \cite{Harada_2013}.

Consequently, in order to compute the PBH formation threshold in the case of a time-dependent $w$ background one should solve numerically \Eq{sound wave equation} and then demand that $\eta_\mathrm{num}(\chi_\mathrm{a}) = \pi$ with $0<\chi_\mathrm{a}<\pi/2$. $\delta_\mathrm{c}$ then is given from \Eq{delta_c w time dependent}. This result generalizes the findings of \cite{Harada_2013} and can be applied in the case of time-dependent $w$ epochs such the preheating epoch during which PBHs can be abundantly produced or the QCD phase transition. 

However, it is important to stress out that the prescription described above for the computation of $\delta_\mathrm{c}$ in the case of a time-dependent equation-of-state parameter, can be  only viewed as an approximate one since it requires the homogeneity of the central overdense core that is not the case when one is met with strong pressure gradients. It is valid then for situations in which $w\ll 1$. As noticed also in \cite{Musco:2018rwt,Escriva:2019phb}, the ``three-zone'' model initially introduced by \cite{Harada_2013} gives $\delta_\mathrm{c}$ for a very sharply peaked homogeneous overdensity profile which eventually collapses into a black hole but it does not take into account the shape dependence of the energy density profile discussed in \Sec{sec:PBH Threshold} and the role of pressure gradients which can potentially disfavor the gravitational collapse and increase the value of $\delta_\mathrm{c}$. For this reason, the PBH formation threshold computed within the ``three-zone'' model can be viewed as a lower bound for $\delta_\mathrm{c}$.

Let us now express the PBH formation threshold in the comoving gauge which is the one which is used mostly in numerical simulations ~\cite{Musco:2004ak,Polnarev:2006aa,Musco:2008hv,Musco:2012au}. In the comoving gauge, the energy density perturbation at horizon crossing,  $\delta^\mathrm{com}_\mathrm{H}$ can be written as ~ \cite{Musco:2018rwt} 
\beq\label{delta_c - comoving gauge}
\delta^\mathrm{com}_\mathrm{H}=Q(t)\frac{1}{3r^2}\frac{\mathrm{d}}{\mathrm{d}r}\left[r^3K(r)\right]r^2_\mathrm{m},
\eeq
where $r_\mathrm{m}$ is the comoving scale of the collapsing overdensity region, $K(r)$ is the curvature profile in the quasi-homogeneous solution regime ~\cite{Musco:2018rwt} and $Q$ is a function of time which is given by
\beq\label{Q}
Q(t) = 1 - \frac{H(t)}{a(t)}\int_{a_\mathrm{ini}}^a\frac{\mathrm{d}a^\prime}{H(a^\prime)}.
\eeq
In the case of a constant equation of state, $Q= \frac{3(1+w)}{5+3w}$. For the case of the ``three-zone'' model considered here, $K(r)=1$ and $r_\mathrm{m}=\sin \chi_\mathrm{a}$ and as a consequence
\beq
\frac{1}{3r^2}\frac{\mathrm{d}}{\mathrm{d}r}\left[r^3K(r)\right]r^2_\mathrm{m} = \sin^2 \chi_\mathrm{a} = \delta^\mathrm{UH}_\mathrm{H}.
\eeq
Therefore, the energy density perturbation at horizon crossing time in the comoving and the uniform Hubble gauge are related as follows
\beq\label{delta_com vs delta_UH}
\delta^\mathrm{com}_\mathrm{H} = Q(t) \delta^\mathrm{UH}_\mathrm{H}.
\eeq

\begin{figure}[h!]
\begin{center}
\includegraphics[width=0.496\textwidth, clip=true]{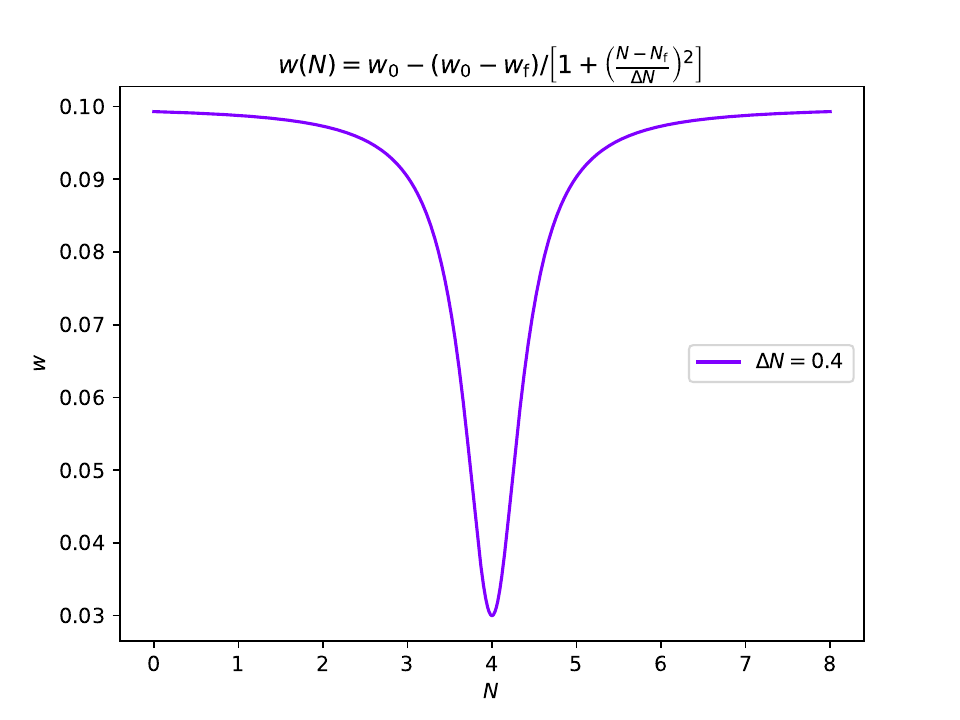}
\includegraphics[width=0.496\textwidth, clip=true] {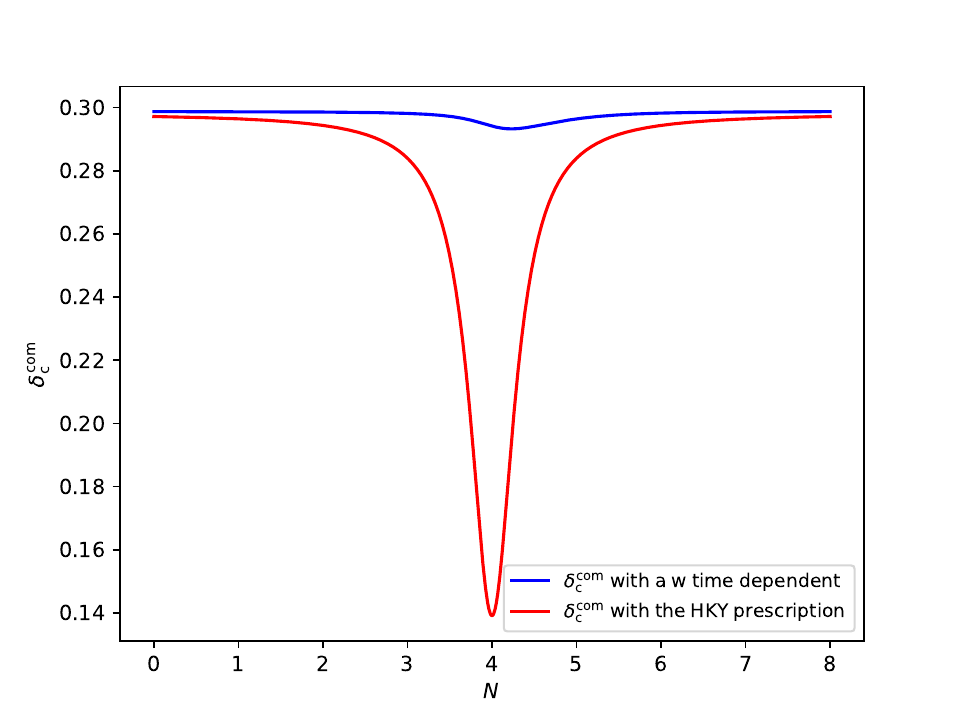}
\caption{ Left Panel: The equation-of-state parameter as a function of the e-fold number $N$ with $w_0=0.1$, $w_\mathrm{f}=0.03$, $\Delta N=0.4$ and $N_\mathrm{f}=4$. Right Panel: The PBH formation threshold, $\delta^\mathrm{com}_\mathrm{c}$, in the comoving gauge, in the case of a time-dependent equation-of-state parameter (blue line) superimposed with $\delta^\mathrm{com}_\mathrm{c}$ computed with the Harada, Kohri, Yoo (HKY) prescription, valid for a constant equation-of-state parameter.}
\label{fig:PBH_threshold_w_time_dependent}
\end{center}
\end{figure}

 In \Fig{fig:PBH_threshold_w_time_dependent}, we plot the evolution of the  PBH threshold in the comoving gauge, $\delta^\mathrm{com}_\mathrm{c}$, the one used mostly in numerical simulations, in the case of a time-dependent equation-of-state parameter varying from $0.03$ to $0.1$ within $2$ e-folds having taken into account the prescription described above. We compare also our prescription with the prescription of ~\cite{Harada_2013} valid for a constant equation-of-state parameter.

As one may see, the PBH formation threshold computed with a time-dependent $w$ prescription is almost constant with a small decrease at $N=4$ which is expected due to the decrease of $w$ at $N\sim 4$.  Interestingly, one can notice that despite the fact with the HKY prescription $\delta_\mathrm{c}$ decreases as $w$, if one takes into account the time-dependence of $w$ this decrease is smoothed presenting a small feature around the minimum of $w$. This effect  can have important consequences for PBH formation since a higher $\delta_\mathrm{c}$  means smaller PBH abundances with possible consequences on the targets of future experiments.

\begin{figure}[h!]
\begin{center}
\includegraphics[width=0.496\textwidth, clip=true]{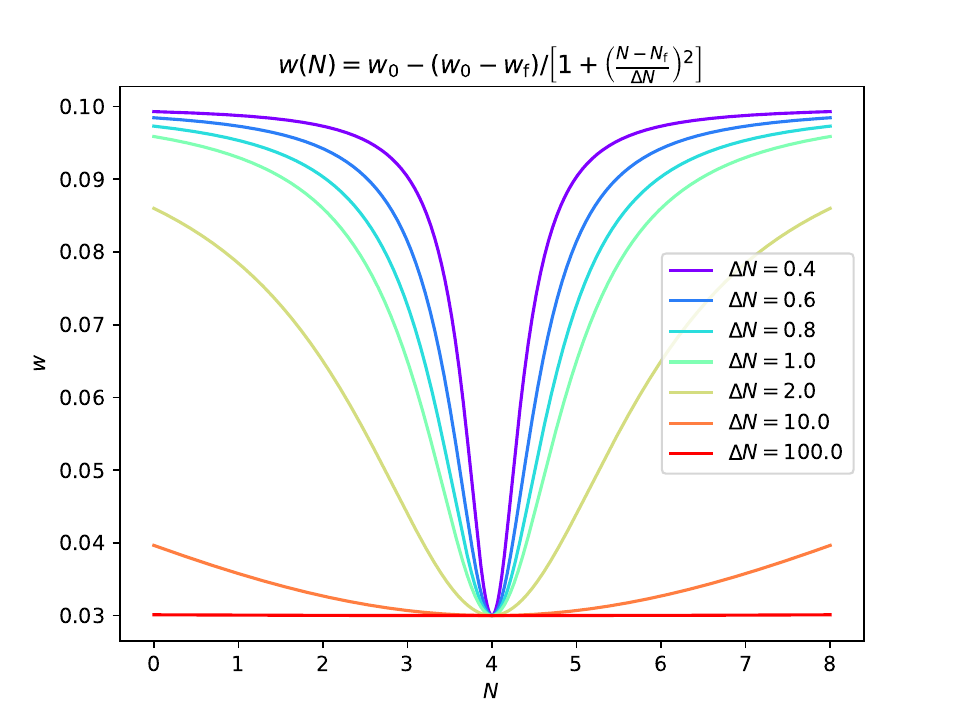}
\includegraphics[width=0.496\textwidth, clip=true] {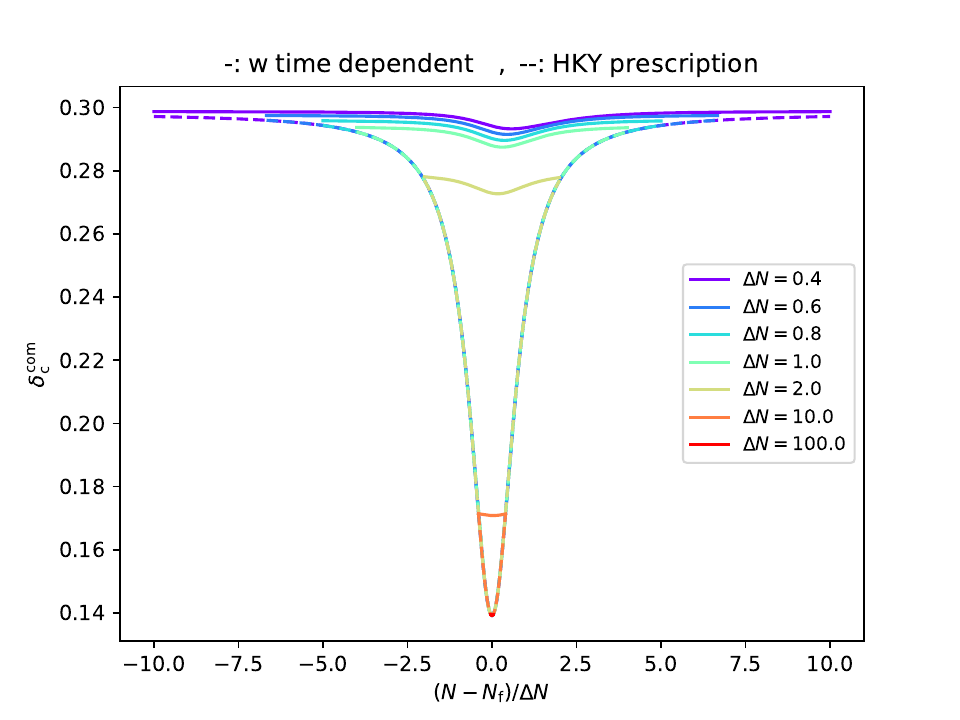}
\caption{ Left Panel: The equation-of-state parameter as a function of the e-fold number $N$ for different values of $\Delta N$ and with $w_0=0.1$, $w_\mathrm{f}=0.03$ and $N_\mathrm{f}=4$. Right Panel: The PBH formation threshold, $\delta^\mathrm{com}_\mathrm{c}$, in the comoving gauge, as a function of $(N-N_\mathrm{f})/\Delta N$ for different values of $\Delta N$, in the case of a time-dependent equation-of-state parameter (solid lines) superimposed with $\delta^\mathrm{com}_\mathrm{c}$ computed with the Harada, Kohri, Yoo (HKY) prescription, valid for a constant equation-of-state parameter (dashed lines).}
\label{fig:PBH_threshold_w_time_dependent_Gamma_study}
\end{center}
\end{figure}

Below, in \Fig{fig:PBH_threshold_w_time_dependent_Gamma_study}, we show as well the dependence on $w$ and $\delta_\mathrm{c}$ on $\Delta N$ which is the width of variation of $w(N)$. 
Interestingly, as it is expected, as one increases $\Delta N$, $w(N)$ approaches a constant value and the $w$ time-dependent prescription described above approaches the one of HKY prescription valid for constant w. 
\section{Observational Constraints on PBHs}
We review here the current observational constraints on the abundance of PBHs distinguishing between PBHs having been evaporated by now and PBHs that are still evaporating, following closely the recent review on the PBH constraints by Carr et al. ~\cite{Carr:2020gox}. Concerning the extraction of the constraints presented below, one assumes a monochromatic PBH mass function (PBHs are produced with the same mass) and that PBHs form during the radiation-dominated era.
\subsection{Evaporated PBHs}
We focus here on evaporated PBHs, which have evaporated by now or they evaporate at the present time. Broadly speaking, the evaporated PBHs are black holes with masses $m_\mathrm{PBH}<10^{15}-10^{16}\mathrm{g}$. The constraints on the abundance of evaporated PBHs are mainly related to BBN constraints, constraints from extra-galactic $\gamma$ rays, constraints from galactic cosmic rays and constraints from CMB distortions. The summarized constraints for the evaporated PBHs are given in \Fig{fig:Non Evaporating PBHs constraints}, taken from ~\cite{Carr:2020gox}. In \Fig{fig:Non Evaporating PBHs constraints}, the rescaled PBH mass function $\beta^\prime(M)$ at formation \footnote{The rescaled PBH mass function $\beta^\prime(M)$ is related to the PBH mass function $\beta(M)$ through the following relation $\beta^\prime(M) = \gamma^{1/2}\left(\frac{g_{*}}{106.75}\right)^{-1/4}\beta(M)$, where $g_*$ is the number of the relativistic degrees of freedom at formation time and $\gamma$ is a parameter of order one associated to the details of the gravitational collapse of an overdensity region to a PBH. For more details see ~\cite{Carr:2020gox}.} is plotted as a function of their mass $M$.  The relevant constraint in the case of absence of Hawking evaporation (black dotted line in \Fig{fig:Non Evaporating PBHs constraints}) are shown as well by requiring that energy density parameter of PBHs today is smaller than one, i.e. $\Omega_\mathrm{PBH}(M) \equiv \frac{\rho_\mathrm{PBH}}{\rho_\mathrm{c}}<1$. Below, we summarize very briefly the main physical mechanisms which give rise to the constraints of the evaporated PBHs depending on their mass.
\begin{figure}[h!]
\begin{center}
\includegraphics[width=0.896\textwidth, clip=true]{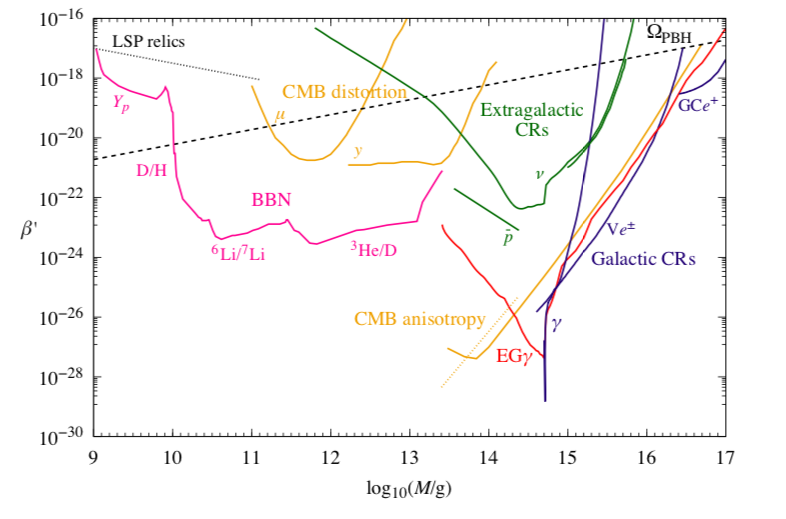}
\caption{The constraints on the abundance of the evaporated PBHs as a function of their mass. With the colored lines one see the constraints on the rescaled PBH mass fraction $\beta^\prime(M)$ at formation due to different physical phenomena explained in the main text and assuming the validity of Hawking radiation mechanism. Otherwise, the constraints are depicted with the black dotted line where there is no Hawking radiation and are obtained by the requirement that $\Omega_\mathrm{PBH}(M)<1$. The allowed regions for $\beta^\prime(M)$ are the ones below the colored lines. Figure credited to ~\cite{Carr:2020gox}.}
\label{fig:Evaporating PBHs constraints}
\end{center}
\end{figure}
\begin{itemize}
\item{\textbf{BBN Constraints} 

The BBN constraints on the PBH abundance are depicted with the magenta solid line in \Fig{fig:Evaporating PBHs constraints}. In particular, PBHs with masses $m_\mathrm{PBH}<10^9\mathrm{g}$ can not be constrained by studying the BBN processes since they evaporate well before the time of the weak freeze-out and thus they are not tractable. For PBHs with masses $m_\mathrm{PBH} \approx 10^9-10^{10}\mathrm{g}$, Hawking radiated mesons and antinucleons induce extra interconversion of protons to neutrons increasing in this way the neutron-to-proton ratio at the time of freeze-out of the weak interaction ~\cite{1978SvA....22..138V} triggering in this way an increase in the final $^4\mathrm{He}$ abundance ~\cite{1978PThPh..59.1012M}. Regarding now the PBHs with masses $m_\mathrm{PBH}\approx 10^{10}-10^{12}\mathrm{g}$,  long-lived high energy hadrons produced out of PBH evaporation, such as pions, kaons and nucleons remain long enough in the ambient medium and trigger dissociation processes of light elements produced during BBN ~\cite{Kohri:1999ex}, reducing in this way  $^4\mathrm{He}$ and increasing $\mathrm{D}$,$^3\mathrm{He}$,$^6\mathrm{Li}$  and $^7\mathrm{Li}$. Finally, for the PBHs with $m_\mathrm{PBH}\approx 10^{12}-10^{13}\mathrm{g}$, photons produced out of the particle cascade process further dissociate $^4\mathrm{He}$, increasing the abundance of light synthesized elements ~\cite{Keith:2020jww,Acharya:2020jbv}. However, it is important to stress out that the BBN constraints carry out some uncertainties regarding the baryon-to-photon ratio, the reaction and the decay rates of the elements produced during the BBN processes.   In \Fig{fig:Evaporating PBHs constraints}, the most conservative constraints are depicted.

}
\item{\textbf{CMB Constraints} 

The CMB constraints on the PBH abundance are depicted with the brown solid line in \Fig{fig:Evaporating PBHs constraints} and as it can be seen these constraints come from CMB spectral distortions as well as from CMB anisotropies. Regarding the CMB anisotropy constraint, which is the dominant constraint on the PBH abundance for PBH masses $m_\mathrm{PBH}\approx 10^{13}-10^{14}\mathrm{g}$, it is related to the damping of the CMB temperature anisotropy power spectrum and a boost in the polarization at small scales. In particular, when high energy electrons, positrons and photons are injected into the baryon-photon plasma around the recombination time ($z\sim 1000$) as products of PBH evaporation they can excite and ionize the neutral hydrogen and helium leading in this way to an increase of scattering processes between CMB photons and free electrons, thereby damping the CMB temperature anisotropy and increasing the polarization at small scales ~\cite{Zhang:2007zzh,Poulin:2016anj}. Concerning now the CMB spectral distortion constraint,  less stringent than the BBN constraint, which applies for the PBH mass range $m_\mathrm{PBH}\approx 10^{11}-10^{14}\mathrm{g}$, it is related to deviations of the CMB spectrum from the spectrum of a black body ~\cite{1977SvAL....3..110Z,Hu:1993gc,Tashiro:2008sf}. Specifically, when the universe is quite young, i.e. before the emission of the CMB, the CMB achieves a black-body spectrum through photon-electron interactions, i.e. Compton and double Compton scatterings, despite of a possible high energy injection. However, as soon as the universe cools down and one is met with the decouplings of these interactions at around $z=10^6$, distortions from a black-body spectrum can be induced via energy injection due to PBH evaporation.
}
\item{\textbf{Galactic/Extragalactic Cosmic Rays Constraints} 

The extragalactic cosmic ray constraints on the PBH abundance is depicted with the green solid line in \Fig{fig:Evaporating PBHs constraints}, it concerns PBHs with masses $m_\mathrm{PBH}\approx 10^{12}-10^{16}\mathrm{g}$ and as it can be seen, it is less stringent than all the other constraints ~\cite{1976ApJ...206....1P,1991ApJ...371..447M,Barrau:2003nj,Ballesteros:2019exr}. With the red solid line we see the constraint as well as from the extragalactic $\gamma$ ray background (EGB) which is the dominant one for PBH masses around $m_\mathrm{PBH}=10^{14}\mathrm{g}$ ~\cite{Carr:2009jm, Arbey:2019vqx}. EGB is related mainly to the primary and secondary emission of photons due to Hawking evaporation of PBHs residing outside of our galaxy. The photons mostly contributing to the EGB due to PBH evaporation are mainly soft $\gamma$ ray and $X$ ray emitted photons which lead to an isotropic background different from the extragalactic cosmic ray background of other astrophysical sources. Regarding now the galactic cosmic ray constraint, it is depicted with the blue solid line in \Fig{fig:Evaporating PBHs constraints} and it is the dominant constraint for PBH masses $m_\mathrm{PBH}\approx 10^{15}-10^{16}\mathrm{g}$. It is related mainly to the anisotropic $\gamma$ ray background emitted from evaporated PBHs clustered inside our galactic halo ~\cite{2009A&A...502...37L, Wright:1995bi,Carr:2016hva} as well as to the $e^{\pm}$  ~\cite{Laha:2019ssq,Dasgupta:2019cae} and $\nu/ \bar{\nu}$~\cite{Wang:2020uvi} emission due to Hawking radiation within the galactic bulge.

}

\end{itemize}

\subsection{Evaporating PBHs}
After having reviewed the constraints on the PBH abundance for PBHs which have evaporated by now we recap here the relevant constraints concerning PBHs which have not completed their evaporation yet. These PBHs are black holes with masses $m_\mathrm{PBH}>10^{15}-10^{16}\mathrm{g}$ and are considered to cluster in the galactic halo in the same way as other forms of dark matter. As in the case of the evaporated PBHs, we consider here as well that PBHs have a monochromatic mass. Historically, due to the assumption that high mass PBHs can constitute a viable candidate for cold dark matter (CDM), the constraints are given in terms of the fraction of PBHs to CDM, $f(M)$ defined as 
\beq\label{f(M) definition}
f(M)\equiv \frac{\Omega_\mathrm{PBH}(M)}{\Omega_\mathrm{CDM}},
\eeq
where $\Omega_\mathrm{CDM} = \Omega^{(0)}_\mathrm{DM}=0.265$ ~\cite{Aghanim:2018eyx} and $\Omega_\mathrm{PBH}(M)$ is the current energy density parameter of PBHs. In the case of PBHs forming during radiation era $f(M)$ is related to $\beta^\prime(M)$ with the following expression ~\cite{Carr:2020gox}
\beq\label{f(M)- beta(M) relation}
f(M) \simeq 10^{8}\beta^\prime(M) \left(\frac{M}{M_\odot}\right)^{-1/2}
\eeq
\begin{figure}[h!]
\begin{center}
\includegraphics[width=0.796\textwidth, clip=true]{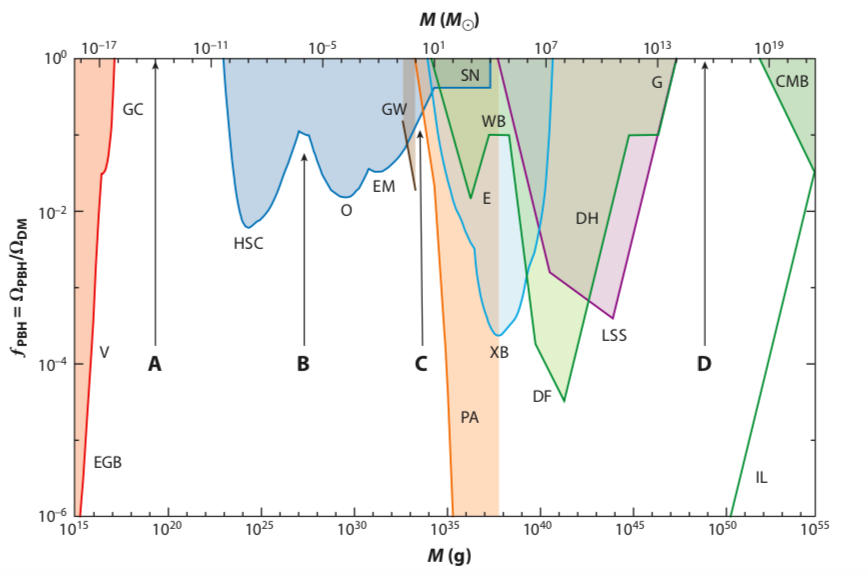}
\caption{The constraints on the fraction of PBHs to cold dark matter as a function of their mass for PBHs with masses $m_\mathrm{PBH}>10^{15}\mathrm{g}$. With the colored regions one see the constraints on the fraction of PBHs to CDM, $f(M)$, due to different physical phenomena explained in the main text. In particular, the colored regions are the ones which are forbidden by observations. There are four possible mass windows A,B,C,D in which PBHs can have an appreciable contribution to dark matter. In the left corner of the figure one can see the constraint from the extra-galactic $\gamma$ ray background described before. Figure credited to ~\cite{Carr:2020xqk}.}
\label{fig:Non Evaporating PBHs constraints}
\end{center}
\end{figure}

At this point point it is important to stress out that the majority of the constraints concerning the evaporating PBHs, presented in  \Fig{fig:Non Evaporating PBHs constraints} are obtained by studying the observational effects that the gravitational potential of PBHs can trigger and can be divided in five large categories depending on the way PBHs induce a gravitational effect: a) gravitational lensing, b) dynamical effect, c) accretion, d) large scale structure and e) gravitational waves. Below, we recap briefly the observational constraints on the fraction of PBHs to cold dark matter depending on the gravitational effect PBHs can trigger.

\begin{itemize}
\item{\textbf{Lensing Constraints}

 When between a source of light, i.e. a star, and an observer there is massive compact object, such as a PBH, one expects, as a prediction of general relativity, to observe light blending. As a consequence, the light source is observed either as an arc segment, either as a ring either as a multimple image depending on the level of alignment between the light source and the massive lensing object as well as on the mass of the lens. Regarding the PBH observational constraints due to gravitational lensing, they are related to microlensing effects in which no distortion in the shape of the light source can be seen but instead one can monitor how the amount of light received from a source change with time ~\cite{Sasaki:2018dmp}. Concerning now the lensing constraints on evaporating black holes, they come from microlensing events of stars observed by MACHO ~\cite{Alcock:2000ph} and EROS ~\cite{Tisserand:2006zx} collaborations (EM) in the Large and Small Magellanic Clouds (LMC and SMC) which probe the fraction of the galactic halo to Massive Compact Halo Objects (MACHOs) as well as from microlensing events of stars in the galactic bulge probed by OGLE (O) ~\cite{Niikura:2019kqi} . They come also from  the lack of lensing events in type Ia Supernovae (SN) ~\cite{Zumalacarregui:2017qqd} as well as from microlensing events of stars in M31 induced by PBHs lying in the halo regions of Milky way and M31 as observed by Subaru Hyper Suprime-Cam (HSC) ~\cite{Niikura:2017zjd}. 
}
\item{\textbf{Dynamical Constraints}

The dynamical constraints on evaporating PBHs are related to the gravitational effect a PBH can have on an astrophysical system through gravitational interactions. Up to now, many astrophysical systems have been studied. Indicatively, we mention the disruption of a white dwarf and the subsequent nuclear fusion triggered due to the passage of PBH in its vicinity ~\cite{Graham:2015apa} as well as the disruption of a neutron star in which a PBH trapped inside it can quickly accrete the surrounding matter and destroy the star ~\cite{Capela:2013yf,Kouvaris:2013kra}. Another interesting dynamical effect of a PBH is the disruption of weakly coupled binaries of stars ~\cite{Yoo:2003fr,Chaname:2003fn} or globular clusters ~\cite{1999ApJ...516..195C} of weakly coupled stars which reside in the galactic halo and which can easily disrupted from encounters of PBHs. Finally, one should mention dynamical constraints imposed to the fraction of PBHs into dark matter from disruption of ultra-faint dwarf galaxies, whose stars move faster due to gravitational interaction with PBHs ~\cite{Brandt:2016aco}, from dynamical friction on PBHs ~\cite{1985ApJ...299..633L} as well as  from the heating of stars in the galactic disk  ~\cite{1994ApJ...437..184X} when a PBH passes through its vicinity. In \Fig{fig:Non Evaporating PBHs constraints} we see in the green colored regions the dominant dynamical constraints which come from wide binaries (WB), star clusters in Eridanus II (E), heating of  stars in the galactic disk (DH), galaxy tidal distortions (G), halo dynamical friction and cosmic microwave dipole (CMB). 
}
\item{\textbf{Accretion Constraints}

The accretion constraints come from accretion of gas into PBHs and the effects of this process. Here we recap briefly the accretion effects to the CMB radiation taking place in the early universe as well as the the electromagnetic radiation emitted from accreted matter to PBHs. Regarding the accreting effects to CMB, one should take into account that the gas of baryonic nature surrounding a PBH is attracted by its gravity, starts to fall into the central region and being in this way ionised either by internal gas collisions or by the outgoing Hawking radiation. This ionisation process heats the gas filling the universe and modifies the CMB black-body spectrum, the time of photon decoupling as well as the ionisation history ~\cite{1981MNRAS.194..639C,Ricotti:2007au,Blum:2016cjs,Ali-Haimoud:2016mbv}. Detecting or not therefore these non-standard features on the CMB radiation one can impose limits on the PBH abundance. Concerning now the electromagnetic radiation from accreted matter to PBHs, one can impose constraints on the PBH abundance by comparing observational data and theoretical predictions of electromagnetic waves, mainly $X$ rays and radio waves, from PBH which attract and accrete their surrounding gas at present time ~\cite{Fujita:1997fh,Gaggero:2016dpq,Kording:2006sa,Mineo:2011id}. In \Fig{fig:Non Evaporating PBHs constraints} we see in the light blue colored region the dominant accretion constraint which comes from X-ray binaries (XB) as well as in the orange colored region the accretion constraint from the Planck satellite CMB measurement (PA).
}

\item{\textbf{Large Scale Structure Constraints}

The large scale structure constraints lie in the fact that the Poissonian fluctuations in the number density of randomnly distributed in space PBHs can enhance the dark matter perturbations in small scales ~\cite{Meszaros:1975ef}. This  enhancement of the dark matter power spectrum on small scales can have an impact on the $\mathrm{Ly\alpha}$ forest observations due to gravitational interactions of the dark matter perturbations with the baryon perturbations. Observing therefore the $\mathrm{Ly\alpha}$ absorption spectra from distant quasars and taking into account the enhacement of dark matter perturbations on small scales one can constrain the PBH abundance ~\cite{Afshordi:2003zb,Croft:2000hs}. In addition, one can constrain the PBH abundance by suggesting PBHs as a possible solution to the tension of the observed near infrared cosmic infrared background (CIB) anisotropies ~\cite{Kashlinsky:2016sdv,Kashlinsky:2004jt}. The large scale structure constraints are shown in \Fig{fig:Non Evaporating PBHs constraints} in the purple colored region. 

}
\item{\textbf{Gravitational-Wave Constraints}

The gravitational-wave constraints on the PBH abundance emanate from the different ways PBHs can lead to gravitational wave (GW) production. In particular, GWs can be induced from the primordial, large curvature perurbations that must have preceded and given rise to the existence of PBHs due to second order gravitational interactions ~\cite{Bugaev:2009zh, Saito:2008jc, Nakama:2015nea, Nakama:2016enz, Cai:2018dig, Yuan:2019udt, Kapadia:2020pir}. In addition, GWs are also expected to be emitted by the mergers PBHs, similarly to the GWs observed through the coalescence of black hole binaries by the LIGO/VIRGO collaboration ~\cite{Nakamura:1997sm, Ioka:1998nz, Eroshenko:2016hmn, Raidal:2017mfl, Zagorac:2019ekv,Hooper:2020evu}. Furthermore, one expects also a gravitational-wave background due Hawking radiated gravitons ~\cite{Anantua:2008am,Dong:2015yjs}. Finally, one should also account for the GWs generated by PBHs themselves due to the gravitational potential they underlie ~\cite{Papanikolaou:2020qtd,Domenech:2020ssp}. By comparing therefore gravitational-wave observations with the theoretical predictions of gravitational waves associated to PBHs one can constrain the PBH abundance. In \Fig{fig:Non Evaporating PBHs constraints}, in the brown colored region one shows the GW constraints on the PBH abundance from the early LIGO results in the range $0.5–30M_\odot$ ~\cite{Raidal:2017mfl}.

}
\end{itemize}
\section{Cosmological Consequences of PBHs}
Given the stringent PBH constraints presented above for a wide range of PBH masses, one may deduce that PBHs are rather unlikely to constitute the totality of dark matter and be detected in the future. However, even if they do not constitute all of dark matter, their cosmological consequences are quite important since by studying them we can learn a lot about the universe state in different times of the cosmic history and probe different physical phenomena depending on the PBH mass.

Specifically, the small mass PBHs ($m_\mathrm{PBH}\leq 10^{15}$g) which have evaporated by now can give access to the early universe physics such as the physics of inflation and the primordial cosmological perturbations ~\cite{Kalaja:2019uju} by probing the matter power spectrum on scales smaller than those of the CMB, the Big Bang Nucleosynthesis (BBN) physics ~\cite{2010arXiv1006.5342S,Keith:2020jww} through Hawking evaporation and the physics of reheating ~\cite{GarciaBellido:1996qt,Hidalgo:2011fj,Martin:2019nuw}, the physics of the cosmic microwave background (CMB) ~\cite{Ali-Haimoud:2016mbv} through distortions of the CMB spectrum, the primordial gravitational wave physics ~\cite{Clesse:2018ogk} by probing the stochastic gravitational background induced at second order in perturbation theory as well as the primordial phase transitions ~\cite{Jedamzik:1999am}. 

%In particular, by studying and setting constraints on small mass PBHs one can probe the matter power spectrum on small scales different from the CMB scales and thus constrain the properties of inflationary models which can potentially give rise to such PBHs. At the same time , given the black hole Hawking evaporation, these small PBHs have evaporated by now leaving a no direct imprint for their existence. However, one can probe indirectly their impact by looking for the products of their evaporation and their effect on the standard BBN picture we have now. 

Regarding the intermediate mass PBHs ($m_\mathrm{PBH}\sim 10^{-15}M_\odot$) which evaporate in our era, one can probe with them high energy astrophysical phenomena. Interestingly, as recently noticed by Carr and al. ~\cite{Carr:2020xqk} intermediate mass PBHs have been conjectured to explain the extragalactic ~\cite{1976ApJ...206....1P} and galactic ~\cite{2009A&A...502...37L} $\gamma$-ray backgrounds through Hawking evaporation, short-period gamma-ray bursts ~\cite{1997ApJ...486..169C}, the reionization of the pregalactic medium ~\cite{Belotsky:2014twa} and antimatter in cosmic rays ~\cite{Barrau:1999sk} pointing out to an increasing interest regarding the connection of PBHs to high energy astrophysics. 

Finally, the higher mass PBHs which still exist today, i.e. ($m_\mathrm{PBH}> 10^{15}$g), can have significant cosmological consequences regarding the gravitational and the dark sector of the universe. Specifically, according to recent arguments, PBHs may indeed constitute a part or all of the dark matter~\cite{Chapline:1975ojl} and they may explain the generation of large-scale structures through Poisson fluctuations~\cite{Meszaros:1975ef,Afshordi:2003zb}. Furthermore, they may provide seeds for supermassive black holes in galactic nuclei~\cite{Carr:1984id, Bean:2002kx} as well as account for the progenitors of the black-hole merging events recently detected by the LIGO/VIRGO collaboration~\cite{LIGOScientific:2018mvr} through their gravitational wave (GW) emission.

%\subsubsection{PBH as Dark Matter}
%\subsubsection{PBHs as seeds of Cosmic Structures}
%\subsubsection{PBHs and the Universe Thermal History}
%\subsubsection{Probing the matter power spectrum at small scales}
%\subsubsection{PBHs and GWs}

\newpage
\chapter{Inflation Theory and PBH Production from Preheating}\label{sec:Inflation Theory and PBH production from Preheating}
In this chapter, we recap the basics of the theory of inflation which solves, as mentioned in \Sec{subsec:HBB Problems}, a range of fundamental problems of the Hot Big Bang theory. In particular, we recap initially the standard single-field slow-roll inflation paradigm and then we briefly present the theory of cosmological inflationary perturbations, which have seeded the PBHs studied in the context of this thesis. Finally,  after reviewing the literature regarding the preheating era we discuss the PBH production in the context of single-field inflationary models theory due to metric preheating a research area to which I contributed with two scientific publications \cite{Martin:2019nuw, Martin:2020fgl}.
 
Regarding my personal contribution to the above mentioned scientific publications,  on the one hand,  in \cite{Martin:2019nuw} I made major contributions by performing a refined calculation for the PBH formation criterion during preheating and writing up the numerical code for the computation and the dynamical evolution of the PBH abundance.  I also produced the figures of the paper,  wrote up the conclusions and proof read the paper.  On the other hand,  in \cite{Martin:2020fgl},  my personal contribution was minor.  In particular,  I checked my PhD advisor's calculation for the computation of the background equation-of-state parameter during preheating and proof read the paper.  I also wrote up some conclusions regarding the effect of the inflaton's radiative decay on PBH formation.

\section{Inflation Theory}
As discussed in \Sec{subsec:HBB Problems}, in order to account for the shortcomings of the Hot Big Bang theory one should require the existence of an early era of accelerated expansion, where $\ddot{a}>0$, which translates to the condition  $p<-\rho/3$ from \Eq{Acceleration Equation}. 
\subsection{Single-Field Inflation}
 This era of negative pressure can be naturally realized in the context of single-field inflation in which a single scalar field $\phi$, the inflaton field,  is minimally coupled to gravity. The action describing such a field reads as 
\beq \label{single-field action} 
 \mathcal{S}_\phi =\int d^{4}x\sqrt{-g}\left[ \frac{1}{2}\partial_{\mu}\phi\partial^{\mu}\phi+V(\phi)\right],  
\eeq
where the first term in the integral is the kinetic energy of the field and the second the potential energy. The stress-energy tensor associated to the above action can be obtained from \Eq{stress-energy tensor} by replacing $\mathcal{S}_\mathrm{matter}$ with $\mathcal{S}_\phi$ and is given by 
\beq\label{single-field inflaton stress-energy tensor}
T^{(\phi)}_\mathrm{\mu\nu}=\partial_\mathrm{\mu}\phi\partial_\mathrm{\nu}\phi +g_\mathrm{\mu\nu}\left[-\frac{1}{2}g^{\mathrm{\rho\sigma}}\partial_\mathrm{\rho}\phi\partial_\mathrm{\sigma}\phi + V(\phi) \right].
\eeq
Given the fact we are working in the context of a flat FLRW background, at the background level the inflaton field $\phi$ should be homogeneous, thereby depending only on time. Thus, in this case, the energy and pressure densities of the inflaton field can be obtained from $T^{(\phi)}_\mathrm{\mu\nu}$ and are given by
\begin{eqnarray}\label{rho + p - single-field inflation}
\rho_{\phi}=T_{00}=\frac{\dot{\phi}^{2}}{2}+V(\phi) \label{rho - single-field inflation}\\
p_{\phi}=T_{ii}=\frac{\dot{\phi}^{2}}{2}-V(\phi). \label{p - single-field inflation}
\end{eqnarray}

The condition then for a homogeneous scalar field to drive a period of accelerated expansion, i.e. $3p + \rho <0$, becomes
\beq\label{Acceleration Condition - Inflation}
V(\phi)>\dot{\phi}^2.
\eeq
The above condition ensures that in order for inflation to take place, the inflaton field should slowly roll down its potential so that its potential energy dominates over its kinetic one. 

Regarding now the background dynamics of the inflaton field, it can be obtained by plugging \Eq{rho - single-field inflation} and \Eq{p - single-field inflation} into the continuity equation \Eq{Continuity Equation}. One then gets the Klein-Gordon equation for $\phi$
\beq \label{Klein-Gordon Equation for phi}
\ddot{\phi}+3H\dot{\phi} +V_{\phi}(\phi)=0
\eeq
where $V_{\phi}(\phi)\equiv \mathrm{d}V(\phi)/\mathrm{d}\phi$. Concerning the Friedmann equation, it reads as
\beq\label{Friedman Equation - Inflation}
3\Mp^2H^2 = V(\phi) + \frac{\dot{\phi}^2}{2}.
\eeq

\subsubsection{The Slow-Roll Regime}\leavevmode\\

The slow-roll regime is defined when the condition \Eq{Acceleration Condition - Inflation} is saturated, i.e. when $V(\phi)\gg\dot{\phi}^{2}$. In such a case, one gets from \Eq{rho - single-field inflation} and \Eq{p - single-field inflation} that $p_{\phi}\simeq-\rho_{\phi}$. In this regime, one obtains from the continuity equation \ref{Continuity Equation}  that $\rho$ is almost constant in time and from the Friedman equation one obtains that $H$ is almost constant in time too. This leads the spacetime to behave as the de Sitter one in which
\beq\label{de Sitter spacetime}
a(t) = a_\mathrm{ini}e^{H(t-t_\mathrm{ini})}.
\eeq
The de Sitter universe is equivalent with a universe dominated by a cosmological constant. See the discussion in \Sec{The constant w}. The slow-roll regime is therefore the limit in which the universe is perturbatively close to the de Sitter one. 
This slow-roll limit is very interesting since there is observational evidence for an almost scale invariant power spectrum on the CMB scales, as the one predicted by the slow-roll single-field inflation ~\cite{Akrami:2018odb}. One can then quantify the deviation from the de Sitter universe, by introducing the so called slow-roll parameters, upon which one can perform a perturbative expansion of the curvature power spectrum ~\cite{Stewart:1993bc,Gong:2001he,Martin:2002vn,Habib:2002yi}. Although there are many possible sets of slow-roll parameters, the mostly used in the literature are the so called Hubble-flow parameters ~\cite{Schwarz:2001vv,Schwarz:2004tz}, $\epsilon_n$, defined iteratively through the following expression
\beq\label{slow-roll parameters}
\epsilon_{n+1} = \frac{\mathrm{d}\ln |\epsilon_{n}|}{\mathrm{d}N},
\eeq
where $N$ is a time variable, called the e-fold number and it is defined as the logarithm of the scale factor, $N\equiv\ln a$. In this parametrisation for the slow-roll parameters, $\epsilon_0$ is defined as $\epsilon_0=\frac{H_\mathrm{ini}}{H}$, where the index $\mathrm{ini}$ denotes an initial time. In the case of a de Sitter universe, $\epsilon_0$ is constant and equal to $1$. Thus, in the slow-roll regime, which describes a quasi de Sitter universe, $\epsilon_0$ should be almost constant in time and close to $1$ and its time derivatives calculated through \Eq{slow-roll parameters} should be small, a fact that makes them very useful to describe pertrurbatively the deviation from the de Sitter expansion. In the language of the slow-roll parameters one is met with the slow-roll inflation as long as $|\epsilon_n|\ll 1$, for all $n>0$.

\subsection{Cosmological Perturbations in the Inflationary Epoch}
 As mentioned in \Sec{sec:early universe}, inflation constitutes the ``standard theory'' for the description of the early moments of the cosmic history since the primordial cosmological perturbations generated during inflation can seed the large scale structures observed today as well as the anisotropies of the relic cosmic microwave background radiation. Therefore, we recap here the theory of cosmological inflationary perturbations, which represents a cornerstone of the modern cosmology.
 
\subsubsection{The Scalar-Vector-Tensor Decomposition}\leavevmode\\

In order to include cosmological perturbations on the top of a homogeneous and isotropic background universe, one should go beyond homogeneity and isotropy. Thus, the most general perturbed metric which can model small perturbations of a FLRW universe can be written as ~\cite{Mukhanov:1990me}
\beq\label{general metric decomposition}
\mathrm{d}s^2 = a^2(\eta)\left[ -(1+2A)\mathrm{d}\eta^2 + 2 B_i \mathrm{d}x^i\mathrm{d}\eta + \left(\gamma_{ij}+h_{ij}\right)\mathrm{d}x^i\mathrm{d}x^j \right],
\eeq
where $A$, $B_i$ and $h_{ij}$ are functions of space and time describing the deviation from a homogeneous and isotropic universe. $\gamma_{ij}$ is the spatial part of the background metric. In our case, since we consider a flat FLRW background universe, $\gamma_{ij}=\delta_{ij}$.

In order now to extract the dynamics of the cosmological perturbations it is very useful to decompose them into scalar, vector and tensor components, thus the name Scalar-Vector-Tensor (SVT) decomposition ~\cite{Bardeen:1980kt}. In particular, any vector field, $B_i$ can be decomposed into the divergence of a scalar field, $B$, and to a vector field , $\bar{B}_i$, with vanishing divergence \footnote{For the scalar, vector and tensor quantities introduced here, the indices are lowered and raised according to the background metric $\gamma_{ij}$.}, that is 
\beq
B_i=\partial_i B + \bar{B}_i, \quad \mathrm{with} \quad \partial^i\bar{B}_i=0.
\eeq
In the same way, any tensor field, $h_{ij}$, can be decomposed into
\beq
h_{ij} = -2\psi\gamma_{ij} + 2\partial_i\partial_jE + 2\partial_{(i}\bar{E}_{j)} + 2\bar{E}_{ij}, \quad \mathrm{with} \quad \partial_i\bar{E}^{ij}=0 \quad \mathrm{and} \quad \bar{E}^i_i=0.
\eeq
The vector perturbations are rapidly supressed during the inflationary stage and therefore they are usually disregarded ~\cite{Peter:2013avv}. Scalar and tensor perturbations are instead studied with a lot of attention. Focusing then for the moment on scalar perturbations at linear order and making use of the SVT decomposition described above, the metric in \Eq{general metric decomposition} reads as
  \beq\label{metric decomposition}
\mathrm{d}s^2 = a^2(\eta)\left\lbrace -(1+2A)\mathrm{d}\eta^2 + 2\partial_i B\mathrm{d}x^i\mathrm{d}\eta + \left[(1-2\psi)\delta_{ij} + 2 \partial_i\partial_j E \right]\mathrm{d}x^i\mathrm{d}x^j\right\rbrace,
\eeq
where we have replaced $\gamma_{ij}=\delta_{ij}$ since as mentioned in \Sec{sec:universe composition} the observed negligible spatial curvature favors a flat FLRW background universe.

\subsubsection{The Gauge Issue} \label{par:gauge issue}\leavevmode\\
 
The study of cosmological perturbations lies in comparing the differences of physical quantities between a background spacetime (here FLRW metric) which is homogeneous and isotropic and the physical spacetime which does not obey necessarily to the cosmological principle. Thus, in order to compute the cosmological perturbations of a physical quantity one should compare the value it assumes at the unperturbed background spacetime with  its value at the perturbed one at the same spacetime point. Since these values live in different spacetime geometries, it is important to find a correspondence that links the same point on the two different spacetimes. This correspondence is called gauge choice and fixing a gauge is equivalent to choosing a threading into lines, corresponding to fixed spatial coordinates, and a slicing into hypersurfaces, corresponding to fixed time. 

Let us consider a generic infinitesimal coordinate/gauge transformation which reads
\beq\label{generic gauge transformation}
x^\mathrm{\mu} \rightarrow \tilde{x}^\mathrm{\mu} = x^\mathrm{\mu} + \xi^\mathrm{\mu},
\eeq
where $\xi^\mathrm{\mu}=(\xi^0,\xi^i)$ is an arbitrary four-vector whose components $\xi^0$ and $\xi^i$ depend on space and time. As discussed in the previous section any vector field $\xi^i$ can be decomposed into the divergence of a scalar field, $\xi$, and to a vector field , $\xi^i_\mathrm{tr}$, with vanishing divergence ($\xi^i_{\mathrm{tr},i}=0$), i.e. $\xi^i = \gamma^{ij}\xi_{,j} + \xi^i_\mathrm{tr}$, where the comma stands for the covariant derivative with respect to the background space coordinates. Therefore, taking into account only the functions $\xi^0$ and $\xi$ which preserve the scalar nature of the metric perturbations one can write \Eq{generic gauge transformation} as
\beq\label{gauge transformation}
\tilde{x}^0 = x^0 +\xi^{0}(x^0,x^i), \quad \bar{x}^i=x^i + \gamma^{ij}\xi_{,j}(x^0,x^i).
\eeq
Under this gauge transformation the scalar perturbations $A$, $B$, $\psi$ and $E$ transform like ~\cite{Mukhanov:1990me}
\beq\label{scalar perturbation transforms}
\tilde{A} = A - \frac{a^\prime}{a}\xi^0 - \xi^{0\prime}, \quad \tilde{\psi} = \psi +\frac{a^\prime}{a}\xi^0, \quad \tilde{B} = B+\xi^0 - \xi^\prime, \quad \tilde{E}=E-\xi,
\eeq
where the prime denotes derivative with respect to the conformal time $\eta$ defined in \Sec{sec:Hubble parameter + z}.

The issue which is risen now with the gauge choice is that there is not a preferred gauge. This is equivalent to the fact that there is not a unique choice of the functions $A$, $B$, $\psi$ and $E$ ~\cite{Bardeen:1980kt}. Therefore, to address this issue one can make two choices:
\begin{itemize}
\item{Make all the calculations in terms of gauge invariant quantities}
\item{Make a gauge choice and perform the calculations in that gauge}
\end{itemize}
Both of these choices have advantages and disadvantages. Making a specific gauge choice may render the computations technically simpler but at the same time it can potentially introduce gauge artifacts. On the other hand, performing a gauge-invariant computation, maybe more technically involved, gives the advantage to work with only physical quantities.  One can construct gauge-invariant quantities by taking combinations of $A$, $B$, $\psi$ and $E$. The simplest gauge-invariant quantities constructed from linear combinations of $A$, $B$, $\psi$ and $E$ that describe the gravitational sector are the so called Bardeen potentials and are defined as follows ~\cite{Bardeen:1980kt}:
\beq\label{Bardeen potentials}
\Phi \equiv A +\frac{1}{a}\left[\left(B-E^\prime\right)a\right]^\prime, \quad \Psi \equiv \psi - \frac{a^\prime}{a}(B-E^\prime).
\eeq
In the absence of anisotropic stress, i.e. in the absence of non diagonal space-space components in the stress-energy tensor, one can prove that $\Phi =\Psi$. Regarding the matter sector, one can construct a gauge-invariant fluctuation for the scalar field $\phi$ as follows
\beq\label{delta phi_gi}
\delta\phi^{(\mathrm{gi})}(\eta,\boldmathsymbol{x}) \equiv \delta\phi + \phi^\prime(B-E^\prime).
\eeq
In the same manner, one can define gauge-invariant scalar fluctuations like $\delta\rho^{(\mathrm{gi})}$ and $\delta p^{(\mathrm{gi})}$. In particular, given a scalar quantity $f$ one can define a gauge-invariant fluctuation $\delta f^{(\mathrm{gi})} $ as
\beq\label{gi scalar fluctuation}
\delta f^{(\mathrm{gi})} \equiv \delta f + f^\prime(B-E^\prime).
\eeq
Here it is important to know that the construction of  $\delta f^{(\mathrm{gi})}$ present in \Eq{gi scalar fluctuation} is the one mostly used in the literature but in fact there are infinitely more possibilities one can construct a gauge-invariant scalar fluctuation.

\subsubsection{The Perturbed Einstein's equations in a gauge-invariant form}\label{par:perturbed Einstein equations}\leavevmode\\

The matter and metric (gravitational) fluctuations are related to each other through the Einstein's equations. Specifically, working with gauge-invariant quantities one can show that in the absence of anisotropic stress, i.e. $\Phi=\Psi$, the perturbed Einstein's equations $\delta G^\mathrm{(gi)}_\mathrm{\mu\nu}=8\pi \delta T^\mathrm{(gi)}_\mathrm{\mu\nu}$ \footnote{The gauge-invariant Einstein tensor and stress-energy tensor perturbations can be constructed as follows:
\begin{eqnarray}
\delta G^{\mathrm{(gi)}0}_0 \equiv \delta G^{0}_0 + G^{0\prime}_0(B-E^\prime), \quad \delta G^{\mathrm{(gi)}i}_j \equiv \delta G^{i}_j + G^{i\prime}_j(B-E^\prime), \quad \delta G^{\mathrm{(gi)}0}_i \equiv \delta G^{0}_i + \left(G^{0}_0-\frac{1}{3}G^k_k\right)\partial_i(B-E^\prime) \\
\delta T^{\mathrm{(gi)}0}_0 \equiv \delta T^{0}_0 + G^{0\prime}_0(B-E^\prime), \quad \delta T^{\mathrm{(gi)}i}_j \equiv \delta T^{i}_j + T^{i\prime}_j(B-E^\prime), \quad \delta T^{\mathrm{(gi)}0}_i \equiv \delta T^{0}_i + \left(T^{0}_0-\frac{1}{3}T^k_k\right)\partial_i(B-E^\prime).
\end{eqnarray} 
}
after a straightforward but tedious calculation take the following form ~\cite{Mukhanov:1990me}
\begin{eqnarray}\label{Perturbed Einstein Equations}
\nabla^{2}\Phi - 3\mathcal{H}\Phi^\prime - 3\mathcal{H}^{2}\Phi &=& \frac{a^2}{2\Mp^2} \delta T^{\mathrm{(gi)}0}_0
\label{Poisson}, \\
\partial_i(\Phi^\prime + \mathcal{H}\Phi) &=& \frac{a^2}{2\Mp^2} \delta T^{\mathrm{(gi)}0}_i, \label{velocity field}\\
\left[\Phi^{\prime\prime}+3\mathcal{H}\Phi^{\prime} + (2\mathcal{H}^\prime + \mathcal{H}^{2})\Phi\right]\delta^i_j &=& - \frac{a^2}{2\Mp^2} \delta T^{\mathrm{(gi)}i}_j\label{pressure gradients}\, .
\end{eqnarray}
Working therefore with the stress-energy tensor of the inflaton field, \Eq{single-field inflaton stress-energy tensor}, one can find that the perturbed stress-energy tensor reads as ~\cite{Mukhanov:1990me}
\beq\label{perturbed T_munu -single field inflation}
\begin{aligned}
& \delta T^{\mathrm{(gi)}0}_0 = a^{-2}\left(-\phi^{\prime 2}\Phi +\phi^\prime \delta\phi^\mathrm{(gi)\prime} +V_{\phi} (\phi)a^2  \delta\phi^\mathrm{(gi)}   \right) \\
& \delta T^{\mathrm{(gi)}0}_i = a^{-2}\phi^\prime \partial_i\delta\phi^\mathrm{(gi)}  \\
& \delta T^\mathrm{(gi)i}_j = a^{-2}\left(\phi^{\prime 2}\Phi -\phi^\prime \delta\phi^\mathrm{(gi)\prime} +V_{\phi} (\phi)a^2  \delta\phi^\mathrm{(gi)}  \right)\delta^i_j
\end{aligned}
\eeq
Consequently, plugging \Eq{perturbed T_munu -single field inflation} into \Eq{Perturbed Einstein Equations} one obtains that
\begin{eqnarray}\label{Perturbed Einstein Equations - Single Field Inflation}
\nabla^{2}\Phi - 3\mathcal{H}\Phi^\prime - 3\mathcal{H}^{2}\Phi &=& \frac{1}{2\Mp^2} \left(-\phi^{\prime 2}\Phi +\phi^\prime \delta\phi^\mathrm{(gi)\prime} +V_{,\phi} (\phi)a^2  \delta\phi^\mathrm{(gi)}   \right), \label{eq:00 component-Einstein equation} \\
(\Phi^\prime + \mathcal{H}\Phi) &=& \frac{1}{2\Mp^2}\phi^\prime\delta\phi^\mathrm{(gi)} , \label{eq:0i component-Einstein equation} \\
\Phi^{\prime\prime}+3\mathcal{H}\Phi^{\prime} + (2\mathcal{H}^\prime + \mathcal{H}^{2})\Phi &=& - \frac{1}{2\Mp^2}\left(\phi^{\prime 2}\Phi -\phi^\prime \delta\phi^\mathrm{(gi)\prime} +V_{,\phi} (\phi)a^2  \delta\phi^\mathrm{(gi)}  \right). \label{eq:ii component-Einstein equation}
\end{eqnarray}

At this point, one can extract an equation for the dynamics of $\Phi$, which describes the gravitational sector. In particular, by substracting \Eq{eq:00 component-Einstein equation} from \Eq{eq:ii component-Einstein equation}, using \Eq{eq:0i component-Einstein equation} to express $\delta\phi^\mathrm{(gi)}$ as a function of $\Phi$ and $\Phi^\prime$ as well as the Klein-Gordon equation \Eq{Klein-Gordon Equation for phi}, one obtains that
\beq\label{eq:Phi equation of motion}
\Phi^{\prime\prime} + 2\left(\mathcal{H}-\frac{\phi^{\prime\prime}}{\phi^\prime}\right)\Phi^\prime - \nabla^2\Phi + 2\left(\mathcal{H}^\prime - \mathcal{H}\frac{\phi^{\prime\prime}}{\phi^\prime}\right)\Phi = 0.
\eeq

\subsubsection{The Curvature Perturbation} \label{par:curvature perturbation}\leavevmode\\

As we saw previously, the matter perturbations, here the scalar field perturbations $\delta\phi^\mathrm{(gi)}$, source metric perturbations $\Phi$ through the Einstein equations, which can be translated to perturbations of the curvature of the spacetime. These perturbations of the curvature of spacetime are of great importance since these are the ones which can explain through the theory of inflation the large scale structure and the CMB anisotropies. In this paragraph, we introduce the curvature perturbation in two gauges, most studied in the literature and then we construct gauge-invariant curvature perturbations starting from these two gauges ~\cite{Riotto:2002yw}.

\begin{itemize}
\item{ \textbf{The comoving curvature perturbation}

The function $\psi$, appearing in the spatial part of the perturbed metric is related to the intrinsic curvature of hypersuperfaces of constant time. In the case of a flat FLRW background one can show that the spatial Ricci scalar, $^{(3)}R$, defined as  $^{(3)}R\equiv \gamma^{ij}R_{ij}$, is related to $\psi$ as follows
\beq 
^{(3)}R=\frac{4}{a^{2}}\nabla^{2}\psi.
\eeq
The \textit{comoving curvature perturbation}, $\mathcal{R}$, is defined as the metric perturbation $\psi$ in the comoving slicing, which is the slicing of free-falling comoving observers for which the expansion is isotropic. In this gauge, there is no energy flux measured by the comoving observers, i.e. $T^{0}_i=0$.  Therefore, $\mathcal{R}$ is defined as
\beq
\mathcal{R} \equiv \psi |_{T^0_i=0}.
\eeq
One then can express $\mathcal{R}$ in a general gauge through a gauge transformation on constant time surfaces $\eta \rightarrow \eta + \delta\eta$. Under this transformation, $\psi$ transforms according to \Eq{gauge transformation} with $\xi^0=\delta\eta$ as follows
\beq \label{psi transform - comoving gauge}
\psi\rightarrow \tilde{\psi}=\psi+\mathcal{H}\delta\eta.
\eeq
In the comoving gauge, $T^\mathrm{0}_i=0$ and given the fact that at the background level $^{(0)}T^{0}_i=0$ as can be checked from \Eq{single-field inflaton stress-energy tensor}, and that $\delta T^{\mathrm{(gi)}0}_i \propto \partial_i \delta\phi \phi^\prime$ as can be seen from \Eq{perturbed T_munu -single field inflation}, one gets that the fluctuation for the scalar field $\phi$ in the comoving gauge, $\delta\phi_\mathrm{com}$ vanishes, i.e. $\delta\phi_\mathrm{com}=0$. To proceed now further and extract the expression for $\mathcal{R}$ in a generic gauge we should identify how a scalar fluctuation, like $\delta\phi$, transforms under a gauge transformation.

Let then $f$  be a scalar quantity and consider the generic coordinate transformation given by \Eq{generic gauge transformation}. Since $f$ is a scalar quantity, its value at a given physical point is the same in all coordinate systems. Thus, $\tilde{f}(\tilde{x}^\mathrm{\mu}) = f(x^\mathrm{\mu})$. In addition, one has as well for the unperturbed background that $f_0(x^\mathrm{\mu}) = \tilde{f}_0(x^\mathrm{\mu})$ where the index $0$ denotes background quantities. Consequently, one can deduce how the scalar fluctuation $\delta f(x^\mathrm{\mu})\equiv f(x^\mathrm{\mu})-f_0(x^\mathrm{\mu})$ transforms under a generic coordinate transformation. In particular, one obtains that 
\beq
\begin{split}
\tilde{\delta f}(\tilde{x}^\mathrm{\mu}) & =  \tilde{f}(\tilde{x}^\mathrm{\mu})-\tilde{f}_0(\tilde{x}^\mathrm{\mu})
 \\ & =  f(x^\mathrm{\mu})-f_0(\tilde{x}^\mathrm{\mu}) 
 \\ & =f(\tilde{x}^\mathrm{\mu}) - \delta x^\mathrm{\mu} \frac{\partial f}{\partial	x^\mathrm{\mu}}(\tilde{x}^\mathrm{\mu})-f_0(\tilde{x}^\mathrm{\mu}) 
 \\ & \simeq  \delta f (\tilde{x}^\mathrm{\mu}) -  \delta x^\mathrm{\mu} \frac{\partial}{\partial x^\mathrm{\mu}} \left(f_0(\tilde{x}^\mathrm{\mu}) + \delta f (\tilde{x}^\mathrm{\mu}) \right),
\end{split}
\eeq
where in the last equality we expanded up to first order the function $f(\tilde{x}^\mathrm{\mu})$. Then, given the fact that for a homogeneous and isotropic scalar field the background function $f_0$ depends only on time and considering up to first order contributions, $\tilde{\delta f}(\tilde{x}^\mathrm{\mu})$ reads as
\beq
\tilde{\delta f}(\tilde{x}^\mathrm{\mu})  = \delta f(x^\mathrm{\mu}) - f^\prime_0 \delta \eta.
\eeq

Thus, following the derivation presented above, $\delta\phi$ will transform as $\delta\phi\rightarrow \delta\phi - \phi^\prime\delta\eta$ and therefore one gets for a transformation on constant time hypersurfaces that 
\beq\label{delta_eta - comoving gauge}
\delta\phi \rightarrow \delta\phi_\mathrm{com} = \delta\phi - \phi^\prime\delta\eta  = 0 \Leftrightarrow \delta\eta = \frac{\delta\phi}{\phi^\prime}.
\eeq
Consequently, given \Eq{psi transform - comoving gauge} and \Eq{delta_eta - comoving gauge}, the comoving curvature perturbation, $\mathcal{R}$ can be defined in a generic gauge as
\beq\label{R definition}
\mathcal{R}\equiv \psi + \mathcal{H}\frac{\delta\phi}{\phi^\prime}.
\eeq
At this point it is important to point out that the above quantity is gauge-invariant by construction. Thus, $\mathcal{R}$ can be viewed as the gravitational potential on hypersurfaces where $\delta\phi=0$
$$\mathcal{R}=\psi|_{\delta\phi=0}.$$ 
Let us also stress out here the usefulness of the comoving curvature perturbation given its constancy on large scales ~\cite{Martin:1997zd}. It can be used then to propagate the inflationary power spectrum from the end of inflation to the post-inflationary era.
}
\item{ \textbf{The uniform energy density curvature perturbation}

In the same way, we can define the \textit{uniform energy density curvature perturbation}, $\zeta$ by considering a slicing with $\delta\rho=0$. Thus, $\zeta$ is defined as 
\beq
\zeta\equiv \psi|_{\delta	\rho=0}.
\eeq
For a transformation on constant time hypersurfaces, given the fact that $\delta\rho$ being a scalar fluctuation it transforms as $\delta\rho\rightarrow \delta\rho-\rho^\prime\delta\eta$, one can find, following the same reasoning as in the case of the comoving slicing, that in the uniform energy density slicing $\delta\eta=\frac{\delta\rho}{\rho^\prime}$. Therefore, in a generic gauge $\zeta$ is defined as 
\beq
\zeta\equiv \psi+\mathcal{H}\frac{\delta\rho}{\rho^\prime}
\eeq
and it is by construction a gauge-invariant quantity.
It is worth to mentioning here that it can be proved that, on superhorizon scales, i.e. $k\ll aH$, the curvature perturbation on the uniform energy density gauge, $\zeta$, is equal to the comoving curvature perturbation, $\cal{R}$ ~\cite{Riotto:2002yw}.
%It is worth to mention that on superhorizon scales $\zeta$ and $\mathcal{R}$ are equal:
%During the inflationary period $\rho=\frac{1}{2}\dot{\phi}^{2}+V(\phi)$, $\dot{\rho}=-3H\dot{\phi}^{2}$ and $\delta\rho=\dot{\phi}\delta\dot{\phi}+V'\delta\phi\simeq V'\delta\phi\simeq -3H\dot{\phi}\delta\phi$. Therefore,
%$$\zeta=\psi+H\frac{\delta\rho}{\dot{\rho}}=\psi+H\frac{\delta\phi}{\dot{\phi}}=\mathcal{R}$$
}
\end{itemize}

\subsubsection{The equation of motion for the scalar perturbations}\label{par:MS equation} \leavevmode\\

In \Sec{par:gauge issue}, we have introduced the Bardeen potentials $\Phi$ and $\Psi$, which are equal in the case of a vanishing anisotropic stress and which describe the gravitational sector as well as the gauge-invariant inflaton perturbation $\delta\phi^\mathrm{(gi)}$ which describes the matter sector. These gauge-invariant quantities are related to each through the Einstein equations as we saw in \Sec{par:perturbed Einstein equations}. This implies that one can construct a gauge-invariant quantity which describes in a unique way the scalar sector, i.e. the gravitational and the matter one. For this reason, the Mukhanov-Sasaki variable is introduced, which is a combination of the Bardeen potential and the gauge-invariant inflaton perturbation $\delta\phi^\mathrm{(gi)}$ and it is defined as ~\cite{Mukhanov:1981xt,Mukhanov:1988jd}
\beq\label{eq:Mukhanov-Sasaki variable definition}
v (\eta, \boldmathsymbol{x}) = a \left(\delta\phi^\mathrm{(gi)}+\phi^\prime\frac{\Phi}{\mathcal{H}}\right).
\eeq
At this point, it is useful to mention that the Mukhanov-Sasaki variable is related to the comoving curvature perturbation $\cal {R}$ as follows 
\beq\label{eq:v-R relation}
v = \frac{a\phi^\prime}{\mathcal{H}}\cal{R},
\eeq
where the last equation is a gauge-invariant equation constructed by starting from the Newtonian gauge in which $E=B=0$ and $\Psi = \psi$ and having taken into account that $\Phi = \Psi$ in the absence of anisotropic stress.

To extract the equation of motion for $u (\eta, \boldmathsymbol{x})$, one should write the total action of the system at hand, which is the sum of the action of the gravitational sector, i.e.  the Einstein-Hilbert action \ref{Actions}, plus the action of the matter sector which is the inflaton scalar field action \ref{single-field action} and write the respective Laplace equation for the Mukhanov-Sasaki equation. To do so, one expands the action of the system up to second order in perturbations to obtain after a rather lengthy calculation that ~\cite{Mukhanov:1990me}
\beq\label{eq:second_order_action}
^{(2)}\delta \mathcal{S} = \frac{1}{2}\int\mathrm{d}^4x \left[ v^{\prime 2} - \delta^{ij}\partial_i v\partial_j v + \frac{\left(a\sqrt{\epsilon_1}\right)^{\prime\prime}}{a\sqrt{\epsilon_1}}u^2\right],
\eeq
where $\epsilon_1$ is the first slow-roll parameter defined through the recursive \Eq{slow-roll parameters}. Then, the next step,  is to write the action in terms of the the Fourier modes of $v(\eta,\boldmathsymbol{x})$, given the fact that in the context of a linear theory each mode evolves independently. Consequently, expanding  $v(\eta,\boldmathsymbol{x})$ in Fourier modes we have that
\beq\label{eq:v Fourier expansion}
v(\eta,\boldmathsymbol{x}) = \frac{1}{\left(2\pi\right)^{3/2}}\int_\mathbb{R}\mathrm{d}^3\boldmathsymbol{k}v_\boldmathsymbol{k}(\eta) e^{i\bmk\cdot\bmx},
\eeq
 with $v_{-\bmk}=v^{*}_\bmk$ since $v(\eta,\boldmathsymbol{x})$ is real. Then, inserting \Eq{eq:v Fourier expansion} into  \Eq{eq:second_order_action} one gets that ~\cite{Mukhanov:1990me}
 \beq\label{eq:second_order_action-Fourier_space}
^{(2)}\delta \mathcal{S} = \int\mathrm{d}\eta \int_{\mathbb{R}^{+}\times\mathbb{R}^2} \mathrm{d}^3\bmk \Bigg\{ v^{\prime}_\bmk v^{*\prime}_\bmk  + v_\bmk v^{*}_\bmk\left[ \frac{\left(a\sqrt{\epsilon_1}\right)^{\prime\prime}}{a\sqrt{\epsilon_1}} - k^2\right] \Bigg\},
\eeq
where we integrate over the half of the Fourier space given the redundancy $v_{-\bmk}=v^{*}_\bmk$. Therefore, the Lagrangian density in Fourier space reads as
\beq
\mathcal{L} \equiv  \int_{\mathbb{R}^{+}\times\mathbb{R}^2} \mathrm{d}^3\bmk \Bigg\{ v^{\prime}_\bmk v^{*\prime}_\bmk  + v_\bmk v^{*}_\bmk\left[ \frac{\left(a\sqrt{\epsilon_1}\right)^{\prime\prime}}{a\sqrt{\epsilon_1}} - k^2\right] \Bigg\},
\eeq
with the conjugate momentum $p_\bmk$ being defined as
\beq
p_\bmk \equiv \frac{\delta\mathcal{L}}{\delta v^{*\prime}_\bmk} = v^\prime_\bmk.
\eeq
Thus, the Laplace equation, which comes out of minimizing the action $^{(2)}\delta \mathcal{S}$, reads as $\partial\mathcal{L}/\partial v^{*}_\bmk = \partial_\eta\left(\partial\mathcal{L}/\partial p^{*}_\bmk \right)$ and leads to the equation of motion for the Mukhanov-Sasaki variable which is given by 
\beq\label{eq:MS_equation}
v^{\prime\prime}_\bmk +\left[k^2- \frac{\left(a\sqrt{\epsilon_1}\right)^{\prime\prime}}{a\sqrt{\epsilon_1}}\right]v_\bmk = 0.
\eeq
From the above equation, one clearly sees that each mode $\bmk$ behaves as a parametric oscillator with a time-dependent frequency $\omega(\eta,\bmk)$ expressed as
\beq
\omega^2(\eta,\bmk)=k^2- \frac{\left(a\sqrt{\epsilon_1}\right)^{\prime\prime}}{a\sqrt{\epsilon_1}}.
\eeq
As one can see, the frequency $\omega(\eta,\bmk)$ depends on the scale factor and its derivatives. Thus, different inflationary potentials, which lead to different dynamics of the scale factor through the Friedman equation, lead to different dynamics of $\omega(\eta,\bmk)$ and subsequently to different dynamics of $v_\bmk(\eta)$.
\newpage
\section{PBHs from the Preheating Instability}\label{sec:metric preheating}
Having introduced before the fundamentals of the inflationary theory and the theory of cosmological perturbations we study here the period of preheating after the end of inflation and the possibility of PBH production during this early era of the cosmic history. Initially, we make a brief introduction of preheating reviewing the relevant literature on the field and then we discuss the PBH production from metric preheating in the context of single-field inflationary models as discussed in the relevant research works completed within my PhD ~\cite{Martin:2019nuw,Martin:2020fgl}.

\subsection{Preheating}\label{sec:preheating general}
When inflation ends, i.e. when the kinetic and the potential energy of the inflaton field become comparable, the inflaton field approaches a local minimum of its potential, which can be approximated in most cases by a quadratic potential of the form $V=m^2\phi^2/2$, where $m$ is a mass scale representing the curvature of the inflationary potential at its minimum. \footnote{In fact, the quadratic form  $V=m^2\phi^2/2$ can be seen as the leading order term of a Taylor expansion of the inflaton potential around its minimum. For potentials with vanishing curvature $m$ at their minimum the leading term is of higher order.} Then, when the inflaton reaches the minimum of its potential, it starts oscillating like $\phi \propto a^{-3/2}\sin (mt)$ driving a decelerated expansion in which the universe's thermal state behaves on average as a matter-domination state with $\langle \rho \rangle \propto a^{-3}$ ~\cite{Turner:1983he}, where $\langle \rangle$ denotes an average over the inflaton's oscillations. This oscillatory era is quoted in the literature as preheating. 

These rapid oscillations can parametrically amplify the quantum fields present during preheating both at the background and at the perturbative level ~\cite{Kofman:1994rk,Kofman:1997yn}. However, the consideration described above does not take into account the coupling of the inflaton to other degrees of freedom, which is necessary to ensure the transition to the radiation era, i.e. the Hot Big Bang phase of the universe. For this reason, more realistic preheating models consider these couplings as well. Below, we recap briefly how one can potentially couple the inflaton field with other degrees of freedom and generate a parametrically resonant amplification regime for the fields present during preheating. 

\subsubsection{The background}\leavevmode\\

At the background level, one can introduce the coupling of the inflaton field with other degrees of freedom by adding an extra friction term ``$\Gamma \dot{\phi}$'' - where $\Gamma$ is a decay rate - in the Klein-Gordon equation \ref{Klein-Gordon Equation for phi}, which basically accounts for the decay of the inflaton to a perfect fluid, usually radiation. Initially, $H\gg \Gamma$ and the effect of the inflaton's decay is negligible. As the universe expands, $H$ decreases and at some point $H\sim\Gamma$, which is the time when the decay of the inflaton starts taking place. The decay rate depends on the specifics of the model considered, i.e. the coupling of the inflaton with the perfect fluid as well as the relevant mass scales and can be calculated within perturbation theory. This perturbative approach, initially studied in the context of reheating after inflation by ~\cite{Dolgov:1982th,Abbott:1982hn}, has however some limitations. It can describe energy transfers due to individual decays of the inflaton to other degrees of freedom and it is efficient only for the last stages of reheating when the energy transfer to the inflaton's decay products is small compared to the very efficient energy transfer taken place during the very rapid coherent oscillations of the inflaton around its minimum during the early stage of preheating ~\cite{Kofman:1994rk,Kofman:1997yn}. 

Therefore, one should consider non perturbative effects when considering the background behavior during the oscillatory phase of preheating. To illustrate this with an example, one can couple the inflaton field $\phi$ with another scalar field $\chi$ with an interaction Lagrangian $\mathcal{L}_\mathrm{int} = -g^2\phi^2\chi^2/2$ where $g$ is a dimensionless coupling constant. Then, the total Lagrangian of the system can be written as $\mathcal{L} = \mathcal{L} _\mathrm{\phi} +\mathcal{L}_\mathrm{\chi} + \mathcal{L} _\mathrm{int}$, with $ \mathcal{L} _\mathrm{\phi}=-m^2\phi^2/2 + \dot{\phi}^2/2$, $ \mathcal{L} _\mathrm{\chi}=-m^2_\mathrm{\chi}\chi^2/2 + \dot{\chi}^2/2$ and the equation of motion for the Fourier mode $\chi_{\bm k}$ can be written as
\beq\label{eq:Perurbative preheating - chi EM}
\ddot{\chi}_{\bm k}+ 3 H \dot{\chi}_{\bm k}+\left[
  \frac{k^2}{a^2(t)}+m_\chi^2+ g^2\phi_0^2(t)\sin^2(m
  t)\right]\chi_{\bm k}=0,
\eeq
with $m_\mathrm{\chi}$ the mass of $\chi$ and $\phi_0(t)$ the decreasing amplitude of the inflaton's oscillations written as $\phi\simeq \phi_0(t) \sin (mt)$.  The above equation can be written in a more compact form by introducing the variable $X_{\bm k} = \chi_{\bm k} a^{3/2}$ and using the time variable $z\equiv mt$,
\beq\label{eq:mathieu}
\frac{\mathrm{d}^2 X_{\bm k}}{\mathrm{d}z^2} + \left[A_{\bm k} - 2q\cos (2z)\right] X_{\bm k} = 0,
\eeq
where the quantities $A_{\bm k}$ and $q$ are defined as
\beq\label{eq:A:q:def}
A_{\bm k} =\frac{k^2}{a^2m^2}+\frac{m_\chi^2}{m^2}
-\frac32 \frac{H^2}{m^2}\left(\frac32-\epsilon_1\right)
+2q, \quad
  q=\frac{g^2\phi_0^2}{4 m^2}.
\eeq
\begin{figure}[h]
\begin{center}
\includegraphics[width=0.6\textwidth]{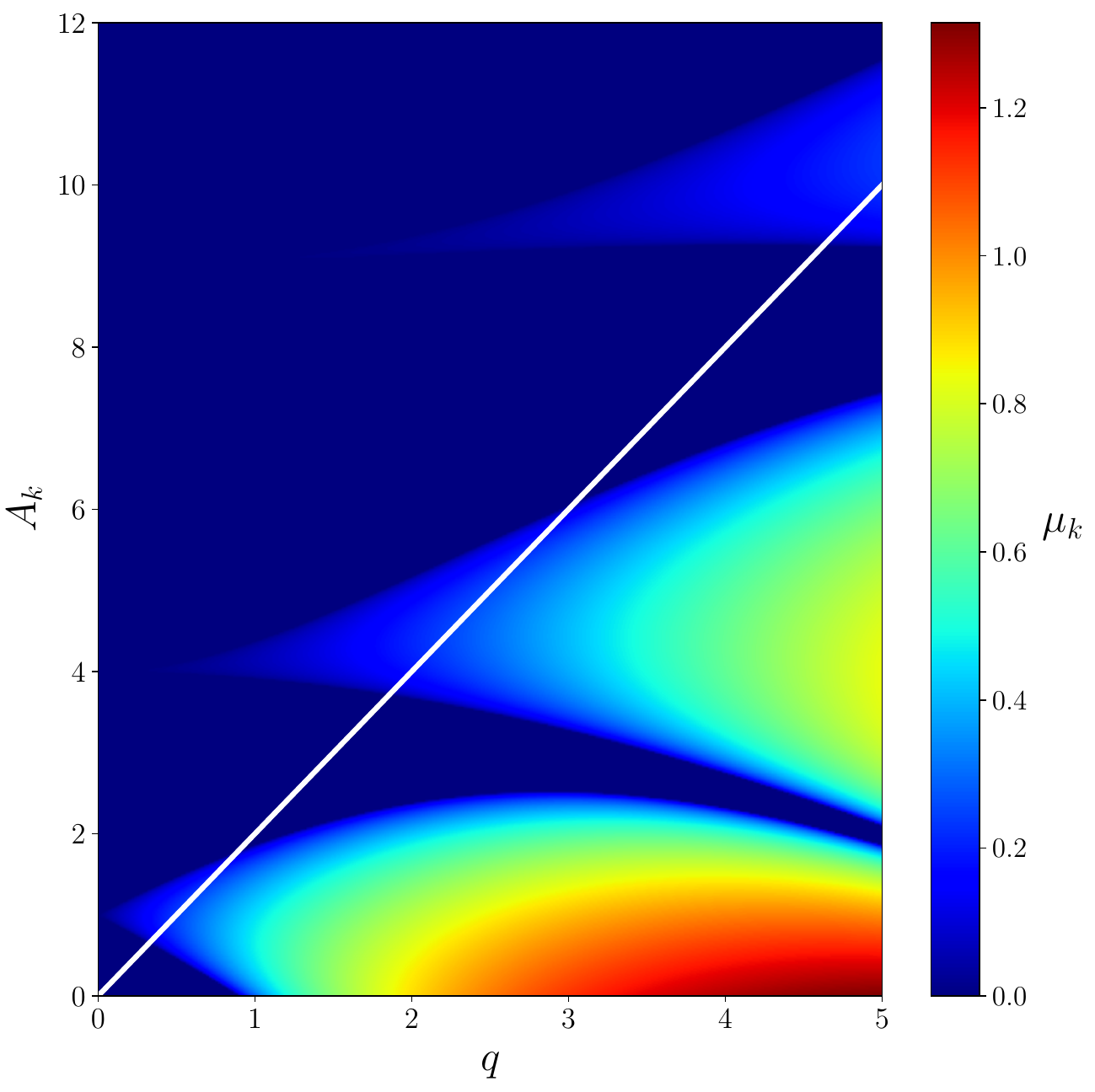}
\caption{Instability chart of the Mathieu equation in the case of a Minkowski background spacetime. The colour code
  represents the value of the Floquet exponent $\mu_{\bm k}$ of the unstable mode. Stable solutions corresponds to $\mu_{\bmk}=0$ and are represented by the dark blue regions while the unstable solutions correspond to $\mu_\bmk>0$ and are structured in different colored bands. The white curve is associated with the line $A_\bmk = 2q$.  Figure credited to ~\cite{Martin:2020fgl}.}
\label{fig:mapmathieu}
\end{center}
\end{figure} 

The above equation gives rise to a parametric resonance structure for the solutions $\chi_{\bm k}$ depending on the range of values of $A_\bmk$ and $q$ as initially noted in the context of reheating after inflation in ~\cite{Dolgov:1989us,Traschen:1990sw} and further studied in ~\cite{Kofman:1994rk,Kofman:1997yn}. To illustrate this resonance structure, we consider here the case of the Minkowski spacetime and for simplicity we assume $m_\mathrm{\chi}=0$. In this regime, $A_{\bm k} = k^2/m^2 + 2q$, $q$ is constant and \Eq{eq:mathieu} becomes a Mathieu equation with unstable, exponentially growing solutions $\chi_{\bm k} \propto e^{\mu_{\bm k} z}$, with $\mu_{\bm k}$ being the Floquet index of the unstable mode ~\cite{Abramovitz:1970aa:hypergeometric}. Since $q>0$, the region of interest is the one with $A_{\bm k} > 2q$. The unstable regions are the ones with $\mu_{\bm k}>0$. In \Fig{fig:mapmathieu}, we plot $\mu_{\bm k}$ as a function of $A_{\bm k}$ and $q$. As one can clearly see from this figure, there are bands at the level of $A_{\bm k} - q$ in which the parametric resonance structure is most pronounced, i.e. $\mu_{\bm k} \sim 1$. The most pronounced band is the one with the smallest value of $A_{\bm k}$. At this point, we should stress out that $q$ is related to the range of modes $k$ being excited. In particular, $\Delta k \sim q^\ell $, where  $\Delta k$ is the range of the excited modes $k$ for a band labelled with the integer $\ell>1$, with $\ell=1$ being the resonance band with the smallest value of $A_{\bm k}$. Thus, when $q\gg 1 $ we are in the so-called ``broad-resonance'' regime in which the range of the excited modes  $k$ is large. On the contrary, when $q\ll 1$ we are in the ``narrow-resonance'' regime in which a small range of modes $k$ are excited. 

If one now takes into account the spacetime expansion then $A_{\bm k}$ and $q$ become functions of time and \Eq{eq:mathieu} is not a Mathieu equation anymore. It is an equation of the Hill type in general ,which also gives rise to a parametric resonance structure for the solutions $\chi_\bmk$ as noted in ~\cite{Traschen:1990sw, Kofman:1994rk, Shtanov:1994ce,Kofman:1997yn}. In this case, in which the spacetime dynamics is restored, a mode $k$ will spend more time inside the wide bands in which $q\gg 1$. Therefore, in an expanding background spacetime the broad-resonance regime is the most important one regarding the amplification of the $\chi$ field. This broad-resonance regime, present when an expanding background is considered, is often quoted as ``stochastic-resonance'' regime in which each growing mode scans different instability bands and the relevant number of produced particles for a specific mode changes in a chaotic way. For more details see ~\cite{Kofman:1997yn}.

\subsubsection{The perturbations: The case of metric preheating}\leavevmode\\

Up to now, we have treated preaheating at the background level only. However, preheating plays an important role at the level of the perturbations as well if one includes the metric and the matter perturbations. In particular, as studied vastly in the literature, resonant amplification of the matter fluctuations are accompanied  with a resonant amplification of the scalar metric fluctuations, responsible for gravitational fluctuations in the curvature, since the two are coupled through the perturbed Einstein’s equations ~\cite{Kofman:1994rk,Kofman:1997yn}. This effect of the resonant amplification of the metric perturbations is often quoted as ``metric preheating" ~\cite{Finelli:1998bu,Bassett:1999mt,Jedamzik:1999um,Bassett:1999cg}. In order to study the effect of metric preheating, one should therefore examine if the Mukhanov-Sasaki equation, \Eq{eq:MS_equation}, governing the dynamics of the scalar perturbations, exhibits a parametric resonance structure like the one present at the background level, when the inflaton oscillates around the minimum of its potential.

Given the oscillations of $\phi(t)$, one expects that $H$ and the slow roll-parameter $\epsilon_1$ should oscillate as well. Therefore, the oscillations of $\epsilon_1$ should induce oscillations on the time-dependent frequency $\omega^2(\eta,\bmk)$ of the mode $v_\bmk$. In such as case, \Eq{eq:MS_equation} could be of the Mathieu or Hill type with the presence of a resonance instability structure regarding the different modes $v_\bmk$. In the context of single field inflation, due to the constancy of the comoving curvature perturbations on super-horizon scales, it was thought initially that there should not be a growth of the scalar perturbations ~\cite{Finelli:1998bu}. However, as realized in ~\cite{Jedamzik:2010dq,Easther:2010mr} one can find a resonance instability structure regarding the metric perturbations  in the context of single-field inflation models but in the narrow-resonance regime, which is immune to the perturbative inflaton decay to radiation ~\cite{Martin:2020fgl}. In the context of multi-field inflation however, the situation is different since there parametric amplification of entropy/isocurvature fluctuations can source the parametric amplification of the adiabatic/curvature fluctuations in the regime of broad resonance ~\cite{Bassett:1998wg, Green:2000he,Bassett:2000ha,Suyama:2004mz}.

\subsection{PBHs from the Preheating Instability}
During the preheating period after inflation, as we saw before, the inflaton oscillates coherently at its ground state, around the minimum of its potential and decays into other degrees of freedom. During this oscillatory phase, resonant amplification of matter and metric perturbations take place ~\cite{Kofman:1994rk,Kofman:1997yn} leading in this way to amplified curvature perturbations which in their turn collapse and form PBHs. Historically, PBHs emanated from the preheating instability were proposed in the context of multi-field inflation and in particular in the context of two-field chaotic inflation ~\cite{Bassett:1998wg, Green:2000he,Bassett:2000ha,Suyama:2004mz} and more recently in ~\cite{Torres-Lomas:2012tna,Torres-Lomas:2014bua} since in this case the parametric amplification of entropy/isocurvature fluctuations can source the parametric amplification of the adiabatic/curvature fluctuations in the regime of broad resonance. In the context of single-field inflation, it is predicted as well ~\cite{Jedamzik:2010dq} [see also ~\cite{Easther:2010mr}] that a pronounced resonant instability structure in the narrow regime this time can amplify metric perturbations inducing in this way the production of PBHs as recently studied in our research work ~\cite{Martin:2019nuw}. 

\subsubsection{Metric preheating in single-field inflation (research article) }\leavevmode\\
After reviewing the basics of preheating before, we discuss here how PBHs can be produced during the preheating instability in the context of single-field inflationary models due to the effect of metric preheating. To do so, one should study the equation of motion for the scalar perturbations, namely the Mukhanov-Sasaki equation \eqref{eq:MS_equation}. In particular, after rescaling the Fourier mode $v_\bmk$ according to $\tilde{v}_\bmk = a^{1/2}v_\bmk$ and using the cosmic time $t$ as the time variable, \Eq{eq:MS_equation} can be written as 
\beq\label{eq:MS equation expanded}
\ddot{\tilde{v}}_\bmk + \Biggl\{\frac{k^2}{a^2}+ V_{\phi\phi}(\phi) + 3\frac{\dot{\phi}^2}{\Mp^2} - \frac{\dot{\phi}^4}{2H^2\Mp^4} + \frac{3}{4\Mp^2}\left[\frac{\dot{\phi}^2}{2}-V(\phi)\right] + \frac{2}{\Mp^2}\frac{\dot{\phi}}{H}V_\phi(\phi) \Biggr\}\tilde{v}_\bmk = 0,
\eeq
where $V_\phi (\phi) \equiv \frac{\mathrm{d}V}{\mathrm{d}\phi}$ and $V_{\phi\phi} (\phi) \equiv \frac{\mathrm{d}^2V}{\mathrm{d}\phi^2}$.
As already said in \Sec{sec:preheating general}, during the rapid oscillations of the inflaton after the end of inflation, the inflaton oscillates more rapidly than the expansion rate of the universe entering the regime where $H\ll m$ and in which the energy density stored in $\phi$ scales in average as matter. In this case, the last term in \Eq{eq:MS equation expanded} is the dominant one scaling like $a^{-3/2}$ while the other terms  with time derivatives of $\phi$ scale like $a^{-3}$. Keeping therefore only the last term in \Eq{eq:MS equation expanded} one can recast \Eq{eq:MS equation expanded} in the following form:
\beq
\label{eq:Mathieu:v}
  \frac{\dd^2\tilde{v}_{\bm k}}{\dd \tilde{z}^2}
  +\left[A_{\bm k}-2q\cos(2\tilde{z})\right]\tilde{v}_{\bm k}=0,
\eeq
with $\tilde{z}$ being defined as $\tilde{z} \equiv mt +\pi/4$ and the coefficients $A_\bmk$ and $q$ being
\beq\label{eq:A_k - q:Mathieu:v}
A_\bmk = 1 + \frac{k^2}{m^2a^2}, \quad q= \frac{\sqrt{6}}{2}\frac{\phi_\mathrm{end}}{\Mp}\left(\frac{a_\mathrm{end}}{a}\right)^{3/2},
\eeq
where $\phi_\mathrm{end}$ and $a_\mathrm{end}$ are the values of the inflaton field and the scale factor at the end of inflation.

\begin{figure}[h!]
\begin{center}
\includegraphics[width=0.8\textwidth]{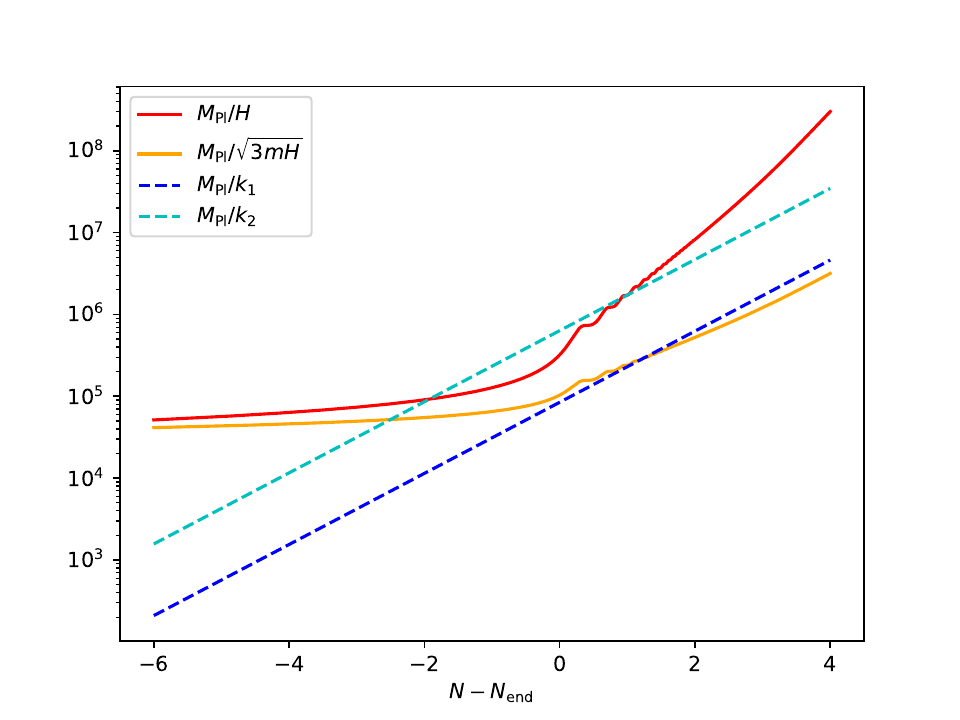}
\caption{Evolution of the relevant physical scales versus the e-folds
  number (counted from the end of inflation). The solid red line stands for the dimensionless Hubble radius $\Mp H^{-1}$, which corresponds to the lower bound in \Eq{eq:excited modes:Mathieu:v} while the solid orange
  line denotes the dimensionless characteristic scale $\Mp/\sqrt{3Hm}$ which corresponds to the
  upper bound in \Eq{eq:excited modes:Mathieu:v}. The dotted lines stand for the time evolution of two
dimensionless scales. Specifically, the ``blue scale'' enters the instability band from below while the ``cyan scale'' enters it from above.}
\label{fig:scaleinf}
\end{center}
\end{figure}

Rigorously, \Eq{eq:Mathieu:v} is not of Mathieu type since $A_\bmk$ and $q$ are time dependent. However, as shown in ~\cite{Jedamzik:2010dq} this time dependence is sufficiently slow that \Eq{eq:Mathieu:v} can be treated using the Floquet analysis for a Mathieu equation. Initially, $q$ is of the order of one since at the end of inflation $\phi_\mathrm{end} = O(\Mp)$ but as the universe expands, it becomes smaller and smaller than one. Thus, after some oscillations of the inflaton field one obtains that $q\ll 1$, being in the narrow-resonance regime and contrary to the case of non-perturbative preheating in which the broad-resonance regime was the more significant one. Considering then the most pronounced first instability band whose bounds are $1-q<A_\bmk<1+q$  ~\cite{Abramovitz:1970aa:hypergeometric}, one has that 
\beq
0<\frac{k}{a}<\sqrt{3Hm}.
\eeq
One then clearly sees the emergence of a new characteristic scale $1/\sqrt{3Hm}$. In order then for the instability to be physical one should consider modes which become subhorizon during the preheating phase i.e $aH<k$. Therefore, the resonance instability structure concerns physical modes which lie inside the range
\beq\label{eq:excited modes:Mathieu:v}
aH<k<a\sqrt{3Hm}.
\eeq
Therefore, the physical modes which are excited are smaller than the Hubble scale $H^{-1}$ and larger than the new characteristic scale $1/\sqrt{3Hm}$. We depict the above range of the excited modes in the following \Fig{fig:scaleinf} in which it is shown the evolution of two relevant physical scales.

Let us now see how the curvature and energy density perturbations evolve for the excited modes satisfying \Eq{eq:excited modes:Mathieu:v}. For the first instability band, the Floquet index of the unstable mode is given by $\mu_\bmk \simeq q/2$ ~\cite{Kofman:1997yn,Lozanov:2019jxc} and $v_\bmk = \tilde{v}_\bmk/a^{1/2} = a^{-1/2} \exp(\int \mu_{\bm{k}} \dd z)\propto a$ ~\cite{Finelli:1998bu,Jedamzik:2010dq}. Therefore, the comoving curvature perturbation $\cal {R}$ related to $v$ through \Eq{eq:v-R relation} reads as
\beq
\mathcal{R}_\bmk = \frac{H}{a\dot{\phi}}v_\bmk = \frac{v_\bmk}{\Mp a \sqrt{2\epsilon_1}} = \mathrm{constant},
\eeq
where in the last equation we have used the fact that on average $\epsilon_1=\frac{\dot{\phi}^2}{2\Mp^2H^2} \simeq \mathrm{constant}$ for $V(\phi)=m^2\phi^2/2$ and that during preheating $\phi \propto a^{-3/2}\sin (mt)$.
One then finds, that the comoving curvature perturbation is a conserved quantity for the modes lying inside the physical instability defined by \Eq{eq:excited modes:Mathieu:v}. Regarding now the evolution of the energy density contrast $\delta_\bmk$, given the fact that the excited modes in study are physical, i.e. they are inside the Hubble radius during preheating, there is no gauge ambiguity in the definition of $\delta_\bmk$. Therefore, we choose to work in the Newtonian gauge, where $E=B=0$ but our results are gauge-independent on subhorizon scales. In the Newtonian gauge, one has ~\cite{Jedamzik:2010dq} that 
\beq\label{eq:delta_k:Newtonian}
\delta_\bmk = -\frac{2}{5}\left(\frac{k^2}{\mathcal{H}^2}+3\right)\mathcal{R}_\bmk.
\eeq
Therefore, on subhorizon scales, i.e. $aH<k$, the first term in the parenthesis is the dominant one and taking into account that $\mathcal{R}_\bmk = \mathrm{constant}$ for $aH<k$ and that during preheating the universe experiences an effective matter domination era in which $a^2H^2\propto a^{-1}$ one gets that for the excited modes inside the physical instability defined by \Eq{eq:excited modes:Mathieu:v}
\beq\label{eq:delta_k:instability}
\delta_\bmk \propto a.
\eeq
One then finds that for the excited modes, the density contrast grows linearly with the scale factor, a fact which have important implications such as early structure formation ~\cite{Jedamzik:2010dq}, gravitational wave emission ~\cite{Jedamzik:2010hq} and PBH formation as studied in our research work ~\cite{Martin:2019nuw}. Regarding PBH formation, as we noted in ~\cite{Martin:2019nuw}, the metric preheating effect studied here can induce formation of PBHs on small scales which exit the Hubble radius a few e-folds before the end of inflation and therefore does not affect the large CMB scales. These PBHs, being formed very early in the cosmic history during the preheating era, can evaporate before BBN and reheat the universe through Hawking radiation ~\cite{Martin:2019nuw}. For more details see our relevant research article attached below.
\includepdf[pages=2]{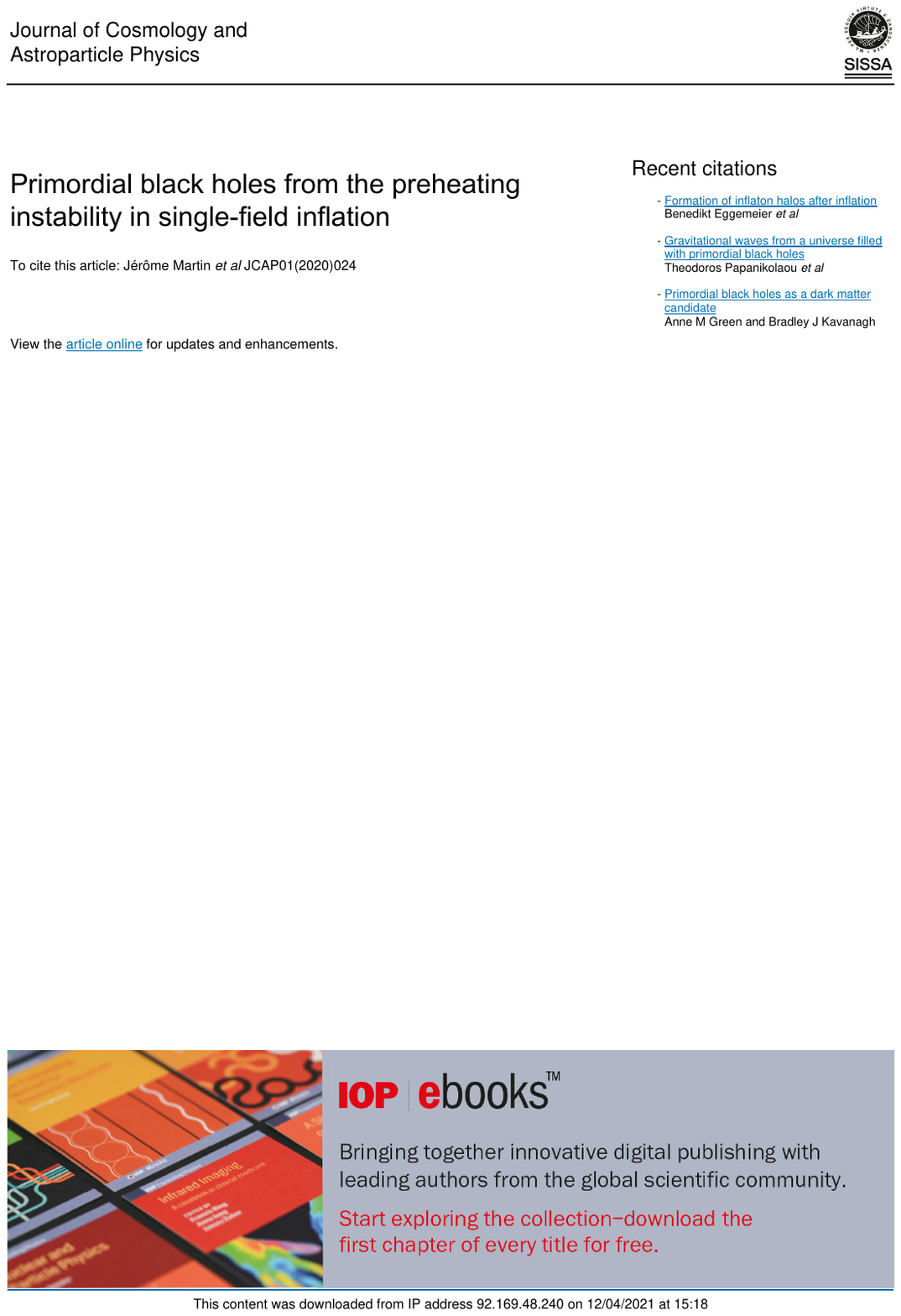}

\subsubsection{Metric preheating in single-field inflation and radiative decay (research article) }\leavevmode\\

We saw before the presence of a resonance instability structure in the context of single-field inflationary models by considering only the self interactions of the inflaton. To ensure therefore the transition to the radiation era one should couple the inflaton to other degrees of freedom, with the first one decaying and reheating the universe. In our work ~\cite{Martin:2019nuw}, the instability phase is abruptly ended either when the produced PBHs after their domination in the energy budget of the universe reheat the universe through Hawking evaporation or only when the inflaton decay products dominate the energy budget of the universe neglecting their effect throughout the preheating instability phase. In this last case, the inflaton oscillating phase was abruptly stopped by assuming instantaneous production of radiation at the end of the instability. However, the production of radiation should be continuous and one would expect that the effect of the inflaton's decay products may destroy the delicate balance which is responsible for the linear growth of the energy density fluctuation for the excited modes lying within the physical instability defined by \Eq{eq:excited modes:Mathieu:v}. 

This question was investigated in our work ~\cite{Martin:2020fgl} in which metric preheating was studied together with radiative decay of the inflaton field. It was also considered the decay of the inflaton to fluids with a generic equation-of-state parameter. As it was found, the perturbative decay effects of the inflaton field do not destroy the metric preheating instability structure since the latter stops only when, at the background level, the energy density of the inflaton's decay products dominate the energy content of the universe. In this way, our initial treatment in which the prehating instability is simply stopped when the universe becomes radiation dominated is found to be quite robust and confirms the unavoidable presence of a metric preheating instability at small scales in the narrow-resonance regime in the context of single-field inflationary models. See attached our relevant research article.
\includepdf[pages=2]{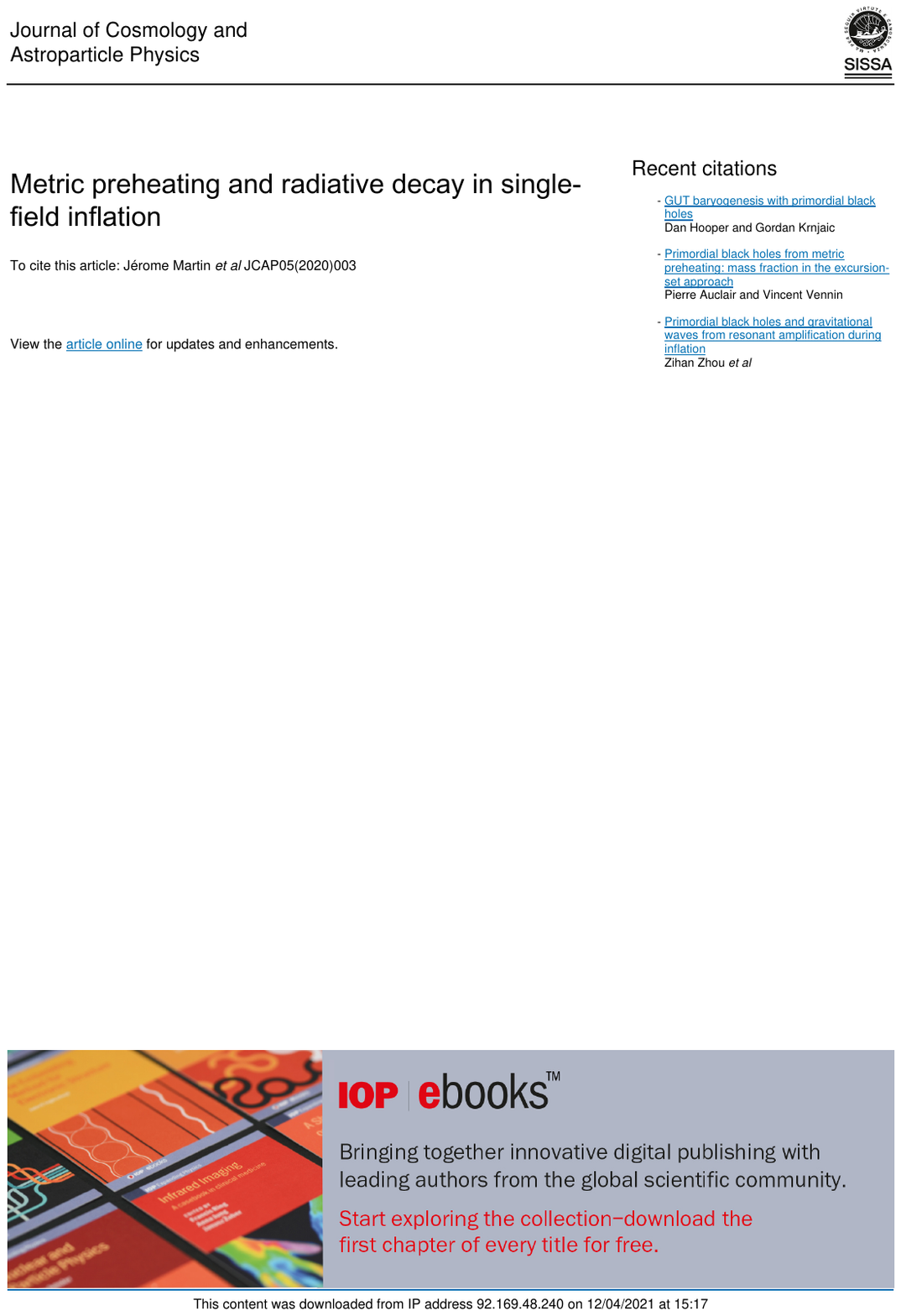}

\newpage
\chapter{PBHs and Induced Gravitational Waves}\label{sec:PBHs and Induced Gravitational Waves}
In this chapter,  we recap briefly the various ways in which PBHs can be connected with gravitational waves and then we present the fundamentals for the calculation of second order induced gravitational waves from first order scalar perturbations in the case of GW production in a radiation-dominated era. Then,  we present the main results of our research work ~\cite{Papanikolaou:2020qtd} in which induced gravitational waves are produced during an era in which ultralight PBHs ($m_\mathrm{PBH}<10^9\mathrm{g}$) drive the universe's expansion.

Regarding my personal contribution to the above mentioned scientific publication ~\cite{Papanikolaou:2020qtd},  I had major contributions by performing an analytic approximative calculation for the extraction of the GW spectral density as well as writing up the numerical code for the calculation of the GW signal and the inference of constraints of the PBH abundances.  In addition,  I produced the figures of the paper,  wrote up the conclusions and the appendices and proof read the paper. 

\section{Gravitational Waves and PBHs}
PBHs can be involved in several ways in the production of gravitational waves giving access to different physical phenomena and contributing as a consequence to a better understanding of the early universe at the time at which they form. 

In particular, as argued in the last decade ~\cite{Bugaev:2009zh, Saito_2009, Nakama_2015, Yuan:2019udt},  PBHs are tightly connected with induced gravitational waves which are sourced by large small-scale primordial curvature perturbations, and which can be potentially detected by future probes like LISA, DECIGO and Einstein Telescope. One then, by studying the stochastic background of induced gravitational waves, can search for PBH ``smoking guns" and constrain therefore the properties of these compact objects, such as their contribution to dark matter ~\cite{Clesse:2018ogk}. 

At the same time, with GW detectors, such as LIGO/VIRGO,  one can probe the coalescence history of compact objects which as recently argued can be potentially explained with the existence of PBHs ~\cite{Sasaki:2016jop}. By studying therefore the gravitational waves emitted out of PBH mergers ~\cite{Nakamura:1997sm, Ioka:1998nz, Eroshenko:2016hmn, Raidal:2017mfl, Zagorac:2019ekv,Hooper:2020evu} one can search for PBH signatures, reconstruct the PBH formation history and  shed light into PBH characteristics such as their mass ~\cite{DeLuca:2020qqa,DeLuca:2020sae} and spin~\cite{Mirbabayi:2019uph,DeLuca:2020bjf}. 

Furthermore, one should stress out the connection between PBHs and the stochastic gravitational wave background of Hawking radiated gravitons emitted out of the PBH evaporation, through which one can constrain the PBH abundances, their masses and their spins ~\cite{Anantua:2008am,Dong:2015yjs}.

In addition, as recently argued in our recent work ~\cite{Papanikolaou:2020qtd}, PBHs can be connected to induced GWs  which are associated to large-scale curvature perturbations underlain by PBHs themselves, not the ones generated from the primordial curvature power spectrum at small scales ~\cite{Papanikolaou:2020qtd}. These GWs can be abundantly produced during an early PBH-dominated era and can be potentially detected by LISA, Einstein Telescope and SKA ~\cite{Inomata:2020yqv}.

Given then all these possibilities of connection of PBHs to GWs, it is evident that, thanks to the recent developments of the gravitational wave astronomy, one can develop theories/models about PBHs, make predictions and test them directly in the laboratory, which in this case is the universe itself.

\section{Induced gravitational waves produced in an radiation-dominated era: The fundamentals}
Having introduced before the different ways PBHs are related to gravitational waves, we focus here on the induced gravitational waves generated at second order from first order scalar perturbations, which was one of the research axes studied within my PhD. Below, we review the calculation of the GW energy density parameter, $\Omega_\mathrm{GW}$, of the stochastic background of induced gravitational waves and we consider the case in which PBHs are produced during an RD era, which is the most studied in the literature. 
\subsection{Gravitational waves at second order}\label{sec:GWs at second order}
\label{sec:2nd:order:GWs}
Before presenting the calculation of the GW energy density parameter of the stochastic background of scalar induced gravitational waves we stress out here a major issue emerging from the study of induced GWs at second order. This issue is actually the fact that while the tensor modes are gauge invariant at first order this does not hold at second order \cite{Hwang_2017,Tomikawa:2019tvi,DeLuca:2019ufz,Yuan:2019fwv,Inomata:2019yww}. This means that, a priori, one needs to specify in which slicing the gravitational waves are detected, \ie which coordinate system is associated to the detection apparatus. 

However, as recently noted in ~\cite{DeLuca:2019ufz, Yuan:2019fwv, Inomata:2019yww,PhysRevLett.113.061301}, this gauge dependence is expected to disappear in the case of induced gravitational waves produced during a radiation era, as the one we review here, due to diffusion damping which exponentially suppresses the scalar perturbations in the late-time limit. In particular, small-scale perturbations, as the ones which seed primordial black holes, decay exponentially like $\propto \exp \left(-k^2/k^2_\mathrm{D}(t)\right)$ ~\cite{1980lssu.book.....P}, within the diffusion scale $k^{-1}_\mathrm{D}(t)$ because the free-streaming length of some species, in particular the one of photons and neutrinos increases as the universe cools down ~\cite{Jeong:2014gna,Hu:1994uz}. This is often quoted as Silk damping ~\cite{1968ApJ...151..459S}. Therefore, the difference between tensor perturbations in two different gauges, which should be written in terms of scalar perturbations should be negligible in the late-time limit. 

In what follows, we choose to perform our calculation by working with the Newtonian coordinate system for the GW detection frame ~\cite{Ananda:2006af,Baumann:2007zm,Kohri:2018awv,Espinosa:2018eve}. Therefore,  by adding to the linearly-perturbed Friedmann-Lema\^itre-Robertson-Walker metric in the Newtonian gauge the second-order tensor perturbation $h_{ij}$ (with a factor $1/2$ as is standard in the literature)\footnote{The contribution from the first-order tensor perturbations is not considered here since we concentrate on gravitational waves induced by scalar perturbations at second order.}, we obtain the total metric
\bea
\label{metric decomposition with tensor perturbations}
\mathrm{d}s^2 = a^2(\eta)\left\lbrace-(1+2\Phi)\mathrm{d}\eta^2  + \left[(1-2\Phi)\delta_{ij} + \frac{h_{ij}}{2}\right]\mathrm{d}x^i\mathrm{d}x^j\right\rbrace \, .
\eea
The tensor perturbation can be Fourier expanded according to
\beq
\label{h_ij Fourier decomposition}
h_{ij}(\eta,\boldmathsymbol{x}) = \int \frac{\mathrm{d}^3\boldmathsymbol{k}}{\left(2\pi\right)^{3/2}} \left[h^{(+)}_\boldmathsymbol{k}(\eta)e^{(+)}_{ij}(\boldmathsymbol{k}) + h^{(\times)}_\boldmathsymbol{k}(\eta)e^{(\times)}_{ij}(\boldmathsymbol{k}) \right]e^{i\boldmathsymbol{k}\cdot\boldmathsymbol{x}},
\eeq
with the polarisation tensors $e^{(+)}_{ij}$ and $e^{(-)}_{ij}$  
defined as
\begin{eqnarray}
e^{(+)}_{ij}(\boldmathsymbol{k}) = \frac{1}{\sqrt{2}}\left[e_i(\boldmathsymbol{k})e_j(\boldmathsymbol{k}) - \bar{e}_i(\boldmathsymbol{k})\bar{e}_j(\boldmathsymbol{k})\right], \\ 
e^{(\times)}_{ij}(\boldmathsymbol{k}) = \frac{1}{\sqrt{2}}\left[e_i(\boldmathsymbol{k})\bar{e}_j(\boldmathsymbol{k}) + \bar{e}_i(\boldmathsymbol{k})e_j(\boldmathsymbol{k})\right],
\end{eqnarray}
where $e_i(\boldmathsymbol{k})$ and $\bar{e}_i(\boldmathsymbol{k})$ are two three-dimensional vectors, such that $\lbrace e_i(\boldmathsymbol{k}), \bar{e}_i(\boldmathsymbol{k}), \boldmathsymbol{k}/k \rbrace$ forms an orthonormal basis. This implies that the polarisation tensors satisfy $e^{(+)}_{ij}e^{(+)}_{ij} = e^{(\times)}_{ij}e^{(\times)}_{ij} =1 , e^{(+)}_{ij}e^{(\times)}_{ij}=0$. Regarding now the equation of motion for the tensor modes, it reads as ~ \cite{Ananda:2006af,Baumann:2007zm,Kohri:2018awv}
\beq
\label{Tensor Eq. of Motion}
h_\boldmathsymbol{k}^{s,\prime\prime} + 2\mathcal{H}h_\boldmathsymbol{k}^{s,\prime} + k^{2}h^s_\boldmathsymbol{k} = 4 S^s_\boldmathsymbol{k}\, ,
\eeq
where 
$s = (+), (\times)$ and the source function $S^s_\boldmathsymbol{k}$ is given by
\beq
\label{eq:Source:def}
S^s_\boldmathsymbol{k}  = \int\frac{\mathrm{d}^3 \boldmathsymbol{q}}{(2\pi)^{3/2}}e^s_{ij}(\boldmathsymbol{k})q_iq_j\left[2\Phi_\boldmathsymbol{q}\Phi_\boldmathsymbol{k-q} + \frac{4}{3(1+w)}(\mathcal{H}^{-1}\Phi_\boldmathsymbol{q} ^{\prime}+\Phi_\boldmathsymbol{q})(\mathcal{H}^{-1}\Phi_\boldmathsymbol{k-q} ^{\prime}+\Phi_\boldmathsymbol{k-q}) \right]\, .
\eeq
As we can see, the source term, $S^s_\boldmathsymbol{k}$, is quadratic in $\Phi$ and therefore it is a second-order quantity. Consequently, the tensor modes, $h^s_\boldmathsymbol{k}$, are second-order quantities as it can be seen from \Eq{Tensor Eq. of Motion}. 

%Moving now into the details of the calculation, in \Eq{eq:Source:def}, the contraction  $e^s_{ij}(\boldmathsymbol{k})q_iq_j \equiv e^s(\boldmathsymbol{k},\boldmathsymbol{q})$ can be expressed in terms of the spherical coordinates $(q,\theta,\varphi)$ of the vector $\bm{q}$ in the basis $\lbrace e_i(\boldmathsymbol{k}), \bar{e}_i(\boldmathsymbol{k}), \boldmathsymbol{k}/k \rbrace$, 
%\beq
%e^s(\boldmathsymbol{k},\boldmathsymbol{q})=
%\begin{cases}
%\frac{1}{\sqrt{2}}q^2\sin^2\theta\cos 2\varphi \mathrm{\;for\;} s= (+)\\
%\frac{1}{\sqrt{2}}q^2\sin^2\theta\sin 2\varphi  \mathrm{\;for\;} s= (\times)
%\end{cases}
%\, .
%\eeq
In the absence of anisotropic stress, if the speed of sound is given by $\cs^2=w$, the equation of motion for the Bardeen potential reads as ~\cite{Mukhanov:1990me}
\bea
\label{Bardeen potential 2}
\Phi_\boldmathsymbol{k}^{\prime\prime} + \frac{6(1+w)}{1+3w}\frac{1}{\eta}\Phi_\boldmathsymbol{k}^{\prime} + wk^2\Phi_\boldmathsymbol{k} =0\, .
\eea
Introducing $x\equiv k \eta$ and $\lambda\equiv (5+3w)/(2+6w)$, this can be solved in terms of the Bessel functions $J_\lambda$ and $Y_\lambda$,
\beq\label{Bardeen Potential - Exact Solution}
\Phi_\boldmathsymbol{k}(\eta) = \frac{1}{x^{\lambda}}\left[C_1(k)J_\mathrm{\lambda}\left(\sqrt{w}x\right) + C_2(k)Y_\mathrm{\lambda}\left(\sqrt{w}x\right)\right],
\eeq
where $C_1(k)$ and $C_2(k)$ are two integration constants. On super sound-horizon scales, \ie when $\sqrt{w}\vert x\vert \ll 1$, this solution features a constant mode and a decaying mode (when $w=0$, this is valid at all scales). By considering the Bardeen potential after it has spent several \efolds~above the sound horizon, the decaying mode can be neglected, and one can write $\Phi_\boldmathsymbol{k}(\eta) = T_\Phi(x) \phi_\boldmathsymbol{k}$, where $\phi_\boldmathsymbol{k}$ is the value of the Bardeen potential at some reference initial time,  $x_0$  and $T_\Phi(x)$ is a transfer function, defined as the ratio of the dominant mode between the times $x$ and $x_0$.  This allows one to rewrite \Eq{eq:Source:def} as
\beq
\label{Source}
S^s_\boldmathsymbol{k}  =
\int\frac{\mathrm{d}^3 q}{(2\pi)^{3/2}}e^{s}(\boldmathsymbol{k},\boldmathsymbol{q})F(\boldmathsymbol{q},\boldmathsymbol{k-q},\eta)\phi_\boldmathsymbol{q}\phi_\boldmathsymbol{k-q}\, ,
\eeq
where one has introduced
\bea
\label{F}
F(\boldmathsymbol{q},\boldmathsymbol{k-q},\eta) & = 2T_\Phi(q\eta)T_\Phi\left(|\boldmathsymbol{k}-\boldmathsymbol{q}|\eta\right) 
\\  & \kern-2em + \frac{4}{3(1+w)}\left[\mathcal{H}^{-1}qT_\Phi^{\prime}(q\eta)+T_\Phi(q\eta)\right]\left[\mathcal{H}^{-1}\vert\boldmathsymbol{k}-\boldmathsymbol{q}\vert T_\Phi^{\prime}\left(|\boldmathsymbol{k}-\boldmathsymbol{q}|\eta\right)+T_\Phi\left(|\boldmathsymbol{k}-\boldmathsymbol{q}|\eta\right)\right],
\eea
which only involves the transfer function $T_\Phi$.
An analytic solution to \Eq{Tensor Eq. of Motion} is obtained with the Green's function formalism,
\bea
\label{tensor mode function}
a(\eta)h^s_\boldmathsymbol{k} (\eta)  =4 \int^{\eta}_{\eta_0}\mathrm{d}\bar{\eta}\,  g_\boldmathsymbol{k}(\eta,\bar{\eta})a(\bar{\eta})S^s_\boldmathsymbol{k}(\bar{\eta}),
\eea
where the Green's function $g_\boldmathsymbol{k}(\eta,\bar{\eta})$ is given by $g_\boldmathsymbol{k}(\eta,\bar{\eta})=G_\boldmathsymbol{k}(\eta,\bar{\eta})\Theta(\eta-\bar{\eta})$. In the previous expression, $\Theta$ is the Heaviside step function, and $G_{\bm{k}}(\eta,\bar{\eta})$ is the solution of the homogeneous equation 
\beq
\label{Green function equation}
G_\boldmathsymbol{k}^{\prime\prime}  + \left( k^{2} - \frac{a^{\prime\prime}}{a}\right)G_\boldmathsymbol{k} = 0\, ,
\eeq
where a prime denotes derivation with respect to the first argument $\eta$, and with initial conditions $ \lim_{\eta\to \bar{\eta}}G_\boldmathsymbol{k}(\eta,\bar{\eta}) = 0$ and $ \lim_{\eta\to \bar{\eta}}G^\prime_\boldmathsymbol{k}(\eta,\bar{\eta})=1$.  The above equation can be solved analytically in terms of Bessel functions and the solution is:
\beq\label{Green function analytical}
kG_\boldmathsymbol{k}(\eta,\bar{\eta}) =\frac{\pi}{2} \sqrt{x\bar{x}}\left[Y_\mathrm{\nu}(x)J_\mathrm{\nu}(\bar{x})  - J_\mathrm{\nu}(x)Y_\mathrm{\nu}(\bar{x})\right],
\eeq
where $\nu = \frac{3(1-w)}{2(1+3w)}$. Since $G_\boldmathsymbol{k}(\eta,\bar{\eta})$ depends only on $k$, from now on it will be noted as $G_k(\eta,\bar{\eta})$.
\subsection{The stress-energy tensor of gravitational waves}
\label{sec:Stress:Energy:GW}
Following closely ~\cite{Maggiore:1999vm}, by considering only the contribution of small scales, where one does not ``feel'' the curvature of the spacetime, the background spacetime can be considered as efffectively flat. Thus, by coarse graining perturbations below the intermediate scale $\ell$ such that $\lambda\ll \ell \ll L_\mathrm{B}$, the effective stress-energy tensor of gravitational waves can be recast in the following form: ~\cite{Maggiore:1999vm}
\beq
\label{t_munu 1}
t_\mathrm{\mu\nu}=-\Mp^2\, \overline{ \left(R^{(2)}_\mathrm{\mu\nu} - \frac{1}{2}\bar{g}_\mathrm{\mu\nu}R^{(2)}\right)},
\eeq
where 
 $\bar{g}_\mathrm{\mu\nu}$ is the background metric, 
$R^{(2)}_\mathrm{\mu\nu}$ is the second-order Ricci tensor and  $R^{(2)} = \bar{g}^\mathrm{\mu\nu}R^{(2)}_\mathrm{\mu\nu}$ its trace. 
The overall bar refers to the coarse-graining procedure.

The physical modes contained in $t_\mathrm{\mu\nu}$ can be extracted either by specifying a gauge, as for instance the transverse-traceless gauge where $\partial_\mathrm{\beta}h^\mathrm{\alpha\beta}=0$ and  $h=\bar{g}^\mathrm{\alpha\beta}h_\mathrm{\alpha\beta}=0$,
 or in a gauge-invariant way by using space-time averages~\cite{Maggiore:1999vm} (see also Appendix of ~\cite{Isaacson:1968zza}). Both approaches coincide on sub-Hubble scales and the 0-0 component of $t_{\mu\nu}$ reads
\bea
\label{rho_GW effective}
 \rhoGW (\eta,\boldmathsymbol{x}) = t_{00} &=  \frac{\Mp^2}{32 a^2}\, \overline{\left(\partial_\eta h_\mathrm{\alpha\beta}\partial_\eta h^\mathrm{\alpha\beta} +  \partial_{i} h_\mathrm{\alpha\beta}\partial^{i}h^\mathrm{\alpha\beta} \right)}\, ,
\eea
which is simply the sum of a kinetic term and a gradient term. 

In the case of a free wave [\ie in the absence of a source term in \Eq{Tensor Eq. of Motion}], these two contributions are identical, since the energy is equipartitioned between its kinetic and gradient components. This is the case in a radiation era where the scalar perturbations due to diffusion damping are in general exponentially supressed and therefore decouple in the late-time limit from the tensor perturbations. Therefore, the source term in the right hand side of \Eq{Tensor Eq. of Motion} can be neglected and considering only sub-horizon scales one can neglect as well the friction term  $2\mathcal{H}h_\boldmathsymbol{k}^{s,\prime}$ in \Eq{Tensor Eq. of Motion}. Therefore, \Eq{Tensor Eq. of Motion} becomes a free-wave equation and one is met with an equipartition between the gradient and the kinetic component in \Eq{rho_GW effective}. Consequently, one obtains that 
\beq\label{rho_GW_effective_averaged}
\begin{split}
\left\langle \rhoGW (\eta,\boldmathsymbol{x}) \right\rangle & = t_{00}  \simeq 2 \sum_{s=+,\times}\frac{\Mp^2}{32a^2}\overline{\left\langle\left(\nabla h^{s}_\mathrm{\alpha\beta}\right)^2\right \rangle }
 \\ & =   \frac{\Mp^2}{16a^2 \left(2\pi\right)^3} \sum_{s=+,\times} \int\mathrm{d}^3\boldmathsymbol{k}_1 \int\mathrm{d}^3\boldmathsymbol{k}_2\,  k_1 k_2 \overline{  \left\langle h^{s}_{\boldmathsymbol{k}_1}(\eta)h^{s,*}_{\boldmathsymbol{k}_2}(\eta)\right\rangle} e^{i(\boldmathsymbol{k}_1-\boldmathsymbol{k}_2)\cdot \boldmathsymbol{x}}\, .
 \end{split}
\eeq
In this expression, the bar denotes averaging over the  sub-horizon oscillations of the tensor field, which is done in order to extract the envelope of the gravitational-wave spectrum at those scales and brackets mean an ensemble average. Defining now $\OmegaGW(\eta,k)$ through the relation
\bea
\label{eq:OmegaGW:def}
\left\langle  \rhoGW (\eta,\boldmathsymbol{x}) \right\rangle \equiv \rho_\mathrm{tot} \int \OmegaGW(\eta,k)  \dd\ln k,
\eea
where $\rho_\mathrm{tot}$ is the total energy density of the universe, one then can compute $\OmegaGW(\eta,k)$ by computing $\left\langle  \rhoGW (\eta,\boldmathsymbol{x}) \right\rangle$. Equivalently, given \Eq{rho_GW_effective_averaged} one can compute $\OmegaGW(\eta,k)$ by computing the two-point correlation function of the tensor field, $\langle h^r_{\boldmathsymbol{k}_1}(\eta)h^{s,*}_{\boldmathsymbol{k}_2}(\eta)\rangle$.
\subsection{The tensor power spectrum at second order}
\label{sec:tensor:power:spectrum}
We  extract now the two-point correlation function of the tensor field, $\langle h^r_{\boldmathsymbol{k}_1}(\eta)h^{s,*}_{\boldmathsymbol{k}_2}(\eta)\rangle$. As we will show later, it is of the form
\bea
\label{tesnor power spectrum definition}
\langle h^r_{\boldmathsymbol{k}_1}(\eta)h^{s,*}_{\boldmathsymbol{k}_2}(\eta)\rangle \equiv \delta^{(3)}(\boldmathsymbol{k}_1 - \boldmathsymbol{k}_2) \delta^{rs} \frac{2\pi^2}{k^3_1}\mathcal{P}_{h}(\eta,k_1),
\eea
where $\mathcal{P}_{h}(\eta,k)$ is the tensor power spectrum. According to \Eq{tensor mode function}, the two-point function of the tensor fluctuation can be written in terms of the two-point function of the source,
\beq\label{h^rh^s analytic 1}
\langle h^r_{\boldmathsymbol{k}_1}(\eta)h^{s,*}_{\boldmathsymbol{k}_2}(\eta)\rangle = \frac{16}{a^2(\eta)}\int_{\eta_0}^{\eta}\mathrm{d}\bar{\eta}_1 G_{k_1}(\eta,\bar{\eta}_1)a(\bar{\eta}_1)\int_{\eta_0}^{\eta}\mathrm{d}\bar{\eta}_2G_{k_2}(\eta,\bar{\eta}_2)a(\bar{\eta}_2)\langle S^r_{\boldmathsymbol{k}_1}(\bar{\eta}_1)S^{s,*}_{\boldmathsymbol{k}_2}(\bar{\eta}_2)\rangle,
\eeq
where the source correlator can be derived from \Eq{Source}, leading to 
\beq\label{Source correlator 1}
\begin{split}
\langle S^r_{\boldmathsymbol{k}_1}(\bar{\eta}_1)S^{s,*}_{\boldmathsymbol{k}_2}(\bar{\eta}_2)\rangle & = \int\frac{\mathrm{d}^3 q_1}{(2\pi)^{3/2}}e^r(\boldmathsymbol{k}_1,\boldmathsymbol{q}_1)F(\boldmathsymbol{q}_1,\boldmathsymbol{k}_1-\boldmathsymbol{q}_1,\bar{\eta}_1) \\ & \kern -3em
\times \int\frac{\mathrm{d}^3 q_2}{(2\pi)^{3/2}}e^s(\boldmathsymbol{k}_2,\boldmathsymbol{q}_2)F^{*}(\boldmathsymbol{q}_2,\boldmathsymbol{k}_2-\boldmathsymbol{q}_2,\bar{\eta}_2)\langle \phi_{\boldmathsymbol{q}_1}\phi_{\boldmathsymbol{k}_1-\boldmathsymbol{q}_1} \phi^{*}_{\boldmathsymbol{q}_2}\phi^{*}_{\boldmathsymbol{k}_2-\boldmathsymbol{q}_2} \rangle .
\end{split}
\eeq

By choosing the initial time $x_0$ well before the horizon entry, one can show that the primordial value $\phi_\bmk$ is related to the comoving curvature perturbation $\zeta_\bmk$ as follows ~\cite{Kohri:2018awv} 
\bea
\label{curvature power spectrum vs phi_k}
\langle\phi_{\boldmathsymbol{k}_1}\phi^{*}_{\boldmathsymbol{k}_2}\rangle = \delta(\boldmathsymbol{k}_1-\boldmathsymbol{k}_2)\frac{2\pi^2}{k^3_1} \mathcal{P}_\Phi(k_1),
\eea
where $\mathcal{P}_\Phi(k)$ is the primordial  power spectrum of the gravitational potential well before the horizon entry.

Combining the above results, and considering the case of a radiation era where $w=1/3$, \Eq{Source correlator 1} gives rise to
\bea
\label{Source correlator 2}
\langle S^r_{\boldmathsymbol{k}_1}(\bar{\eta}_1)S^{s,*}_{\boldmathsymbol{k}_2}(\bar{\eta}_2)\rangle & = \pi \delta^{(3)}(\boldmathsymbol{k}_1 -\boldmathsymbol{k}_2)  
 \int \mathrm{d}^3 \boldmathsymbol{q}_1e^r(\boldmathsymbol{k}_1,\boldmathsymbol{q}_1)e^s(\boldmathsymbol{k}_1,\boldmathsymbol{q}_1) 
  \\ &
  F(\boldmathsymbol{q}_1,\boldmathsymbol{k}_1-\boldmathsymbol{q}_1,\bar{\eta}_1) F^{*}(\boldmathsymbol{q}_1,\boldmathsymbol{k}_1-\boldmathsymbol{q}_1,\bar{\eta}_2) \frac{\mathcal{P}_\Phi(q_1)}{q^3_1}\frac{\mathcal{P}_\Phi(|\boldmathsymbol{k}_1-\boldmathsymbol{q}_1|)}{|\boldmathsymbol{k}_1-\boldmathsymbol{q}_1|^3}.
\eea
Rewriting then the above integral in terms of the two auxiliary variables $u = |\boldmathsymbol{k}_1 - \boldmathsymbol{q}_1|/k_1$ and $v = q_1/k_1$ and plugging \Eq{Source correlator 2} into \Eq{h^rh^s analytic 1} 
%In the orthonormal basis $\lbrace e_i(\boldmathsymbol{k}_1), \bar{e}_i(\boldmathsymbol{k}_1), \boldmathsymbol{k}_1/k_1 \rbrace$, let $(q_1,\theta,\phi)$ be the spherical coordinates of the vector $\boldmathsymbol{q}_1$. Applying the law of cosines (also known as Al Kashi's theorem) to  the triangle formed of the vectors $\boldmathsymbol{k} _1$, $\boldmathsymbol{q} _1$ and $\boldmathsymbol{k} _1 - \boldmathsymbol{q} _1$, one finds $\cos\theta=(1+v^2-u^2)/2v$, while one simply has $q_1=k_1 v$. The integral over $\boldmathsymbol{q}_1$ can thus be written as
%\beq
%\int_{\mathbb{R}^3}\mathrm{d}^3\boldmathsymbol q_1 =  k^3_1\int_{0}^{\infty}\mathrm{d}v\, v^2  \int_{|1-v|}^{1+v}\mathrm{d}u\,   \frac{u}{v}   \int_0^{2\pi}\mathrm{d}\phi.
%\eeq
%Then, noticing that $F(\boldmathsymbol{q},\boldmathsymbol{k-q},\eta)$ depends only on the modulus of its first two arguments, i.e. $F(\boldmathsymbol{q},\boldmathsymbol{k-q},\eta)=F(q,|\boldmathsymbol{k-q}|,\eta)$ [see \Eq{F}] and given that, by construction, $\vert \boldmathsymbol{q}_1-\boldmathsymbol{k}_1\vert = k_1 u$ does not depend on $\phi$, the integral over $\phi$ in \Eq{Source correlator 2} can be performed independently, and one finds
%\beq
%\int_0^{2\pi}\mathrm{d}\phi \, e^r(\boldmathsymbol{k}_1,\boldmathsymbol{q}_1)e^s(\boldmathsymbol{k}_1,%\boldmathsymbol{q}_1) = \frac{k^4_1}{2}v^4\left[1-\frac{(1+v^2-u^2)^2}{4v^2}\right]^2\pi\,  \delta^{rs}.
%\eeq
one obtains after a straightforward but lengthy calculation [See ~\cite{Papanikolaou:2020qtd} for more details] that the two-point function of the tensor field can be cast in the form of \Eq{tesnor power spectrum definition}, where
 the tensor power spectrum is given by
\bea
\label{Tensor Power Spectrum}
\mathcal{P}_h(\eta,k) = 4\int_{0}^{\infty} \mathrm{d}v\int_{|1-v|}^{1+v}\mathrm{d}u \left[ \frac{4v^2 - (1+v^2-u^2)^2}{4uv}\right]^{2}I^2(u,v,x)\mathcal{P}_\Phi(kv)\mathcal{P}_\Phi(ku)\,,
\eea
with
\bea
\label{I function}
I(u,v,x)=\int_{x_0}^{x} \mathrm{d}\bar{x}\, \frac{a(\bar{x})}{a(x)}\, k\, G_{k}(x,\bar{x}) F_k(v,u,\bar{x}).
\eea
In this expression, $x=k\eta$ and we use the notation  $F_{k}(v,u,x)\equiv  F(q ,|\boldmathsymbol{k}-\boldmathsymbol{q}|,\eta)$ given the fact that $x=k\eta$, $q=vk$ and $|\boldmathsymbol{k}-\boldmathsymbol{q}|=uk$.
Having extracted therefore an analytic formula for the tensor power spectrum defined through \Eq{tesnor power spectrum definition} one can compute $\OmegaGW(\eta,k)$ by combining \Eq{rho_GW_effective_averaged} and \Eq{eq:OmegaGW:def}. At the end, one gets that in the free-wave approximation, where one can assume equipartition of the kinetic and gradient energies, $\OmegaGW(\eta,k)$ reads as
\bea
\label{Omega_GW_RD}
\OmegaGW (\eta,k) =  \frac{1}{24}\left[\frac{k}{\calH(\eta)}\right]^{2}\overline{\mathcal{P}}_{h}(\eta,k).
\eea

With the above formula one can compute the energy contribution of induced GWs at a reference time during the RD era. Here we choose this time as the time of PBH formation, which, as explained in \Eq{sec:PBH basics}, is considered to be the time at which the mode $k$ related to the PBH scale crosses the horizon. To compute then the contribution of the induced GWs to the energy budget of the universe at present epoch one should evolve $\OmegaGW(\eta_\mathrm{f},k)$ computed at PBH formation time up to today. To do so, one has that 
\beq
\OmegaGW(\eta_0,k) = \frac{\rhoGW(\eta_0,k)}{\rho_\mathrm{c}(\eta_0)} = \frac{\rhoGW(\eta_\mathrm{f},k)}{\rho_\mathrm{c}(\eta_\mathrm{f})}\left(\frac{a_\mathrm{f}}{a_\mathrm{0}}\right)^4 \frac{\rho_\mathrm{c}(\eta_\mathrm{f})}{\rho_\mathrm{c}(\eta_0)} = \OmegaGW(\eta_\mathrm{f},k)\Omega^{(0)}_\mathrm{r}\frac{\rho_\mathrm{r,f}a^4_\mathrm{f}}{\rho_\mathrm{r,0}a^4_0},
\eeq
where we have taken into account that $\Omega_\mathrm{GW}\sim a^{-4}$. The index $0$ refers to the present time. Then, taking into account that the energy density of radiation can be recast as $\rho_r = \frac{\pi^2}{15}g_{*\mathrm{\rho}}T_\mathrm{r}^4$ and that the temperature of the radiation bath, $T_\mathrm{r}$, scales as  $T_\mathrm{r}\propto g^{-1/3}_{*\mathrm{S}}a^{-1}$ one finds that 
\beq\label{Omega_GW_RD_0}
\Omega_\mathrm{GW}(\eta_0,k) = \Omega^{(0)}_r\frac{g_{*\mathrm{\rho},\mathrm{f}}}{g_{*\mathrm{\rho},0}}\left(\frac{g_{*\mathrm{S},\mathrm{0}}}{g_{*\mathrm{S},\mathrm{f}}}\right)^{4/3}\OmegaGW(\eta_\mathrm{f},k),
\eeq
where $g_{*\mathrm{\rho}}$ and $g_{*\mathrm{S}}$ stand for the energy and entropy relativistic degrees of freedom.

\subsection{The GW energy density parameter in a RD era}
Considering now GW emission during RD era, from \Eq{F}, the function $F_k(u,v,\bar{x})$ in a RD era reads
\bea
\label{f function - w =1/3}
F_\mathrm{RD}(v,u,x) & = 2T_\Phi(vx)T_\Phi(ux) + \left[vxT_\Phi^{\prime}(vx)+T_\Phi(vx)\right]\left[uxT_\Phi^{\prime}(ux)+T_\Phi(ux)\right],
\eea
where we have used the fact that $\mathcal{H} = aH = 1/\eta$ during an RD era.

\begin{figure}[h!]
\begin{center}
\includegraphics[width = 0.6\textwidth, height = 0.5\textwidth, clip=true]{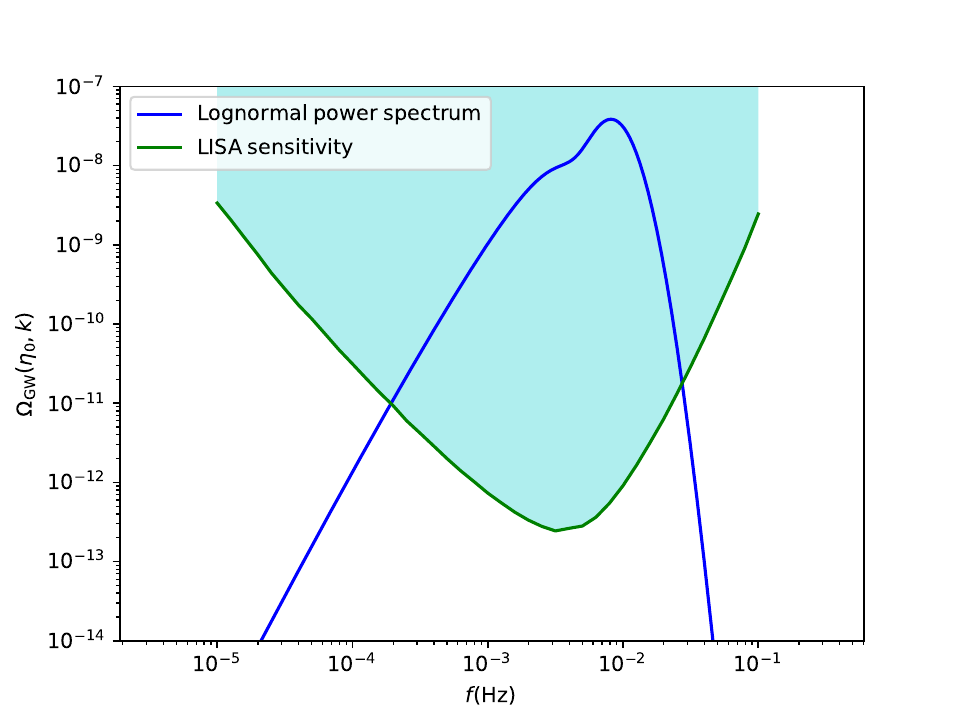}
\caption{The energy density parameter $\OmegaGW(\eta_0,k)$ for a lognormal  curvature power spectrum [see \Eq{eq:log P_zeta}] superimposed to the gravitational wave sensitivity curve of LISA ~\cite{Caprini:2015zlo}.}
\label{fig:PBH_GW_LISA}
\end{center}
\end{figure}

Regarding the evolution of $T_\Phi$, having written before $\Phi_\bmk = T_\Phi \phi_\bmk$ one can write \Eq{Bardeen Potential - Exact Solution} for the transfer function $T_\Phi$ by specifying the initial conditions at $x_0$ for the $T_\phi$ and for $T^\prime_\Phi$. At early times, i.e. $x_0\rightarrow 0$, all modes can be considered super-horizon leading to a constant value for $T_\Phi$ according to the discussion after \Eq{Bardeen potential 2}. One then can choose that $T_\Phi(x_0\rightarrow 0)=1$ and $T^\prime(x_0\rightarrow 0) = 0$ and $T_\Phi(x)$ reads as 
\beq\label{eq:T_Phi - w=1/3}
T_\Phi(x) = \frac{9}{x^2}\left[\frac{\sin\left(x/\sqrt{3}\right)}{x/\sqrt{3}}-\cos\left(x/\sqrt{3}\right)\right]
\eeq
Thus, $F_\mathrm{RD}(v,u,x)$ becomes
\bea
\label{f function  expanded- w =1/3}
\begin{split}
F_\mathrm{RD}(v,u,x)  & = \frac{18}{u^3v^3x^6}\biggl[ 18uvx^2\cos\left(\frac{ux}{\sqrt{3}}\right)\cos\left(\frac{vx}{\sqrt{3}}\right) + \\ & (54-6(u^2+v^2)x^2+u^2v^2x^4)\sin\left(\frac{ux}{\sqrt{3}}\right)\sin\left(\frac{vx}{\sqrt{3}}\right)
 \\ & + 2\sqrt{3}ux(v^2x^2-9)\cos\left(\frac{ux}{\sqrt{3}}\right)\sin\left(\frac{vx}{\sqrt{3}}\right) \\ & + 2\sqrt{3}vx(u^2x^2-9)\sin\left(\frac{ux}{\sqrt{3}}\right)\cos\left(\frac{vx}{\sqrt{3}}\right)\biggr]
\end{split}
\eea
and from \Eq{I function} one obtains after a straightforward but long calculation that 
\begin{align*}
 I_\mathrm{RD}(u,v,x) & =\frac{3}{4u^3v^3x}\Bigg\{ -\frac{4}{x^3}\biggl [ uv(u^2+v^2-3)x^3\sin x - 6uvx^2\cos\frac{ux}{\sqrt{3}}\cos\frac{vx}{\sqrt{3}} + \\ &
+ 6\sqrt{3}ux\cos\frac{ux}{\sqrt{3}}\sin\frac{vx}{\sqrt{3}} + 6\sqrt{3}vx\sin\frac{ux}{\sqrt{3}}\cos\frac{vx}{\sqrt{3}} 
\\ & -3\Bigl(6+(u^2+v^2-3)x^2\Bigr)\sin\frac{ux}{\sqrt{3}}\sin\frac{vx}{\sqrt{3}}\biggr]   \\ &
+(u^2+v^2-3)^{2}\Biggl( \sin x\bigg\{ \mathrm{Ci}\left[\left(1-\frac{v-u}{\sqrt{3}}\right)x\right]+ \mathrm{Ci}\left[\left(1+\frac{v-u}{\sqrt{3}}\right)x\right] \\ & - \mathrm{Ci}\left[\left(1-\frac{v+u}{\sqrt{3}}\right)x\right] - \mathrm{Ci}\left[\left(1+\frac{v+u}{\sqrt{3}}\right)x\right] + \ln \left|\frac{3-(u+v)^{2}}{3-(u-v)^2}\right| \bigg\}  \\ &
+ \cos x \bigg\{ -\mathrm{Si}\left[\left(1-\frac{v-u}{\sqrt{3}}\right)x\right]-\mathrm{Si}\left[\left(1+\frac{v-u}{\sqrt{3}}\right) x \right]  + \\& 
+ \mathrm{Si}\left[\left(1-\frac{v+u}{\sqrt{3}}\right) x \right] +\mathrm{Si}\left[\left(1+\frac{v+u}{\sqrt{3}}\right) x \right]\bigg\} \Biggr) \Bigg\}, \label{eq:I_RD}
\end{align*}
where the $\mathrm{Ci}(x)$ and $\mathrm{Si}(x)$ functions are defined as
\beq
\mathrm{Si}(x) = \int_0^x\mathrm{d}\bar{x}\frac{\sin\bar{x}}{\bar{x}}, \quad \mathrm{Ci}(x) = -\int_x^\infty\frac{\cos\bar{x}}{\bar{x}}.
\eeq
Taking now the oscillation average of $I^2_\mathrm{RD}(u,v,x)$ in the late-time limit, i.e. $x\rightarrow \infty$ one obtains that 
\beq\label{eq:I_RD_late_time}
\begin{split}
\bar{I^2_\mathrm{RD}}(v,u,x\rightarrow\infty) & = \frac{1}{2}\left[\frac{3(u^2+v^2-3)}{4u^3v^3x}\right]^{2}\biggl\{\biggl[-4uv + (u^2+v^2-3)\ln \left| \frac{3 - (u+v)^{2}}{3-(u-v)^{2}}\right|\biggr]^2  \\ & + \pi^2(u^2+v^2-3)^2\Theta(v+u-\sqrt{3})\biggr\}
\end{split}
\eeq
To compute now the energy density parameter $\OmegaGW (\eta,k)$ we consider a log-normal curvature power spectrum which peaks at the frequency, $f_{*}$, at which the LISA experiment exhibits its maximum sensitivity, i.e. $f_{*}=f_\mathrm{LISA}=3.4\mathrm{mHz}$.
\beq\label{eq:log P_zeta}
\mathcal{P}_\zeta(k) = A_\zeta e^{-\frac{\ln^2\left(\frac{k}{k_{*}}\right)}{2\sigma^2}},
\eeq
where $k_{*}=2\pi f_{*}$ and we have chosen $A_\zeta = 0.029$ and $\sigma=0.5$. Then, plugging \Eq{eq:log P_zeta} into \Eq{Tensor Power Spectrum} we perform numerically the double integral \eqref{Tensor Power Spectrum} and by combining \Eq{Omega_GW_RD} and \Eq{Omega_GW_RD_0} we show in \Fig{fig:PBH_GW_LISA} $\OmegaGW (\eta_0,k)$ superimposed to the gravitational wave sensitivity curve of LISA ~\cite{Caprini:2015zlo}. 
\section{Induced gravitational waves produced in a PBH-dominated era (research article)}
Having presented before the basics of the calculation of second order induced gravitational waves from first order scalar perturbations in the case of GW production in a radiation-dominated era, we recap here the main results of our work~\cite{Papanikolaou:2020qtd}  in which gravitational waves are produced in an era of domination of ultralight PBHs with masses $m_\mathrm{PBH}<10^{9}\mathrm{g}$ by emphasizing the differences with the case of induced GWs produced in a RD universe.

Firstly, we should stress out that contrary to the induced GWs produced in a RD era where the gauge-dependence of GWs mentioned in \Sec{sec:GWs at second order} is expected to disappear in the late-time limit due to diffusion damping ~\cite{DeLuca:2019ufz, Yuan:2019fwv, Inomata:2019yww,PhysRevLett.113.061301} this does not happen for GWs produced in a matter-dominated era, like the one driven by PBHs. However, in ~ \cite{Papanikolaou:2020qtd} we do not derive observable predictions, but we rather investigate a GW backreaction problem, which we assume bears little dependence on the gauge. In particular, if the energy density carried by gravitational waves overcomes the one of the background, one expects perturbation theory to break down in any gauge. 

In addition, we should underline the nature of the induced GWs studied in~\cite{Papanikolaou:2020qtd} as well. In the majority of the literature, induced GWs are sourced be primordial scalar perturbations which have preceded and given rise to PBHs. However, in our work~\cite{Papanikolaou:2020qtd} the induced GWs  are sourced by scalar perturbations underlain by PBHs themselves. For this reason, by considering monochromatic PBHs produced in a RD universe with their initial spatial distribution being of Poisson type (unclustered), we treated the PBH energy density perturbations as isocurvature perturbations and we extracted at the end the power spectrum of the gravitational potential $\mathcal{P}_\Phi(k)$ underlain by a gas of PBHs during the subsequent PBH domination era. Finally, by plugging our expression for $\mathcal{P}_\Phi(k)$ in \Eq{Tensor Power Spectrum} and making use of \Eq{Omega_GW_RD} we computed the GW spectrum $\Omega_\mathrm{GW}(\eta,k)$.

Another aspect which should be emphasized here is that in our work ~\cite{Papanikolaou:2020qtd}, we considered GWs produced during a PBH-dominated era, where $\Phi=\mathrm{constant}$, a fact which forces the source term \eqref{eq:Source:def} to be constant and as a consequence the equation of motion \eqref{Tensor Eq. of Motion} for the tensor modes is not anymore a free-wave equation. The amplitude of gravitational waves converges then to a solution constant in time, which highly suppresses the kinetic contribution compared to the gradient contribution in \Eq{rho_GW effective}. Consequently, in the case of induced GWs produced during a PBH-dominated era the result \eqref{Omega_GW_RD} should be divided by two.

Finally, by making use of  \Eq{Omega_GW_RD}, we extracted both numerically and analytically $\Omega_\mathrm{GW}(\eta_\mathrm{evap},k)$ at evaporation time $\eta_\mathrm{evap}$ and we found that it crucially depends on two parameters, namely the PBH mass, $m_\mathrm{PBH}$ and the initial abundance of PBHs at formation time $\Omega_\mathrm{PBH,f}$. Typically, we found that the amount of gravitational waves increases with $m_\mathrm{PBH}$, since heavier black holes live  longer, hence dominate the universe for a longer period before they evaporate, and with $\Omega_\mathrm{PBH,f}$, since more abundant black holes dominate the universe earlier, hence for a longer period too. 

Subsequently, by integrating $\Omega_\mathrm{GW}(\eta_\mathrm{evap},k)$ over the relevant modes $k$ we required that the induced GWs produced during the PBH-dominated era are not overproduced, i.e. $\Omega_\mathrm{GW,tot}(\eta_\mathrm{evap})<1$, deriving in this way both analytically and numerically the following upper bound constraint on the initial abundance of PBHs, $\Omega_\mathrm{PBH,f}$, as a function of their mass $m_\mathrm{PBH}$:
\bea
 \label{Omega_f constraints}
\Omega_\mathrm{PBH,f} < 1.4\times 10^{-4}\left(\frac{10^9\mathrm{g}}{ m_\mathrm{PBH} }\right)^{1/4} .
 \eea
At this point, let us stress out that since PBHs with masses smaller than $10^9\mathrm{g}$ evaporate before BBN, they cannot be directly constrained (at least without making further assumption, see \eg \cite{Dai:2019eei}). Therefore, to the best of our knowledge, the above constraint is the first one ever derived on ultra-light PBHs. We should also underline that given the fact that we have not assumed a specific PBH production mechanism - we just only assumed that initially PBHs are unclustered and they all have the same mass - the constraint quoted in \Eq{Omega_f constraints} is rather model independent. For more details see our relevant research article attached below.
\includepdf[pages=2]{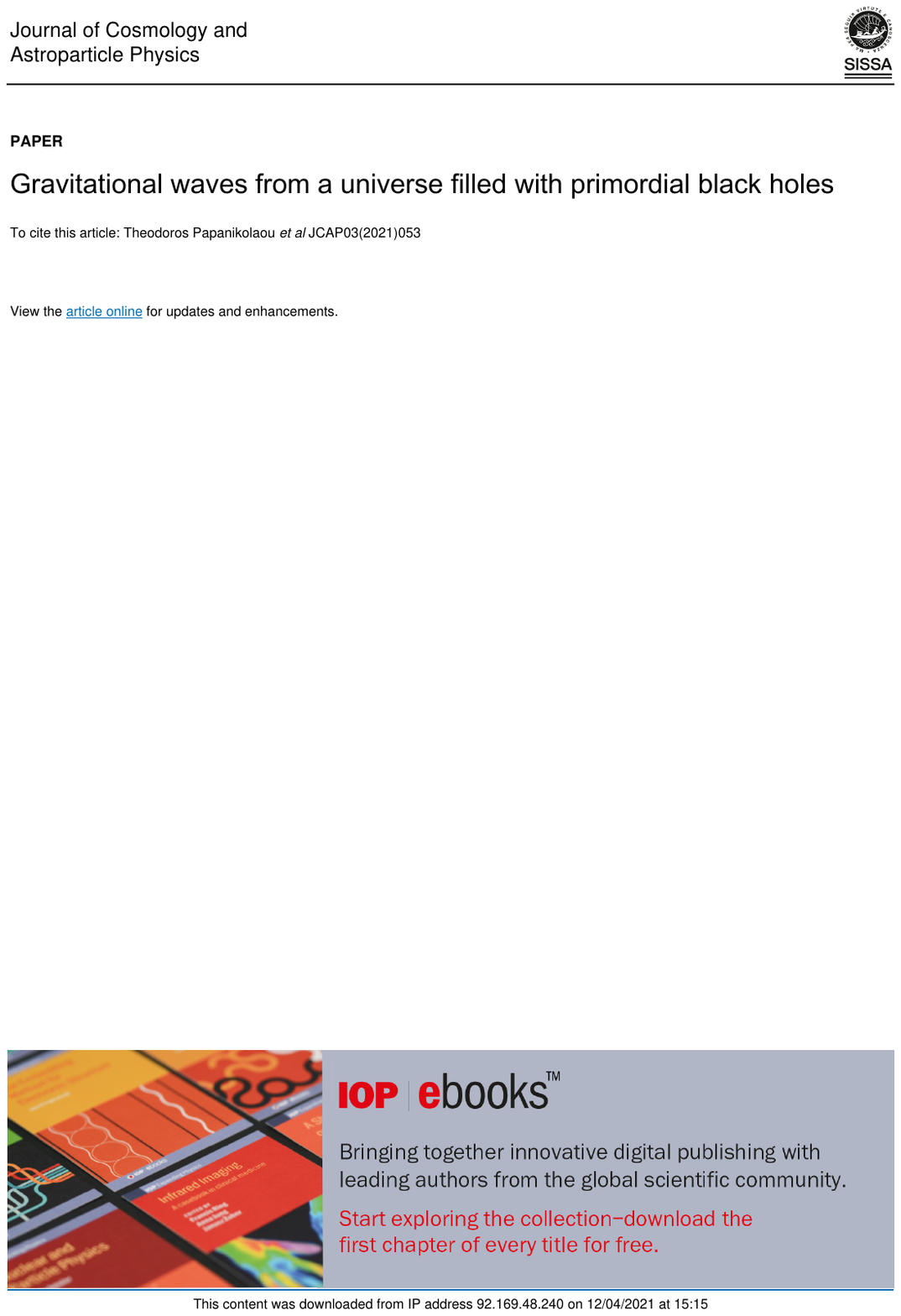}
\newpage
\chapter{PBH formation for an anisotropic perfect fluid}\label{sec:Anisotropic Collapse of PBHs}
Up to now, most of the research works in the literature model the PBH gravitational collapse as spherical and isotropic due to its calculational simplicity \footnote{During a spherically symmetric gravitational collapse the fluid elements are moving only in the radial direction, leading to a spherically symmetric object. In the particular case we treat here, we consider an additional degree of freedom of the stress energy tensor which accounts for the anisotropy of the collapse and which is compatible with spherical symmetry, i.e. the radial and tangential pressures appearing in the radial and angular diagonal parts of the stress-energy tensor, which in our case are not the same ~\cite{Bayin:1985cd}. This regime can be physically realized within theories with scalar fields and multifluids ~\cite{1982NCimB..69..145L}}. In particular, in the context of peak theory, within which PBHs are often studied, rare large peaks which collapse to form PBHs are expected to be quasi spherical ~\cite{Bardeen:1985tr}. However, one should go beyond these assumptions to model realistically the PBH gravitational collapse process. Regarding the spherical symmetry hypothesis, there were some early studies going beyond and adopting the “pancake” collapse ~\cite{1965ApJ...142.1431L,1970Ap......6..320D, 1970A&A.....5...84Z, 1980PhLB...97..383K} as well as some recent ones focusing on a non-spherical collapse of PBHs in a matter domination universe ~\cite{Harada:2015ewt} and on the ellipsoidal collapse of PBHs ~\cite{Kuhnel:2016exn}. 

Regarding the anisotropic nature of the gravitational collapse, to the best of our knowledge there is no systematic treatment of this topic in the context of PBHs. In general, one expects to have anisotropies in theories with scalar fields and multifluids where the anisotropy is described in spherical symmetry as a difference between the radial and the tangential pressure ~\cite{1982NCimB..69..145L}. In addition, in the astrophysics context, there has been done a huge progress in the study of anisotropic star solutions in GR which can be obtained both analytically~\cite{1974ApJ...188..657B,1980PhRvD..22..807L,Bayin:1982vw,Mak:2001eb,Dev:2004ss,Herrera:2004xc,Veneroni:2018gfm} and numerically ~\cite{Doneva:2012rd,Biswas:2019gkw} but which do not treat the anisotropic character of the collapse in a covariant way. In a recent study ~\cite{Raposo:2018rjn} a covariant formulation has been proposed to study anisotropic stars as ultracompact objects mimicking the dynamical behavior of a black-hole . 

Inspired by this work, we study here the anisotropic formulation of the initial conditions for the collapse of cosmological perturbations into PBHs, giving at the end some results for the possible effect of the anisotropy on the PBH formation threshold, $\delta_\mathrm{c}$. Initially, we review the Misner-Sharp and Misner-Hernadez hydrodynamic equations used to describe the dynamical evolution of a spherically symmetric configuration. Then, we discuss a covariant form for the equation of state of an anisotropic radiation fluid describing the evolution of cosmological perturbations in the early universe. Afterwards, by making a gradient expansion approximation on superhorizon scales we extract the initial conditions for the hydrodynamic and metric perturbations as well as their dependence on the degree of the anisotropy of the gravitational collapse. Finally, we give a synthetic overview regarding the numerical calculation of the PBH formation threshold, $\delta_\mathrm{c}$, and adopting a perturbative approach based on the assumption that $\delta_\mathrm{c}$ depends on the shape of the initial energy density profile in the same way as in the isotropic case, we give an estimation regarding the dependence of $\delta_\mathrm{c}$ on the degree of anisotropy of the gravitational collapse. 

The results and conclusions of this chapter regarding PBH formation for an anisotropic perfect fluid were sumbitted on arXiv ~\cite{Musco:2021sva} on October 12th after the submission of the manuscript. 

\section{The Misner-Sharp Equations for an anisotropic perfect fluid}\label{sec:Misner-Sharp Equations}
Working in spherical symmetry, the space-time metric can be written in the Misner-Sharp form ~\cite{Misner:1964je}
\beq\label{MS-Metric}
\mathrm{d}s^2=-A^2(t,r)\mathrm{d}t^2+B^2(t,r)\mathrm{d}r^2+R^2(t,r)\mathrm{d}\Omega^2
\eeq
where $R$ is the areal radius, $A$ the lapse function, $B$ a function related to the spatial curvature of the space time, $r$ the radial comoving coordinate, $t$ the cosmic time coordinate and $\mathrm{d}\Omega^2$ the solid line infinitesimal element of a unit $2$-sphere, i.e. $\mathrm{d}\Omega^2 = \mathrm{d}\theta^2 + \sin^2\theta\mathrm{d}\phi^2$. This is the so called cosmic time slicing, corresponding to a FLRW metric when the Universe is homogeneous and isotropic. Introducing then the differential operators $D_t$ and $D_t$ 
\beq\label{D_r and D_t in Misner-Sharp}
 D_t\equiv \left. \frac{1}{A}\frac{\partial}{\partial t}\right\vert_r \mathrm{\quad and \;\;\;\;} D_r\equiv \left.\frac{1}{B}\frac{\partial}{\partial r}\right\vert_t,
\eeq 
which are basically the derivatives with respect to the proper time and proper space respectively, one can define two auxiliary quantities $U$ and $\Gamma$ 
\beq\label{U and Gamma}
U\equiv D_tR \mathrm{\;\;\;\;and\;\;\;\;} \Gamma\equiv D_r R,
\eeq
where $U$ is the radial component of the four-velocity in an ``Eulerian" (non comoving) frame and $\Gamma$ is the so called generalized Lorentz factor. In the background homogeneous and isotropic FLRW Universe, $R(t,r)=a(t)r$, $U=H(t)R(t,r)$ and $\Gamma^2 = 1-Kr^2$, where $a$ is the scale factor, $H$ is the Hubble parameter and $K$ is the spatial curvature present of the FLRW metric (\ref{FRLW metric - cosmic time}).

The quantities $U$ and $\Gamma$ are related through the Misner-Sharp mass $M$ which for spherical symmetric spacetimes is defined as ~\cite{Misner:1964je, Hayward:1994bu} 
\beq\label{Misner Sharp Mass}
M(t,r)\equiv \frac{R(t,r)}{2}\left[1-\nabla_\mathrm{\mu}R(t,r)\nabla^\mathrm{\mu}R(t,r)\right],
\eeq
%The above quantity, within spherical symmetry is a well defined and unique quantity ~\cite{Faraoni:2016xgy}. 
From the above definition one can get the constraint equation
\beq\label{constraint equation}
\Gamma^2=1+U^2-\frac{2M}{R}
\eeq
obtained by integrating the 00-component of the Einstein equations. 

Regarding the form of the stress-energy tensor for an anisotropic perfect fluid, it can be written in a covariant way ~\cite{Raposo:2018rjn} as
\beq\label{Stress Energy Tensor}
T_{\mathrm{\mu\nu}}= \rho u_\mathrm{\mu} u_\mathrm{\nu} +p_\mr k_\mathrm{\mu}k_\mathrm{\nu} +p_\mt \Pi_\mathrm{\mu\nu},
\eeq
where $p_r$ and $p_t$ are the radial and tangential pressure respectively, $u_\mathrm{\mu}$ is the fluid four-velocity and $k_\mathrm{\mu}$ is a unit spacelike vector orthogonal to $u_\mathrm{\mu}$, i.e $u_\mathrm{\mu}u^{\mathrm{\mu}}=-1 = - k_\mathrm{\mu}k^\mathrm{\mu}$ and $u^{\mathrm{\mu}}k_\mathrm{\mu}=0$. $\Pi_\mathrm{\mu\nu} = g_\mathrm{\mu\nu}+u_\mathrm{\mu}u_\mathrm{\nu}-k_\mathrm{\mu}k_\mathrm{\nu}$ is a projection onto a two surface orthogonal to $u^\mathrm{
\mu}$ and $k^\mathrm{\mu}$. Working now in the comoving frame of the fluid we get that $u_\mathrm{\mu} = (-A,0,0,0)$ and $k_\mathrm{\mu} = (0,B,0,0)$. For an anisotropic spherically symmetric fluid, one has $p_\mr\neq p_\mt$.
%In addition, we should stress out that we  introduce here two equation-of-state parameters, $w_\mathrm{r}(t,r)$ and $w_\mathrm{t}(t,r)$, which in general depend on $t$ and $r$, such as that 
%\begin{eqnarray}
%p_\mathrm{r}=w_\mathrm{r}(t,r)\rho \\
%p_\mathrm{t}=w_\mathrm{t}(t,r)\rho
%\end{eqnarray}

Considering now the Einstein field equations $G^\mathrm{\mu}_\mathrm{\nu}=8\pi T^\mathrm{\mu}_\mathrm{\nu}$ \footnote{In this chapter, we work in a unit system where $c=G=1$.} and the conservation of the stress energy tensor $\nabla_\mathrm{\mu}T^\mathrm{\mu\nu}=0$, one can obtain the Misner-Sharp hydrodynamic set equations \cite{Misner:1964je, May:1966zz} for an anisotropic spherically symmetric fluid with $p_\mr \neq p_\mathrm{t}$:
\beq \label{MS-Equations}
\begin{aligned}
& D_tU = -\frac{\Gamma}{\rho+p_\mathrm{r}} \left[ D_r p_\mathrm{r} +  \frac{2\Gamma}{R} \left( p_\mathrm{r}-p_\mathrm{t} \right) \right] - \frac{M}{R^2} - 4\pi Rp_\mathrm{r}   \\
& \frac{D_t\rho_0}{\rho_0} = - \frac{1}{R^2\Gamma}D_r\left(R^2U\right) \\
& \frac{D_t\rho}{\rho+p_\mathrm{r}} = \frac{D_t\rho_0}{\rho_0} + \frac{2U}{R} \frac{p_\mathrm{r}-p_\mathrm{t}}{\rho+p_\mathrm{r}}  \\ 
& \frac{D_rA}{A} = -\frac{1}{\rho+p_\mathrm{r}} \left[ D_r p_\mathrm{r} + \frac{2\Gamma}{R} \left(p_\mathrm{r}-p_\mathrm{t}\right) \right] \\ 
& D_rM = 4\pi R^2\Gamma\rho  \\
& D_tM=-4\pi R^2 Up_\mathrm{r}  \\ 
& D_t\Gamma = -\frac{U}{\rho+p_\mathrm{r}} \left[ D_rp_\mathrm{r} + \frac{2\Gamma}{R} \left(p_\mathrm{r}-p_\mathrm{t}\right) \right],
\end{aligned}
\eeq
where one can appreciate the additional terms appearing when $p_\mr \neq p_\mathrm{t}$.
This system of differential equations, combined with the constraint equation given by \Eq{constraint equation}, can be solved once the equations of state for $p_\mr$ and $p_\mt$ are specified. In the context of PBH formation with a massless radiation fluid, one of these equations of state can be obtained assuming, as it looks reasonable, the conservation of the trace, i.e. $T^\mathrm{\mu}_\mathrm{\mu}=0$ ~\cite{Ellis:1971pg}, which gives an additional constraint relation between $p_\mr$ and $p_\mt$, 
\beq\label{trace constraint}
\rho - p_\mathrm{r} - 2p_\mathrm{t} = 0.
\eeq
%%%%%%%%%%%%%%%%%%%%%%%%%%%%%% %%% %%%%%%%%%%%%%%%%%%%%%%%%%%%%%%

\section{The Misner-Hernadez Equations for an anisotropic perfect fluid}\label{sec:Misner-Sharp-Hernadez Equations}
When using the Misner-Sharp equations presented in the previous section to study the gravitational collapse process leading to the formation of a black hole, one is facing a well known problem associated to the cosmic time slicing. Because the observer is comoving with the fluid, it is possible to follow the evolution of the region both inside and outside the apparent horizon \footnote{ The formation of an apparent horizon in spherical symmetry, in a collapsing or expanding medium, is reached when the condition for a marginally trapped surface $R(r,t) = 2M(r,t)$ is satisfied \cite{Faraoni:2016xgy}.} up to the formation of the singularity. However, in this slicing this is reached when the matter outside the horizon is still collapsing into the black hole. Therefore without doing something, the formation of the singularity is going to prevent following the rest of the outer evolution, computing also the final mass of the black hole, which is one of the fundamental outcomes of such simulations.
%The Misner-Sharp formulation of the Einstein's equations presented in \Sec{sec:Misner-Sharp Equations} has an important drawback regarding the black hole gravitational collapse. In particular, in this formulation, in which the time variable is the cosmic time, singularities appear after a finite time before an apparent horizon has formed and as a consequence the time evolution can not be continued. Thus, once the singularity has been formed, parts of the evolution, which could  potentially been tracked by a distant observer, cannot be followed using the Misner-Sharp formulation of the field equations~\cite{Misner:1964je,May:1966zz}. 
For this reason, despite the simplicity and intuitiveness of the cosmic time slicing one should consider a null foliation of spacetime, characterized by a far distant observer, in order to track consistently the gravitational collapse process \footnote{An alternative approach is to make a numerical excision of the central region, where the singularity is formed, and continue the evolution in the cosmic time slicing. This technique however requires some care. Making a coordinate transformation in spherical symmetry and going to a null slicing, does not allow to follow the full evolution of the region inside the apparent horizon, but is a very good choice in order to have a full description of the collapsing region outside.}. To do so, we revise here the Misner-Hernadez formulation of the Einstein's equations in which the time variable is now the ``observer time'' defined as the time at which an outgoing radial null ray emitted from an event reaches a distant observer. In this way, the formation of the singularity is screened by the asymptotic formation of the apparent horizon because of the infinite redshift associated to signals emitted from the region where the apparent horizon forms ~\cite{Hernandez:1966zia}, and all the evolution of the region outside the apparent horizon can be followed.

An outgoing null ray is described by the equation
\beq\label{photon path}
A\mathrm{d}t=B\mathrm{d}r
\eeq
and the observer time $u$ is defined by
\beq\label{observer time}
G(r,u)\mathrm{d}u = A(r,t)\mathrm{d}t - B(r,t)\mathrm{d}r,
\eeq
which inserted into the metric \ref{MS-Metric} allows to obtain the following form:
\beq\label{Misner Hernadez Spherical symmetric metric}
\mathrm{d}s^2 = -G^2\mathrm{d}u^2 -2GB\mathrm{d}u\mathrm{d}r +R^2\mathrm{d}\Omega^2,
\eeq
where the function $G(r,u)$ plays the role of the lapse in the null slicing and satisfies the following useful relation [See Appendix \ref{External Derivative}]:
\beq\label{f constraint equation}
\frac{D_kG}{G}=\frac{D_rA}{A}+\frac{D_tB}{B}= \frac{D_rA}{A} + \frac{D_rU}{\Gamma}.
\eeq
The operators defined in \Eq{D_r and D_t in Misner-Sharp} are given by
\beq\label{D_r and D_t in Misner-Hernadez}
 D_k\equiv \left. \frac{1}{B}\frac{\partial}{\partial r}\right\vert_u = D_r + D_t, \quad D_t\equiv \left. \frac{1}{G}\frac{\partial}{\partial u}\right\vert_r,
\eeq 
$\Gamma = D_rR = D_kR-U$  and the equations seen in 
\Eq{MS-Equations} take the following form, derived for the first time, in the isotropic limit, by Misner-Hernandez:
\beq\label{Misner-Hernadez Equations without lambda}
\begin{aligned}
& D_tU  = -\frac{\Gamma}{\rho+p_\mathrm{r}} \left[ D_kp_\mathrm{r} - D_t p_\mr + \frac{2\Gamma}{R} ( p_\mathrm{r}-p_\mathrm{t})  \right] - \frac{M}{R^2} - 4\pi Rp_\mathrm{r}  \\
& \frac{D_t \rho_0}{\rho_0} = \frac{1}{\Gamma} \left( D_k U - D_t U  \right) - \frac{2U}{R} \\
& \frac{D_t \rho}{\rho+p_\mr}   = \frac{D_t \rho_0}{\rho_0} + \frac{2U}{R} \frac{p_\mr-p_\mt}{\rho+p_\mr} \\  
& \frac{D_kG}{G}  =\frac{1}{\Gamma}\left[D_kU+\frac{M}{R^2}+4\pi R p_\mathrm{r}\right] \Leftrightarrow D_k\left(\frac{\Gamma+U}{G}\right)=-\frac{4\pi R}{G}(\rho+p_\mathrm{r}) \\
& D_kM  =4\pi R^2(\Gamma\rho - p_\mathrm{r}U) \\
& D_tM  =-4\pi R^2 Up_\mathrm{r} \\ 
& D_t\Gamma  = - \frac{U}{\rho+p_\mr}\left[ D_k p_\mathrm{r} - D_t p_\mr + \frac{2\Gamma}{R} ( p_\mathrm{r}-p_\mathrm{t})  \right].
\end{aligned} 
\eeq

At the computational level, the strategy adopted is the following: first, we perturb the Misner-Sharp equations by performing the gradient expansion approximation on the superhorizon regime in order to specify the initial conditions on a space-like slice at constant cosmic time in terms of a time-independent curvature profile. We do so because in such slicing we know how to specify a consistent set of cosmological perturbations. These initial conditions are then evolved with the Misner-Sharp equations \eqref{MS-Equations} in order to generate a second set of initial data on a null slice at constant observer time (outgoing null ray). Finally, this second set of the initial data can then be evolved with the Misner-Hernadez equations \eqref{Misner-Hernadez Equations without lambda}, following the full evolution of the perturbations until an apparent horizon is formed in case of a perturbation with an amplitude larger than the threshold, or seeing that the perturbation bounces and disperses if the amplitude is below the threshold. For more details regarding the numerical scheme see ~\cite{Musco:2005bua,Musco:2004ak}.

%%%%%%%%%%%%%%%%%%%%%%%%%%%%%% %%% %%%%%%%%%%%%%%%%%%%%%%%%%%%%%%

\section{The equation of state of an anisotropic fluid}
After having derived the Einstein equations for an anisotropic fluid in the cosmic and null time slicing, we introduce here a covariant formulation modeling the difference between the radial and tangential pressures of the collapsing fluid in terms of pressure or energy density gradients. In particular, following \cite{Raposo:2018rjn, Bowers:1974tgi} the difference $p_\mt- p_\mr$ can be expressed, to a certain degree of arbitrariness, in a covariant form as
\begin{eqnarray}
p_\mt & = & p_\mr + \lambda f(r,t) k^\mu \nabla_\mu p_\mr \quad \textrm{(pressure gradients)} \label{D_rp_r} \\ 
& & \quad \quad \quad \textrm{or} \nonumber \\
p_\mt & = & p_\mr + \lambda f(r,t) k^\mu \nabla_\mu \rho \quad \ \textrm{(energy density gradients)}  \label{D_rrho},
\end{eqnarray}
where $f(r,t)$ is a generic function of $r$ and $t$ while $\lambda$ is a parameter tuning the level of the anisotropy of the collapse. Using the metric \eqref{MS-Metric} one can show that $k^\mu \nabla_\mu = D_r$. At this point, one should mention that the Minser-Sharp equations \eqref{MS-Equations} should be regularized in the following way at $R=0$ ~\cite{Bowers:1974tgi}: 
\beq\label{boundary condition}
\lim_{R\rightarrow 0} \frac{p_\mathrm{r}-p_\mathrm{t}}{R} = 0.
\eeq
A possible choice for $f(r,t)$ satisfying the boundary condition \eqref{boundary condition} and keeping the parameter $\lambda$ dimensionless, without introducing an additional characteristic scale into the problem, is $f(r,t) = R(r,t)$. In this case, using \Eq{D_rp_r} and \Eq{D_rrho} combined with \Eq{trace constraint}, the equations of state for $p_\mr$ and $p_\mt$ read as
\begin{eqnarray}\label{eq:p_r+p_t - f= R}
p_\mathrm{r} & = & \frac{1}{3}\left[\rho-2\lambda R D_rp_\mr \right],\quad p_\mathrm{t} = \frac{1}{3}\left[\rho+\lambda R D_rp_\mr \right]  \quad \textrm{(pressure gradients)}  \label{eq:p_r+p_t - f= R-D_rp_r}  \\
p_\mathrm{r} & = & \frac{1}{3}\left[\rho-2\lambda R D_r\rho \right],\quad p_\mathrm{t} = \frac{1}{3}\left[\rho+\lambda RD_r\rho\right]  \quad \textrm{(energy density gradients).}  \label{eq:p_r+p_t - f= R-D_rrho} 
\end{eqnarray}
Another interesting possibility is to choose $f(r,t)=\rho^n(r,t)$, where $n$ is an integer. In this last case, the anisotropy parameter $\lambda$ is in general dimensionful but the equations of state for $p_\mr$ and $p_\mt$ depend only on thermodynamic quantities, namely on $p_\mr$ and $\rho$, which are all local quantities of the comoving fluid element, a key difference with respect  to the previous model. Using this choice, one obtains that 
\begin{eqnarray}\label{eq:p_r+p_t - f= rho^n}
p_\mathrm{r} & = &\frac{1}{3}\left[\rho-2\lambda \rho^n D_rp_\mr \right],\quad p_\mathrm{t} = \frac{1}{3}\left[\rho+\lambda \rho^n D_rp_\mr \right]\quad\textrm{(pressure gradients)}  \label{eq:p_r+p_t - f= rho^n-D_rp_r} \\
p_\mathrm{r} & = & \frac{1}{3}\left[\rho-2\lambda \rho^nD_r\rho \right],\quad p_\mathrm{t} = \frac{1}{3}\left[\rho+\lambda \rho^n D_r\rho\right]\quad\textrm{(energy dens. gradients).} \label{eq:p_r+p_t - f= rho^n-D_rrho}
\end{eqnarray}
As one can see, from \Eq{eq:p_r+p_t - f= R} and \Eq{eq:p_r+p_t - f= rho^n}, in the limit $\lambda=0$ one reproduces the isotropic limit in which $p_\mathrm{r}=p_\mathrm{t}=\rho/3$. Below, we give the form of the Misner-Sharp equations (\ref{MS-Equations}) in the anisotropic regime where $p_\mr\neq p_\mt$ for $f(r,t)=R(r,t)$ and $f(r,t)=\rho^n(r,t)$.
\begin{itemize}

\item
{
\textbf{$f(r,t)=R(r,t), p_\mathrm{r}-p_\mathrm{t} = -\lambda R D_rp_\mathrm{r}$}

\beq \label{Misner-Sharp Equations with Drpr}
\begin{aligned}
& D_t U = -\frac{\Gamma(1 -2\lambda\Gamma)}{\rho+p_\mathrm{r}}D_r p_\mathrm{r} - \frac{M}{R^2}-4\pi Rp_\mathrm{r} \\
%& \frac{D_\mathrm{t}\rho_0}{\rho_0} = - \frac{1}{\Gamma R^2} \frac{1}{R^2\Gamma}D_\mathrm{r}\left(R^2U\right) \\
& \frac{D_t\rho}{\rho+p_\mathrm{r}} = \frac{D_t\rho_0}{\rho_0} -  2\lambda U \frac{ D_r p_\mathrm{r}}{\rho+p_\mr} \\ 
& \frac{D_rA}{A} = - \frac{(1 -2\lambda\Gamma)}{\rho+p_\mathrm{r}}D_rp_\mathrm{r}  \\ 
& D_t\Gamma = - \frac{U(1 -2\lambda\Gamma)}{\rho+p_\mathrm{r}}D_rp_\mathrm{r} 
\end{aligned}
\eeq

}
\item
{
\textbf{$f(r,t)=R(r,t), p_\mathrm{r}-p_\mathrm{t} = -\lambda R D_r \rho$}

\beq \label{Misner-Sharp Equations with Drrho}
\begin{aligned}
& D_tU = -\frac{\Gamma}{\rho+p_\mathrm{r}} (D_r p_\mathrm{r} -2\lambda\Gamma D_r\rho) - \frac{M}{R^2}-4\pi Rp_\mathrm{r} \\
& \frac{D_t\rho}{\rho+p_\mathrm{r}} =  \frac{D_t\rho_0}{\rho_0} - 2\lambda U \frac{D_r\rho}{\rho+p_\mr}  \\ 
& \frac{D_r A}{A}=-\frac{1}{\rho+p_\mathrm{r}}(D_r p_\mathrm{r}-2\lambda\Gamma D_r\rho) \\ 
& D_t\Gamma = -\frac{U}{\rho+p_\mathrm{r}} (D_rp_\mathrm{r}-2\lambda\Gamma D_r\rho)
\end{aligned}
\eeq
}
\item
{
\textbf{$f(r,t)=\rho^n(r,t), p_\mathrm{r}-p_\mathrm{t} = -\lambda \rho^n(r,t) D_rp_\mr$}

\beq \label{Misner-Sharp Equations with f=rho^n-D_rp_r}
\begin{aligned}
& D_tU = -\frac{\Gamma}{\rho+p_\mathrm{r}} \left(1 -\frac{2\lambda\Gamma\rho^n}{R}\right) D_rp_\mathrm{r} - \frac{M}{R^2}-4\pi Rp_\mathrm{r} \\
& \frac{D_t\rho}{\rho+p_\mathrm{r}} =  \frac{D_t\rho_0}{\rho_0} - \frac{2\lambda U\rho^n}{R} \frac{D_r p_\mr}{\rho+p_\mr}  \\ 
& \frac{D_rA}{A}=-\frac{D_rp_\mathrm{r}}{\rho+p_\mathrm{r}}\left(1-\frac{2\lambda\Gamma\rho^n}{R} \right) \\ 
& D_t\Gamma = -\frac{U}{\rho+p_\mathrm{r}} \left(1-\frac{2\lambda\Gamma\rho^n}{R}\right)D_r p_\mathrm{r}
\end{aligned}
\eeq
}

\item
{
\textbf{$f(r,t)=\rho^n(r,t), p_\mathrm{r}-p_\mathrm{t} = -\lambda \rho^n(r,t) D_r\rho$}

\beq \label{Misner-Sharp Equations with f=rho^n-D_rrho}
\begin{aligned}
& D_tU = -\frac{\Gamma}{\rho+p_\mathrm{r}} \left(D_r p_\mr -\frac{2\lambda\Gamma\rho^n}{R} D_r\rho\right) - \frac{M}{R^2}-4\pi Rp_\mathrm{r} \\
& \frac{D_t \rho}{\rho+p_\mathrm{r}} =  \frac{D_t\rho_0}{\rho_0} - \frac{2\lambda U\rho^n}{R} \frac{D_r\rho}{\rho+p_\mr}  \\ 
& \frac{D_rA}{A}=-\frac{1}{\rho+p_\mathrm{r}}\left(D_r p_\mr-\frac{2\lambda\Gamma\rho^n}{R}D_r\rho \right) \\ 
& D_t\Gamma = -\frac{U}{\rho+p_\mathrm{r}} \left(D_r p_\mr-\frac{2\lambda\Gamma\rho^n}{R}D_r\rho\right)
\end{aligned}
\eeq
}

\end{itemize}

%%%%%%%%%%%%%%%%%%%%%%%%%%%%%%%%  SECTION 3 %%%%%%%%%%%%%%%%%%%%%%%%%%%%

\section{The quasi homogeneous solution }\label{sec:The Quasi Homogeneous Solution}
Having introduced a covariant formulation of the equation of state for a spherically symmetric anisotropic fluid, one should specify the initial conditions for all the relevant quantities describing a cosmological perturbation on superhorizon scales. To do so, let us consider the asymptotic solution of the Einstein’s equations in the limit of $t\rightarrow 0$. This corresponds to a FLRW metric with an $r$ dependent curvature profile $K(r)$ which does not depend on time,
\beq\label{FRLW metric-K(r) form}
\mathrm{d}s^2_\mathrm{AQH} = -\mathrm{d}t^2 + a^2(t)\left[\frac{\mathrm{d}r^2}{1-K(r)r^2}+r^2\mathrm{d}\Omega^2\right].
\eeq

The above solution is often quoted as the Asymptotic Quasi Homogeneous solution (AQH) (as $t\rightarrow 0$)~\cite{Lifshitz:1963ps} and within this formulation of the metric, $K(r)$ can be seen as an initial curvature profile specified on superhorizon scales \footnote{$K(r)$ can be directly linked to the comoving curvature perturbation $\mathcal{R}$ defined in \ref{R definition}. In particular, $\mathcal{R} = \frac{r^2_\mathrm{m}}{2r^2}\left[r^3K(r)\right]^\prime$ ~\cite{Polnarev:2006aa}. %It represents as well the only independent source of cosmological adiabatic perturbations ~\cite{Liddle:2000cg}
}. At this point, it is important to stress out that $K(r)$ corresponds to arbitrarily large metric perturbations while the hydrodynamic perturbations, i.e. energy density and velocity ones, vanish asymptotically as $t\rightarrow 0$ and therefore on such regime they can be treated as small perturbations. One can then solve analytically the hydrodynamic equations and write self-consistently the initial conditions for the energy density, the velocity perturbations and all the relevant variables of the equations as a function of the curvature profile $K(r)$ at a time when the quasi-homogeneous solution is valid at certain order.

Considering perturbations well outside the horizon, all the hydrodynamic and metric quantities are nearly homogeneous and their perturbations are small deviations away from their background value. To parametrise these deviations, we introduce a fictitious parameter $\epsilon$  defined as the ratio between the Hubble radius $H^{-1}$ and the characteristic physical scale, $L$, of the collapsing region,
\beq\label{epsilon definition}
\epsilon(t) = \frac{H^{-1}}{L} = \frac{1}{H(t)a(t)r_\mathrm{m}} \ll 1 \footnote{It is important to mention here that the gradient expansion in terms of a fictitious parameter $\epsilon\ll 1$ is valid only for superhorizon scales and it is used in order to determine the initial conditions for all the metric and hydrodynamic quantities. When later the characteristic scale of the overdensity reenters the Hubble radius and $\epsilon \sim 1$ the gradient expansion breaks down. Thus, once the initial conditions are written by the gradient expansion approximation, they could be evolved using the Misner-Sharp and Misner-Hernadez equations, which can describe the details of the non linear gravitational collapse process.},
\eeq
where $r_\mathrm{m}$ is the comoving characteristic scale of the collapsing region. In this way, all quantities can be written as a power series in $\epsilon$. Thus, considering only the growing mode which is of $O(\epsilon^2)$ in the first non zero term of the expansion ~\cite{Lyth:2004gb,Musco:2018rwt}, at first order one has for the hydrodynamic variables $\rho$, $U$, $p_\mr$, $p_\mt$ and $M$ that~\cite{Polnarev:2006aa}
\begin{equation}\label{Perturbed Hydrodynamical Variables}
\begin{split}
\rho & = \rho_\mathrm{b}(t)\left[1 + \epsilon^2 \tilde{\rho}(r,t)\right] \\
p_\mathrm{r} & = \frac{\rho_\mathrm{b}(t)}{3}\left[1+\epsilon^2\tilde{p}_\mathrm{r}(r,t)\right] \\ 
p_\mathrm{t} & = \frac{\rho_\mathrm{b}(t)}{3}\left[1+\epsilon^2\tilde{p}_\mathrm{t}(r,t)\right] \\ 
U & = H(t)R\left[1+\epsilon^2\tilde{U}(r,t)\right]\\
M & = \frac{4\pi}{3}\rho_\mathrm{b}(t)R^3\left[1+\epsilon^2\tilde{M}(r,t)\right].
\end{split}
\end{equation}
This approach is known in the literature as the long wavelength approach ~\cite{Shibata_1999}, or gradient expansion~\cite{Salopek:1990jq}, or separate universe hypothesis ~\cite{Wands:2000dp,Lyth:2004gb} and reproduces the time evolution of the linear perturbation theory. %allowing for accounting of non linear curvature perturbations as well as long as the spacetime is sufficiently smooth on scales larger than $L_\mathrm{m}$ ~\cite{Lyth:2004gb}. 

Regarding the metric components $A$, $B$ and $R$, which are coupled to the matter ones through Einstein's equations, one can write in the same way that
\begin{equation}\label{Perturbed Metric Components}
\begin{split}
A & = 1 + \epsilon^2 \tilde{A}(r,t) \\
B & = \frac{R^\prime}{\sqrt{1-K(r)r^2}}\left[1+\epsilon^2\tilde{B}(r,t)\right] \\
R & = a(t)r\left[1+\epsilon^2\tilde{R}(r,t)\right].
\end{split}
\end{equation}

The next step is to perform now the perturbative analysis and extract the initial conditions for the hydrodynamic and metric perturbations as a function of the curvature profile $K(r)$. To do so, by performing the gradient expansion at the level of the Misner-Sharp equations (\ref{MS-Equations}), we extract below the equations for the metric and the hydrodynamic perturbations without specifying a specific model describing the difference $p_\mr-p_\mt$. 

We start initially with the metric perturbations $\tilde{A}$, $\tilde{B}$ and $\tilde{R}$. From the definition of $U$ one has that $\dot{R}=AU$. Perturbing this equation by keeping only first order terms, i.e. $\sim O(\epsilon^2)$, one gets for $\tilde{R}$ that it should obey the following equation:
\beq\label{R_tilde cosntraint}
\begin{split}
\dot{a}(1+\epsilon^2\tilde{R})+a\left(\epsilon^2\tilde{R}\right)^{.} & = (1+\epsilon^2\tilde{A})HR(1+\epsilon^2\tilde{U})\Leftrightarrow \\
2\epsilon\dot{\epsilon}\tilde{R}+\epsilon\dot{\tilde{R}} & = \epsilon H(\tilde{A}+\tilde{U})\Leftrightarrow \\ 
2\tilde{R}+\frac{\partial\tilde{R}}{\partial N }  & = \tilde{A}+\tilde{U},
\end{split}
\eeq
where $N=\ln(a/a_\mathrm{ini})$ is the number of e-folds and $a_\mathrm{ini}$ is the scale factor at an initial time. In the last step, we used the fact that $\dot{\epsilon}/\epsilon=H$. %because $\epsilon$ is defined with respect to homogeneous quantities. 
Then, from the $01$ Einstein equation one can easily get that $\frac{\dot{B}}{B}=A\frac{U^\prime}{R^\prime}$. Combining the above equation with $\dot{R}=AU$ we obtain that $\frac{\dot{B}}{B}  - \frac{\dot{R}^\prime}{R^\prime} = - \frac{A^\prime U}{R^\prime}$, which, once perturbed by keeping orders up to $O(\epsilon^2)$, gives the following equation for $\tilde{B}$:
\beq\label{B_tilde constraint}
2\tilde{B}+\frac{\partial\tilde{B}}{\partial N}  = -r \tilde{A}^\prime.
\eeq
The prime $^\prime$ denotes differentiation with respect to the radial comoving coordinate.
Finally, regarding the perturbation of the lapse function $\tilde{A}$, perturbing the equation $\frac{D_rA}{A} = -\frac{1}{\rho+p_\mathrm{r}} \left[ D_r p_\mathrm{r} + \frac{2\Gamma}{R} \left(p_\mathrm{r}-p_\mathrm{t}\right) \right]$ one obtains that
\beq\label{A_tilde constraint}
\tilde{A}^\prime = - \frac{1}{4}\left[\tilde{p}^\prime_\mathrm{r}+\frac{2}{r} \left(\tilde{p}_\mathrm{r}-\tilde{p}_\mathrm{t}\right)\right].
\eeq

We continue the perturbative gradient expansion scheme considering the hydrodynamic perturbations $\tilde{U}$, $\tilde{\rho}$ and $\tilde{M}$. Regarding $\tilde{\rho}$ one perturbs the equation $D_rM=4\pi R^2\Gamma\rho$ which gives
\beq
\frac{1}{3}(1+\epsilon^2\tilde{M})\left[3\frac{R^\prime}{R}+\epsilon^2\tilde{M}^\prime\right]  = (1+\epsilon^2\tilde{\rho})\frac{R^\prime}{R}\Leftrightarrow 
\eeq
\beq\label{rho_tilde constraint}
\tilde{\rho} =\frac{1}{3r^2}\left(r^3\tilde{M}\right)^\prime.
\eeq
Then, the equation for $\tilde{M}$, obtained by perturbing $D_tM=-4\pi R^2 Up_\mathrm{r}$, reads as
\beq
(1+\epsilon^2\tilde{M})\left[\frac{\dot{\rho}_\mathrm{b}}{\rho_\mathrm{b}} + 3\frac{\dot{R}}{R} + \left(\epsilon^2\tilde{M}\right)^{.}\right] = -\frac{\dot{R}}{R}(1+\epsilon^2\tilde{p}_\mathrm{r})\Leftrightarrow
\eeq
\beq\label{M_tilde constraint}
\tilde{M}+\frac{\partial\tilde{M}}{\partial N}=-4\tilde{U}-4\tilde{A}-\tilde{p}_\mathrm{r}.
\eeq
Regarding $\tilde{U}$, by perturbing \Eq{constraint equation} one gets that
\beq\label{U_tilde constraint}
\tilde{U}=\frac{1}{2}\left[\tilde{M}-K(r)r_\mathrm{m}^2\right].
\eeq

To summarize, the differential equations describing the behavior of the metric and hydrodynamic perturbations in the gradient expansion approximation in which $\epsilon\ll 1$ are:
\beq\label{eq:metric+hydrodynamic perturbations}
\begin{aligned}
2\tilde{R}+\frac{\partial\tilde{R}}{\partial N }  & =  \tilde{A}+\tilde{U} \\
2\tilde{B}+\frac{\partial\tilde{B}}{\partial N} & =  -r \tilde{A}^\prime \\
\tilde{A}^\prime & =  - \frac{1}{4}\left[\tilde{p}^\prime_\mathrm{r}+\frac{2}{r} \left(\tilde{p}_\mathrm{r}-\tilde{p}_\mathrm{t}\right)\right] \\
\tilde{\rho} & = \frac{1}{3r^2}\left(r^3\tilde{M}\right)^\prime \\
\tilde{M}+\frac{\partial\tilde{M}}{\partial N} & = -4\tilde{U}-4\tilde{A}-\tilde{p}_\mathrm{r} \\
\tilde{U} & = \frac{1}{2}\left[\tilde{M}-K(r)r_\mathrm{m}^2\right].
\end{aligned}
\eeq
As one may see from the above equations, the only place, in which the dependence on the prescription modeling the difference $p_\mr-p_\mt$ enters, is at the level of the differential equation for the lapse function perturbation, $\tilde{A}$.

\section{The initial conditions in presence of anisotropies}
We extract below the initial conditions of the hydrodynamic and metric perturbations as a function of the time-independent curvature profile $K(r)$ by specifying our choice for the EoS of an anisotropic radiation dominated medium. 
\subsection{Equation of state in terms of pressure gradients}\label{sec:Equation of state in terms of pressure gradients}
We choose here the EoS where the difference $p_\mr-p_\mt$ is proportional to pressure gradients [See \Eq{D_rp_r}]. Regarding the free function $f(r,t)$ we choose it to be either $f(r,t)=R(r,t)$ (which is a non local quantity) or $f(r,t)=\rho^n(r,t)$ (which is a local quantity). 

\subsubsection{$f(r,t)=R(r,t)$ } \label{sec:Drpr-f=R}
In this case, the anisotropy parameter $\lambda$ is dimensionless and therefore one does not need to introduce a characteristic scale at the level of the equation of state. Using the EoS \eqref{eq:p_r+p_t - f= R-D_rp_r}, with a straightforward calculation one has that the constraint equation for $\tilde{p}_\mr$ reads as
\beq\label{p_tilde-rho_tilde}
\tilde{p}_\mathrm{r}-\tilde{\rho}=-\frac{2\lambda r}{3}\sqrt{1-K(r)r^2}\tilde{p}^{\prime}_\mathrm{r} = -\frac{2\lambda r}{3}F^\prime(r),
\eeq
where 
\beq\label{F(r) definition}
F(r)\equiv \int_{0}^{r}\sqrt{1-K(r^\prime)r^{\prime2}}\tilde{p}^{\prime}_\mathrm{r}(r^\prime)\mathrm{d}r^\prime = -\frac{3}{2\lambda}\int^{r}_{0}\frac{\tilde{p}_\mathrm{r}-\tilde{\rho}}{r^\prime}\mathrm{d}r^\prime,
\eeq
and the corresponding equation for the lapse perturbation $\tilde{A}$ reads as
\beq\label{A_tilde constraint - Drpr}
\tilde{A}^\prime = - \frac{1}{4}\left(\tilde{p}^\prime_\mathrm{r} - 2\lambda\sqrt{1-K(r)r^2}\right).
\eeq

These equations coupled with \Eq{eq:metric+hydrodynamic perturbations} allows to find the explicit dependence of the initial perturbation profiles on the curvature profile $K(r)$. Let us start with the metric perturbations $\tilde{A}$, $\tilde{R}$ and $\tilde{B}$. Integrating \Eq{A_tilde constraint - Drpr} and using the fact that $\tilde{\rho}(0)=\tilde{p}_\mathrm{r}(0)$, as it can be seen by \Eq{p_tilde-rho_tilde}, one can infer that
\beq\label{A_tilde}
\tilde{A} - \tilde{A}(0)=-\frac{1}{4}\tilde{p}_\mathrm{r} + \frac{\tilde{\rho}(0)}{4}+\frac{\lambda}{2}F(r).
\eeq
At $r=\infty$, where $\tilde{p}_\mathrm{r}(\infty)=0$, from the above equation one has that 
\beq\label{A tilde boundary condition}
\tilde{A}(\infty) = 0 = \tilde{A}(0)+\frac{\tilde{\rho}(0)}{4}+\frac{\lambda}{2}F(\infty)\Leftrightarrow  \tilde{A}(0)= -\frac{1}{4}\left[\tilde{\rho}(0)+2\lambda F(\infty)\right].
\eeq
Thus, plugging \Eq{A tilde boundary condition} into \Eq{A_tilde} and taking into account \Eq{p_tilde-rho_tilde}, one gets that
\beq\label{A_tilde in tems of F}
\tilde{A} =-\frac{\tilde{\rho}}{4} + \frac{\lambda}{2}\left[\frac{r\mathcal{F}^\prime(r)}{3}+\mathcal{F}(r)\right],
\eeq
where we have introduced the function $\mathcal{F}(r)$ defined as
\beq \label{F_r}
\mathcal{F}(r)\equiv F(r)- F(\infty)=-\int_r^{\infty}\frac{\partial\tilde{p}_\mathrm{r}(r^\prime)}{\partial r^\prime}\sqrt{1-K(r^\prime)r^{\prime 2}}\mathrm{d}r^\prime.
\eeq
Considering now \Eq{R_tilde cosntraint} for $\tilde{R}$, one can crearly see that the right-hand side is time-independent because of \Eq{A_tilde in tems of F} and \Eq{U_tilde constraint} allowing $\tilde{R}$ to be written as
\beq\label{R_tilde}
\tilde{R} =\frac{1}{2} \left(\tilde{A}+\tilde{U}\right)=- \frac{1}{2}\left\{\frac{\tilde{\rho}}{4} - \frac{\lambda}{6}\left[r\mathcal{F}^\prime(r) + \mathcal{F}(r)\right]+\frac{1}{6}K(r)r_\mathrm{m}^2\right\}
\eeq
As for $\tilde{B}$, following the same reasoning, one can see that 
\beq\label{B_tilde}
\tilde{B}=-\frac{r\tilde{A}^\prime}{2}=\frac{r}{8}\tilde{p}^\prime_\mathrm{r}(1-2\lambda\sqrt{1-K(r)r^2}) = \frac{r}{8}F^\prime(r)\left[\frac{1}{\sqrt{1-K(r)r^2}}-2\lambda\right].
\eeq

To work out the expression for $\tilde{M}$ one can combine Eqs. (\eqref{M_tilde constraint}, \eqref{A_tilde} , \eqref{p_tilde-rho_tilde} and \eqref{U_tilde constraint} ) and obtain that
\beq\label{M_tilde_1}
\tilde{M} +\frac{1}{3}\frac{\partial\tilde{M}}{\partial N}=\frac{2}{3}K(r)r_\mathrm{m}^2 - \frac{2}{3}\lambda\mathcal{F}(r).
\eeq
Following the same reasoning as in the case of $\tilde{R}$, given the time-independent nature of the right-hand side of \Eq{M_tilde_1} one can deduce that 
\beq\label{M_tilde}
\tilde{M} = \frac{2}{3}K(r)r_\mathrm{m}^2 - \frac{2}{3}\lambda\mathcal{F}(r).
\eeq
Considering the behavior of $\tilde{U}$, one can plug \Eq{M_tilde} into \Eq{U_tilde constraint} to get that
\beq\label{U_tilde}
\tilde{U}=-\frac{1}{6}K(r)r_\mathrm{m}^2 - \frac{\lambda}{3}\mathcal{F}(r).
\eeq
Finally, by plugging \Eq{M_tilde} into \Eq{rho_tilde constraint} one obtains the expression for the energy density,
\beq\label{rho_tilde}
\tilde{\rho}=\frac{2}{3}\left\{\frac{\left[r^3K(r)\right]^\prime}{3r^2}r^2_\mathrm{m} -\lambda\left[\frac{r}{3}\mathcal{F}^\prime(r)+  \mathcal{F}(r)\right]\right\}.
\eeq

One can further simplify the expressions for $\tilde{A}$, $\tilde{R}$, $\tilde{B}$, $\tilde{M}$, $\tilde{\rho}$ and $\tilde{U}$ writing them in a more compact form. To do so, we introduce the following effective curvature profile,
\beq\label{Effective Curvature Profile-Drpr}
\mathcal{K}(r) \equiv K(r) - \frac{\lambda}{r^2_\mathrm{m}}\mathcal{F}(r).
\eeq
This allows to write the quasi-homogeneous solution, in a similar form as the isotropic case ($\lambda=0$), introducing the effective energy density and velocity perturbations $\tilde{\rho}_\mathrm{eff}$ and  $\tilde{U}_\mathrm{eff}$ defined as
\beq\label{rho_tilde_eff}
\tilde{\rho}_\mathrm{eff} = \tilde{\rho}-2\lambda\left[\frac{r\mathcal{F}^\prime(r)}{3}+\mathcal{F}(r)\right],
\eeq
\beq\label{U__tilde_eff}
\tilde{U}_\mathrm{eff} = \tilde{U}+\frac{\lambda}{2}\mathcal{F}(r) = -\frac{1}{6}\mathcal{K}(r)r^2_\mathrm{m}
\eeq

In this way, the metric and the hydrodynamic perturbations are given by the following expressions:
\begin{eqnarray}\label{Perturbations}
\begin{split}
\tilde{A} & = - \frac{\tilde{\rho}}{4} +\frac{\lambda}{2}\left[\frac{r\mathcal{F}^\prime(r)}{3}+\mathcal{F}(r) \right]  = -\frac{\tilde{\rho}_\mathrm{eff}}{4}\\
\tilde{B} &  = \frac{r}{8}\mathcal{F}^\prime(r)\left[\frac{1}{\sqrt{1-K(r)r^2}}-2\lambda\right]  = \frac{r}{8}\tilde{\rho}^\prime_\mathrm{eff}\\
\tilde{\rho} & =\frac{2}{3}\frac{\left[r^3\mathcal{K}(r)\right]^\prime}{3r^2}r^2_\mathrm{m}  \\
\tilde{U} & =-\frac{1}{6}\mathcal{K}(r)r_\mathrm{m}^2 - \frac{\lambda}{2}\mathcal{F}(r) = \tilde{U}_\mathrm{eff} - \frac{\lambda}{2}\mathcal{F}(r)\\ 
\tilde{M} &=\frac{2}{3}\mathcal{K}(r)r_\mathrm{m}^2 = -4\tilde{U}_\mathrm{eff}  \\
\tilde{R} & = -\frac{\tilde{\rho}_\mathrm{eff}}{8} + \frac{\tilde{U}}{2} .
\end{split}
\end{eqnarray}

To complete this derivation we need to find the function $\mathcal{F}(r)$ defined in \Eq{F_r}, which is the integral of the pressure-gradient profile $\frac{\partial\tilde{p}_\mathrm{r}}{\partial r}$, corrected by $\Gamma \simeq \sqrt{1-K(r)r^2}$ on super horizon scales, which is measuring the geometrical curvature of the space time. In particular, by combining Eqs. (\ref{p_tilde-rho_tilde}) and (\ref{rho_tilde}) one obtains the following integral equation for $\tilde{p}_\mathrm{r}$
\beq\label{p_r integral}
\tilde{p}_\mathrm{r}=\frac{2}{3}\left\{\frac{\left[r^3K(r)\right]^\prime}{3r^2}r^2_\mathrm{m}  -\lambda\left[ \mathcal{F}(r)  +\frac{4}{3} r f(r)\right]\right\}.
\eeq
where 
\[ f(r)\equiv\frac{\partial\tilde{p}_\mathrm{r}}{\partial r}\sqrt{1-K(r)r^2}\,. \]
By differentiating the above equation we can write the following differential equation,
\beq\label{p_r ODE compact form}
\begin{aligned}
\frac{8\lambda}{9}r\sqrt{1-K(r)r^2}f^\prime(r) &+ \left[\frac{14\lambda}{9}\sqrt{1-K(r)r^2} +1\right]f(r)  \\
&- \frac{2}{3}\left\{\frac{\left[r^3K(r)\right]^\prime}{3r^2}\right\}^\prime r^2_\mathrm{m} \sqrt{1-K(r)r^2} = 0,
\end{aligned}
\eeq
with $f(0)=0$ and for $\lambda=0$ we recover the quasi-homogeneous limit,
\beq\label{f(r) - lambda =0}
f_{\lambda=0}(r)= \frac{2}{3}\left\{\frac{\left[r^3K(r)\right]^\prime}{3r^2}\right\}^\prime r^2_\mathrm{m} \sqrt{1-K(r)r^2}.
\eeq
Solving \Eq{p_r ODE compact form} one can extract the profile of the pressure gradients $\frac{\partial\tilde{p}_\mathrm{r}}{\partial r}$ that inserted into \Eq{F_r} allows us to modify appropriately the metric and hydrodnamical perturbations in the case of an anisotropic fluid described by the equation of state given by \Eq{eq:p_r+p_t - f= R-D_rp_r}.

Below, we show the pressure gradient profiles for both positive and negative values of $\lambda$, as well the behavior of the the energy density and velocity perturbations' profiles. In the figures below, we make the simplest choice specifying $K(r)$ with a Gaussian profile of the form,
\beq
K(r) = \mathcal{A}e^{-(r/r_\mathrm{m,0})^2},
\eeq
with $r_\mathrm{m,0}=1$ and $\mathcal{A}=\frac{3e}{4r^2_\mathrm{m,0}}$ \footnote{The value of $\mathcal{A}=\frac{3e}{4r^2_\mathrm{m,0}}$ chosen corresponds to the threshold for the isotropic case, where the averaged perturbation amplitude defined in \Eq{delta_m definition} is equal to $\delta_\mathrm{m,iso}=0.5$ ~\cite{Musco:2018rwt}.}. Regarding the value of $r_\mathrm{m}$, we take $r_\mathrm{m}=r_\mathrm{m,0}=1$, since $r_\mathrm{m}\simeq r_\mathrm{m,0}$ for any value of $\lambda$ considered here. 
\begin{figure}[h!]
\begin{center}
\includegraphics[height = 0.35\textwidth, width=0.496\textwidth, clip=true]{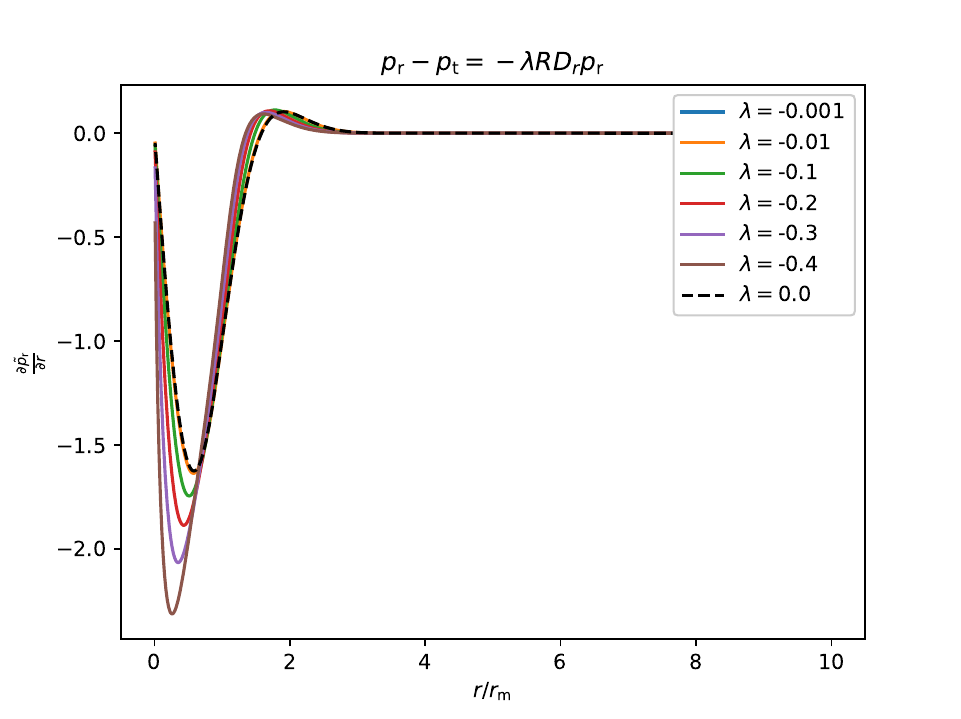}
\includegraphics[height = 0.35\textwidth, width=0.496\textwidth, clip=true]{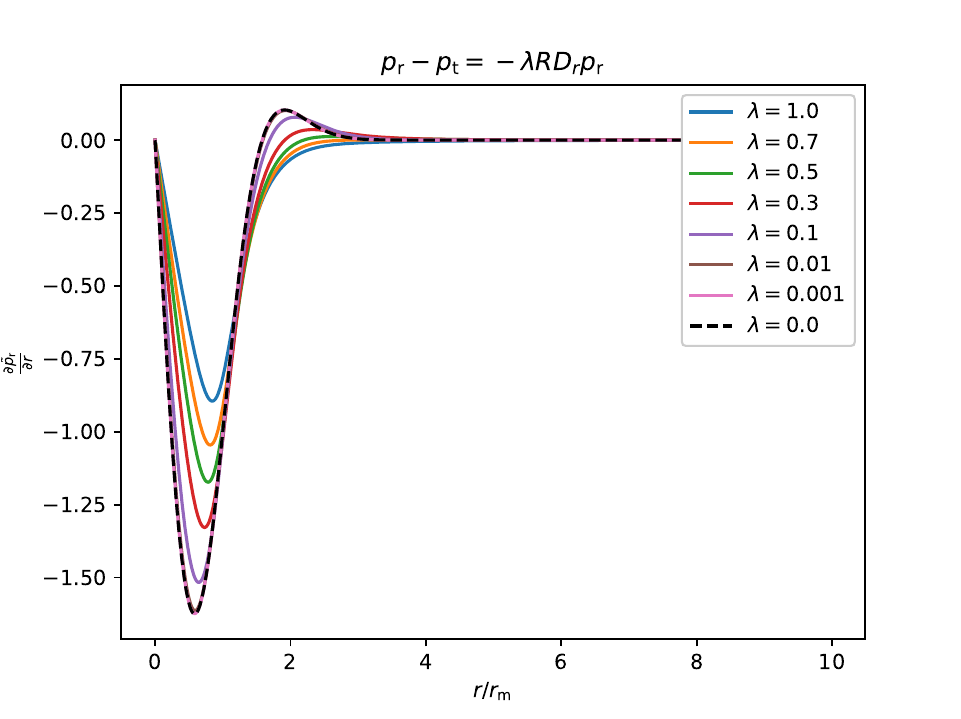}
\caption{In this figure, we show the behavior of $\frac{\partial\tilde{p}_\mr}{\partial r}$ against $r/r_\mathrm{m}$. In the left panel we are considering negative values of the anisotropy parameter $\lambda$ whereas in the right panel we account for positive values of $\lambda$.}
\label{fig:dpr_dr_pr_f=R}
\end{center}
\end{figure}
\begin{figure}[h!]
\begin{center}
\includegraphics[height = 0.35\textwidth, width=0.496\textwidth, clip=true]{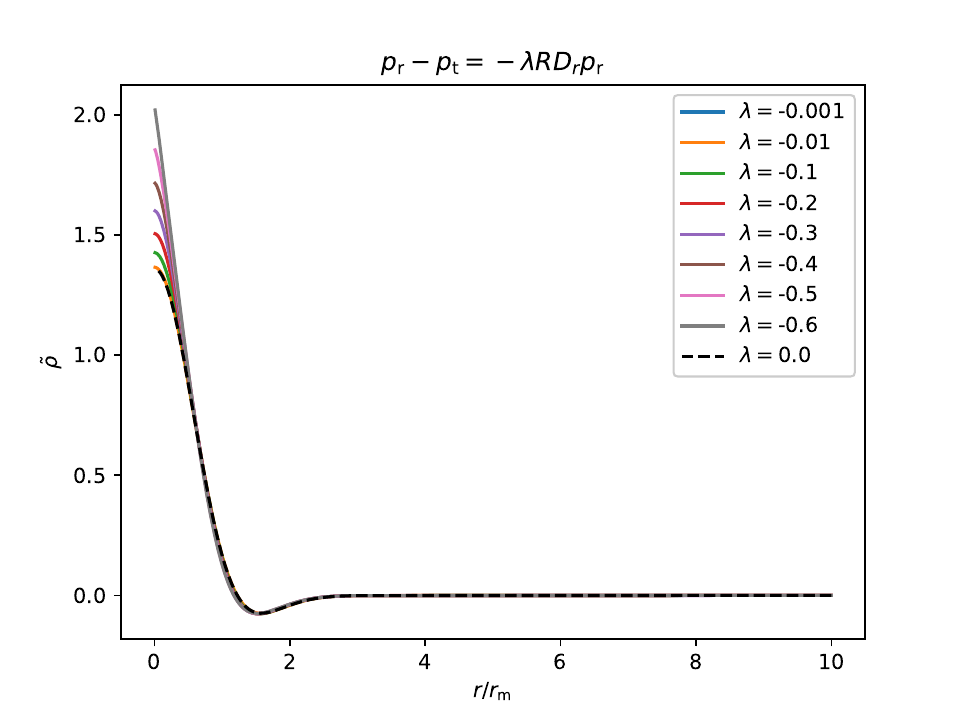}
\includegraphics[height = 0.35\textwidth, width=0.496\textwidth, clip=true]{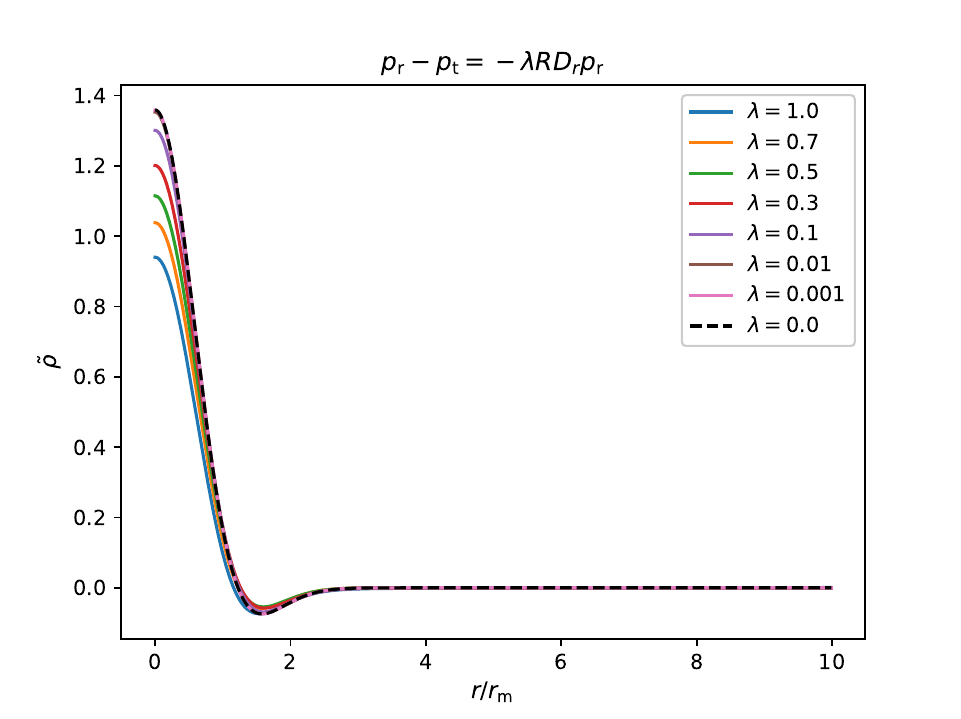}
\caption{In this figure, we show the behavior $\tilde{\rho}$ against $r/r_\mathrm{m}$. In the left panel we are considering negative values of the anisotropy parameter $\lambda$ whereas in the right panel we account for positive values of $\lambda$.}
\label{fig:rho_tilde_pr_f=R}
\end{center}
\end{figure}
\begin{figure}[h!]
\begin{center}
\includegraphics[height = 0.35\textwidth, width=0.496\textwidth, clip=true]{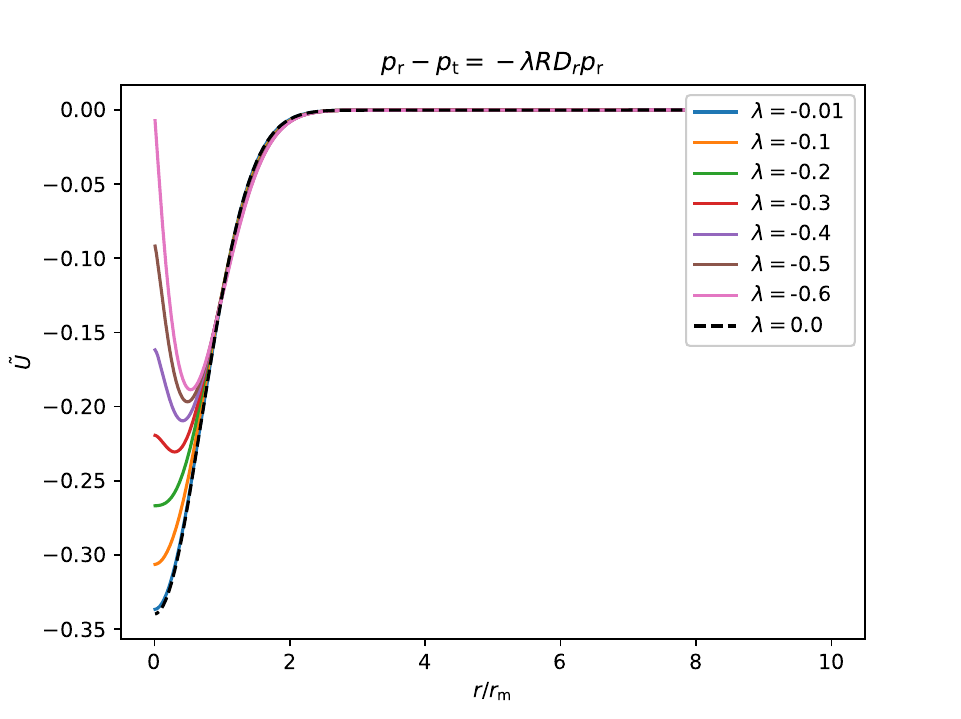}
\includegraphics[height = 0.35\textwidth, width=0.496\textwidth, clip=true]{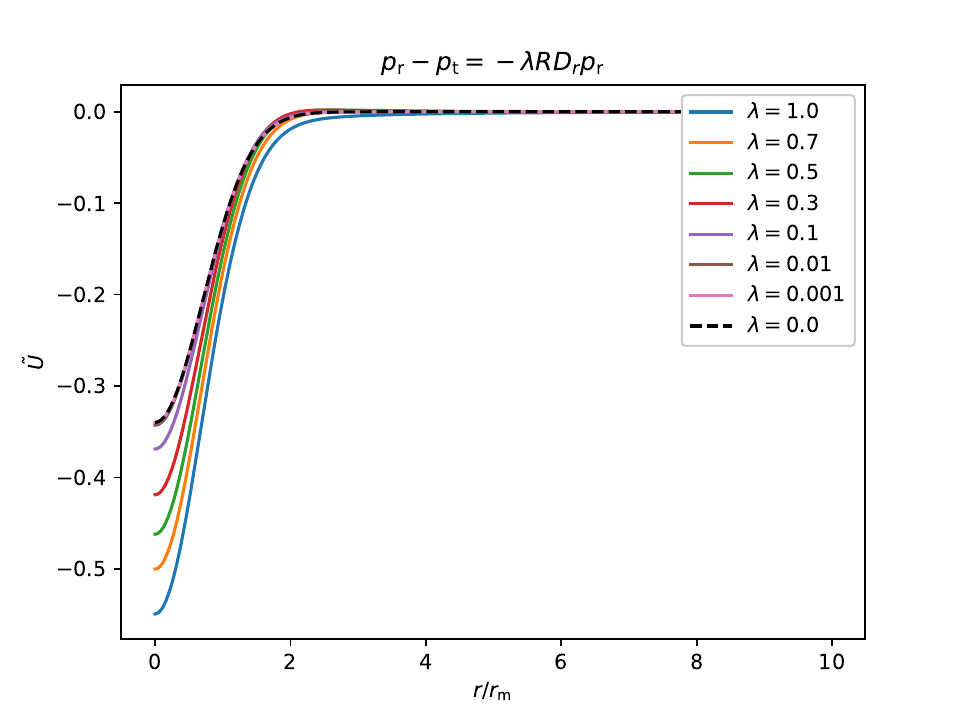}
\caption{In this figure, we show the behavior of $\tilde{U}$ against $r/r_\mathrm{m}$. In the left panel we are considering negative values of the anisotropy parameter $\lambda$ whereas in the right panel we account for positive values of $\lambda$.}
\label{fig:U_tilde_pr_f=R}
\end{center}
\end{figure}
\begin{figure}[h!]
\begin{center}
\includegraphics[height = 0.35\textwidth, width=0.496\textwidth, clip=true]{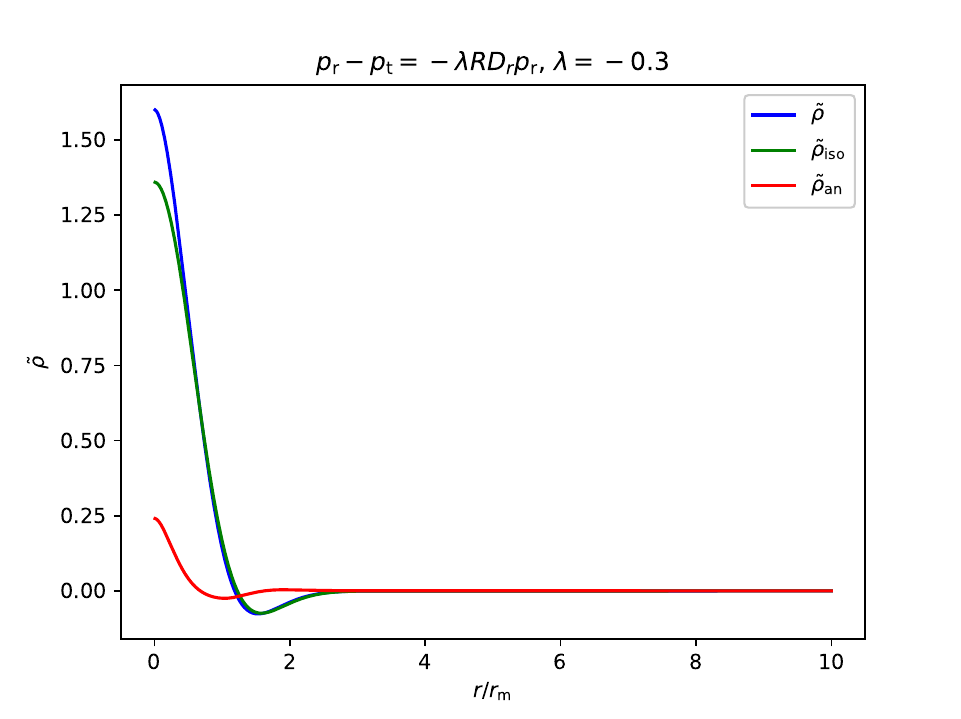}
\includegraphics[height = 0.35\textwidth, width=0.496\textwidth, clip=true]{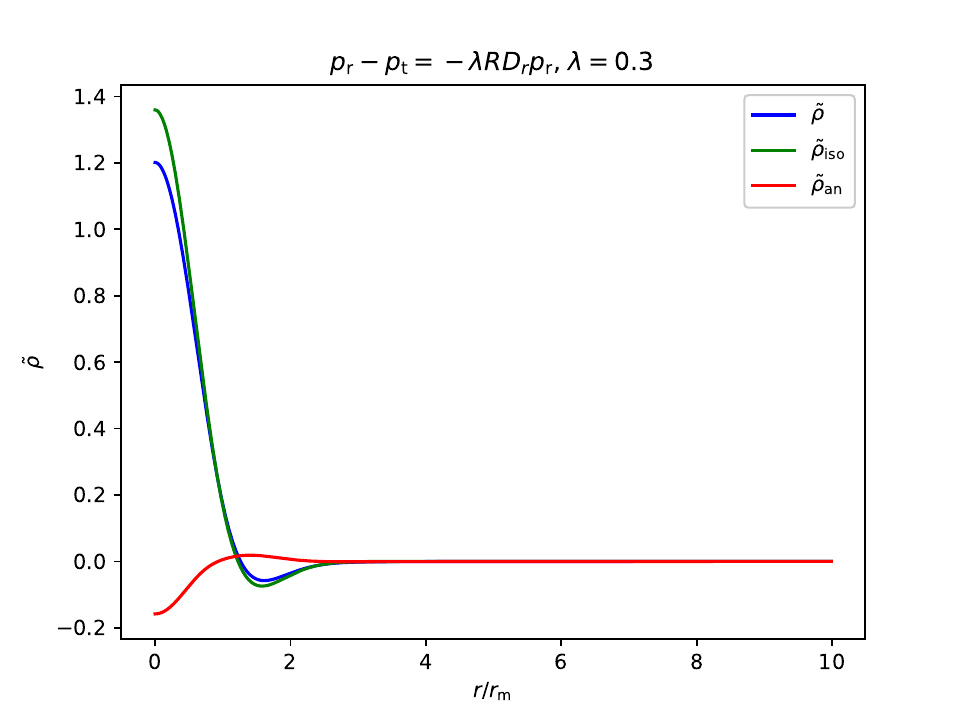}
\caption{The isotropic and anisotropic contributions to the energy density perturbation. In the left panel, we show the case for $\lambda=-0.3$ whereas in the right panel we show the case for $\lambda=0.3$.}
\label{fig:rho_tilde - lambda= constant - pr - f=R}
\end{center}
\end{figure}

As one may see from the above figures, as $\lambda\to 0$ one can clearly see a convergence to the isotropic case and discriminate between two regimes, corresponding to positive and negative values of $\lambda$. When $\lambda<0$, given the fact that the pressure gradient profile is mainly negative, then from $p_\mathrm{r}-p_\mathrm{t}=-\lambda RD_\mathrm{r}p_\mathrm{r}$, one has that $p_\mr<p_\mt$. In this case, since the radial pressure is reduced compared to the tangential one,  one would expect to be easier for a cosmological perturbation to collapse along the radial direction with respect to what one has in the isotropic case. Consistently, one sees that the peaks of the energy density and velocity perturbations are enhanced compared to the case when $\lambda=0$. This behavior can be clearly seen from \Fig{fig:dpr_dr_pr_f=R} and \Fig{fig:U_tilde_pr_f=R}. On the other hand, when $\lambda>0$, we have $p_\mr>p_\mt$ with this larger value of the radial component of the pressure acting against the gravitational collapse compared to the isotropic case. In this case, the amplitude of the energy density and velocity perturbations are therefore reduced with respect to the isotropic ones [See \Fig{fig:dpr_dr_pr_f=R} and \Fig{fig:U_tilde_pr_f=R}]. 

To see this effect more explicitly, we consider a positive and a negative value of $\lambda$, displaying the isotropic and the anisotropic contributions of the energy density perturbation profile in \Fig{fig:rho_tilde - lambda= constant - pr - f=R}. As it can be seen from this figure, when $\lambda=-0.3$ the anisotropy has a positive contribution to the overall amplitude of the energy density perturbation, enhancing it with respect to the isotropic case. On the other hand, when $\lambda=0.3$, the anisotropic contribution reduces the amplitude of the energy density perturbation.

Finally, one should also notice from the left panel of \Fig{fig:dpr_dr_pr_f=R} that for a value of $\lambda$ which is smaller than a critical value, the pressure gradient profile does not converge to zero at the origin as expected from the boundary condition given by \Eq{boundary condition}. This behavior is explained in detail in \App{Lower Limit on lambda<0}, where one can see that due to the mathematical structure of \Eq{p_r ODE compact form}, the radial derivative of $\tilde{p}_\mr$ diverges at $r=0$ when $\lambda<\lambda_\mathrm{c}=-9/14$.

\newpage
\subsubsection{$f(r,t)=\rho^n(r,t)$} \label{sec:Drpr-f=rho^n}
We are going to consider now the EoS when $f(r,t)=\rho^n(r,t)$, as given by \Eq{eq:p_r+p_t - f= rho^n-D_rp_r}. The only difference compared to the case where $f(r,t)=R(r,t)$ is at the level of the differential equation for $\tilde{A}$ and $\tilde{p_\mathrm{r}}-\tilde{\rho}$. In particular, one finds that
\begin{eqnarray}
\tilde{A}^\prime &=& - \frac{1}{4}\tilde{p}^\prime_\mathrm{r}\left[1-\frac{2\lambda\rho^n_\mathrm{b}}{ar}\sqrt{1-K(r)r^2}\right]\label{A_tilde constraint - f=rho^n}\\
\tilde{p}_\mathrm{r}-\tilde{\rho}&=&-\frac{2\lambda\rho^n_\mathrm{b}}{3ar_\mathrm{m}}\sqrt{1-K(r)r^2}\tilde{p_\mr}^{\prime} = -\frac{2\lambda\rho^n_\mathrm{b}}{3ar_\mathrm{m}} r H^\prime(r)\label{p_tilde-rho_tilde - f=rho^n}, \\
\end{eqnarray}
where
\beq\label{H(r) definition}
H(r)\equiv r_\mathrm{m}\int_{0}^{r}\frac{\sqrt{1-K(r^\prime)r^{\prime2}}}{r^\prime}\tilde{p_\mr}^{\prime}(r^\prime)\mathrm{d} r^\prime.
\eeq
Following the same reasoning as before one obtains for $\tilde{A}$ that
\beq\label{A_tilde - f=rho^n}
\tilde{A}=-\frac{\tilde{\rho}}{4}+\frac{\lambda\rho^n_\mathrm{b}}{2ar_\mathrm{m}}\left[\mathcal{H}(r) + \frac{r}{3}\mathcal{H}^\prime(r)\right],
\eeq
where $\mathcal{H}(r)$ is defined as
\beq\label{curly H definition}
\mathcal{H}(r) \equiv -r_\mathrm{m}\int_{r}^{\infty}\frac{\sqrt{1-K(r^\prime)r^{\prime2}}}{r^\prime}\tilde{p_\mr}^{\prime}(r^\prime)\mathrm{d}r^\prime.
\eeq

Considering $\tilde{M}$, by plugging \Eq{U_tilde constraint}  and \Eq{A_tilde - f=rho^n} into \Eq{M_tilde constraint} and taking into account \Eq{p_tilde-rho_tilde - f=rho^n} one finds that $\tilde{M}$ satisfies the following differential equation:
\beq
\tilde{M}  + \frac{1}{3}\frac{\partial\tilde{M}}{\partial N} =  \frac{2}{3}K(r)r^2_\mathrm{m} -\frac{2\lambda}{3}\frac{\rho^n_\mathrm{b}(N)}{a(N)r_\mathrm{m}}\mathcal{H}(r).
\eeq

From the above equation, given the time-dependence of the right-hand side, one can separate the variables $(r,N)$ and the explicit profile of $\tilde{M}$ can be recast in the following way: 
\beq\label{M_tilde-f=rho^n-(r,t) separated}
\tilde{M} = \frac{2}{3}K(r)r^2_\mathrm{m} + \Phi_{p_\mr}(N)\mathcal{H}(r),
\eeq
where the time-dependent function $\Phi_{p_\mr}(N)$ obeys the following equation:
\beq
\Phi_{p_\mr}^\prime(N) + 3\Phi_{p_\mr}(N) = -2\lambda\frac{\rho^n_\mathrm{b}(N)}{a(N)r_\mathrm{m}}.
\eeq
At this point, one should stress out that $\tilde{M}$ written in the form of \Eq{M_tilde-f=rho^n-(r,t) separated} is the sum of a time-independent isotropic term, $\frac{2}{3}K(r)r^2_\mathrm{m}$, plus a time-dependent anisotropic one, $\Phi_{p_\mr}(N)\mathcal{H}(r)$. The differential equation for $\Phi_{p_\mr}$ can be integrated analytically and it has the following solution:
\beq
\Phi_{p_\mr}(N) = e^{-3N}\left[c -\frac{\lambda\rho^n_\mathrm{b,ini}}{a_\mathrm{ini}r_\mathrm{m}(1-2n)}e^{-(4n-2)N} \right].
\eeq
By demanding that $\Phi_{p_\mr}(N=0)=0$, since one expects that at initial time the anisotropy vanishes, one gets that $c= \frac{\lambda\rho^n_\mathrm{b,ini}}{a_\mathrm{ini}r_\mathrm{m}(1-2n)}$. Here one should notice that $a=a_\mathrm{ini}\Leftrightarrow N=0$ corresponds to an initial time when the perturbations are generated and at which one can consider that the medium is still  isotropic as one expects at the end of inflation Thus, hereafter we consider that $a_\mathrm{ini}=a_\mathrm{inf}$ . At the end, taking into account the definition of $N=\ln\left(a/a_\mathrm{ini}\right)$ measuring the number of e-folds, $\Phi_{p_\mr}$ can be written  in terms of the scale factor $a$ as
\beq\label{Phi analytical}
\Phi_{p_\mr}(a) = \frac{\lambda\rho^n_\mathrm{b,inf}}{a_\mathrm{inf}r_\mathrm{m}(1-2n)}\left(\frac{a}{a_\mathrm{inf}}\right)^{-3}\left[1-\left(\frac{a}{a_\mathrm{inf}}\right)^{-(4n-2)}\right],
\eeq
where the index ``$\mathrm{inf}$'' stands for the time at the end of inflation.
From \Eq{Phi analytical} one may say that for $n=1/2$ there is an indefinite value for $\Phi_{p_\mr}$ of the form $(0/0)$. However, this is not the case as one can see in \App{Phi_pr+I_1+1_2} in which we give $\Phi_{p_\mr}$ in the limit $n\rightarrow 1/2$.

Consequently, plugging \Eq{M_tilde-f=rho^n-(r,t) separated} into \Eq{rho_tilde constraint} and \Eq{U_tilde constraint}, one obtains for the energy density, velocity and lapse perturbations the following form:
\begin{align}
\tilde{\rho} & = \Phi_\mathrm{iso}\frac{1}{3r^2}\left[r^3K(r)\right]^\prime r^2_\mathrm{m} + \Phi_{p_\mr}(a)\frac{1}{3r^2}\left[r^3\mathcal{H}(r)\right]^\prime \label{rho_tilde_full-f=rho^n}\\
\tilde{U} & = \frac{1}{2}\left[ \left(\Phi_\mathrm{iso} -1 \right)K(r)r^2_\mathrm{m} + \Phi_{p_\mr}(a)\mathcal{H}(r)\right]  \label{U_tilde_full-f=rho^n}\\
\tilde{A} & =  -\frac{1}{4}\left\{ \Phi_\mathrm{iso}\frac{1}{3r^2}\left[r^3K(r)\right]^\prime r^2_\mathrm{m} + \Phi_{p_\mr}(a)\frac{1}{3r^2}\left[r^3\mathcal{H}(r)\right]^\prime \right\} +\frac{\lambda\rho^n_\mathrm{b}(a)}{2ar_\mathrm{m}}\frac{\left[r^3\mathcal{H}(r)\right]^\prime}{3r^2}\label{A_tilde_full-f=rho^n},
\end{align}
where $\Phi_\mathrm{iso}=2/3$ and one  can clearly identify the isotropic and anisotropic contributions. 

Considering now the behavior of $\tilde{R}$ and $\tilde{B}$, by plugging \Eq{A_tilde_full-f=rho^n} and \Eq{U_tilde_full-f=rho^n} into \Eq{R_tilde cosntraint} and \Eq{B_tilde constraint} one gets the following differential equations for $\tilde{R}$ and $\tilde{B}$.
\begin{align}%\label{tilde R and B equations - f=rho^n}
\begin{split}
2\tilde{R}+\frac{\partial\tilde{R}}{\partial N }  & =- \frac{\Phi_\mathrm{iso}}{4}\frac{1}{3r^2}\left[r^3K(r)\right]^\prime r^2_\mathrm{m} +  \left(\Phi_\mathrm{iso}-1\right)\frac{K(r)r^2_\mathrm{m}}{2} \\ & - \left[\frac{\Phi_{p_\mr}(N)}{4}-\frac{\lambda\rho^n_\mathrm{b}(N)}{2a(N)r_\mathrm{m}}\right] \frac{\left[r^3\mathcal{H}(r)\right]^\prime}{3r^2}  +  \frac{\Phi_{p_\mr}(N)}{2}\mathcal{H}(r)
\end{split}  \label{tilde R equation - f=rho^n}\\
\begin{split}
2\tilde{B}+\frac{\partial\tilde{B}}{\partial N} & = \frac{r}{4} \Phi_\mathrm{iso} \left\{\frac{1}{3r^2}\left[r^3K(r)\right]^\prime r^2_\mathrm{m} \right\}^\prime +  r \left[\frac{\Phi_{p_\mr}(N)}{4}-\frac{\lambda\rho^n_\mathrm{b}(N)}{2a(N)r_\mathrm{m}}\right] \left\{\frac{1}{3r^2}\left[r^3\mathcal{H}(r)\right]^\prime\right\}^\prime.
\end{split} \label{tilde B equation - f=rho^n}
\end{align}
The above system of equations can be solved defining two new free functions $I_{1,p_\mr}(N)$ and $I_{2,p_\mr}(N)$ and writing the solutions for $\tilde{R}$ and $\tilde{B}$ as following:
\begin{align}
\tilde{R} & = - I_{1,\mathrm{iso}}\frac{1}{3r^2}\left[r^3K(r)\right]^\prime r^2_\mathrm{m} + I_{2,\mathrm{iso}}\frac{K(r)r^2_\mathrm{m}}{2}  - I_{1,p_\mr}(N) \frac{\left[r^3\mathcal{H}(r)\right]^\prime}{3r^2}  + I_{2,p_\mr}(N)\mathcal{H}(r) \label{tilde R parametrisation - f=rho^n} \\
\tilde{B} & = I_{1,\mathrm{iso}}r\left\{\frac{1}{3r^2}\left[r^3K(r)\right]^\prime r^2_\mathrm{m}\right\}^\prime  + I_{1,p_\mr}(N)r \left\{\frac{\left[r^3\mathcal{H}(r)\right]^\prime}{3r^2}\right\}^\prime  \label{tilde  B parametrisation - f=rho^n} ,
\end{align}
where $I_{1,\mathrm{iso}} = 1/12$ and $I_{2,\mathrm{iso}}=-1/6$. By plugging now \Eq{tilde R parametrisation - f=rho^n} and \Eq{tilde  B parametrisation - f=rho^n} into \Eq{tilde R equation - f=rho^n} and  \Eq{tilde B equation - f=rho^n} one obtains the differential equations for $I_{1,p_\mr}(N)$ and $I_{2,p_\mr}(N)$ which read as
\begin{align}
I^\prime_{1,p_\mr}(N) + 2I_{1,p_\mr}(N) & = \frac{\Phi_{p_\mr}(N)}{4}-\frac{\lambda\rho^n_\mathrm{b}(N)}{2a(N)r_\mathrm{m}} \\ 
I^\prime_{2,p_\mr}(N) + 2I_{2,p_\mr}(N) & = \frac{\Phi_{p_\mr}(N)}{2}.
\end{align}
The above equations can be solved analytically imposing the initial conditions $I_{1,p_\mr}(N=0)=I_{2,p_\mr}(N=0)=0$ and their solutions in terms of the scale factor $a$ read as
\begin{align}\label{I1+I2 solutions}
\begin{split}
I_{1,p_\mr}(a)& =  \frac{\lambda\rho^n_\mathrm{b,ini}}{4a_\mathrm{inf}r_\mathrm{m}(2n-1)(4n-1)}\left(\frac{a}{a_\mathrm{inf}}\right)^{-3} \\  & \times\left[4n-1+4(1-2n)\frac{a}{a_\mathrm{inf}} -(4n-3)\left(\frac{a}{a_\mathrm{inf}}\right)^{-(4n-2)}\right]
\end{split} \\
I_{2,p_\mr}(a)& = -\frac{\lambda\rho^n_\mathrm{b,inf}}{2a_\mathrm{ini}r_\mathrm{m}(2n-1)(4n-1)}\left(\frac{a}{a_\mathrm{inf}}\right)^{-3}\left[1-4n +2(2n-1)\frac{a}{a_\mathrm{inf}}+\left(\frac{a}{a_\mathrm{inf}}\right)^{-(4n-2)}\right].
\end{align}
Interestingly, here as well, when $n=1/2$ and $n=1/4$, the behavior $I_{1,p_\mr}$ and $I_2$ should be treated carefully. For this reason, we take the corresponding limits as it can be seen in \App{Phi_pr+I_1+1_2}

Finally, in order to determine explicitly the initial conditions for the hydrodynamic and metric perturbations one should compute the modulating function $\mathcal{H}(r)$, which is analogous to function $\mathcal{F}(r)$ defined earlier.To do so, one can combine \Eq{p_tilde-rho_tilde - f=rho^n} and \Eq{rho_tilde_full-f=rho^n} and obtain after a straightforward calculation the following differential equation for the rescaled pressure gradient profile $h(r)$:
\beq\label{p_r ODE compact form - f=rho^n}
\begin{aligned}
\sqrt{1-K(r)r^2} \left[\frac{2\lambda\rho^n_\mathrm{b}(a_0)}{3a_0r_\mathrm{m}} -\frac{ \Phi_{p_\mr}(a_0)}{3}\right]rh^\prime(r)  & + \left\{\left[\frac{2\lambda\rho^n_\mathrm{b}(a_0)}{3a_0r_\mathrm{m}} -\frac{4\Phi_{p_\mr}(a_0)}{3}\right]\sqrt{1-K(r)r^2}+r\right\}h(r) \\ &  - \Phi_\mathrm{iso}\left[\frac{\left(r^3K(r)\right)^\prime}{3r^2}\right]^\prime r^2_\mathrm{m} \sqrt{1-K(r)r^2} =0
\end{aligned},
\eeq
where $h(r)\equiv r_\mathrm{m}\frac{\partial\tilde{p}_\mathrm{r}}{\partial r}\frac{\sqrt{1-K(r)r^2}}{r}$ and the anisotropy modulating terms $\Phi_{p_\mr}$ and $\frac{\lambda\rho^n_\mathrm{b}(a)}{ar_\mathrm{m}}$ should be computed at an initial time when the gradient expansion is still valid up to a certain order $\epsilon_0 = \epsilon(t_0)$. The above differential equation should satisfy the analogous to $\mathcal{F}(r)$ boundary condition $\lim_{r\rightarrow 0}h(r)=0$ as imposed by \Eq{boundary condition}. Therefore, one can solve the above differential equation for the rescaled pressure gradient profile $h(r)$,  and then integrate it  in order to compute $\mathcal{H}(r)$ which can be finally inserted into \Eq{rho_tilde_full-f=rho^n}, \Eq{U_tilde_full-f=rho^n}, \Eq{A_tilde_full-f=rho^n}, \Eq{tilde R parametrisation - f=rho^n} and \Eq{tilde  B parametrisation - f=rho^n} to obtain the full expressions of the initial conditions for the hydrodynamic and metric perturbations. 

At this point, given the value of $n$, one can define a new dimensionless anisotropy parameter $\tilde{\lambda}$ as 
\beq\label{tilde_lambda}
\tilde{\lambda}=\frac{\lambda\rho^n_\mathrm{b,inf}}{r_\mathrm{m}},
\eeq
where $\rho_\mathrm{b,inf}$ is the background energy density the end of inflation. With the above definition of the anisotropy parameter, the equation of state \eqref{eq:p_r+p_t - f= rho^n-D_rp_r} can be recast as
\beq\label{p_r-p_t-tilde_lambda-f=rho^n}
p_\mathrm{r} = \frac{1}{3}\left[\rho-2\tilde{\lambda}r_\mathrm{m}\left(\frac{\rho}{\rho_\mathrm{b,inf}}\right)^nD_\mathrm{r}p_\mathrm{r}\right].
\eeq

We should also stress out here that  our problem at hand requires  to specify five input parameters in order to fully specify the initial conditions for the hydrodynamic and metric perturbations. In particular, these parameters are a) the index $n$ appearing in the equation of state \Eq{p_r-p_t-tilde_lambda-f=rho^n}, b) the dimensionless anisotropy parameter $\tilde{\lambda}$, c) the ratio between the energy scales at horizon crossing ($\epsilon_\mathrm{HC}=1$) and the energy scale at the end of inflation, i.e. $q = \left(\frac{\rho_\mathrm{b,HC}}{\rho_\mathrm{b,inf}}\right)^{1/4}$, d) the small initial parameter $\epsilon_0 = \frac{H^{-1}_0}{a_\mathrm{0}r_\mathrm{m}}$ and e) the background energy density, $\rho_\mathrm{b,inf}$, measured at the end of inflation, which depends on the underlying inflationary model generating the hydrodynamic and metric perturbations. At the end, the initial conditions for $\tilde{A}$, $\tilde{R}$, $\tilde{B}$, $\tilde{M}$, $\tilde{\rho}$ and $\tilde{U}$ could be written in a compact form as follows:
\begin{eqnarray}\label{Perturbations-Drpr-f=rho^n}
\begin{split}
\tilde{M} & = \frac{2}{3}K(r)r^2_\mathrm{m} + \Phi_{p_\mr}(N)\mathcal{H}(r) \\
\tilde{\rho} & = \Phi_\mathrm{iso}\frac{1}{3r^2}\left[r^3K(r)\right]^\prime r^2_\mathrm{m} + \Phi_{p_\mr}(a_0)\frac{1}{3r^2}\left[r^3\mathcal{H}(r)\right]^\prime \\
\tilde{U} & = \frac{1}{2}\left[ \left(\Phi_\mathrm{iso} -1 \right)K(r)r^2_\mathrm{m} + \Phi_{p_\mr}(a_0)\mathcal{H}(r)\right] \\
\tilde{A} & =  -\frac{1}{4}\left\{ \Phi_\mathrm{iso}\frac{1}{3r^2}\left[r^3K(r)\right]^\prime r^2_\mathrm{m} + \Phi_{p_\mr}(a_0)\frac{1}{3r^2}\left[r^3\mathcal{H}(r)\right]^\prime \right\} +\frac{\lambda\rho^n_\mathrm{b}(a_0)}{2a_0r_\mathrm{m}}\frac{\left[r^3\mathcal{H}(r)\right]^\prime}{3r^2}\\
\tilde{R} & = - I_{1,\mathrm{iso}}\frac{1}{3r^2}\left[r^3K(r)\right]^\prime r^2_\mathrm{m} + I_{2,\mathrm{iso}}\frac{K(r)r^2_\mathrm{m}}{2}  - I_{1,p_\mr}(a_0) \frac{\left[r^3\mathcal{H}(r)\right]^\prime}{3r^2}  + I_{2,p_\mr}(a_0)\mathcal{H}(r)  \\
\tilde{B} & = I_{1,\mathrm{iso}}r\left\{\frac{1}{3r^2}\left[r^3K(r)\right]^\prime r^2_\mathrm{m}\right\}^\prime  + I_{1,p_\mr}(a_0)r \left\{\frac{\left[r^3\mathcal{H}(r)\right]^\prime}{3r^2}\right\}^\prime,
\end{split}
\end{eqnarray}
where the modulating functions $\Phi_{p_\mr}$, $\frac{\lambda\rho^n_\mathrm{b}(a)}{ar_\mathrm{m}}$, $I_{1,p_\mr}$ and $I_{2,p_\mr}$ should be computed at initial time $t_0$ when the gradient expansion is valid up to a certain order $\epsilon_0$. Below, we give their explicit dependence on $q$, $\epsilon_0$, $n$ and $\tilde{\lambda}$.
\begin{align}\label{anisotropy modulators-Drpr-f=rho^n}
\frac{\lambda\rho^n_\mathrm{b}(a_0)}{a_0r_\mathrm{m}} & = \frac{\tilde{\lambda}}{q}\left(\frac{q}{\epsilon_0}\right)^{4n+1} \\
\Phi_{p_\mr}(a_0) & =  \frac{\tilde{\lambda}}{q(1-2n)}\left(\frac{q}{\epsilon_0}\right)^{3}\left[1-\left(\frac{q}{\epsilon_0}\right)^{(4n-2)}\right] \\
\begin{split}
I_{1,p_\mr}(a_0)& =  \frac{\tilde{\lambda}}{4q(2n-1)(4n-1)}\left(\frac{q}{\epsilon_0}\right)^{3} \\ & \times\left[4n-1+4(1-2n)\frac{\epsilon_0}{q} -(4n-3)\left(\frac{q}{\epsilon_0}\right)^{(4n-2)}\right]
\end{split} \\
I_{2,p_\mr}(a_0)& =  -\frac{\tilde{\lambda}}{2q(2n-1)(4n-1)}\left(\frac{q}{\epsilon_0}\right)^{3}\left[1-4n +2(2n-1)\frac{\epsilon_0}{q}+\left(\frac{q}{\epsilon_0}\right)^{(4n-2)}\right] .
\end{align}

Here, one should point out that at the level of the equation of state we identify three main contributions. First, the dimensionless parameter $\tilde{\lambda}$ accounting for the anisotropy of the medium. Second, the ratio $\left(\frac{\rho}{\rho_\mathrm{b,inf}}\right)^n$ which is measuring for the effect of cosmic expansion and finally the term $D_\mathrm{r}p_\mathrm{r}$ which accounts for the effect of the pressure gradients. Regarding the possible values of $n$, one can assume based on physical arguments that at the infinite time limit the pressure gradient contributions disappear, implying that $n\geq 0$. The values of the anisotropic parameter $\tilde{\lambda}$ can in principle take any value. 

In the following figures, we show the behavior of the initial conditions of the pressure gradients, the energy density and  velocity perturbations for a specific toy-model, i.e. $n=0$, considering positive values of the dimensionless anisotropy parameter $\tilde{\lambda}$ and taking into account that, due to the structure of \Eq{p_r ODE compact form - f=rho^n} which describes the behavior of the pressure gradients, one is facing a divergence at $r=0$ for negative values of $\tilde{\lambda}$. [See also the discussion in \App{Lower Limit on lambda<0}.]

\begin{figure}[h!]
\begin{center}
\includegraphics[height = 0.35\textwidth, width=0.496\textwidth, clip=true]{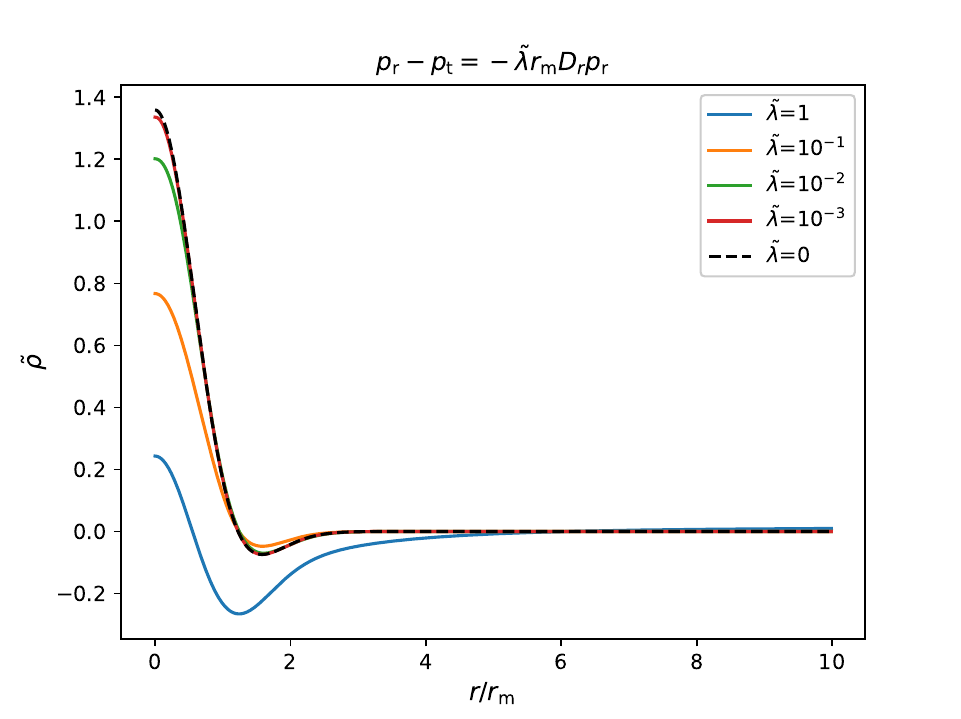}
\includegraphics[height = 0.35\textwidth, width=0.496\textwidth, clip=true]{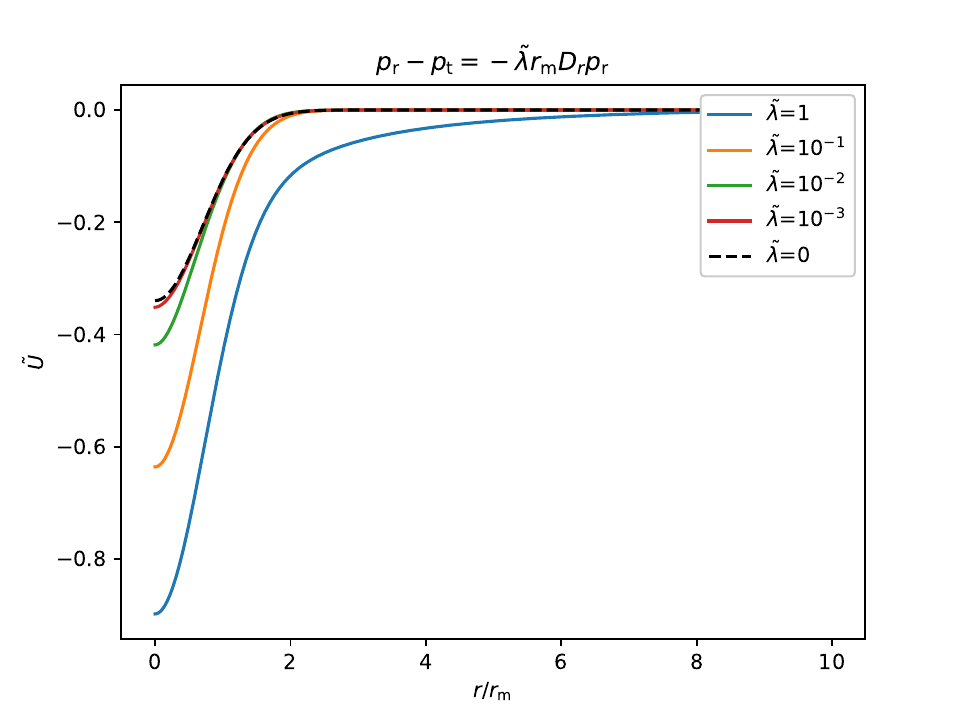}
\caption{In the left panel we show $\tilde{\rho}$ against $r/r_\mathrm{m}$ for different values of $\tilde{\lambda}>0$ while in the right one we show $\tilde{U}$ against $r/r_\mathrm{m}$, considering different values of $\tilde{\lambda}>0$. We have chosen $n=0$, $q=10^{-10}$ and $\epsilon_0 = 10^{-1}$.}
\label{fig:rho_tilde+U_tilde_f=rho^n-Drpr}
\end{center}
\end{figure}

As it can be seen from the figures below,where we choose $q=10^{-10}$ and $\epsilon_0 = 10^{-1}$, the behavior of the initial conditions for the hydrodynamic and metric perturbations is similar to the case where $f(r,t)=R(r,t)$ with a positive value of $\tilde{\lambda}$ enhancing the radial pressure compared to the tangential one and leading to a lower amplitude of the matter perturbations $\tilde{\rho}$ and $\tilde{U}$ with respect to the isotropic case [See \Fig{fig:rho_tilde+U_tilde_f=rho^n-Drpr}].

\begin{figure}[h!]
\begin{center}
\includegraphics[height = 0.40\textwidth, width=0.60\textwidth, clip=true]{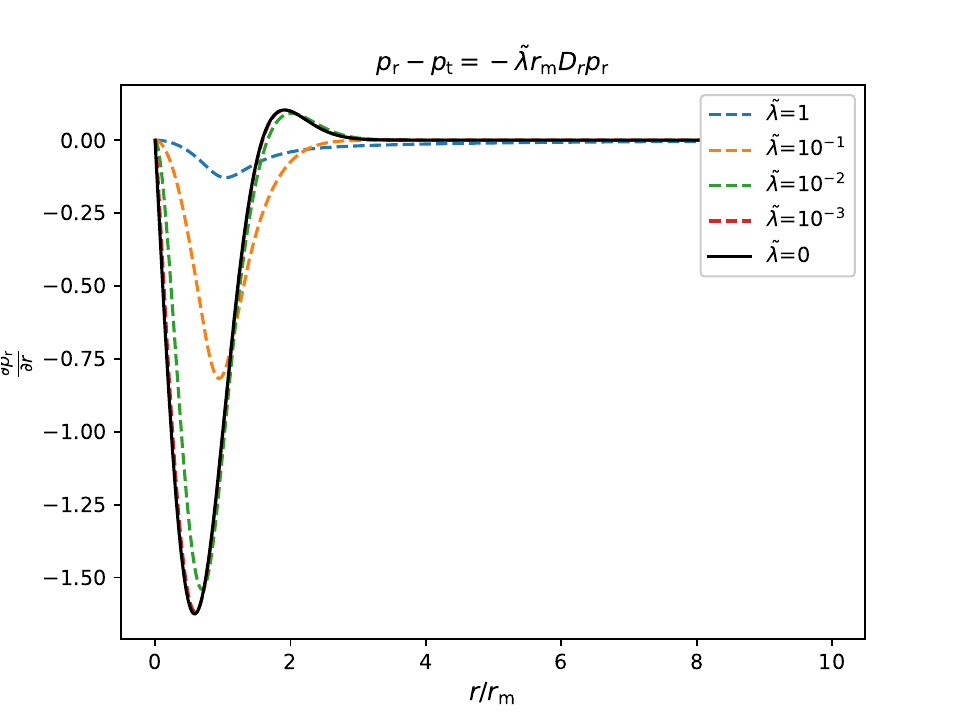}
\caption{In this figure, we plot $\frac{\partial\tilde{p}_\mr}{\partial r}$ against $r/r_\mathrm{m}$ by considering positive values of $\tilde{\lambda}$. We have chosen $n=0$, $q=10^{-10}$ and $\epsilon_0 = 10^{-1}$.}
\label{fig:dpr_dr_pr_f=rho^n-Drpr}
\end{center}
\end{figure}

\subsection{Equation of state in terms of energy density gradients}\label{sec:Equation of state in terms of energy density gradients}
We study here the quasi-homogeneous solution when the EoS is given by \Eq{D_rrho}, where the function $f(r,t)$ is chosen to be either $f(r,t)=R(r,t)$ or $f(r,t)=\rho^n(r,t)$. 

\subsubsection{$f(r,t)=R(r,t)$ } \label{sec:Drrho-f=R}
Following the same reasoning as before,  the only place in which we see a difference compared to the previous cases is in the differential equations for $\tilde{A}$ and $\tilde{p}_\mathrm{r}-\tilde{\rho}$ which now read as
\begin{eqnarray}
\tilde{A}^\prime &=& - \frac{1}{4}\left[\tilde{p}^\prime_\mathrm{r}-6\lambda\sqrt{1-K(r)r^2}\tilde{\rho}^\prime\right]\label{A_tilde constraint - Drrho} \\
\tilde{p}_\mathrm{r}-\tilde{\rho}&=&-2\lambda r\sqrt{1-K(r)r^2}\tilde{\rho}^{\prime} = -2\lambda r G^\prime(r)\label{p_tilde-rho_tilde - Drrho},
\end{eqnarray}
where
\beq\label{G(r) definition}
G(r)\equiv \int_{0}^{r}\sqrt{1-K(r^\prime)r^{\prime2}}\tilde{\rho}^{\prime}(r^\prime)\mathrm{d} r^\prime.
\eeq
 
Then, in order to put the expressions for $\tilde{A}$, $\tilde{R}$, $\tilde{B}$, $\tilde{M}$, $\tilde{\rho}$ and $\tilde{U}$ in a compact form as in \Sec{sec:Drpr-f=R}, we introduce an effective curvature profile similar to the one defined in \Eq{Effective Curvature Profile-Drpr} as follows:
\beq\label{Effective Curvature Profile - Drrho}
\mathcal{K}(r) = K(r) - \frac{3\lambda}{r^2_\mathrm{m}}\mathcal{G}(r),
\eeq
where 
\beq \label{G_r}
\mathcal{G}(r)\equiv G(r)- G(\infty)=-\int_r^{\infty}\frac{\partial\tilde{\rho}(r^\prime)}{\partial r^\prime}\sqrt{1-K(r^\prime)r^{\prime 2}}\mathrm{d} r^\prime.
\eeq 
We define as well the effective energy density and velocity perturbations $\tilde{\rho}_\mathrm{eff}$ and  $\tilde{U}_\mathrm{eff}$ such that the quasi-homogeneous solution is written in a similar form as in the isotropic case ($\lambda=0$),
\beq\label{rho_tilde_eff - Drrho}
\tilde{\rho}_\mathrm{eff} = \tilde{\rho}-2\lambda\left[r\mathcal{G}^\prime(r)+3\mathcal{G}(r)\right]
\eeq
\beq\label{U__tilde_eff - Drrho}
\tilde{U}_\mathrm{eff} = \tilde{U}+\frac{3\lambda}{2}\mathcal{G}(r) = -\frac{1}{6}\mathcal{K}(r)r^2_\mathrm{m}.
\eeq
At the end, one gets the hydrodynamic and metric perturbations in the following compact form:
\begin{eqnarray}\label{Perturbations - Drrho}
\begin{aligned}
\tilde{A} & = -\frac{\tilde{\rho}}{4} + \frac{\lambda}{2}\left[r\mathcal{G}^\prime(r)+3\mathcal{G}(r)\right] = -\frac{\tilde{\rho}_\mathrm{eff}}{4}\\
\tilde{B} &  = \frac{r}{8}\left[\tilde{\rho}^\prime-2\lambda r\mathcal{G}^{\prime\prime}(r) -8\lambda \mathcal{G}^\prime(r)\right]  = \frac{r}{8}\tilde{\rho}^\prime_\mathrm{eff}\\
\tilde{\rho} & =\frac{2}{3}\frac{\left[r^3\mathcal{K}(r)\right]^\prime}{3r^2}r^2_\mathrm{m} \\
\tilde{U} & =-\frac{1}{6}\mathcal{K}(r)r_\mathrm{m}^2 - \frac{3\lambda}{2}\mathcal{G}(r) = \tilde{U}_\mathrm{eff} - \frac{\lambda}{2}\mathcal{G}(r)\\ 
\tilde{M} &=\frac{2}{3}\mathcal{K}(r)r_\mathrm{m}^2 = -4\tilde{U}_\mathrm{eff}  \\
\tilde{R} & = -\frac{\tilde{\rho}_\mathrm{eff}}{8} + \frac{\tilde{U}}{2}  
\end{aligned}
\end{eqnarray}

Finally, in order to express explicitly the initial conditions for the hydrodynamic and metric perturbations, one should compute the behavior of the energy density gradient profile $\frac{\partial \tilde{\rho}}{\partial r}$. Combining then \Eq{p_tilde-rho_tilde - Drrho} and \Eq{rho_tilde constraint} one gets the following equation for 
\[ g(r)\equiv\frac{\partial\tilde{\rho}}{\partial r}\sqrt{1-K(r)r^2} \]
analogous to \Eq{p_r ODE compact form}:
\beq\label{rho ODE compact form}
\begin{aligned}
\frac{2\lambda}{3}r\sqrt{1-K(r)r^2}g^\prime(r) &+ \left[\frac{8\lambda}{3}\sqrt{1-K(r)r^2} + 1\right]g(r) +\\ 
&-  \frac{2}{3}\left\{\frac{\left[r^3K(r)\right]^\prime}{3r^2}\right\}^\prime r^2_\mathrm{m} \sqrt{1-K(r)r^2} = 0,
\end{aligned}
\eeq
with the boundary condition $g(0)=0$. By solving for $g(r)$ the above equation, we integrate $g(r)$ in order to get the anisotropic modulating function $\mathcal{G}(r)$ modifying all the hydrodynamic and metric perturbations as one can see from \Eq{Perturbations - Drrho}.

Below, we show the energy density gradient profiles considering both positive and negative values of $\lambda$, together the behavior of the energy density and velocity perturbation's profiles. For the figures below, we chose as before a curvature profile for $K(r)$ having a Gaussian form $K(r) = \mathcal{A}e^{-(r/r_\mathrm{m,0})^2}$ with $r_\mathrm{m,0}=1$ and $\mathcal{A}=\frac{3e}{4r^2_\mathrm{m,0}}$. Regarding the value of $r_\mathrm{m}$, we take as before $r_\mathrm{m}=r_\mathrm{m,0}=1$, since also in this case $r_\mathrm{m}\simeq r_\mathrm{m,0}$ independently of the value of $\lambda$ we have considered. 
\begin{figure}[h!]
\begin{center}
\includegraphics[height = 0.35\textwidth, width=0.496\textwidth, clip=true]{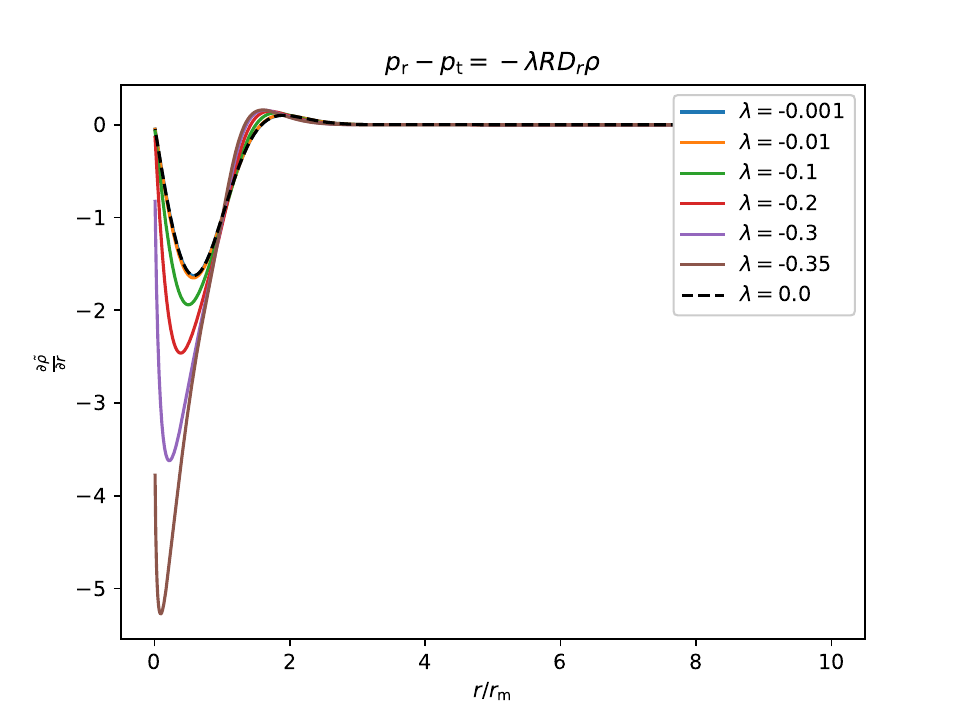}
\includegraphics[height = 0.35\textwidth, width=0.496\textwidth, clip=true]{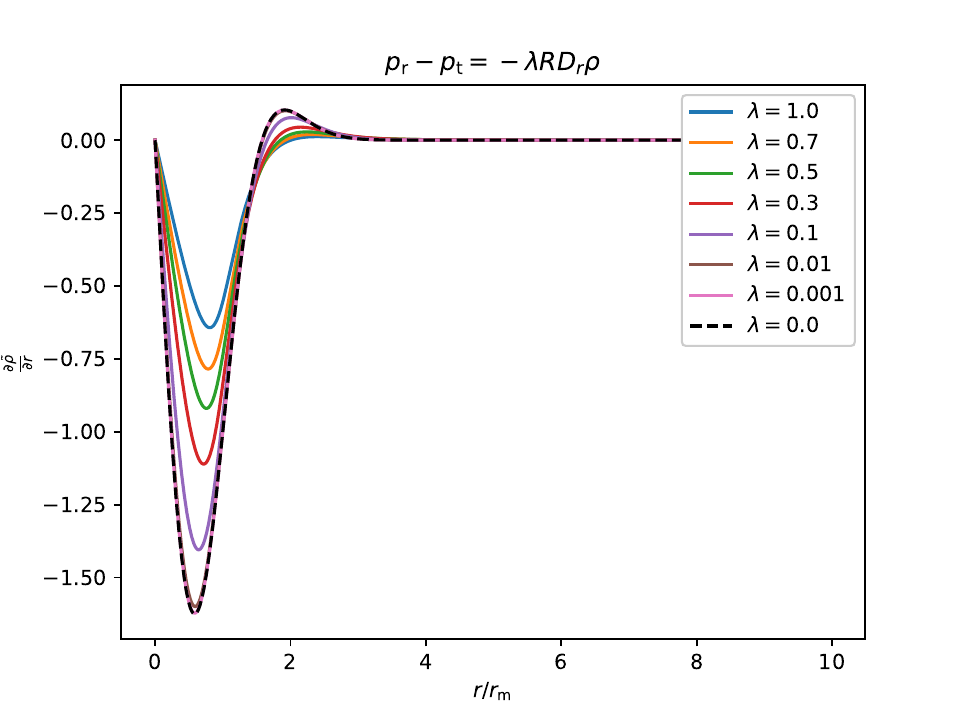}
\caption{In this figure, we show the $\frac{\partial\tilde{\rho}}{\partial r}$ against $r/r_\mathrm{m}$. In the left panel, we consider negative values of $\lambda$ whereas in the right panel we account for positive values of $\lambda$. }
\label{fig:drho_dr_pr_f=R}
\end{center}
\end{figure}
\begin{figure}[h!]
\begin{center}
\includegraphics[height = 0.35\textwidth, width=0.496\textwidth, clip=true]{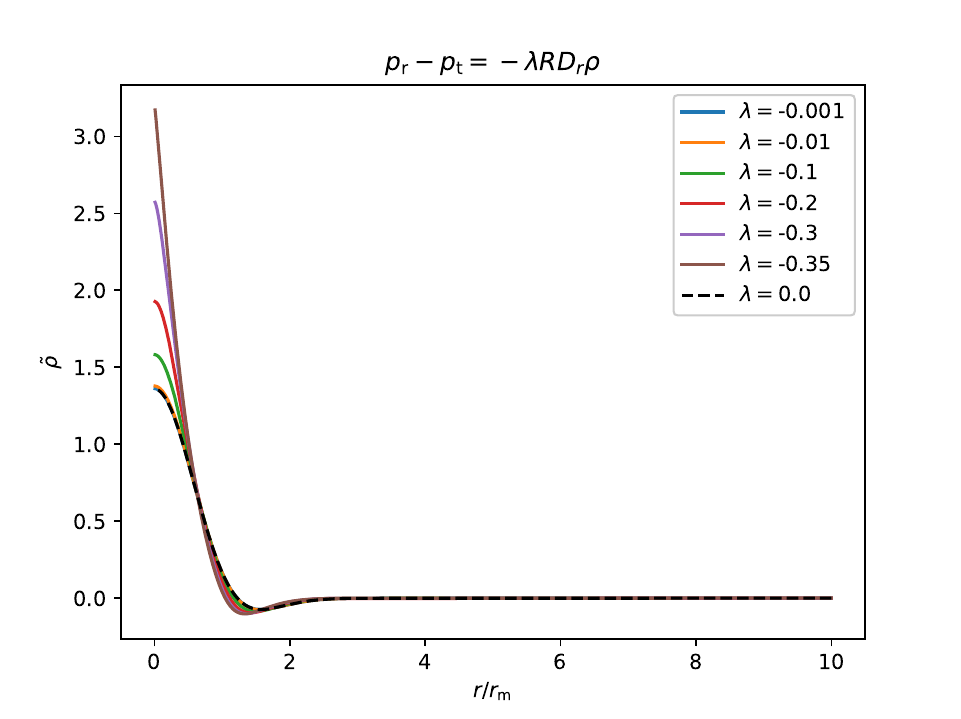}
\includegraphics[height = 0.35\textwidth, width=0.496\textwidth, clip=true]{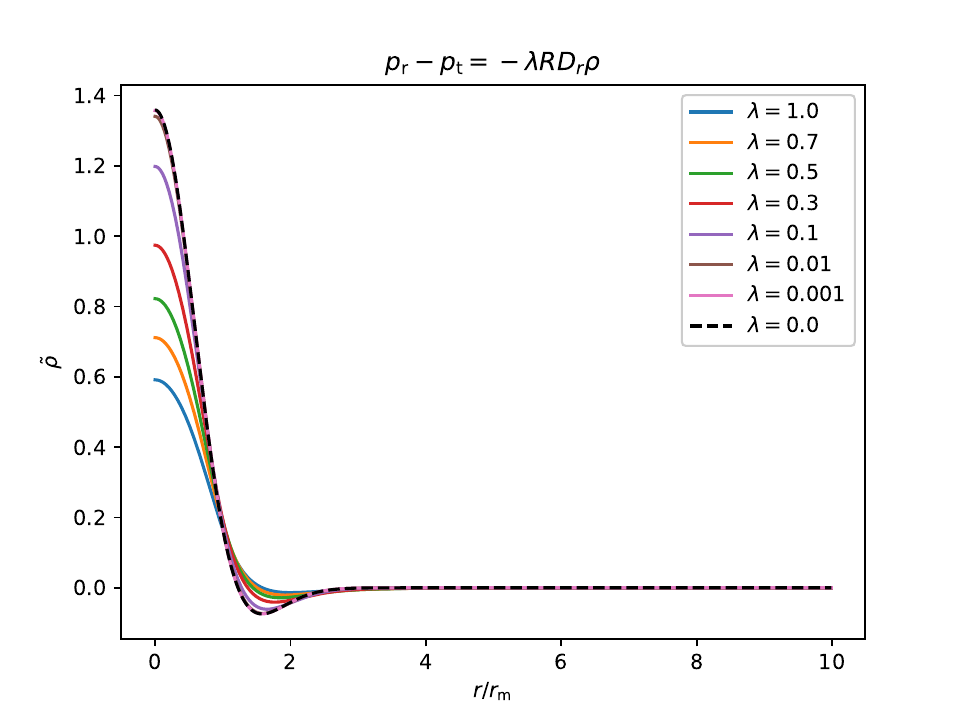}
\caption{In this figure, we show the $\tilde{\rho}$ against $r/r_\mathrm{m}$. In the left panel, we consider negative values of $\lambda$ whereas in the right panel we account for positive values of $\lambda$. }
\label{fig:rho_tilde_rho_f=R}
\end{center}
\end{figure}
\begin{figure}[h!]
\begin{center}
\includegraphics[height = 0.35\textwidth, width=0.496\textwidth, clip=true]{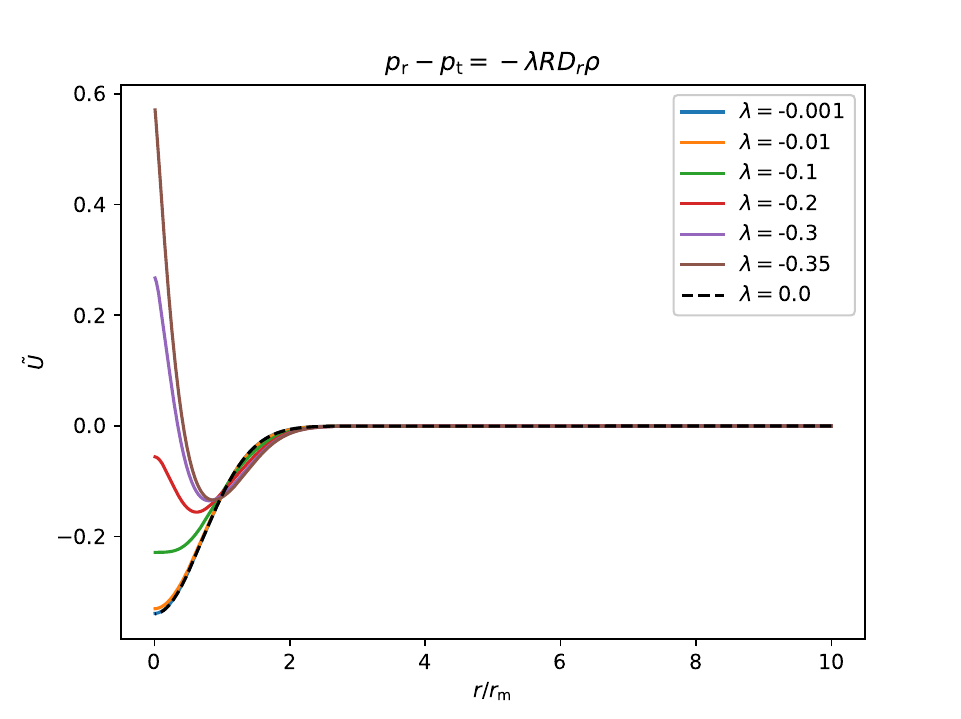}
\includegraphics[height = 0.35\textwidth, width=0.496\textwidth, clip=true]{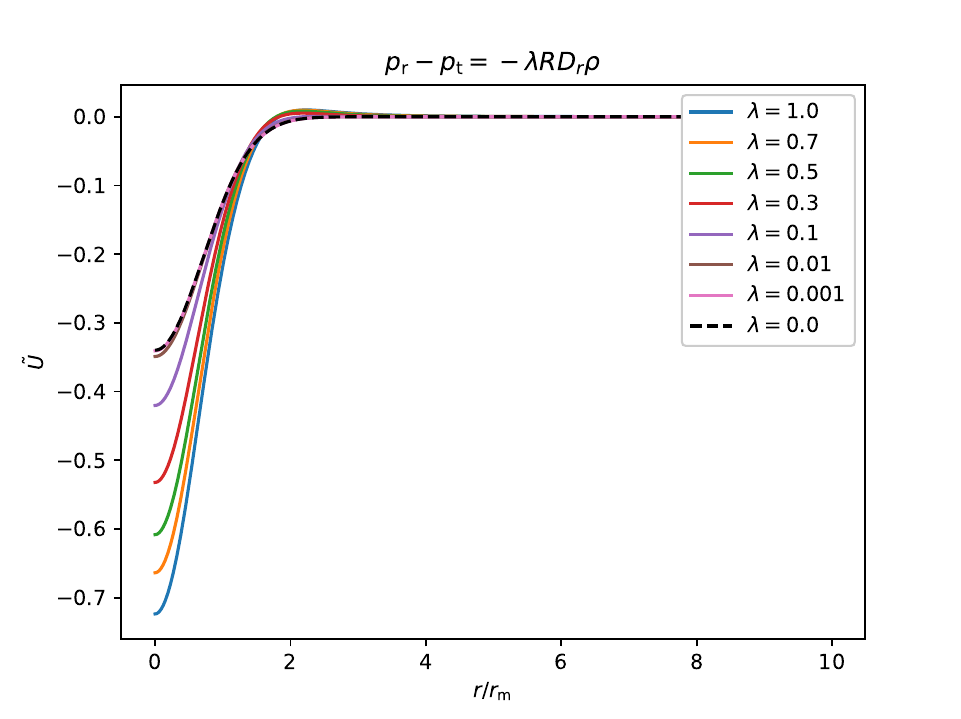}
\caption{In this figure, we show the $\tilde{U}$ against $r/r_\mathrm{m}$. In the left panel, we consider negative values of $\lambda$ whereas in the right panel we account for positive values of $\lambda$. }
\label{fig:U_tilde_rho_f=R}
\end{center}
\end{figure}

As in the case where $p_\mathrm{r}-p_\mathrm{t}=-\lambda RD_\mathrm{r}p_\mathrm{r}$, we identify two regimes: a) when $\lambda<0$, characterized by the fact that $\frac{\partial\tilde{\rho}}{\partial r}$ is mainly negative and one obtains that $p_\mr<p_\mt$ suggesting in this way the possibility for a cosmological perturbation to collapse more easily and b) when $\lambda>0$ where $p_\mr>p_\mt$ and it is more difficult for a perturbation to collapse.

Regarding the behavior of the energy density gradient, similarly to the case where $p_\mathrm{r}-p_\mathrm{t}=-\lambda RD_\mathrm{r}p_\mathrm{r}$, one can identify from \Fig{fig:drho_dr_pr_f=R} a lower bound in $\lambda$, which can be explained from the mathematical structure of \Eq{rho ODE compact form} describing the behavior of $\frac{\partial\tilde{\rho}}{\partial r}$. [See also the \App{Lower Limit on lambda<0}].
\subsubsection{$f(r,t)=\rho^n(r,t)$}\label{sec:Drrho-f=rho^n}

We now consider the  EoS with $f(r,t)=\rho^n(r,t)$, given by \Eq{eq:p_r+p_t - f= rho^n-D_rrho}. Following the same strategy as in \Sec{sec:Drpr-f=rho^n}, the differential equations for $\tilde{A}$ and $\tilde{p}_\mathrm{r}-\tilde{\rho}$ read as 
\begin{eqnarray}
\tilde{A}^\prime &=& - \frac{1}{4}\left[\tilde{p}^\prime_\mathrm{r}-\frac{6\lambda\rho^n_\mathrm{b}}{ar}\sqrt{1-K(r)r^2}\tilde{\rho}^\prime\right]\label{A_tilde constraint - f=rho^n-D_rrho} \\
\tilde{p}_\mathrm{r}-\tilde{\rho}&=&-\frac{2\lambda\rho^n_\mathrm{b}}{a}\sqrt{1-K(r)r^2}\tilde{\rho}^{\prime} = -\frac{2\lambda\rho^n_\mathrm{b}}{ar_\mathrm{m}} r J^\prime(r)\label{p_tilde-rho_tilde - f=rho^n-D_rrho},
\end{eqnarray}
where
\beq\label{J(r) definition}
J(r)\equiv r_\mathrm{m}\int_{0}^{r}\frac{\sqrt{1-K(r^\prime)r^{\prime2}}}{r^\prime}\tilde{\rho}^{\prime}(r^\prime)\mathrm{d}r^\prime.
\eeq
The differential equations for $\tilde{R}$, $\tilde{B}$, $\tilde{M}$, $\tilde{\rho}$ and $\tilde{U}$, taking into account that these quantities do not depend on the prescription modeling the difference $p_\mr-p_\mt$, will be given by Eqs. (\ref{eq:metric+hydrodynamic perturbations}).
Integrating them using the above equations, similarly as in \Sec{sec:Drpr-f=rho^n}, one can write the profiles of $\tilde{A}$, $\tilde{R}$, $\tilde{B}$, $\tilde{M}$, $\tilde{\rho}$ and $\tilde{U}$ as follows:
\begin{eqnarray}\label{Perturbations-Drrho-f=rho^n}
\begin{split}
\tilde{M} & = \frac{2}{3}K(r)r^2_\mathrm{m} + \Phi_\rho(a_0)\mathcal{J}(r)\label{M_tilde_full-f=rho^n-D_rrho} \\
\tilde{\rho} & = \Phi_\mathrm{iso}\frac{1}{3r^2}\left[r^3K(r)\right]^\prime r^2_\mathrm{m} + \Phi_{\rho}(a_0)\frac{1}{3r^2}\left[r^3\mathcal{J}(r)\right]^\prime \label{rho_tilde_full-f=rho^n-D_rrho}\\
\tilde{U} & = \frac{1}{2}\left[ \left(\Phi_\mathrm{iso} -1 \right)K(r)r^2_\mathrm{m} + \Phi_{\rho}(a_0)\mathcal{J}(r)\right]  \label{U_tilde_full-f=rho^n-D_rrho} \\
\tilde{A} & =  -\frac{1}{4}\left\{ \Phi_\mathrm{iso}\frac{1}{3r^2}\left[r^3K(r)\right]^\prime r^2_\mathrm{m} + \Phi_{\rho}(a_0)\frac{1}{3r^2}\left[r^3\mathcal{J}(r)\right]^\prime \right\} +\frac{3\lambda\rho^n_\mathrm{b}(a_0)}{2a_0r_\mathrm{m}}\frac{\left[r^3\mathcal{J}(r)\right]^\prime}{3r^2}\label{A_tilde_full-f=rho^n-D_rrho} \\
\tilde{R} & = - I_{1,\mathrm{iso}}\frac{1}{3r^2}\left[r^3K(r)\right]^\prime r^2_\mathrm{m} + I_{2,\mathrm{iso}}\frac{K(r)r^2_\mathrm{m}}{2}  - I_{1,\rho}(a_0) \frac{\left[r^3\mathcal{J}(r)\right]^\prime}{3r^2}  + I_{2,\rho}(a_0)\mathcal{J}(r) \label{tilde R parametrisation - f=rho^n-D_rrho} \\
\tilde{B} & = I_{1,\mathrm{iso}}r\left\{\frac{1}{3r^2}\left[r^3K(r)\right]^\prime r^2_\mathrm{m}\right\}^\prime  + I_{1,\rho}(a_0)r \left\{\frac{\left[r^3\mathcal{J}(r)\right]^\prime}{3r^2}\right\}^\prime  \label{tilde  B parametrisation - f=rho^n-D_rrho},
\end{split}
\end{eqnarray}
where $\mathcal{J}(r)$ is defined as
\beq\label{curly J definition}
\mathcal{J}(r) \equiv -r_\mathrm{m}\int_{r}^{\infty}\frac{\sqrt{1-K(r^\prime)r^{\prime2}}}{r^\prime}\tilde{\rho}^{\prime}(r^\prime)\mathrm{d}r^\prime.
\eeq
The anisotropy modulating functions $\Phi_\rho$, $I_{1,\rho}$ and $I_{2,\rho}$, calculated at initial time $t_0$ when the gradient expansion is valid up a certain order $\epsilon_0$, satisfy the following differential equations 
\begin{align}
\Phi_\rho^\prime(N) + 3\Phi_\rho(N) & = -6\lambda\frac{\rho^n_\mathrm{b}(N)}{a(N)r_\mathrm{m}}\\
I^\prime_{1,\rho}(N) + 2I_{1,\rho}(N) & = \frac{\Phi_{\rho}(N)}{4}-\frac{3\lambda\rho^n_\mathrm{b}(N)}{2a(N)r_\mathrm{m}} \\ 
I^\prime_{2,\rho}(N) + 2I_{2,\rho}(N) & = \frac{\Phi_{\rho}(N)}{2},
\end{align}
with initial conditions $\Phi(N=0)=I_{1,\rho}(N=0)=I_{2,\rho}(N=0)=0$. After solving the above differential equations, the solutions of for $\Phi_\rho$, $I_{1,\rho}$ and $I_{2,\rho}$ in terms of the scale factor read as
\begin{align}
\Phi_{\rho}(a) & =  3 \frac{\lambda\rho^n_\mathrm{b,inf}}{a_\mathrm{inf}r_\mathrm{m}(1-2n)}\left(\frac{a}{a_\mathrm{inf}}\right)^{-3}\left[1-\left(\frac{a}{a_\mathrm{inf}}\right)^{-(4n-2)}\right]\label{Phi analytical-D_rrho}\\
\begin{split}
I_{1,\rho}(a)& = \frac{3\lambda\rho^n_\mathrm{b,inf}}{4a_\mathrm{inf}r_\mathrm{m}(2n-1)(4n-1)}\left(\frac{a}{a_\mathrm{inf}}\right)^{-3} \\ & \times\left[4n-1+4(1-2n)\frac{a}{a_\mathrm{inf}} -(4n-3)\left(\frac{a}{a_\mathrm{inf}}\right)^{-(4n-2)}\right]
\end{split} ]\label{I_1 analytical-D_rrho} \\
I_{2,\rho}(a)& = -\frac{3\lambda\rho^n_\mathrm{b,inf}}{2a_\mathrm{inf}r_\mathrm{m}(2n-1)(4n-1)}\left(\frac{a}{a_\mathrm{inf}}\right)^{-3}\left[1-4n +2(2n-1)\frac{a}{a_\mathrm{inf}}+\left(\frac{a}{a_\mathrm{inf}}\right)^{-(4n-2)}\right]\label{I_2 analytical-D_rrho}.
\end{align}

As in \Sec{sec:Drpr-f=rho^n}, $N=0$ refers to $a=a_\mathrm{inf}$, the time when inflation ends, corresponding to the epoch when the cosmological perturbations are generated. For this reason, when $N=0$ we are essentially still in the isotropic regime. The limits for $n\rightarrow 1/2$ and $n\rightarrow 1/4$ for the modulating functions $\Phi_\rho$, $I_{1,\rho}$ and $I_{2,\rho}$ are given in \App{Phi_pr+I_1+1_2}.

Finally, in order to specify explicitly the initial conditions for the hydrodynamic and metric pertubations one should compute the energy density gradient profile, $\frac{\partial\tilde{\rho}}{\partial r}$ and then integrate it to obtain the modulating function $\mathcal{J}(r)$. Doing so, one can combine \Eq{p_tilde-rho_tilde - f=rho^n-D_rrho} and \Eq{rho_tilde constraint} and obtain after a straighforward calculation the following differential equation for the rescaled energy density gradient profile $j(r)$:
\beq\label{p_r ODE compact form - f=rho^n-D_rrho}
\begin{aligned}
\frac{\Phi_{\rho}(a_0)\sqrt{1-K(r)r^2}}{3}rj^\prime(r)  & + \left\{\frac{4\Phi_{\rho}(a_0)}{3}\sqrt{1-K(r)r^2}-r\right\}j(r) \\ &  + \Phi_\mathrm{iso}\left[\frac{\left(r^3K(r)\right)^\prime}{3r^2}\right]^\prime r^2_\mathrm{m} \sqrt{1-K(r)r^2} =0
\end{aligned},
\eeq
where $j(r)\equiv r_\mathrm{m}\frac{\partial\tilde{\rho}}{\partial r}\frac{\sqrt{1-K(r)r^2}}{r}$. 

At this point, as we have seen in \Sec{sec:Drpr-f=rho^n}, one can introduce the rescaled anisotropy parameter $\tilde{\lambda}$ defined exactly as before through \Eq{tilde_lambda} and express the anisotropic modulating functions  $\frac{\lambda\rho^n_\mathrm{b}(a_0)}{a_0r_\mathrm{m}}$, $\Phi_{\rho}(a_0)$, $I_{1,\rho}(a_0)$ and$I_{2,\rho}(a_0)$ in terms of $\tilde{\lambda}$, $n$, $q$ and $\epsilon_0$, given below by the following explicit expressions:
\begin{align}\label{anisotropy modulators-Drpr-f=rho^n}
\frac{\lambda\rho^n_\mathrm{b}(a_0)}{a_0r_\mathrm{m}} & = \frac{\tilde{\lambda}}{q}\left(\frac{q}{\epsilon_0}\right)^{4n+1} \\
\Phi_{\rho}(a_0) & =  \frac{3\tilde{\lambda}}{q(1-2n)} \left(\frac{q}{\epsilon_0}\right)^{3}\left[1-\left(\frac{q}{\epsilon_0}\right)^{(4n-2)}\right] \\
\begin{split}
I_{1,\rho}(a_0)& =  \frac{3\tilde{\lambda}}{4q(2n-1)(4n-1)}\left(\frac{q}{\epsilon_0}\right)^{3}\\ & \times\left[4n-1+4(1-2n)\frac{\epsilon_0}{q} -(4n-3)\left(\frac{q}{\epsilon_0}\right)^{(4n-2)}\right]
\end{split} \\
I_{2,\rho}(a_0)& =  -\frac{3\tilde{\lambda}}{2q(2n-1)(4n-1)}\left(\frac{q}{\epsilon_0}\right)^{3}\left[1-4n +2(2n-1)\frac{\epsilon_0}{q}+\left(\frac{q}{\epsilon_0}\right)^{(4n-2)}\right].
\end{align}
 
In the following figures, we have chosen as before $q=10^{-10}$ and $\epsilon_0 = 10^{-1}$, in order to compute the behavior of the initial profiles of the energy density gradients, the energy density and velocity perturbations in the special case of $n=0$. As discussed in the \App{Lower Limit on lambda<0}, negative values of $\tilde{\lambda}$ lead to a divergent behavior of the energy density gradient profile at $r=0$. Therefore, we consider only positive values for the dimensionless anisotropic parameter $\tilde{\lambda}$. 

The effect of the anisotropy is similar to the case where the difference between the radial and the tangential pressure is proportional to pressure gradients with $\tilde{\lambda}>0$ leading to an enhancement of the radial pressure and therefore decreasing the amplitude energy density perturbation.

\begin{figure}[h!]
\begin{center}
\includegraphics[height = 0.40\textwidth, width=0.60\textwidth, clip=true]{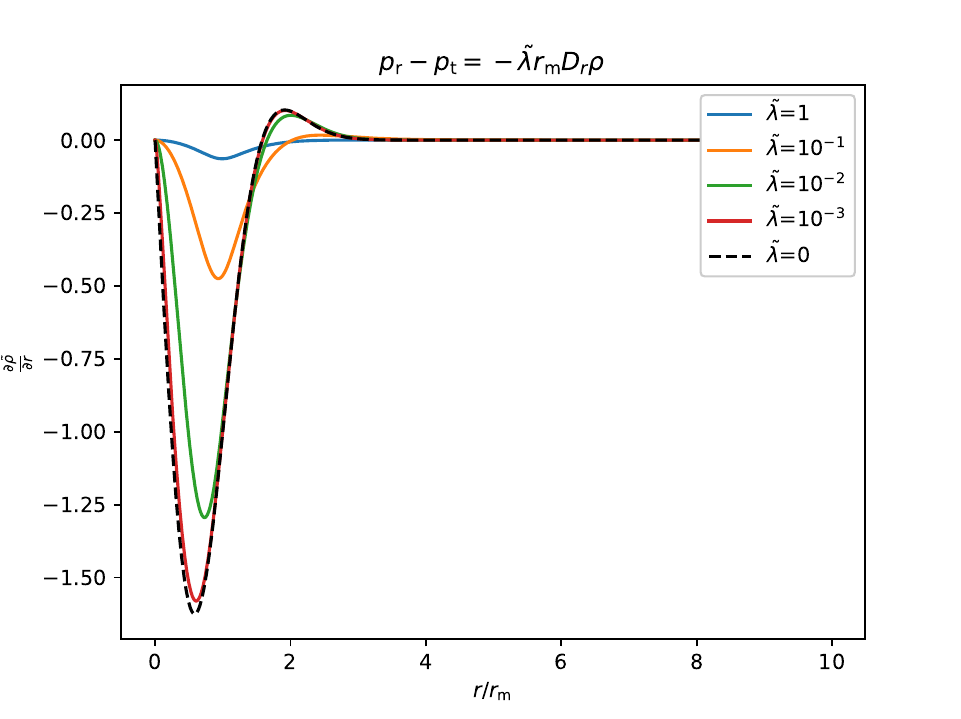}
\caption{In this figure, we plot $\frac{\partial\tilde{\rho}}{\partial r}$ against $r/r_\mathrm{m}$ by considering positive values of $\tilde{\lambda}$. We have chosen $n=0$, $q=10^{-10}$ and $\epsilon_0 = 10^{-1}$. }
\label{fig:drho_dr_pr_f=rho^n-Drrho}
\end{center}
\end{figure}

\begin{figure}[h!]
\begin{center}
\includegraphics[height = 0.35\textwidth, width=0.496\textwidth, clip=true]{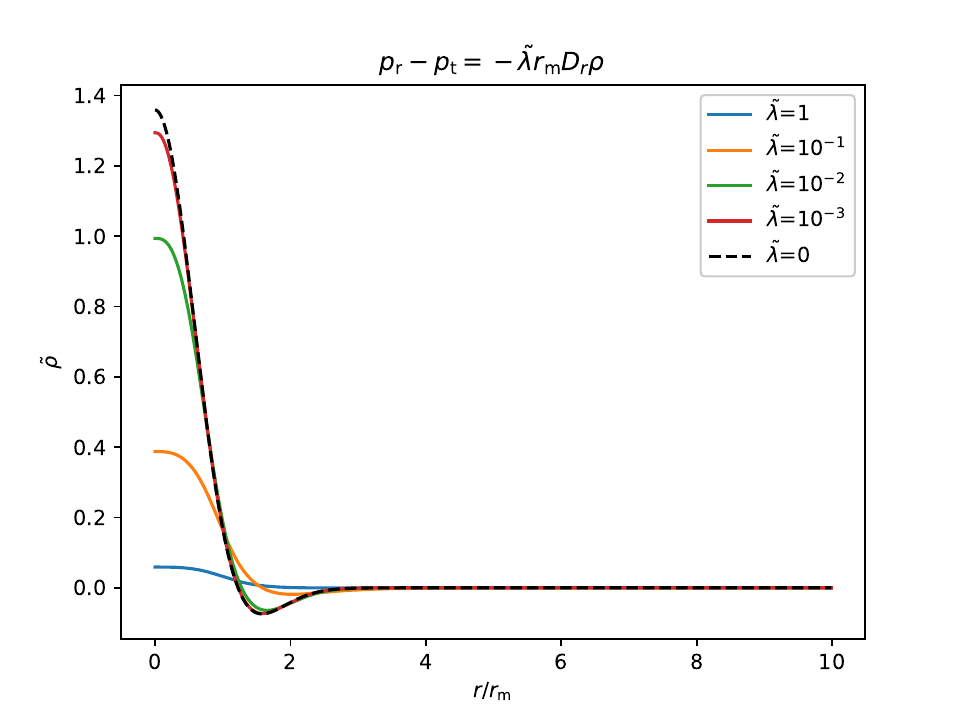}
\includegraphics[height = 0.35\textwidth, width=0.496\textwidth, clip=true]{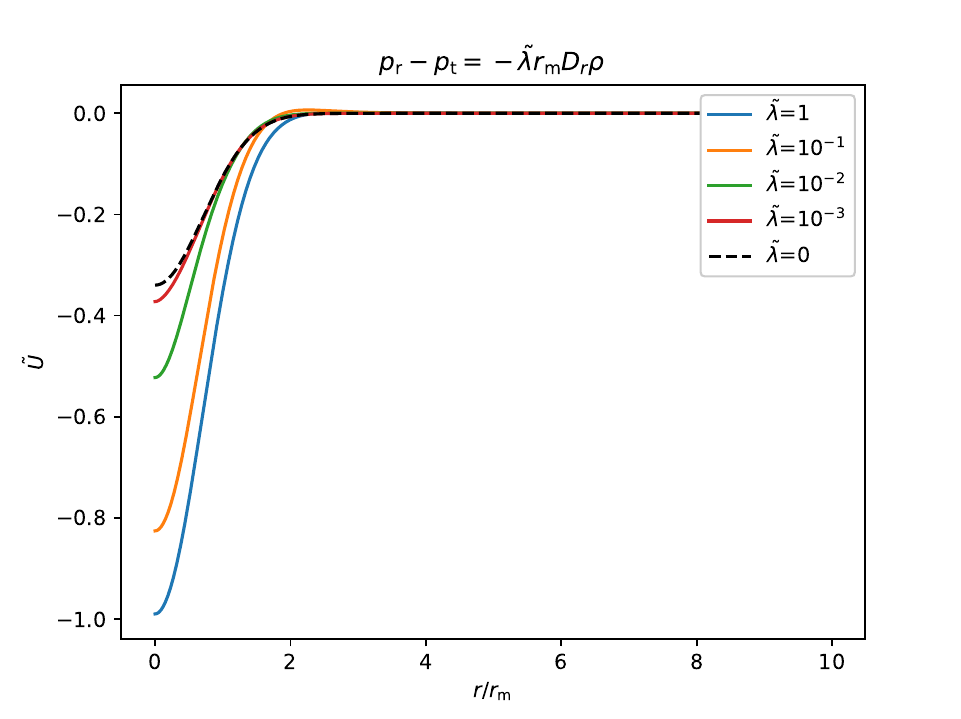}
\caption{In the left panel we show $\tilde{\rho}$ against $r/r_\mathrm{m}$ for different values of $\tilde{\lambda}>0$ while in the right one we show $\tilde{U}$ against $r/r_\mathrm{m}$ for different values of $\tilde{\lambda}>0$. We have chosen $n=0$, $q=10^{-10}$ and $\epsilon_0 = 10^{-1}$.}
\label{fig:rho_tilde+rho_tilde=fixed_f=rho^n-Drrho}
\end{center}
\end{figure}

\newpage
\section{Towards the PBH formation thershold}
After specifying the initial conditions for the hydrodynamic and metric perturbations, the next step is to compute numerically the PBH formation threshold $\delta_\mathrm{c}$ as a function of the degree of the anisotropy. Given the fact that we have not yet fully completed the numerical computation of $\delta_\mathrm{c}$ (work in progress) we give below a synthetic overview about the prescription one should follow to make such computation.

1. Given the anisotropy parameters $\lambda$ or $\tilde{\lambda}$ and $n$ and the amplitude of the curvature profile $\mathcal{A}$, one can compute the characteristic scale of the collapsing overdensity region $r_\mathrm{m}$, corresponding to the location where the compaction function defined in \Eq{C definition} has a maximum. This is done on the superhorizon regime when the compaction function is time-independent. This allows to compute the averaged perturbation amplitude of the collapsing overdensity $\delta_\mathrm{m}$, defined as the integral of the energy density perturbation over a spherical volume of radius $r_\mathrm{m}$ in \Eq{delta_m definition}.

2. Then, one should evolve the non linear hydrodynamic equations and check if a black hole apparent horizon is formed, i.e. when the condition $ R(r,t) = 2M(r,t)$ is fulfilled. The threshold $\delta_c$ is obtained then as the limiting intermediate case between overcritical perturbations ($\delta_\mathrm{m} > \delta_c$) collapsing into a black hole, and subcritical ones ($\delta_\mathrm{m} < \delta_c$), where the perturbation bouncing back into the medium without forming a black hole. In practice this is computed with a bisection method up to a certain precision. 

To estimate here what one would obtain performing numerical simulations of our anisotropic models, we compute $\delta_\mathrm{c}$ by adopting a perturbative approach based on the assumption that $\delta_\mathrm{c}$ depends on the shape of the initial energy density profile in the same way as it happens in the isotropic case. If so, one can compute the dependence of $\delta_\mathrm{c}$ in terms of the degree of anisotropy of the gravitational collapse, computing how the shape of the initial perturbation is changed by the amplitude of the anisotropy. 

To find this dependence, we use the analytic relation for the threshold of PBH formation as a function of the shape parameter, $\alpha$, defined as
\beq\label{alpha}
\alpha \equiv -\frac{r^2_\mathrm{m}\mathcal{C}^{\prime\prime}(r_\mathrm{m})}{4\mathcal{C}(r_\mathrm{m})}.
\eeq
For our analytic estimation of $\delta_\mathrm{c}$, we use the numerical fit given by Eq. (44) of ~\cite{Musco:2020jjb} where the threshold for PBH formation is given as a polynomial function of $\alpha$ as follows,
\beq\label{delta_c_analytical_Musco}
\delta_\mathrm{c}= 
\begin{cases}
\alpha^{0.047}-0.50 \quad 0.1 \lesssim\alpha\lesssim 7 \\
\alpha^{0.035}-0.475 \quad 7 \lesssim\alpha\lesssim 13 \\
\alpha^{0.026}-0.45 \quad 13 \lesssim\alpha\lesssim 30.
\end{cases}
\eeq
Fixing the amplitude of the anisotropy measured by $\lambda$ or $\tilde{\lambda}$ one should firstly compute $r_\mathrm{m}$ by requiring $\mathcal{C}(r_\mathrm{m})=0$ and subsequently compute at $r_\mathrm{m}$ the compaction function and its second derivative, which allows to compute the shape parameter, $\alpha$ using \Eq{alpha}. By inserting then this into \Eq{delta_c_analytical_Musco} one can compute the threshold for PBH formation. Below, we give the dependence of $\alpha$ and $\delta_\mathrm{c}$ in terms of $\lambda$ or $\tilde{\lambda}$ when $f(r,t) = R(r,t)$ or $f(r,t)=\rho^n(r,t)$. In particular, when $f(r,t)=\rho^n(r,t)$, we choose $n=0$ , $q=10^{-10}$ and $\epsilon_0=10^{-1}$. For our practical purposes, we fix as well the perturbation amplitude $\delta_\mathrm{m}=0.5$, a condition from which one can extract the amplitude $\mathcal{A}$ of the curvature profile for a fixed value of the amplitude of the anisotropy.

\begin{figure}[h!]
\begin{center}
\includegraphics[height = 0.40\textwidth, width=0.496\textwidth, clip=true]{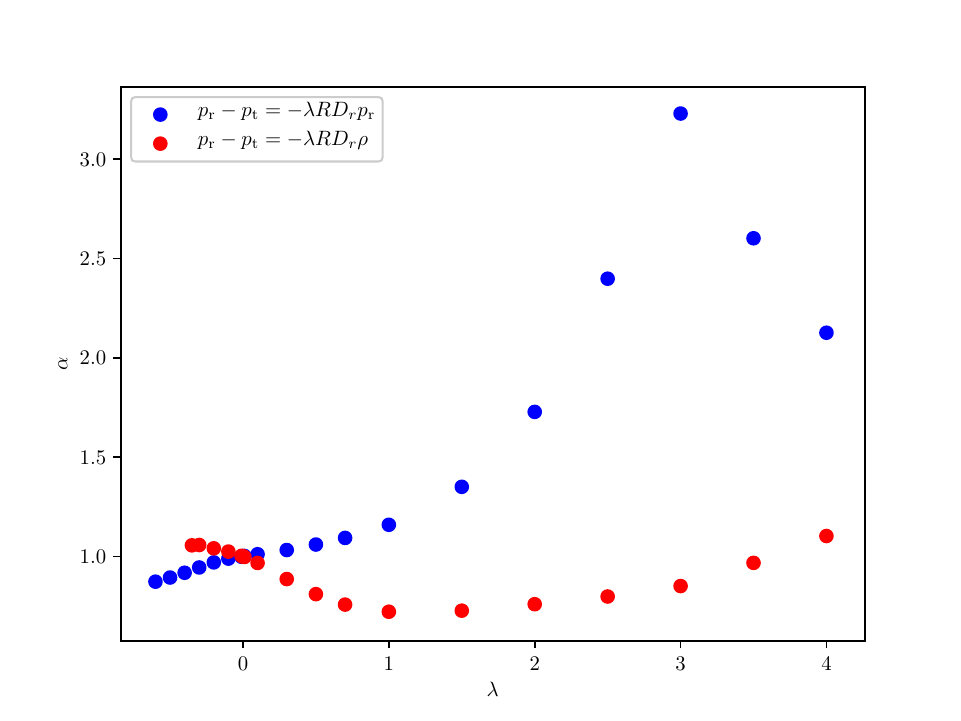}
\includegraphics[height = 0.40\textwidth, width=0.496\textwidth, clip=true]{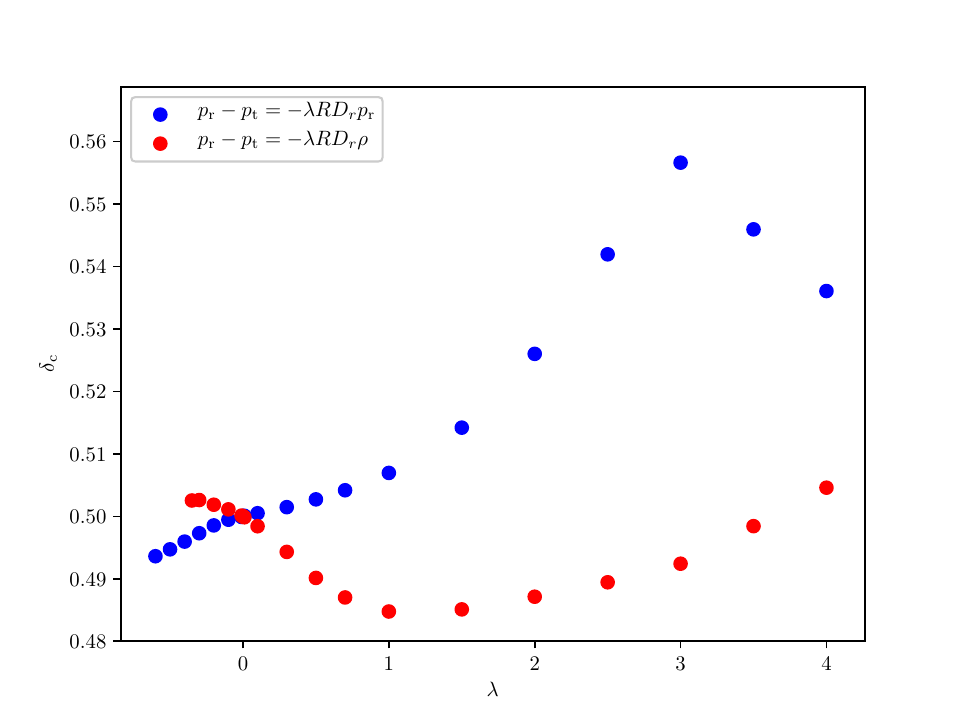}
\caption{The shape parameter (left panel) and the PBH formation threshold (right panel) as a function of the degree of anisotropy $\lambda$. The blue circles stand for the prescription in which  $p_\mathrm{r}-p_\mathrm{t}=-\tilde{\lambda}R(r,t) D_\mathrm{r}p_\mr$ while the red ones correspond to the prescription in which $p_\mathrm{r}-p_\mathrm{t}=-\tilde{\lambda}R(r,t) D_\mathrm{r}\rho$.}
\label{fig:delta_c_lambda_no_dim}
\end{center}
\end{figure}
\begin{figure}[h!]
\begin{center}
\includegraphics[height = 0.40\textwidth, width=0.496\textwidth, clip=true]{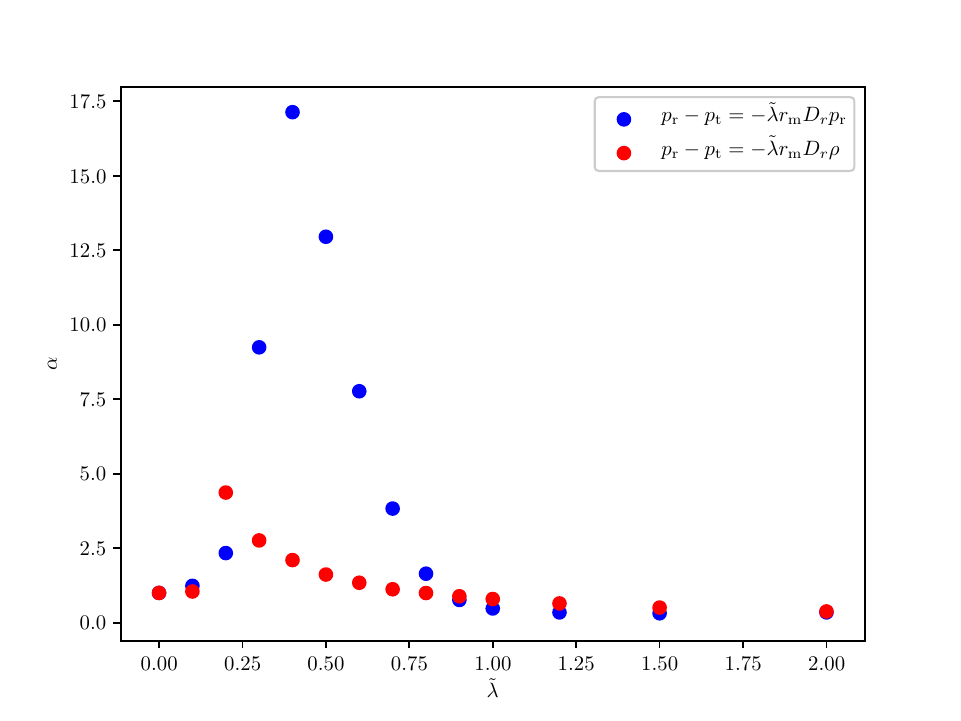}
\includegraphics[height = 0.40\textwidth, width=0.496\textwidth, clip=true]{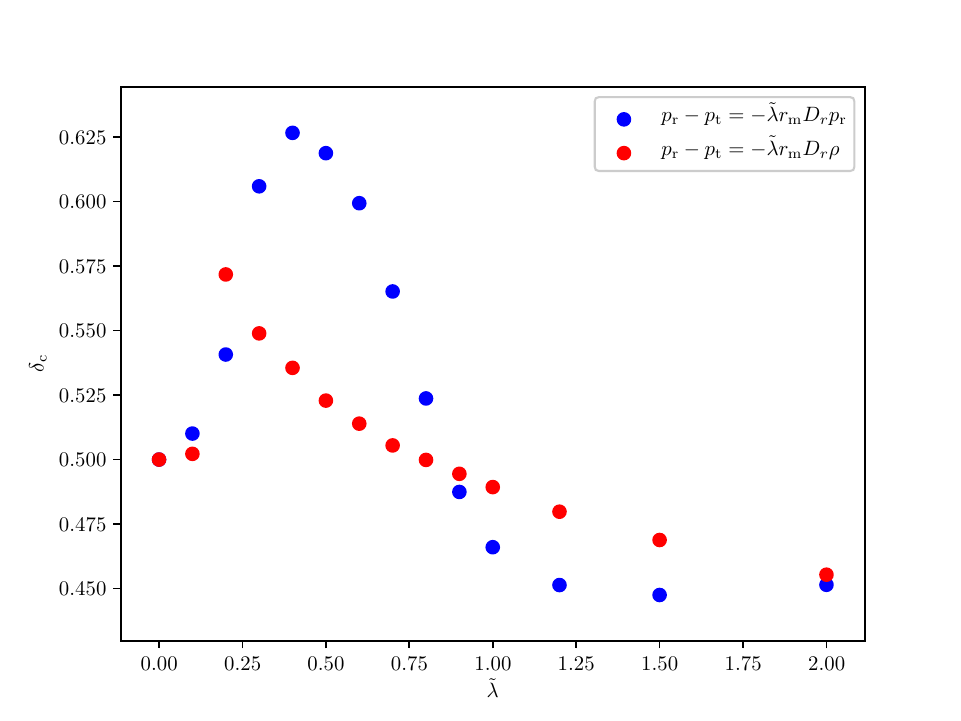}
\caption{The shape parameter (left panel) and the PBH formation threshold (right panel) as a function of the degree of anisotropy $\tilde{\lambda}$. The blue circles stand for the prescription in which $p_\mathrm{r}-p_\mathrm{t}=-\tilde{\lambda}r_\mathrm{m} \left(\frac{\rho}{\rho_\mathrm{b,inf}}\right)^n D_\mathrm{r}p_\mr$  while the red ones correspond to the prescription in which $p_\mathrm{r}-p_\mathrm{t}=-\tilde{\lambda}r_\mathrm{m} \left(\frac{\rho}{\rho_\mathrm{b,inf}}\right)^n D_\mathrm{r}\rho$. In both prescriptions we have chosen $n=0$, $q=10^{-10}$ and $\epsilon_0=10^{-1}$.}
\label{fig:delta_c_lambda_dim}
\end{center}
\end{figure}

As one can see from the figures below, an initial increase of $\delta_\mathrm{c}$ is observed, which is somehow expected for an increasing amplitude of the anisotropy, as mentioned in the previous sections. This is because in this regime, the radial pressure is enhanced with respect to the tangential one, an effect that works against the gravitational collapse. On the contrary, after a critical value of $\lambda$ or $\tilde{\lambda}$, $\delta_\mathrm{c}$ follows a decreasing behavior, an effect that could not be predicted. However, if one makes a careful analysis on the dependence of $\alpha$ in terms of  $\lambda$ or $\tilde{\lambda}$ it is possible to identify a dependence of  $\alpha\propto (\lambda\mathrm{\;or\;}\tilde{\lambda})\left.\frac{\partial(\tilde{p}_\mr \mathrm{\;or\;} \tilde{\rho})}{\partial r}\right\vert{r_\mathrm{m}}$, with two competing terms: $\lambda$ or $\tilde{\lambda}$, and $\left.\frac{\partial(\tilde{p}_\mr \mathrm{\;or\;} \tilde{\rho})}{\partial r}\right\vert{r_\mathrm{m}}$ which decreases in absolute value with respect to $\lambda$ or $\tilde{\lambda}$ as it can be seen numerically from Figs. \ref{fig:dpr_dr_pr_f=R}, \ref{fig:dpr_dr_pr_f=rho^n-Drpr}, \ref{fig:drho_dr_pr_f=R} and \ref{fig:drho_dr_pr_f=rho^n-Drrho}. Consequently, one expects a critical turning point in the behavior of $\delta_\mathrm{c}$. The explicit dependence of $\delta_\mathrm{c}$ in terms of $\lambda$ or $\tilde{\lambda}$ has not been deduced yet and it is part of an ongoing research work.

At this point, one should comment the fact that in the case where $f(r,r) = R(r,t)$ the behavior of $\alpha$ and $\delta_\mathrm{c}$ when the difference $p_\mr-p_\mt$ is modeled as proportional to pressure gradients is quite different with respect to the case when $p_\mr-p_\mt$ is modeled as proportional to energy density gradients. In the first case, $\delta_\mathrm{c}$ initially increases with $\lambda$ and then after a critical point decreases. However, in the second case, $\delta_\mathrm{c}$ initially decreases and then increases with $\lambda$. This behavior could be caused from the fact that the EoS for $f(r,t)=R(r,t)$ is given not only in terms of local quantities such as $p_\mr$ and $\rho$ but also in terms of non local quantities as the areal radius $R(r,t)$. This is an indication that this model is not so well physically motivated. This claim however needs to be confirmed by performing full numerical simulations.

Finally, it is important to stress out that \Eq{delta_c_analytical_Musco} gives just an estimation of $\delta_\mathrm{c}$ for values of $\lambda$ or $\tilde{\lambda}$ in which $p_\mt-p_\mt\ll 1$ and cannot be trusted if one wants to find the exact value of $\delta_\mathrm{c}$ in presence of anisotropies. For this reason it is important to perform in the future  the full numerical analysis and evolve the non-linear hydrodynamic equations. Despite this fact, the results obtained here give a reasonable estimation of the effect of the anisotropy.

\chapter{Conclusions - Outlook}
Since '70s, when PBHs were initially introduced by  Novikov and Zeldovich ~\cite{1967SvA....10..602Z} and  Stephen Hawking ~\cite{1971MNRAS.152...75H}, they have been attracting an increasing attention within the scientific community. Already in ~\cite{Chapline:1975ojl}, PBHs were proposed to contribute to dark matter and seed the supermassive black holes we see in the center of galaxies ~\cite{1984MNRAS.206..315C, Bean:2002kx}. In '90s, the first formation scenarios made their appearance ranging from inflationary production mechanisms ~\cite{Dolgov:1992pu,Carr:1994ar}, primordial phase transitions ~\cite{Crawford:1982yz,Kodama:1982sf} up to gravitational collapse of topological defects ~\cite{Hawking:1987bn,Polnarev:1988dh,MacGibbon:1997pu}. In the following decades, a huge progress took place regarding the analytical ~\cite{Harada_2013} and numerical ~\cite{1980SvA....24..147N,Niemeyer:1997mt,Shibata_1999} methods describing the PBH gravitational collapse process. In addition, the wide range of masses of PBHs gave access to different physical phenomena, a fact which gave the possibility to make significant progress on the constraints of the abundance of PBHs at different mass ranges by studying data from different observational probes ~\cite{Carr:2020gox}.

In the view of this significant progress on the field of PBHs physics, both at the theoretical and the observational level, a first goal of this thesis was to constrain parameters of the early universe through PBH physics and vice-versa to constrain PBHs by studying aspects of the early universe. In particular, in a first part of the thesis we set constraints on parameters of the early universe, namely the energy scale at end of inflation and the energy scale at the onset of the radiation era, by studying PBHs produced from the preheating instability in the context of single-field inflationary models ~\cite{Martin:2019nuw, Martin:2020fgl}. Interestingly, we find that PBHs can be so abundantly produced that reheating can proceed from their evaporation. By taking also into account the decay of the inflaton field to a radiation fluid, we show that the resonant instability structure of preheating responsible for the PBH production is not disrupted by the presence of the radiative products of the inflaton, a fact which points out to the presence of a generic PBH production mechanism from the preheating instability in the context of single-field inflation.

Regarding future prespectives of this first research part of this thesis, one should point out the possibility to narrow down the observational predictions of the CMB. Particularly, for a fixed single-field inflationary potential, the only theoretical uncertainty in the observational predictions of the CMB is on the number of e-folds elapsed between the time when the CMB pivot scale exits the Hubble radius and the end of inflation, namely $\Delta N_{*}$. This number depends on the energy scale at the end of inflation, $\rho_\mathrm{inf}$, which is given by the inflationary model under consideration, the  energy density at the onset of the radiation era $\rho_\mathrm{rad}$, and the averaged equation-of-state parameter between the end of inflation and the onset of the radiation era, $\bar{w}_\mathrm{rad}$ and reads ~\cite{Martin:2006rs} 
\begin{align}\label{DN_*}
\Delta N_{*} & = \frac{1-3\bar{w}_\mathrm{rad}}{12(1+\bar{w}_\mathrm{rad})}\ln \left(\frac{\rho_\mathrm{rad}}{\rho_\mathrm{inf}}\right)+\frac{1}{4}\ln\left(\frac{\rho_*}{9\Mp^4}\frac{\rho_*}{\rho_\mathrm{inf}}\right) \\&-\ln\left(\frac{k_\mathrm{P}/a_\mathrm{now}}{\tilde{\rho}^{1/4}_\mathrm{\gamma,now}}\right),
\end{align}
where $\bar{w}_\mathrm{rad}\equiv \frac{\int_{N_\mathrm{inf}}^{N_\mathrm{rad}}w(N)dN}{\Delta N_\mathrm{rad}}$ is the mean equation of state during reheating with $\Delta N_\mathrm{rad} = N_\mathrm{rad}-N_\mathrm{inf}$ , $\rho_{*}$ is the energy scale at the time when the CMB pivot scale exits the Hubble radius, $a_\mathrm{now}$ is the present value of the scale factor, and $\tilde{\rho}_\mathrm{\gamma,now}$ is the the energy density of radiation today rescaled by the number of relativistic degrees of freedom. Taking the pivot scale $k_\mathrm{P} /a_\mathrm{now}$ to be 0.05Mpc$^{-1}$ and $\tilde{\rho}_\mathrm{\gamma}$  to its measured value, the last term is $N_0 \equiv -\ln\left(\frac{k_\mathrm{P}/a_\mathrm{now}}{\tilde{\rho}^{1/4}_\mathrm{\gamma,now}}\right)\simeq 61.76$.

\begin{figure}[h!]
\begin{center}
\includegraphics[width=0.49\textwidth, clip=true]{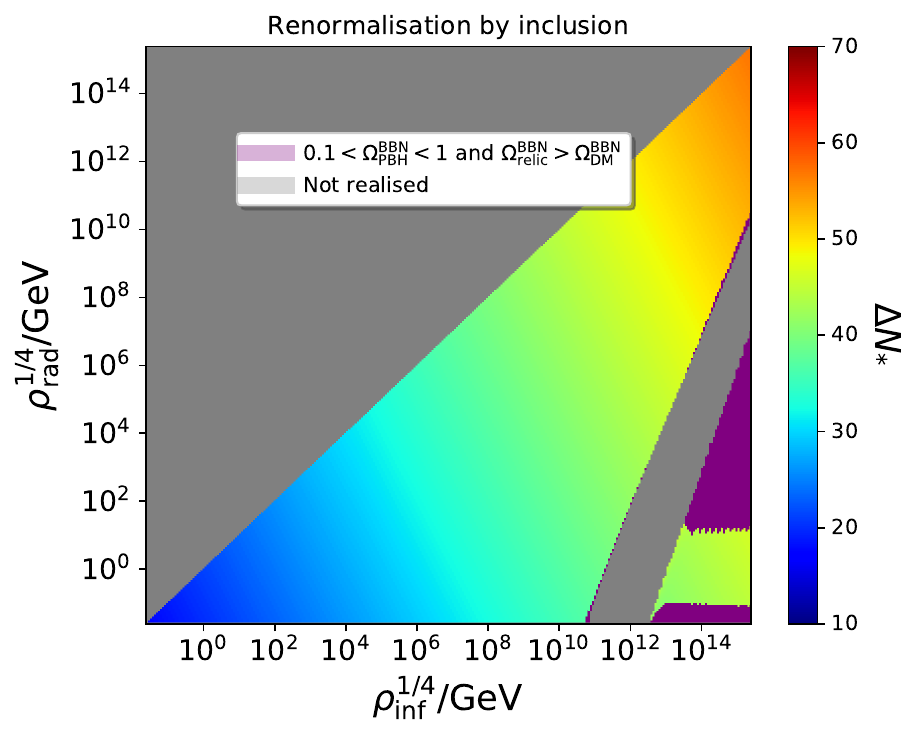}
 \includegraphics[width=0.49\textwidth, clip=true]{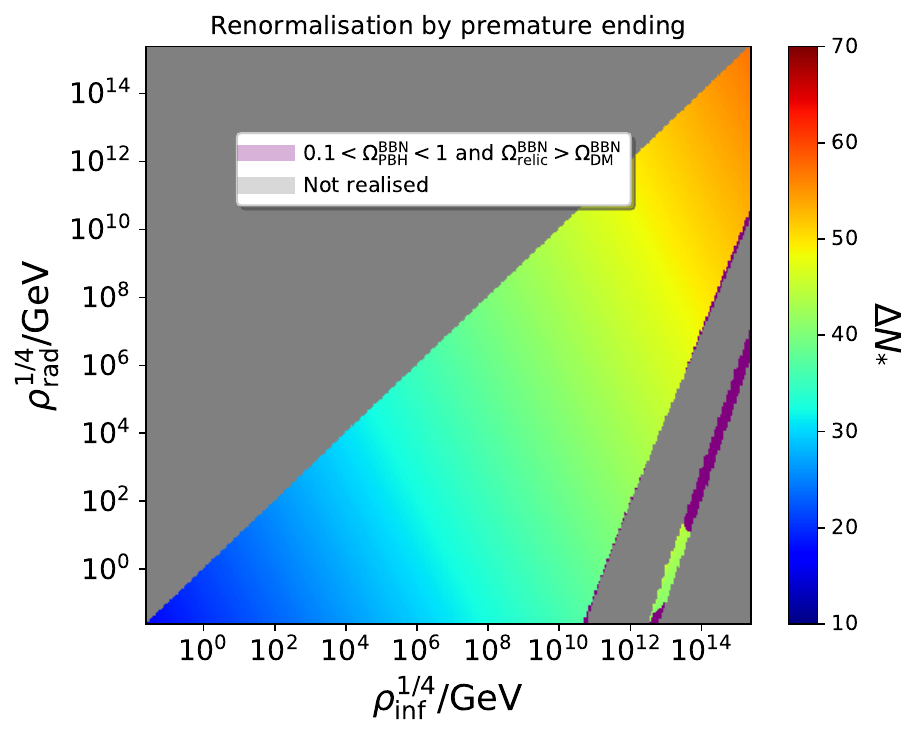}
\caption{$\Delta N_{*}$ as a function of $\rho_\mathrm{inf}$ and $\rho_\mathrm{rad}$, when the PBH mass fraction is renormalised by inclusion (left panel) and by
  premature ending (right panel).}
\label{fig:DN_constraints}
\end{center}
\end{figure}

Considering then constraints on the energy scale at the end of inflation and on energy scale at the onset of the radiation from PBHs produced during preheating, one can constrain $\Delta N_{*}$. In particular, by using the theoretical setup introduced ~\cite{Martin:2019nuw} one can compute $\bar{w}_\mathrm{rad}$ and from \Eq{DN_*} constrain $\Delta N_{*}$. In 
\Fig{fig:DN_constraints} we give the constraints on $\Delta N_{*}$ as a function of $\rho_\mathrm{inf}$ and $\rho_\mathrm{rad}$.

Vice-versa, in a second part of the thesis, we study a backreaction problem of the gravitational waves induced by energy density perturbations underlain by a gas of PBHs. In particular, by requiring  that the induced gravitational waves associated to PBHs are not overproduced in an era when ultralight PBHs ($m_\mathrm{PBH}<10^9\mathrm{g}$) dominate the energy budget of the universe we set constraints on the abundance of PBHs at the time they form as a function of their mass ~\cite{Papanikolaou:2020qtd}. To the best of our knowledge, these constraints are actually the first solid model-independent constraints in the literature regarding ultralight PBHs, which are poorly constrained since they evaporate before BBN and they do not leave a direct observational imprint apart from rather speculative Planckian relics produced as leftovers of the PBH Hawking evaporation. 

\begin{figure}[h!]
\begin{center}
  \includegraphics[width=0.68\textwidth,  clip=true]
                  {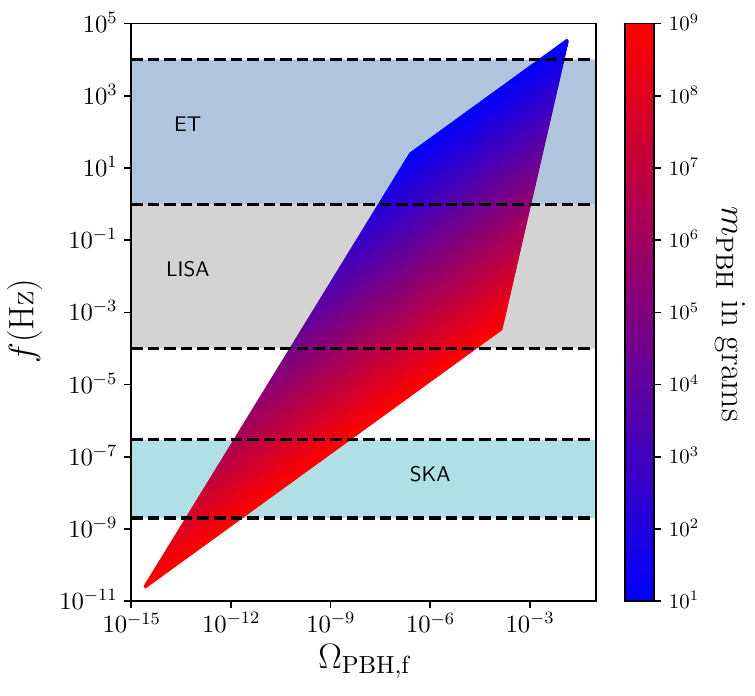}
\caption{The peak frequency of the stochastic gravitational wave background induced by a dominating gas of primordial black holes, as a function of  
their abundance at the time they form, $\Omega_\mathrm{PBH,f}$ (horizontal axis), and their mass $m_\mathrm{PBH}$ (colour coding).
The displayed region of parameter space corresponds to values of $m_\mathrm{PBH}$ and $\Omega_\mathrm{PBH,f}$ such that black holes dominate the universe content for a transient period, they form after inflation and Hawking evaporate before BBN and that the induced gravitational waves are not overproduced, see \Eq{Omega_f constraints}. In practice, \Eq{GW frequency} is displayed with $g_\ueff=100$, $z_\mathrm{eq}=3387$ and $H_0=70\,\mathrm{km}\,\mathrm{s}^{-1}\,\Mpc^{-1}$. For comparison, the detection frequency bands of ET, LISA and SKA are also shown. Figure credited ~\cite{Papanikolaou:2020qtd}.}
\label{fig:GW frequency}
\end{center}
\end{figure}

At this point, one should point out the potential detectability of the gravitational-wave signal of the stochastic background of induced gravitational waves produced from a universe filled with ultralight PBHs. As we found in ~\cite{Papanikolaou:2020qtd}, the peak frequency of the relevant gravitational-wave spectrum, given in \Eq{GW frequency}, depends crucially on the initial PBH abundance when the PBH forms, $\Omega_\mathrm{PBH,f}$, and the PBH mass, $m_\mathrm{PBH}$, and lies within the frequency band of future gravitational-wave experiments like the Einstein Telescope (ET)~\cite{Maggiore:2019uih}, the Laser Interferometer Space Antenna (LISA)~\cite{Audley:2017drz,Caprini:2015zlo} and  the Square Kilometre Array (SKA) facility~\cite{Janssen:2014dka}, 
\bea
\label{GW frequency}
\frac{f}{\mathrm{Hz}} \simeq \frac{1}{\left(1+z_\mathrm{eq}\right)^{1/4}}\left(\frac{H_0}{70\mathrm{kms^{-1}Mpc^{-1}}}\right)^{1/2}\left(\frac{g_\mathrm{eff}}{100}\right)^{1/6}\Omega^{2/3}_\mathrm{PBH,f} \left(\frac{m_\mathrm{PBH} }{10^9\mathrm{g}}\right)^{-5/6} ,
\eea
where $H_0$ is the value of the Hubble parameter today, $g_\mathrm{eff}$ is the effective number of relativistic degrees of freedom at PBH formation time and $z_\mathrm{eq}$ is the redshift at matter-radiation equality. See also the following \Fig{fig:GW frequency} in which the peak frequency is plotted as a function of the initial PBH abundance at formation time and the PBH mass.

This is very important, since one can potentially further constrain ultralight PBHs from the upcoming data of future gravitational-wave observational probes. However, in order to give an explicit answer on whether this signal can be detected, one should take into account the dynamical evolution of the gravitational-wave spectrum from the end of the PBH-dominated era up to our epoch by resolving the gradual transition from the PBH-dominated era up to the subsequent radiation-dominated era which is rather subtle ~\cite{Inomata:2019zqy,Domenech:2020ssp}.

Finally, in the last part of the thesis, which is a work in progress and not published yet, we study aspects of the gravitational collapse of a radiation fluid to PBHs in the presence of anisotropies. In particular, we model in a covariant way the difference between the radial, $p_\mr$, and the tangential pressure, $p_\mt$, as proportional to either to pressure or energy density gradients with a proportionality factor $\lambda$ which accounts for the anisotropic nature of the gravitational collapse.
Then, by performing a gradient expansion approximation on superhorizon scales at the level of the Einstein's equations we extract the initial conditions for the hydrodynamic and metric perturbations in the presence of anisotropies. At the end, we  deduce the dependence of the PBH formation threshold in terms of the anisotropy parameter $\lambda$ adopting a perturbative approach based on the assumption that $\delta_\mathrm{c}$ depends on the shape of the initial energy density profile in the same way as in the isotropic case. Although this is something that requires a numerical investigation which is an ongoing work, the results obtained here give a reasonable estimation of what is the effect of the anisotropy.

The next step is to evolve in time the non-linear hydrodynamic equations and study numerically the PBH apparent horizon formation and the explicit dependence of the PBH formation threshold on the anisotropy parameter $\lambda$. In this way, we will be able to derive the dependence as well of the PBH mass fraction on the anisotropy parameter $\lambda$. In this way, one can make use of the observational constraints on the PBH abundances and set tight constraints on the anisotropy parameter of our model. In addition, in the case where the anisotropy parameter $\lambda$ becomes dimensionful, it depends on the intrinsic energy scale of the problem, which in our case is the energy scale when the perturbations are generated, namely the energy scale at the end of inflation. Consequently, in this regime one can translate the potential observational constraints on $\lambda$ to constraints on the energy scale at the end of inflation giving access in this way to the inflationary landscape.

To summarize, with this thesis we studied aspects of the early universe through PBH physics. In particular, we set constraints on cosmological parameters of the early universe by studying PBHs produced during the preheating instability in the context of single-field inflation and we constrained PBHs by studying induced gravitational waves produced in an early PBH-dominated era. In addition, we studied some facets of the PBH gravitational collapse in the presence of anisotropies. The research findings related to this thesis can potentially open up new directions in the field of PBH physics as stressed out above shedding light at the same time in the understanding of the physics of the early universe.

\newpage
\selectlanguage{french}
\chapter{Compte Rendu Français}
Ce compte rendu contient une description courte des résultats principaux de cette thèse sur articles.
\section{Contexte Scientifique}
Les trous noirs primordiaux (TNP), proposés initialement dans les années '70 ~\cite{1967SvA....10..602Z, Carr:1974nx,1975ApJ...201....1C}, attirent de plus en plus l'attention de la communauté scientifique, vu qu'ils peuvent adresser un grand nombre de problèmes de la cosmologie contemporaine. D'après des arguments récents, ils peuvent potentielllement constituer une partie ou la totalité de la matière noire en expliquant en même temps la génération des structures gravitationnels de grande échelle à travers les fluctuations Poissoniennes ~\cite{Meszaros:1975ef,Afshordi:2003zb}. De plus, les TNP pourraient fournir les graines des trous noirs supemassifs au milieu des noyaux galactiques \cite{Carr:1984id, Bean:2002kx} aussi bien que consituer les ancêtres des évenements de coalescence des trous noirs recemment  détectés  par les missions LIGO/VIRGO ~\cite{LIGOScientific:2018mvr} à travers l'émission des ondes gravitationnels.
 
Malgré le fait que les TNP ne sont pas encore détectés, ces objets astrophysiques jouent un rôle cardinal sur la cosmologie étant donné le fait que avec eux, dépendamment de leur masse, on peut explorer et contraindre une grande variété de phénomènes physiques. En particulier, les TNP de pétite masse ($m_\mathrm{PBH}\leq 10^{15}\mathrm{g}$) qui se sont évaporés maintenant peuvent donner accès à la physique de l'univers primordial comme la physique de l'inflation et des perturbations cosmologiques primordiales~\cite{Kalaja:2019uju}, la physique du rechauffement de l'univers et de la nucléosynthèse après le Big Bang (BBN) \cite{2010arXiv1006.5342S,Keith:2020jww} , la physique des ondes gravitationnels primordiaux ~\cite{Clesse:2018ogk}  et des transitions de phase primordiales  ~\cite{Jedamzik:1999am} aussi bien que la physique du fond cosmologique (Cosmic Microwave Background)~\cite{Ali-Haimoud:2016mbv}. De l'autre côté, avec les TNP d'une masse intérmediaire qui s'évaporent à notre époque cosmique, on peut investiguer des phénomènes de l'astrophysique de haute énérgie comme le fond des rayons cosmiques par le biais de l'évaporation Hawking de TNP~\cite{Carr:2016hva}. Enfin, les TNP d'une grande masse qui existent encore aujourd'hui, i.e. ($m_\mathrm{PBH}> 10^{15}\mathrm{g}$), peuvent donner accès à des phénomènes de la physique gravitationnelle, comme les lentilles gravitationnelles ~\cite{Niikura:2019kqi,Zumalacarregui:2017qqd}, à la formation des structures de grande échelle ~\cite{Carr:2018rid} aussi bien que à la physique du secteur noir de l'univers, à savoir la matière noire ~\cite{Carr:2020xqk}  et l'énergie noire~\cite{Nesseris:2019fwr}.

\section{Recherche effectuée pendant la thèse}
Après avoir suscité avant le contexte scientifique sur le domaine de TNP on résume ici les résultats de la recherche effectuée pendant mes études doctorales, pendant laquelle on a combiné des aspects de l'univers primordial et de la physique des ondes gravitationnels avec la physique des TNP. On a étudié aussi des facettes du processus de l’effondrement gravitationnel des TNP en présence des anisotropies. 

\subsection{TNP de l'instabilité de préchauffement}
L'inflation constitue la théorie cardinale de la cosmologie primordiale qui peut décrire d'une manière concordante les conditions initiales de l'univers primordial et résoudre les problèmes fondamentaux de l'époque du Hot Big Bang, à savoir les problème de l'horizon et de la platitude. De plus, l'inflation peut générer naturellement les petrurbations cosmologiques primordiales qui ont engendré les structures de grande échelle à l'univers aussi bien que le fond cosmologique.

Pour adrésser ainsi à tous ces aspects mentionnés avant, l'inflation postule une phase initiale pendant laquelle l'univers s'étend avec un rythme accéléré et son budget énergétique est dominé par un champ scalaire, $\phi$, rapporté comme l'inflaton, qui est associé à un potentiel inflationnaire $V(\phi)$. Après avoir passé d'une phase de roulement lent tout au long de son potentiel, l'infaton commence à osciller à l'origine de son potentiel, un comportement oscillatoire qui conduit inévitablement à l'émergence d'une structure d'instabilité résonante au niveau de l'équation de mouvement des perturbations scalaires. C'est alors pendant cette phase oscillatoire de l'inflaton, souvant citée comme préchauffement, qu'on étudie la production des TNP dans le contexte des modèles inflationnaires avec un champ scalaire ~\cite{Martin:2019nuw, Martin:2020fgl}.

\begin{figure}[t]
\begin{center}
\includegraphics[width=0.496\textwidth, clip=true]{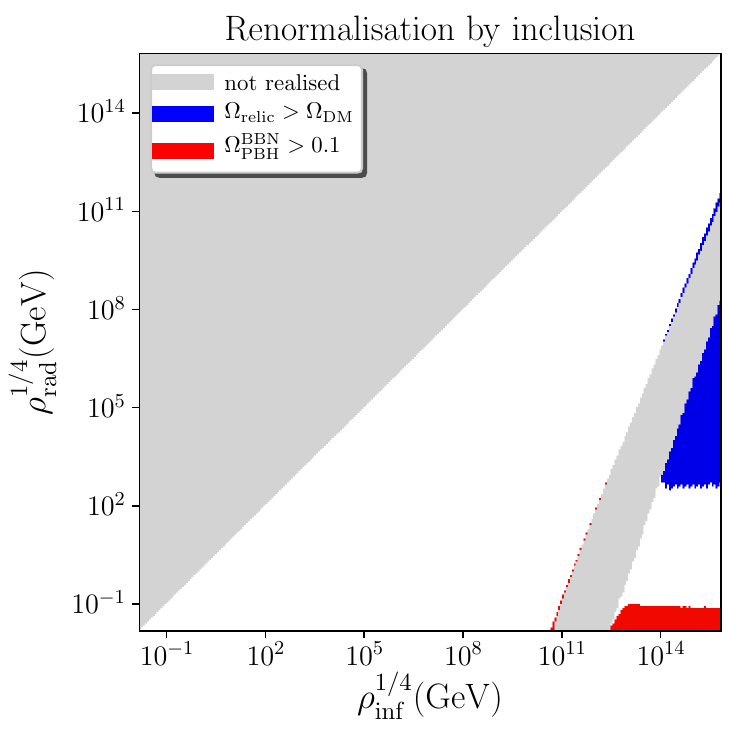}
\includegraphics[width=0.496\textwidth, clip=true]{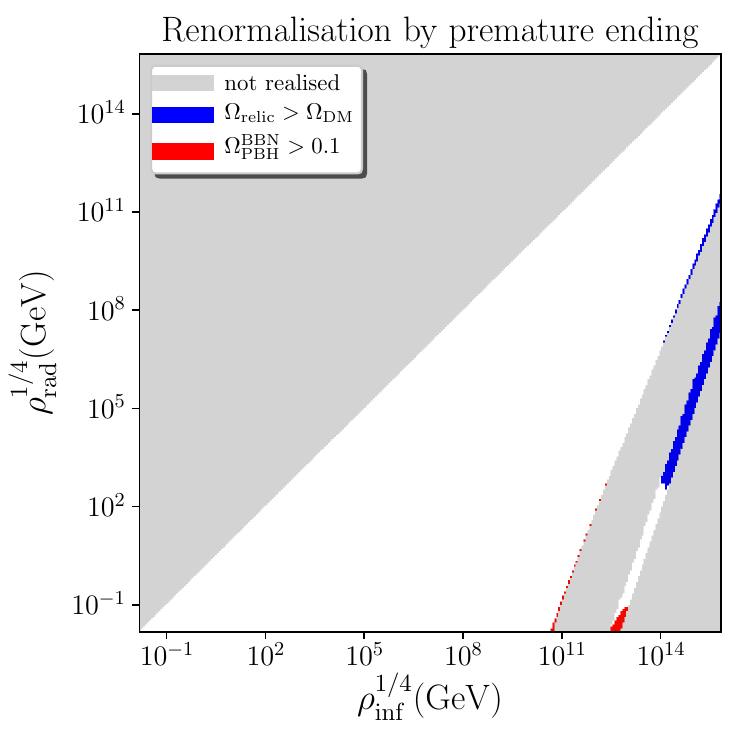}
\caption{Contraintes combinées au niveau des paramètres $\rho_\mathrm{rad}$ et $\rho_\mathrm{inf}$ quand la fonction de masse de TNP se renormalise par inclusion
  (panneau gauche) et par l'arrêt prématuré de l'instabilité de préchauffement (panneau droit). Les régions grises
  sont exclues vu qu'elles correspondent aux valeurs de $\rho_\mathrm{rad}$, qui ne peuvent pas être réalisés naturellement tandis que les régions rouges sont aussi exclues parce qu'elles conduisent aux abondances très grandes de TNP à l'époque de BBN. Les régions bleues sont exclues vu que dans lequelles quelq'un se confronte avec une surpoduction des reliques Planckiennes. La région qui reste, démontrée en blanc, est celle qui est permise. Figure créditée à ~\cite{Martin:2019nuw}.}
\label{fig:combined_constraints}
\end{center}
\end{figure}

En particulier, on a trouvé que les TNP produits pendant la période du prechauffement peuvent potentiellement dominer le contenu énergétique de l'univers et conduire au réchauffement de l'univers, pendant lequel les particules élémentaires du Modèle Standard se produisent, à travers leur évaporation. Par conséquent, en exigeant que les TNP ne dominent pas le budget énergétique de l'univers pendant la période de la nucléosynthèse après le Big Bang et qu'ils ne surproduisent pas des reliques Planckiennes, on a imposé des contraintes concernant l’échelle d’énergie de l’univers au commencement de l’époque du Hot Big Bang, $\rho_\mathrm{rad}$ aussi bien l’échelle d’énergie de l’univers à la fin de la période d’inflation, $\rho_\mathrm{inf}$ ~\cite{Martin:2019nuw}. Dans la \Fig{fig:combined_constraints} ci-dessous on montre les contraintes combinées au niveau des paramètres $\rho_\mathrm{rad}$ et $\rho_\mathrm{inf}$ pour les deux prescriptions de renormalisation de la fonction de masse des TNP étant introduites pour adresser le problème de la surproduction de TNP. Pour plus de détails à voir sur ~\cite{Martin:2019nuw}.

Ensuite, vu que sur ~\cite{Martin:2019nuw} on a considéré que les auto-interactions de l'inflaton, on a avancé notre recherche en couplant d'une manière covariante le champ inflaton avec un fluide de radiation afin d'assurer la transition de la période de l'inflation à la période du Hot Big Bang. Enfin, on a trouvé que la desintégration de l'inflaton en radiation ne change pas la structure de l'instabilité résonante du préchauffement, responsable pour la production de TNP, jusqu'à le moment où le fluide de radiation domine énergétiquement l'univers. Par suite, l'émergence de l'instabilité de résonance qui favorise la formation des TNP est encore présente pendant la période du préchauffement, en indiquant de cette façon un mécanisme de production de TNP général dans le contexte des modèles inflationnaires avec un champ scalaire~\cite{Martin:2020fgl}.

\subsection{Ondes gravitationnelles d'un univers rempli des TNP}
Au sein de cette thèse, on combiné aussi des aspects de la physique des ondes gravitationnelles avec la physique des TNP. \'Evidemment, il y a pleins de canaux connectant la physique des TNP avec la physique des ondes gravitationnnelles. Parmi eux, on peut mettre en exergue les trois plus significatifs. Tout d'abord, il faut se référer aux ondes gravitationnelles de second ordre qui se sont induites par les perturbations de courbure primordiales qui ont précédé et ont donné naissance aux TNP. Un second canal possible de connection de TNP avec les ondes gravitationnelles est le fond stochastique gravitationnel des gravitons émis par le biais de l'évaporation Hawking des TNP. Enfin, le troisième canal, le plus étudié par rapport aux autres, est le fond stochastique des ondes gravitationnelles produites à travers des événements de coalescence des TNP. 

Pendant le temps de déroulement de cette thèse, on s'était concentré sur les ondes gravitationnellles induites par des perturbations scalaires associés aux TNP eux-mêmes, et pas aux perturbations de courbure primordiales qui ont donné naissance aux TNP. En particulier, on a étudié un gas des TNP qui crée son propre potentiel gravitationnel et qui induit inévitablement un fond des ondes gravitationnelles à travers des effets gravitationnels de seconde ordre. Plus spécifiquement, on a traité des régimes où les TNP constituent la composante principale du budget énergétique de l’univers pendant une période avant BBN. En demandant alors que ces ondes gravitationnelles induites ne se produisent pas en excès à la fin de la période de domination énergétique des TNP,  on a imposé des contraintes indépedentes du modèle de production de TNP sur leur abondance au moment de leur formation en fonction de leur masse.  Ci-dessous, on donne notre approximation analytique des contraintes extraites au niveau de l'abondance des TNP,
\bea
 \label{Omega_f constraints_french}
\Omega_\mathrm{PBH,f} < 1.4\times 10^{-4}\left(\frac{10^9\mathrm{g}}{ m_\mathrm{PBH} }\right)^{1/4}, 
 \eea
où $\Omega_\mathrm{PBH,f}$ et $m_\mathrm{PBH}$ sont l'abondance initiale des TNP le moment de leur formation et la masse des TNP respectivement.

\begin{figure}[h!]
\begin{center}
  \includegraphics[width=0.796\textwidth,  clip=true]
                  {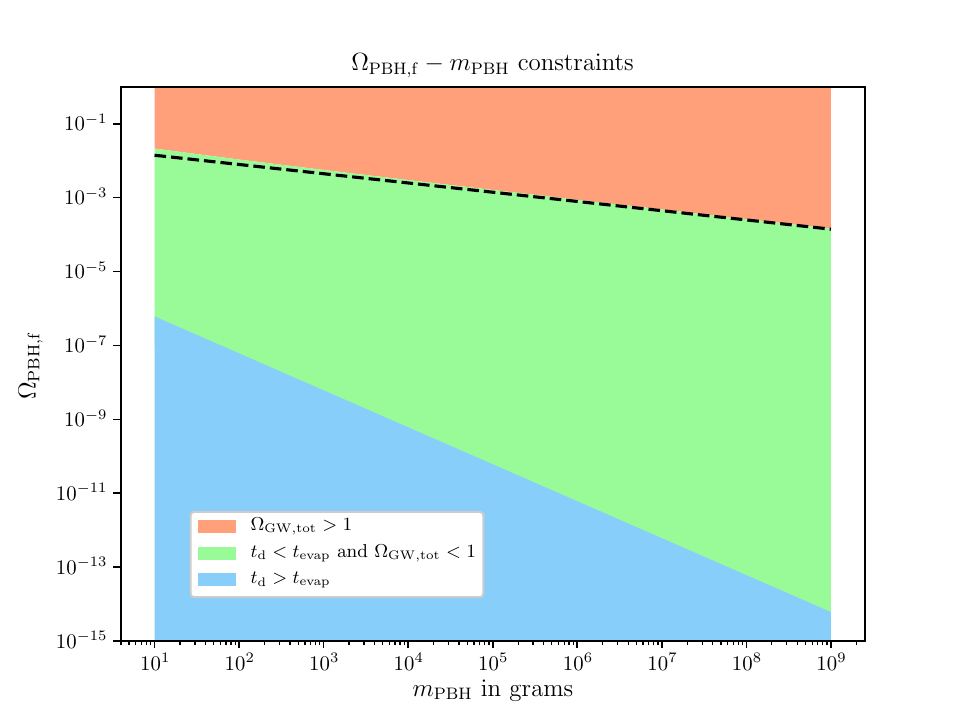}
\caption{L'énergie de densité fractionnelle aux ondes gravitationnelles, $\Omega_\mathrm{GW,tot}$, au moment d'évaporation de TNP en fonction de deux paramètres, à savoir la masse des TNP, $m_\mathrm{PBH}$,  et leur abondance, $\Omega_\mathrm{PBH,f}$, le moment ils se forment. La région orange correspond aux régimes où $\Omega_\mathrm{GW}(t_\mathrm{evap})>1$ et conduit à la surproducion des ondes gravitationnelles. Cette région est alors exclue. La région bleue correspond aux régimes où les TNP ne dominent jamais le budget énergétique de l'univers tandis que la région verte correspond aux cas où on se confronte avec des phases transitoires de domination des TNP et où les ondes gravitationnelles ne se produisent pas en excès. La ligne noire en tirés représente notre approximation analytique ~\eqref{Omega_f constraints_french}, par rapport à la contrainte supérieure au niveau de $\Omega_\mathrm{PBH,f}$. \'Evidemment, on peut voir qu'il nous fournit avec une approximation très bonne de la frontière entre les régions orange et verte. }
\label{fig:Omega_f_newtonian_gauge_RD_to_MD_constraints}
\end{center}
\end{figure}

Vue que ces TNP se forment et s'évaporent avant BBN, leurs masses sont miniscules, à savoir $m_\mathrm{PBH}\in\left[10\mathrm{g},10^{9}\mathrm{g}\right]$ et d'après la bibliographie internationale sur le sujet, ils ne sont pas bien contraints. En étudiant alors les ondes gravitationnelles induites produites dans une époque de domination énergétique des TNP  on a pu imposer les premières contraintes solides sur les TNP ultra-légers ~\cite{Papanikolaou:2020qtd}. On voit aussi ci-dessous dans la \Fig{fig:Omega_f_newtonian_gauge_RD_to_MD_constraints} les contraintes extraites au niveau de paramètres $\Omega_\mathrm{PBH,f}$ et $m_\mathrm{PBH}$.

\section{Effondrement anisotrope des TNP}
Enfin, pendant cette thèse, on a abordé aussi une autre facette de la physique des TNP, réliée à l'effondrement gravitationnel en présence des anisotropies. En particulier, en adoptant la symétrie sphérique, on a introduit une pression anisotrope afin de tenir en compte le caractère anisotrope de l'éffondrement gravitationnel d'un fluide de radiation. Plus spécifiquement, étant inspiré par le comportement des objets ultra-compacts qui imitent le comportement dynamique des trous noirs, on modélé d'une façon covariante la difference entre la densité de la pression radiale, $p_\mr$, et la densité de la pression tangentielle, $p_\mt$, en postulant que la différence $p_\mr-p_\mt$ est proportionnelle soit aux gradients de pression soit aux gradients de densité d'énergie, 
\begin{eqnarray}
p_\mt & = & p_\mr + \lambda f(r,t) k^\mu \nabla_\mu p_\mr \quad \textrm{(gradients de pression)} \label{D_rp_r_french} \\ 
& & \quad \quad \quad \textrm{or} \nonumber \\
p_\mt & = & p_\mr + \lambda f(r,t) k^\mu \nabla_\mu \rho \quad \ \textrm{(gradients de densité d'énergie)},
\end{eqnarray}
où $k^\mu$ est un quadrivecteur spatial unitaire orthogonal à la quadrivitesse du fluide, $\nabla_\mu$ correspond à la dérivative covariante, $\lambda$ est un paramètre qui correspond au degré de l'anisotropie et $f(r,t)$ est une fonction libre. Ensuite, en utilisant la méthode de l'éxpansion de gradients au niveau des équations hydrodynamiques, on a extrait les conditions initiales pour les perturbations hydrodynamiques et metriques en fonction du profile de courbure $K(r)$, rélié à la géométrie de l'éspace-temps. 

Enfin, on a extrait la dépendence du seuil de formation de TNP, $\delta_\mathrm{c}$, en fonction du degré de l'anisotropie $\lambda$ en supposant que $\delta_\mathrm{c}$ dépend de la forme du profil initial de la densité d'énergie de la même manière que dans le cas isotrope. Malgré le fait que cela demande une exploration numérique, constituant un travail en cours, les resultats obtenus ici donnent une estimation raisonable de l'effet de l'anisotropie. Dans la figure ci-dessous, on voit la dépendence du  $\delta_\mathrm{c}$ en fonction du degré de l'anisotropie $\lambda$ ou $\tilde{\lambda}$, dépendant de la modelisation de la différence $p_\mr-p_\mt$.

\begin{figure}[h!]
\begin{center}
\includegraphics[height = 0.35\textwidth, width=0.496\textwidth, clip=true]{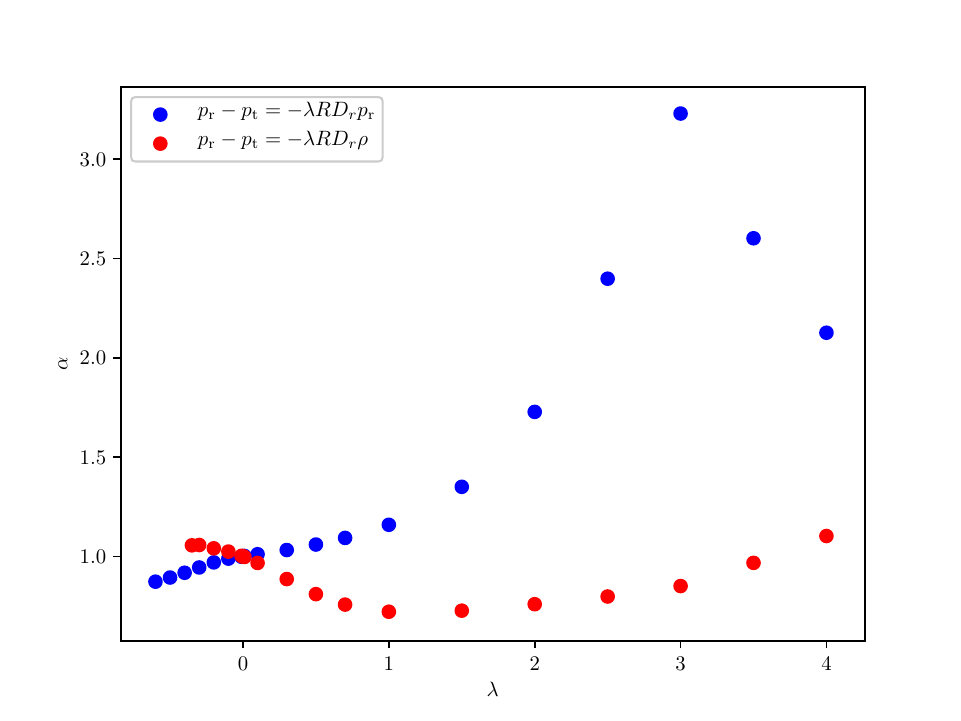}
\includegraphics[height = 0.35\textwidth, width=0.496\textwidth, clip=true]{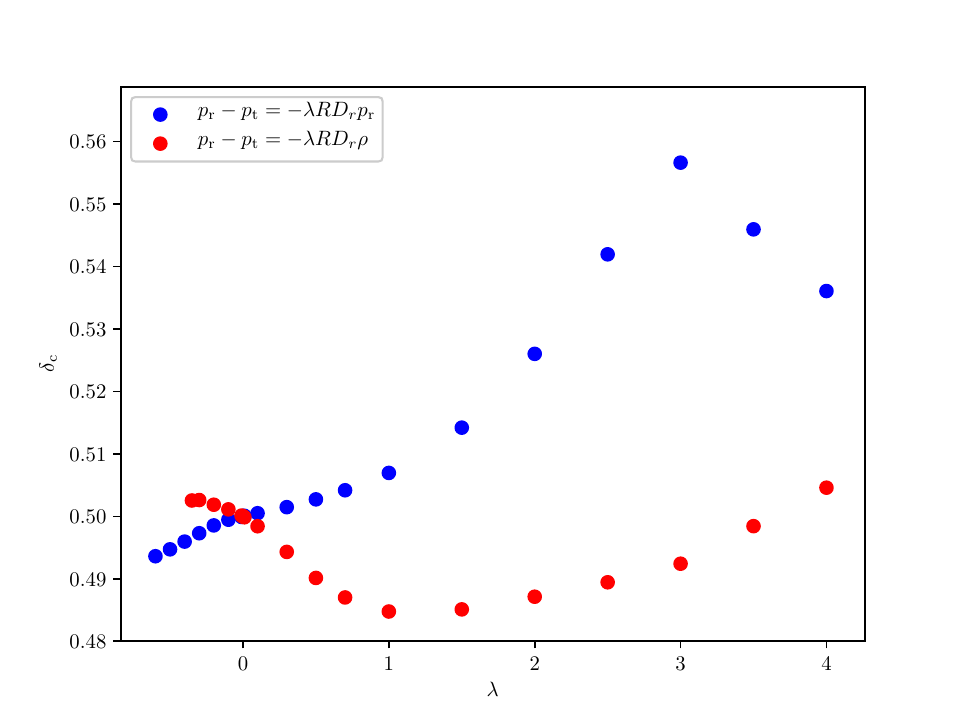}
\caption{Le paramètre de forme (panneau gauche) et le seuil de formation de TNP (panneau droit) en fonction du degré de l'anisotropie $\tilde{\lambda}$. Les circles bleus correspondent à la modelisation où $p_\mathrm{r}-p_\mathrm{t}=-\tilde{\lambda}R(r,t) D_\mathrm{r}p_\mr$ tandis que les circles rouges correspondent à la modelisation où $p_\mathrm{r}-p_\mathrm{t}=-\tilde{\lambda}R(r,t) D_\mathrm{r}\rho$.}
\label{fig:delta_c_lambda_no_dim_french}
\end{center}
\end{figure}

\begin{figure}[h!]
\begin{center}
\includegraphics[height = 0.35\textwidth, width=0.496\textwidth, clip=true]{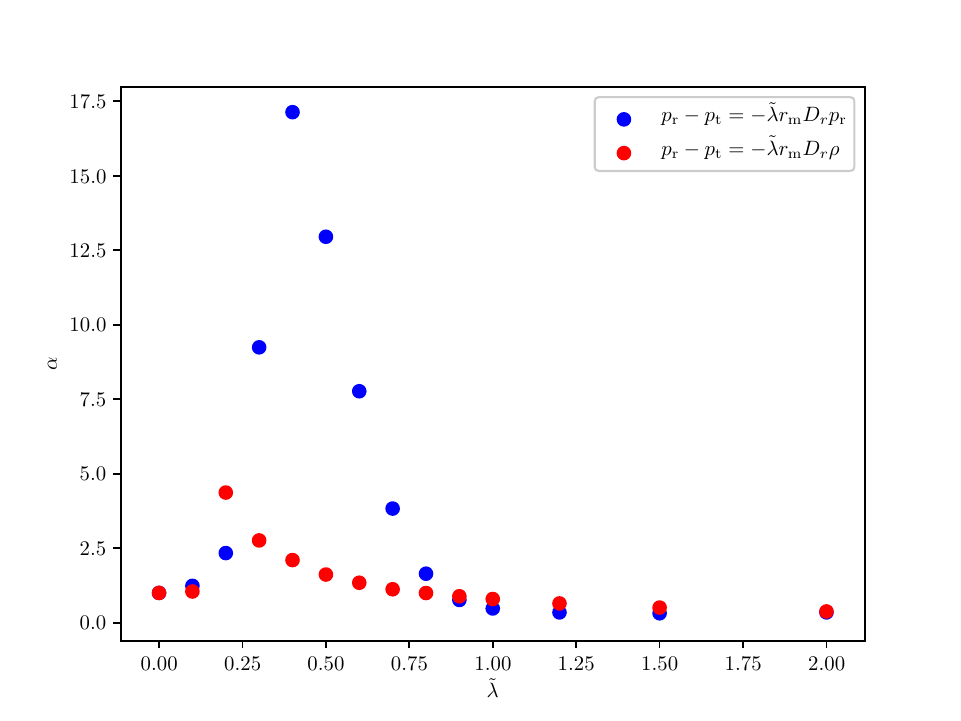}
\includegraphics[height = 0.35\textwidth, width=0.496\textwidth, clip=true]{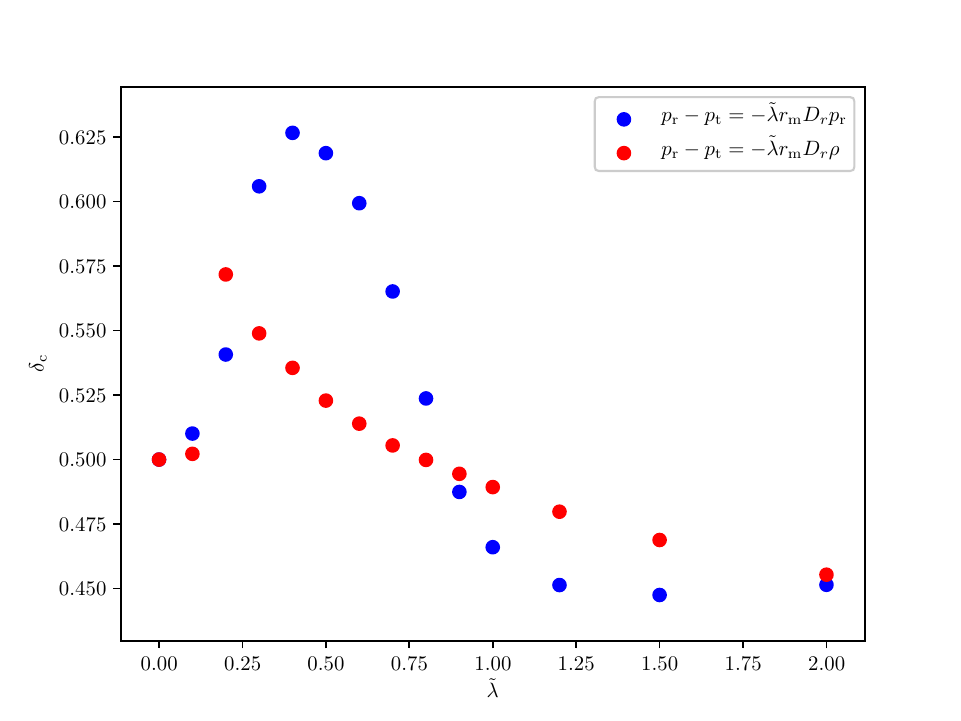}
\caption{Le paramètre de forme (panneau gauche) et le seuil de formation de TNP (panneau droit) en fonction du degré de l'anisotropie $\tilde{\lambda}$. Les circles bleus correspondent à la modelisation où $p_\mathrm{r}-p_\mathrm{t}=-\tilde{\lambda}r_\mathrm{m} \left(\frac{\rho}{\rho_\mathrm{b,inf}}\right)^n D_\mathrm{r}p_\mr$ tandis que les circles rouges correspondent à la modelisation où $p_\mathrm{r}-p_\mathrm{t}=-\tilde{\lambda}r_\mathrm{m} \left(\frac{\rho}{\rho_\mathrm{b,inf}}\right)^n D_\mathrm{r}\rho$. Dans les deux modélisations on a choisi $n=0$, $q=10^{-10}$ and $\epsilon_0=10^{-1}$.}
\label{fig:delta_c_lambda_dim_french}
\end{center}
\end{figure}

\section{Conclusions - Perspectives}
Depuis les années '70, quand les TNP s'étaient initialement introduits par  Novikov and Zeldovich ~\cite{1967SvA....10..602Z} et Stephen Hawking ~\cite{1971MNRAS.152...75H}, ils attirent de plus en plus l'attention de la communauté scientifique. Déjà, dans ~\cite{Chapline:1975ojl} les TNP sont proposés afin de contribuer au budget énergétique de la matière noire et engender les trous noirs supermassifs qu'on voit au centre des galaxies  ~\cite{1984MNRAS.206..315C, Bean:2002kx}. Dans les années '90, les premiers mécanismes de production de TNP se sont apparus, se variant des modèles inflattionaires \cite{Dolgov:1992pu,Carr:1994ar} et des transitions de phase primordiales ~\cite{Crawford:1982yz,Kodama:1982sf} jusqu'à l'effondrement gravitationnel des défauts topologiques ~\cite{Hawking:1987bn,Polnarev:1988dh,MacGibbon:1997pu}. Dans les  decennies suivantes, un grand progrès s'était manifesté par rapport aux méthodes analytiques ~\cite{Harada_2013} et numériques ~\cite{1980SvA....24..147N,Niemeyer:1997mt,Shibata_1999} décrivant le processus de l'effondrement gravitationnel des TNP. De plus, la grande gamme de masse des TNP nous a permis d'avoir accès aux phénomènes physiques différents donnant de cette façon la possibilité de contraindre l'abondance des TNP en étudiant les données observationnelles des différentes éxperiences ~\cite{Carr:2020gox}.

Vu ce progrès significatif au domaine de la physique des TNP, au niveau théorique et à la fois observationnel, un premier but de cette thèse était de poser des contraintes aux paramètres de l'univers primordial à travers la physique des TNP et vice-versa de contraindre les TNP en étudiant quelques aspects de l'univers primordial. En particulier, dans la première partie de la thèse, on a posé des contraintes  sur des paramètres cosmologiques de l'univers primordial, à savoir l'échelle d'énergie à la fin de la période de l'inflation et l'échelle de l'énergie au commencement de l’époque du Hot Big Bang en étudiant des TNP se produisant pendant la période de préchauffement au contexte de la théorie de l'inflation avec un champ scalaire ~\cite{Martin:2019nuw, Martin:2020fgl}. Il est intéressant de noter que les TNPs se produisant pendant la période de préchauffement se produisent si abondamment qu'ils peuvent conduire au réchauffement de l'univers par le biais de leur évaporation.

En ce qui concerne les perspectives futures de cette première partie de la thèse il faut souligner les contraintes potentielles que quelq'un peut poser sur les prévisions observationnelles du CMB. Plus spécifiquement, pour un potentiel inflattionaire avec un champ scalaire fixe, la seule incertitude au niveau des prévisions observationnelles du CMB se repose au nombre d'e-folds passés entre le temps dans lequel l'échelle pivot du CMB franchit le rayon Hubble et la fin de la période l'inflation, à savoir $\Delta N_{*}$. Ce nombre, dépendant de l'échelle d'énergie à la fin de la période de l'inflation, $\rho_\mathrm{inf}$, qui est donné par le modèle d'inflation sous considération, de l'échelle de l'énergie au commencement de l’époque du Hot Big Bang, $\rho_\mathrm{rad}$, et du paramètre de l'équation d'état moyen entre la fin de l'inflation et le commencement de l’époque du Hot Big Bang, $\bar{w}_\mathrm{rad}$, s'écrit comme ~\cite{Martin:2006rs} 
\begin{align}\label{DN_*_compte_rendu}
\Delta N_{*} & = \frac{1-3\bar{w}_\mathrm{rad}}{12(1+\bar{w}_\mathrm{rad})}\ln \left(\frac{\rho_\mathrm{rad}}{\rho_\mathrm{inf}}\right)+\frac{1}{4}\ln\left(\frac{\rho_*}{9\Mp^4}\frac{\rho_*}{\rho_\mathrm{inf}}\right) \\&-\ln\left(\frac{k_\mathrm{P}/a_\mathrm{now}}{\tilde{\rho}^{1/4}_\mathrm{\gamma,now}}\right),
\end{align}
où $\bar{w}_\mathrm{rad}\equiv \frac{\int_{N_\mathrm{inf}}^{N_\mathrm{rad}}w(N)dN}{\Delta N_\mathrm{rad}}$ est le paramètre de l'équation d'état moyen pendant la période du réchauffement, $\Delta N_\mathrm{rad} = N_\mathrm{rad}-N_\mathrm{inf}$ , $\rho_{*}$ est l'échelle d'énergie au moment où l'échelle pivot du CMB franchit le rayon Hubble, $a_\mathrm{now}$ est le facteur d'échelle aujourd'hui et $\tilde{\rho}_\mathrm{\gamma,now}$ est la densité d'énergie de radiation aujourd'hui. En prenant alors l'échelle pivot du CMB $k_\mathrm{P} /a_\mathrm{now}$ égal à 0.05Mpc$^{-1}$ et $\tilde{\rho}_\mathrm{\gamma}$ à son valeur mesuré aujourd'hui, le dernier terme en \Eq{DN_*_compte_rendu} devient $N_0 \equiv -\ln\left(\frac{k_\mathrm{P}/a_\mathrm{now}}{\tilde{\rho}^{1/4}_\mathrm{\gamma,now}}\right)\simeq 61.76$.

\begin{figure}[h!]
\begin{center}
\includegraphics[width=0.49\textwidth, clip=true]{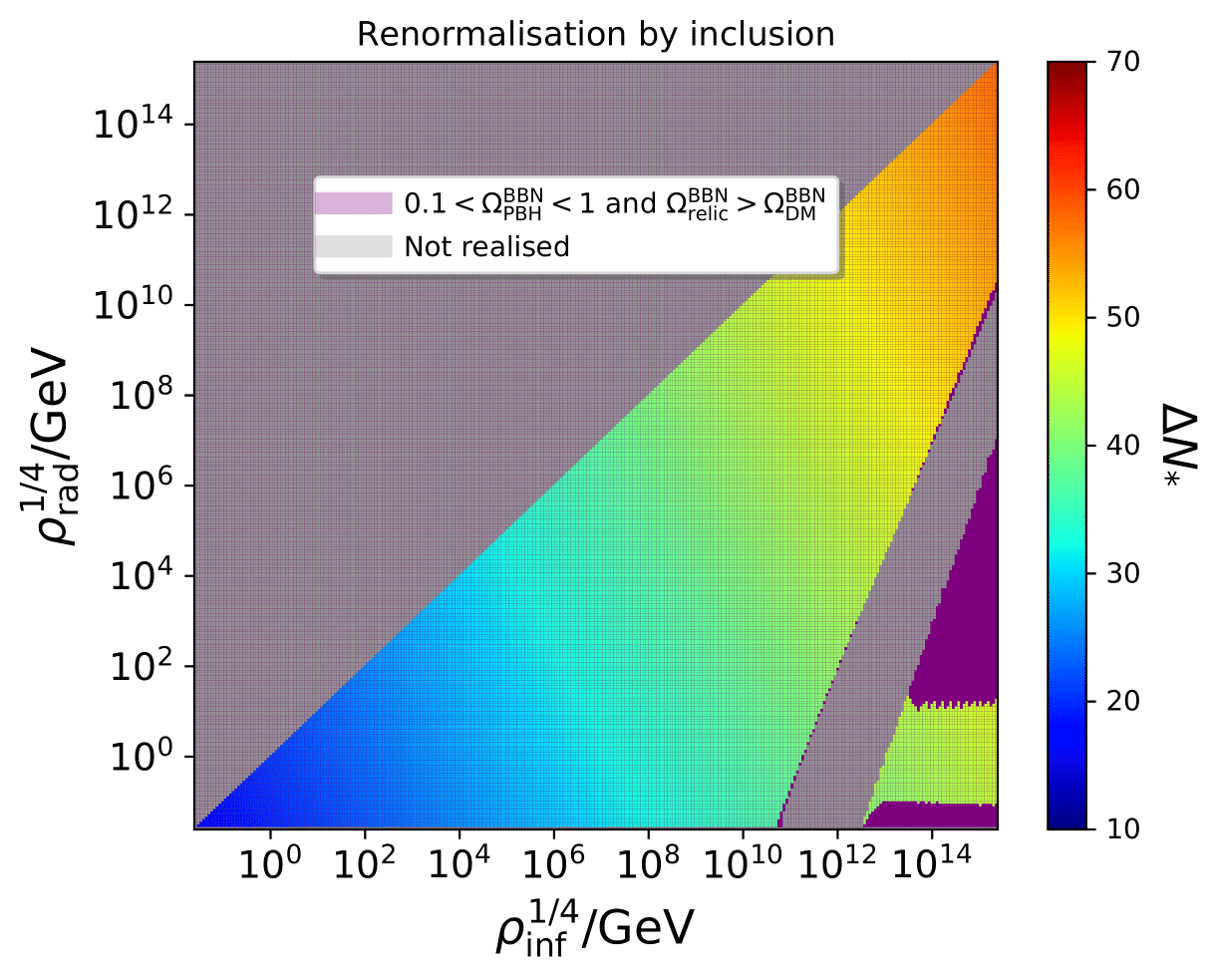}
 \includegraphics[width=0.49\textwidth, clip=true]{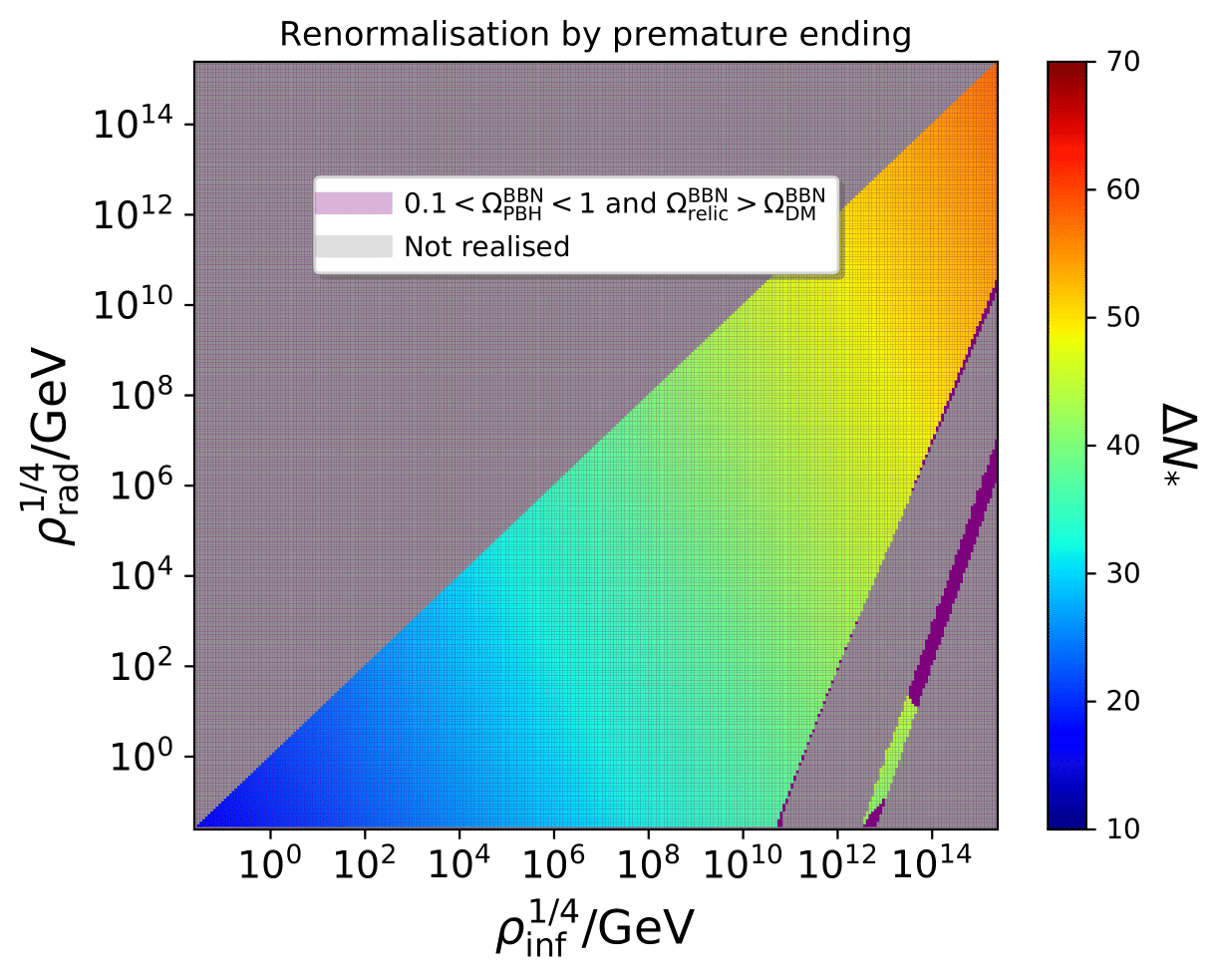}
\caption{$\Delta N_{*}$ en fonction de $\rho_\mathrm{inf}$ et $\rho_\mathrm{rad}$, quand la fonction de masse des TNP est renormalisée avec le schéma de l'inclusion des TNP (panneau gauche) aussi bien qu'avec le shéma de l'arrêt premature de l'instabilité du préchauffement (panneau droit). Pour plus de détails à voir  sur ~\cite{Martin:2019nuw}.}
\label{fig:DN_constraints_compte_rendu}
\end{center}
\end{figure}

Par suite, en considérant les contraintes sur l'échelle d'énergie à la fin de la période de l'inflation et l'échelle de l'énergie au commencement de l’époque du Hot Big Bang en étudiant des TNP se produisant pendant la période de préchauffement, on peut contraindre la quantité $\Delta N_{*}$. En particulier, en utilisant le setup théorique developpé en  ~\cite{Martin:2019nuw}, quelq'un peut calculer $\bar{w}_\mathrm{rad}$ et de l'\Eq{DN_*_compte_rendu} poser des contraintes sur $\Delta N_{*}$. Dans la \Fig{fig:DN_constraints_compte_rendu},on donne les contraintes au niveau de $\Delta N_{*}$ en fonction de $\rho_\mathrm{inf}$ et $\rho_\mathrm{rad}$.

Vice-versa, dans la seconde partie de la thèse, on a étudié un problème de réaction en retour des ondes gravitationnelles induites par des perturbations d'énergie de densité sous-tendues par un gaz des TNP. Plus spécifiquement, en demandant que les ondes gravitationnelles induites associées aux TNP ne se produisent pas en excès pendant une période cosmique où des TNP ultralégers ($m_\mathrm{PBH}<10^9\mathrm{g}$) dominent le budget énergétique de l'univers on a posé des contraintes sur l'abondance des TNP au moment où ils se forment en fonction de leur masse ~\cite{Papanikolaou:2020qtd}. Au meilleur de notre connaissance, elles se sont les premières contraintes solides indépendentes du modèle de production des TNP ultalégers dans la littérature internationale, étant donné le fait que les TNP ultralégers sont très pauvrement contraintes vu qu'ils s'évaporent avant BBN et ils ne laissent pas une empreinte observationnelle directe à part des reliques Planck spéculatives qui se produisent comme un vestige après l'évaporation Hawking des TNP.

\begin{figure}[h!]
\begin{center}
  \includegraphics[width=0.68\textwidth,  clip=true]
                  {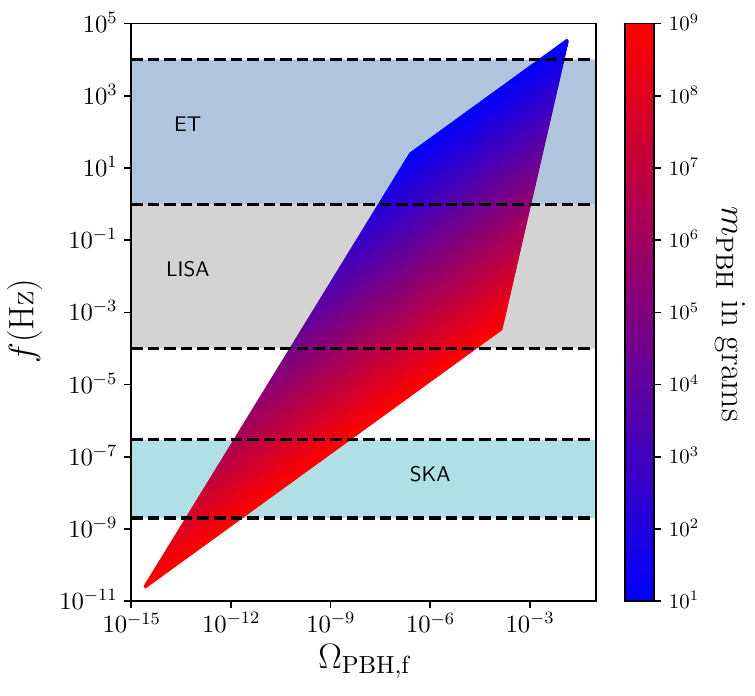}
\caption{La fréquence de crête du fond stochastique des ondes gravitationnelles induites par un gaz des TNP dominant le budget énergétique de l'univers, en fonction de l'abondance des TNP au moment où ils se forment, $\Omega_\mathrm{PBH,f}$ (axe horizontal), et de leur masse, $m_\mathrm{PBH}$ (code des couleurs).
La région de l'éspace des paramètres démontrée correspond aux valeurs de $m_\mathrm{PBH}$ et de $\Omega_\mathrm{PBH,f}$ tels que les TNP dominent le budget énergétique de l'univers pendant une période transitoire, se forment après la fin de la période de l'inflation et s'évaporent avant BBN et que les ondes gravitationnelles induites ne se produisent pas en excès, à voir \Eq{Omega_f constraints_french}. En pratique, \Eq{GW frequency} est démontrée avec $g_\ueff=100$, $z_\mathrm{eq}=3387$ et $H_0=70\,\mathrm{km}\,\mathrm{s}^{-1}\,\Mpc^{-1}$. Pour comparaison, les bandes des fréquences détectables  par ET, LISA et SKA sont aussi montrées. Figure creditée à ~\cite{Papanikolaou:2020qtd}.}
\label{fig:GW frequency_french}
\end{center}
\end{figure}

Sur ce point, il faut mettre en exergue la détectabilité potentielle du signal stochastique des ondes gravitationnelles induites produites dans un univers étant dominé par des TNP ultralégers. De façon interessante, on a trouvé dans ~\cite{Papanikolaou:2020qtd} que la fréquence de crête du spectrum respective, donné par l'\Eq{GW frequency} dépend de manière décisive de l'abondance initiale des TNP au moment de leur formation, $\Omega_\mathrm{PBH,f}$, et de la masse du TNP, $m_\mathrm{PBH}$, et se trouve de la gamme de fréquences des expériences des ondes gravitationnelles comme l'Einstein Telescope (ET) ~\cite{Maggiore:2019uih}, le Laser Interferometer Space Antenna (LISA)~\cite{Audley:2017drz,Caprini:2015zlo} et le Square Kilometre Array (SKA)~\cite{Janssen:2014dka}, 
\bea
\label{GW frequency_french}
\frac{f}{\mathrm{Hz}} \simeq \frac{1}{\left(1+z_\mathrm{eq}\right)^{1/4}}\left(\frac{H_0}{70\mathrm{kms^{-1}Mpc^{-1}}}\right)^{1/2}\left(\frac{g_\mathrm{eff}}{100}\right)^{1/6}\Omega^{2/3}_\mathrm{PBH,f} \left(\frac{m_\mathrm{PBH} }{10^9\mathrm{g}}\right)^{-5/6} ,
\eea
où $H_0$ est le valeur du paramètre Hubble aujourd'hui, $g_\mathrm{eff}$ est le nombre effective des degrés de liberté relativistes au moment de fomation des TNP et $z_\mathrm{eq}$ est le décalage vers le rouge au moment de l'équilibre énergétique entre la matière et la radiation. À voir aussi la \Fig{fig:GW frequency_french} dans laquelle la fréquence de crête se montre en fonction de $\Omega_\mathrm{PBH,f}$ et de $m_\mathrm{PBH}$.

Cette perspective est très importante vu que quelq'un peut potentiellement contraindre plus les TNP ultralégers en étudiant les donnés qui vont arriver par les expériences observationnelles futures des ondes gravitationnelles. Par contre, afin de donner une réponse finale si ce signal stochastique des ondes gravitationnelles induites associées aux TNP ultralégers peut être détecté ou non, il faut tenir en compte l'évolution dynamique du spectrum des ondes gravitationnelles de la fin de l'époque de domination des TNP jusq'à notre époque. Et pour cela, il est nécessaire de résoudre la transition graduelle de l'époque cosmique dominée par des TNP à l'époque cosmique successive dominée par la radiation, une étude assez subtile ~\cite{Inomata:2019zqy,Domenech:2020ssp}.

Enfin, dans la dernière partie de la thèse, qui est un travail en cours pas encore publié, on étudie des aspects de l'effondrement gravitationnel d'un fluide radiative aux TNP en présence des anisotropies. En particulier, on modèle d'une manière covariante la différence entre la pression radiale, $p_\mr$, et la pression tangentielle, $p_\mt$,  en postulant que la différence $p_\mr-p_\mt$ est proportionnelle soit aux gradients de pression soit aux gradients de densité d'énergie avec un paramètre de proportionalité $\lambda$ qui est équivelent à un paramètre d'anisotropie. En réalisant alors le programme perturbative de l'expansion des gradients au niveau des équations d'Einstein on déduit les conditions initiales des perturbations hydrodynamiques et metriques en présence des anisotropies. De plus, on déduit la dépendence du seuil de formation de TNP, $\delta_\mathrm{c}$, en fonction de  $\lambda$ en supposant que $\delta_\mathrm{c}$ dépend de la même manière que dans le cas isotrope de la forme du profil initial de densité d'énergie.

Le prochain pas à faire est d'évoluer les équations hydrodynamiques non-linéaires et étudier numériquement la formation de l'horizon du TNP et la dépendence explicite du seuil de formation des TNP avec le degré de l'anisotropie $\lambda$. Par conséquent, on pourra calculer le seuil de formation des TNP en présence des anisotropies et dériver aussi la dépendence de la fonction de masse des TNP avec le paramètre $\lambda$. Et cela va ouvrir des perspectives sur notre recherehce vu qu'en utilisant les contraintes des abondances des TNP quelq'un peut contraindre le degré de l'anisotropie $\lambda$ de notre modèle. De plus, dans le cas où le paramètre $\lambda$ a des dimensions, il se dépend lui-même de l'échelle d'énergie intrinsèque du processus de l'effondrement gravitationnel, qui est dans notre cas l'échelle d'énergie au moment où les perturbations sont générées, à savoir  l'échelle d'énergie à la fin de la période de l'inflation. Par conséquent, quelq'un peut traduire les contraintes observationnelles potentielles au niveau du degré de l'anisotropie  $\lambda$  aux contraintes au niveu de l'échelle d'énergie à la fin de la période de l'inflation donnant accès de cette façon au panorama inflationnaire.

Pour résumer, avec cette thèse on a étudié des aspects de l'univers primordial par le biais de la physique des TNP. Plus particulièrement, on a posé des contraintes sur des paramètres cosmologiques de l'univers primordial en étudiant les TNP se produisant pendant la période du préchauffement au contexte de la théorie de l'inflation avec un champ scalaire. De plus, en étudiant aussi les ondes gravitationnelles induites produites dans une époque cosmique où des TNP constituent la composante dominante du budget énergétique de l'univers, on a posé des contraintes indépendentes du modèle de production des TNP sur l'abondance des TNP au moment où ils se forment en fonction de leur masse. De surcroît, on a étudié des facettes de l'effondrement gravitationnel des TNP en présence des anisottropies. Comme il était mis en évidence avant, les résultats de la recherche effectuée au sein de cette thèse peuvent potentiellement ouvrir des directions nouvelles sur le domaine de la physique des TNP et porter un éclairage sur la compréhension de la physique de l'univers primordial.

\selectlanguage{english}
\begin{appendix}
\chapter{Appendix}
\section{The sound speed in a time-dependent $\textbf{$w$}$ background}\label{app:sound speed calculation}
Here, we extract the sound speed of a general adiabatic fluid, $c^2_\mathrm{s}$, with a time-dependent equation-of-state parameter, $w$. In a general system, the pressure density $p$ is a function of the energy density $\rho$ as well as of the entropy density $S$, i.e. $p=p(\rho,S)$. Consequently, one can write the following equation
\beq\label{sound speed definition general}
\delta p =c^2_\mathrm{s}  \delta\rho + \left( \frac{\partial p}{\partial S}\right)_{\mathrm{\rho}} \delta S,
\eeq
where the sound speed $c^2_\mathrm{s}$ is defined as $c^2_\mathrm{s} \equiv \left(\frac{\partial p}{\partial\rho}\right)_{\mathrm{S}}$. If one considers then an adiabatic system then they should require that $\left(\frac{\partial p}{\partial S}\right)_{\mathrm{\rho}} = 0$, i.e. there is no entropy production. Consequently, for such a system $c^2_\mathrm{s}$ becomes
\beq\label{c^2_s definition}
 c^2_\mathrm{s} =\frac{\delta p}{\delta \rho}, 
 \eeq
Given then the fact that the background pressure and energy densities of an adiabatic fluid system, $p$ and $\rho$ depend only on time, one can rewrite \Eq{c^2_s definition} by introducing the derivation with respect to the conformal time and using the chain rule, as
\beq
c^2_\mathrm{s} = \frac{p^\prime}{\rho^\prime},
\eeq
where the prime denotes differentiation with respect to the conformal time, $\eta$. Using therefore the continuity equation (\ref{Continuity Equation}) written with the conformal time as the time variable as well as time differentiating the equation of state for a time dependent equation-of-state parameter $w$, one can straightforwardly obtain that
\beq\label{c^2_s with w time dependent}
c^2_\mathrm{s}(\eta) = w(\eta) -\frac{1}{3\left[1+w(\eta)\right]\mathcal{H}(\eta)}\frac{\mathrm{d}w}{\mathrm{d}\eta} ,
\eeq
where $\mathcal{H}\equiv \frac{a^\prime}{a}$ is the conformal Hubble parameter. 
\section{The External Derivative}\label{External Derivative}
Below we derive the relation $ D_\mathrm{k}f = D_\mathrm{r}f+D_\mathrm{t}f$ as well \Eq{f constraint equation} by making use of the external derivative which is defined to be the unique $R$-linear mapping from $k$-forms to $(k + 1)$-forms satisfying the following properties ~\cite{Spivak:1971}: 

1. $\mathrm{d}f$ is the differential of $0$-forms (smooth functions) f.

2. $\mathrm{d}(\mathrm{d}f)=0$ for every $k$-form $f$.

3. $\mathrm{d}(a\wedge b) = \mathrm{d}a\wedge b + (-1)^{p}a\wedge \mathrm{d}b$, where $b$ is a $p$-form.

Having that into our mind, we can prove here the relation $ D_\mathrm{k}f = D_\mathrm{r}f+D_\mathrm{t}f$ for a general function $f$. If we multiply (\ref{observer time}) with a function f ($0$-form) and take the external derivative we have that 
\[ \begin{aligned}\label{Derivatives relation}
& G\mathrm{d}f\wedge \mathrm{d}u + f \mathrm{d}G\wedge \mathrm{d}u = A \mathrm{d}f\wedge \mathrm{d}t + f\mathrm{d}A\wedge \mathrm{d}t - B \mathrm{d}f\wedge \mathrm{d}r - f\mathrm{d}B\wedge \mathrm{d}r\Leftrightarrow \\ 
& G\mathrm{d}f\wedge \mathrm{d}u =A \mathrm{d}f\wedge \mathrm{d}t - B \mathrm{d}f\wedge \mathrm{d}r \Leftrightarrow \\
& G\left(\frac{\partial f}{\partial r}\right)_\mathrm{u} \mathrm{d}r\wedge \mathrm{d}u = A\left(\frac{\partial f}{\partial r}\right)_\mathrm{t} \mathrm{d}r\wedge \mathrm{d}t + B \left(\frac{\partial f}{\partial t}\right)_\mathrm{r} \mathrm{d}r\wedge \mathrm{d}t \Leftrightarrow \\ 
& G\left(\frac{\partial f}{\partial r}\right)_\mathrm{u} \mathrm{d}r\wedge \mathrm{d}u = A\frac{G}{A}\left(\frac{\partial f}{\partial r}\right)_\mathrm{t} \mathrm{d}r\wedge \mathrm{d}u + B \frac{G}{A} \left(\frac{\partial f}{\partial t}\right)_\mathrm{r} \mathrm{d}r\wedge \mathrm{d}u \Leftrightarrow\\ 
& D_\mathrm{k}f = D_\mathrm{r}f + D_\mathrm{t}f,
\end{aligned} \]
where from the first to the second equality, we used the fact that the external derivative applied to (\ref{observer time}) gives $\mathrm{d}G\wedge \mathrm{d}u = \mathrm{d}A\wedge \mathrm{d}t -\mathrm{d}B\wedge \mathrm{d}r$ and from the third to the forth equality we expressed $\mathrm{d}t$ in terms of $\mathrm{d}u$ and $\mathrm{d}r$ through  (\ref{observer time}).

Regarding now the derivation of \Eq{f constraint equation}, one obtains from (\ref{observer time}) by applying the external derivative that 
\[ \begin{aligned}
& \mathrm{d}G\wedge \mathrm{d}u = \mathrm{d}A\wedge \mathrm{d}t -\mathrm{d}B\wedge \mathrm{d}r \Leftrightarrow \\
& \left(\frac{\partial G}{\partial r}\right)_\mathrm{u} \mathrm{d}r\wedge \mathrm{d}u = \left(\frac{\partial A}{\partial r}\right)_\mathrm{t} \mathrm{d}r\wedge \mathrm{d}t + \left(\frac{\partial B}{\partial t}\right)_\mathrm{r} \mathrm{d}r\wedge \mathrm{d}t \Leftrightarrow \\
& \left(\frac{\partial G}{\partial r}\right)_\mathrm{u} \mathrm{d}r\wedge \mathrm{d}u = \frac{G}{A}\left(\frac{\partial A}{\partial r}\right)_\mathrm{t} \mathrm{d}r\wedge \mathrm{d}u +\frac{G}{A} \left(\frac{\partial B}{\partial t}\right)_\mathrm{r} \mathrm{d}r\wedge \mathrm{d}u \Leftrightarrow \\
& \frac{D_\mathrm{k}G}{G} = \frac{D_\mathrm{r}A}{A} + \frac{D_\mathrm{t}B}{B} \Leftrightarrow \\
& \frac{D_\mathrm{k}G}{G} = \frac{D_\mathrm{r}A}{A} + \frac{D_\mathrm{r}U}{\Gamma}
\end{aligned} \]
where in the last equality we replace $\mathrm{t}$ from (\ref{observer time}) and we used the fact that from $01$ Einstein equation, $\frac{D_\mathrm{t}B}{B} =  \frac{D_\mathrm{r}U}{\Gamma}$.
\newpage
\section{Lower limit on the anisotropic parameter $\lambda/\tilde{\lambda}$}\label{Lower Limit on lambda<0}
As we checked numerically for values of $\lambda<0$ less than a critical value the pressure and energy density gradient profiles were diverging at zero for both of the cases $f(r,t)=R(r,t)$ and $f(r,t)=\rho^n(r,t)$. This divergence can be explained by developing the pressure/energy density gradient profile around zero. We will work here with the energy density gradients. The same arguments apply also for the pressure gradients.

Starting with the case of $f(r,t)=R(r,t)$ and working with the differential equation \ref{rho ODE compact form}  regarding the energy density gradients, one can develop $g(r)$ around zero as 
\[ g(r)=g_0 + g_1 r + g_2 r^2/2, \] 
where
\[ g_0 =g(0)= \tilde{\rho}^\prime(0) \,, \quad g_1=g^\prime(0) \quad \textrm{and} \quad g_2=g^{\prime\prime}(0) \]  
Then, after a straightforward calculation, we get that
\[ \label{rho ODE compact form around zero}
\frac{2\lambda r}{3}\left(1-\frac{\mathcal{A}r^2}{2}\right)\left(g_1 +g_2 r\right) +\left[\frac{8\lambda}{3}\left(1-\frac{\mathcal{A}r^2}{2}\right)+1 \right]\left(g_0 + g_1 r + \frac{g_2 r^2}{2}\right)-\tilde{\rho}^\prime_\mathrm{iso}\left(1-\frac{\mathcal{A} r^2}{2}\right) = 0
\]
Considering now only the $O(r^0)$ terms we have that
\[ \label{g_0}
g_0 =  \tilde{\rho}^\prime(0) = \lim_{r\to 0} \frac{\tilde{\rho}^\prime_\mathrm{iso}(r)}{1+\frac{8\lambda}{3}},
\]
where $\rho^\prime_\mathrm{iso}=\frac{2}{3}\left\{\frac{\left[r^3K(r)\right]^\prime}{3r^2}\right\}^\prime r^2_\mathrm{m}$

If $\lambda<-3/8$ and since $\tilde{\rho}^\prime_\mathrm{iso}(0)=0^{-}$ one gets that $\tilde{\rho}^\prime(0) = 0^+$ which is not consistent since $\tilde{\rho}^\prime(0)$  should approach zero from negative values. If however $\lambda >-3/8$ then one obtains the consistent result that $\tilde{\rho}^\prime(0) = 0^-$. For the critical value $\lambda=-3/8$ then one has that 
\[ 
\tilde{\rho}^\prime(0) = \lim_{r\to 0} \frac{\tilde{\rho}^\prime_\mathrm{iso}(r)}{1+\frac{8\lambda}{3}} = \frac{0}{0} = \frac{\tilde{\rho}^{\prime\prime}_\mathrm{iso}(0)}{0} = -\infty \neq 0^-,
\]
after applying the De l'Hopital theorem and considering the fact that $\tilde{\rho}^{\prime\prime}_\mathrm{iso}(0)<0$.

Consequently, in the case of $p_\mathrm{r}-p_\mathrm{t} = -\lambda RD_\mathrm{r}\rho$ with $\lambda<0$ one gets that $\lambda$ should be larger than a critical value, namely $\lambda>\lambda_\mathrm{c}=-3/8$. In the case of $\lambda>0$ we do not have this problem since the expression $1+\frac{8\lambda}{3}$ is always positive, making thus $\tilde{\rho}^\prime(0) = 0^-$ always for $\lambda>0$. As one may notice, $\lambda=\lambda_\mathrm{c}=-3/8$ is the value which makes zero the prefactor in front of $g(r)$ in \Eq{rho ODE compact form} and which determines the sign of $\tilde{\rho}^\prime(0)$ at order $O(r^0)$.

In the case of $p_\mathrm{r}-p_\mathrm{t} = -\lambda RD_\mathrm{r}p_\mathrm{r}$, following the same procedure one can conclude that $\lambda>\lambda_\mathrm{c}=-9/14$ in order not to confront divergent pressure gradient profiles at $r=0$.

Regarding now the case of $f(r,t)=\rho^n(r,t)$ one can apply the same gradient expansion around zero for the differential equation governing the pressure gradient profile's behavior,  i.e. \Eq{p_r ODE compact form - f=rho^n} and derive the necessary condition for which the pressure gradient does not diverge at $r=0$. At this point, we should point out that given that in the formulation where $f(r,t)=\rho^n(r,t)$ we have more than one anisotropy parameters, the divergence condition should be given in terms of $\tilde{\lambda}$, $n$, $q$ and $\epsilon_0$.

Writing then $h(r)$ as 
\[ h(r)=h_0 + h_1 r + h_2 r^2/2, \] 
where
\[ h_0 =h(0)= \tilde{p_\mr}^\prime(0) \,, \quad h_1=h^\prime(0) \quad \textrm{and} \quad h_2=h^{\prime\prime}(0) \]  
and applying the same procedure as before one gets that
\[ \label{h_0}
h_0 =  \tilde{p_\mr}^\prime(0) = \lim_{r\to 0} \frac{\tilde{p}^\prime_{\mathrm{r,iso}}(r)}{\frac{2}{9(1-2n)}\frac{\tilde{\lambda}}{q}\left(\frac{q}{\epsilon_0}\right)^3\left[(5-6n)\left(\frac{q}{\epsilon_0}\right)^{4n-2} -2\right]},
\]
where $\tilde{p}^\prime_\mathrm{r,iso}=\frac{2}{3}\left\{\frac{\left[r^3K(r)\right]^\prime}{3r^2}\right\}^\prime r^2_\mathrm{m}$. Consequently, the necessary condition in order not obtain a divergence at $r=0$ is 
\beq\label{lambda_c - f=rho^n - Drpr}
\frac{2}{9(1-2n)}\frac{\tilde{\lambda}}{q}\left(\frac{q}{\epsilon_0}\right)^3\left[(5-6n)\left(\frac{q}{\epsilon_0}\right)^{4n-2} -2\right]>0.
\eeq
From the above expression if one fixes $q$ and $\epsilon_0$, one may identify two regimes, namely when $n>1/2$ and when $n<1/2$. In particular when $n>1/2$ given the fact that $q/\epsilon_0\ll 1$ one obtains that 
the second term in the brackets in \Eq{lambda_c - f=rho^n - Drpr} is the dominant one and that $\tilde{\lambda}$ should be positive, i.e. $\tilde{\lambda}>0$. If one the other hand $n<1/2$ the first term in the brackets is the dominant one and one can see straightforwardly that again $\tilde{\lambda}$ should be positive. Therefore, if $q/\epsilon_0\ll 1$ which is in general the case, $\tilde{\lambda}>\tilde{\lambda}_\mathrm{c}=0$.

In the case now where the difference between the radial and the tangential pressure is modeled as proportional to energy density gradients, following the same procedure as before and make a gradient expansion around $r=0$ at the level of \ref{p_r ODE compact form - f=rho^n-D_rrho} one gets the following necessary condition to avoid divergences around $r=0$.
\beq\label{lambda_c - f=rho^n - Drrho}
\frac{3}{1-2n}\frac{\tilde{\lambda}}{q}\left(\frac{q}{\epsilon_0}\right)^3\left[ 1- \left(\frac{q}{\epsilon_0}\right)^{4n-2}\right]<0.
\eeq
From the above equation, one can easily check that given the fact that $q/\epsilon_0\ll 1$, $\tilde{\lambda}$ should always be positive for any value of $n$, i.e. $\tilde{\lambda}>\tilde{\lambda}_\mathrm{c}=0$.
\newpage
\section{The limits $n\to 1/2$ and $n\to 1/4$ of $\Phi_{p_\mr}$, $\Phi_{\rho}$, $I_{1,p_\mr}$, $I_{2,p_\mr}$, $I_{1,\rho}$ and $I_{2,\rho}$}\label{Phi_pr+I_1+1_2}
Below we give the limits $n\to 1/2$ and $n\to 1/4$  of $\Phi_{p_\mr}$, $\Phi_{\rho}$, $I_{1,p_\mr}$,$I_{2,p_\mr}$, $I_{1,\rho}$, $I_{2,\rho}$ which are given respectively by equations \eqref{Phi analytical}, \eqref{Phi analytical-D_rrho}, \eqref{I1+I2 solutions}, \eqref{I_1 analytical-D_rrho} and \eqref{I_2 analytical-D_rrho}.

\begin{align}
\lim_{n\rightarrow 1/2}\Phi_{p_\mr}(a) & = -\frac{2\lambda\sqrt{\rho_\mathrm{b,inf}}}{a_\mathrm{inf}r_\mathrm{m}}\left(\frac{a}{a_\mathrm{inf}}\right)^{-3}\ln\left(\frac{a}{a_\mathrm{inf}}\right) \\ 
\lim_{n\rightarrow 1/2} I_{1,p_\mr}(a) & = \frac{\lambda\sqrt{\rho_\mathrm{b,inf}}}{2a_\mathrm{inf}r_\mathrm{m}}\left(\frac{a}{a_\mathrm{inf}}\right)^{-3}\left[2-2\frac{a}{a_\mathrm{inf}}+\ln\left(\frac{a}{a_\mathrm{inf}}\right)\right] \\
\lim_{n\rightarrow 1/4} I_{1,p_\mr}(a) & = -\frac{\lambda\rho^{1/4}_\mathrm{b,ini}}{2a_\mathrm{inf}r_\mathrm{m}}\left(\frac{a}{a_\mathrm{inf}}\right)^{-3}\left\{1+\frac{a}{a_\mathrm{inf}}\left[2\ln\left(\frac{a}{a_\mathrm{inf}}\right)-1\right]\right\}\\
\lim_{n\rightarrow 1/2} I_{2,p_\mr}(a) & =  \frac{\lambda\sqrt{\rho_\mathrm{b,inf}}}{a_\mathrm{inf}r_\mathrm{m}}\left(\frac{a}{a_\mathrm{inf}}\right)^{-3}\left[\ln\left(\frac{a}{a_\mathrm{inf}}\right)-\frac{a}{a_\mathrm{inf}}+1 \right]\\
\lim_{n\rightarrow 1/4} I_{2,p_\mr}(a) & =-\frac{\lambda\rho^{1/4}_\mathrm{b,inf}}{a_\mathrm{inf}r_\mathrm{m}}\left(\frac{a}{a_\mathrm{inf}}\right)^{-3}\left\{1+\left[\ln\left(\frac{a}{a_\mathrm{inf}}\right)-1\right]\frac{a}{a_\mathrm{inf}}\right\}
\end{align}

\begin{align}
\lim_{n\rightarrow 1/2}\Phi_{\rho}(a) & = -\frac{6\lambda\sqrt{\rho_\mathrm{b,inf}}}{a_\mathrm{inf}r_\mathrm{m}}\left(\frac{a}{a_\mathrm{inf}}\right)^{-3}\ln\left(\frac{a}{a_\mathrm{inf}}\right) \\
\lim_{n\rightarrow 1/2} I_{1,\rho}(a) & = \frac{3\lambda\sqrt{\rho_\mathrm{b,inf}}}{2a_\mathrm{inf}r_\mathrm{m}}\left(\frac{a}{a_\mathrm{inf}}\right)^{-3}\left[2-2\frac{a}{a_\mathrm{inf}}+\ln\left(\frac{a}{a_\mathrm{inf}}\right)\right] \\
\lim_{n\rightarrow 1/2} I_{1,\rho}(a) & = -\frac{3\lambda\rho^{1/4}_\mathrm{b,inf}}{2a_\mathrm{inf}r_\mathrm{m}}\left(\frac{a}{a_\mathrm{inf}}\right)^{-3}\left\{1+\frac{a}{a_\mathrm{inf}}\left[2\ln\left(\frac{a}{a_\mathrm{inf}}\right)-1\right]\right\}\\
\lim_{n\rightarrow 1/2}I_{2,\rho}(a) & =  \frac{3\lambda\sqrt{\rho_\mathrm{b,inf}}}{a_\mathrm{inf}r_\mathrm{m}}\left(\frac{a}{a_\mathrm{inf}}\right)^{-3}\left[\ln\left(\frac{a}{a_\mathrm{inf}}\right)-\frac{a}{a_\mathrm{inf}}+1 \right]\\
\lim_{n\rightarrow 1/4} I_{2,\rho}(a) & =-\frac{3\lambda\rho^{1/4}_\mathrm{b,inf}}{a_\mathrm{inf}r_\mathrm{m}}\left(\frac{a}{a_\mathrm{inf}}\right)^{-3}\left\{1+\left[\ln\left(\frac{a}{a_\mathrm{inf}}\right)-1\right]\frac{a}{a_\mathrm{inf}}\right\}.
\end{align}

\end{appendix}

\bibliographystyle{JHEP} 
\bibliography{PBH}
\end{document}